\documentclass[12pt]{article}
\usepackage{mathrsfs}
\makeatletter \@addtoreset{equation}{section} \makeatother
\renewcommand{\theequation}{\thesection.\arabic{equation}}
\newcommand{\ba}{\begin{array}}
\newcommand{\ea}{\end{array}}
\newcommand{\beq}{\begin{equation}}
\newcommand{\eeq}{\end{equation}}
\newcommand{\bea}{\begin{eqnarray}}
\newcommand{\eea}{\end{eqnarray}}
\def\bce{\begin{center}}
\def\ece{\end{center}}

\def\nonu{\nonumber}

\def\pa{\partial}
\def\al{\alpha}
\def\be{\beta}
\def\ga{\gamma}

\def\de{\delta}

\def\la{\lambda}

\def\si{\sigma}

\def\eps6{{\displaystyle \mathop{\epsilon}^{6}}{}}
\def\g6{{\displaystyle \mathop{g}^{6}}{}}

\def\nab6{{\displaystyle \mathop{\nabla}^{6}}{}}
\def\0{{\sst{(0)}}}
\def\1{{\sst{(1)}}}
\def\2{{\sst{(2)}}}
\def\3{{\sst{(3)}}}
\def\4{{\sst{(4)}}}
\def\5{{\sst{(5)}}}
\def\6{{\sst{(6)}}}
\def\7{{\sst{(7)}}}
\def\8{{\sst{(8)}}}

\def\ba{\begin{array}}
\def\ea{\end{array}}
\def\beq{\begin{equation}}
\def\eeq{\end{equation}}
\def\be{\begin{equation}}
\def\ee{\end{equation}}

\def\la{\lambda}
\def\eps{\epsilon}

\def\ba{\begin{array}}
\def\ea{\end{array}}
\def\beq{\begin{equation}}
\def\eeq{\end{equation}}
\def\be{\begin{equation}}
\def\ee{\end{equation}}

\def\la{\lambda}
\def\eps{\epsilon}

\def\eps6{{\displaystyle \mathop{\epsilon}^{6}}{}}

\def\nab6{{\displaystyle \mathop{\nabla}^{6}}{}}

\newcommand{\bean}{\begin{eqnarray*}}
\newcommand{\eean}{\end{eqnarray*}}

\begin{document}
\thispagestyle{empty} \addtocounter{page}{-1}
   \begin{flushright}
%PUPT-2395 \\
%CALT-68-nnnn \\
%{\tt hep-th/yymmnnn}\\
\end{flushright}

\vspace*{1.3cm}
  
\centerline{ \Large \bf
A Charged and Neutral Spin-$4$ Currents
 }
\vspace*{0.3cm}
\centerline{ \Large \bf  
in  the Grassmannian-like  Coset Model } 
%and }
\vspace*{1.0cm}
\centerline{ {\bf
Changhyun Ahn$^{\dagger,\star,\ast,}$\footnote{CA is a visitor to
Seoul
National University of Science and Technology
and a professor emeritus at
Kyungpook National University} and Minsu Kang$^{\star,\ddagger}$}
}
\vspace*{1.0cm} 
\centerline{\it
$\dagger$
Institute for Convergence Fundamental Studies,}
%\centerline{\it College of Liberal Arts,}
\centerline{\it
Seoul
National University of Science and Technology, Seoul
01811, Korea} 
\vspace*{1.0cm} 
\centerline{\it
$\star$
Department of Physics, Kyungpook National University, Daegu
41566, Korea} 
\vspace*{1.0cm} 
\centerline{$\ast$ {\tt  ahn@knu.ac.kr}}
%\qquad
\centerline{$\ddagger$  {\tt 
d1565@naver.com}} 
\vskip2cm

\centerline{\bf Abstract}
\vspace*{0.5cm}

By calculating the second order pole in the operator product
expansion (OPE) of the charged spin-$3$ current
with the neutral spin-$3$ current in the
Grassmannian-like coset model, we determine
the primary charged spin-$4$ current.
Similarly, by computing
the second order pole in the OPE of
the neutral spin-$3$ current
with itself,
we obtain the primary neutral spin-$4$ current.
We determine the OPE
of  the charged spin-$2$ current
with the charged spin-$3$ current
for generic parameters and the large
$k$ (one of the parameters) limit is also obtained for this OPE.
In particular, the above
primary charged spin-$4$ current appears in the first order pole
of this OPE for generic parameters.
We also check that
the above
primary charged and
neutral spin-$4$ currents occur at the second order pole
in the OPE of the charged spin-$3$ current
with itself for fixed parameters.

%\vspace*{4cm}
% \begin{flushright}
%{\it On the occasion of my sixtieth birthday}
%\end{flushright}

\baselineskip=18pt
\newpage
\renewcommand{\theequation}
{\arabic{section}\mbox{.}\arabic{equation}}

\tableofcontents

%%%%%%%%%%%%%%%%%%%%%%%%%%%%%%%%%%%%%%%%%%%%%%%%%%%%%%%%%%%%%%%%%%%%%
%%%%%%%%%%%%%%%%%%%%%%%%%%%%%%%%%%%%%%%%%%%%%%%%%%%%%%%%%%%%%%%%%%%%%%

%%%%%%%%%%%%%%%%%%%%%%%%%%%%%%%%%%%%%%%%%%%%%%%%%%%%%%%%%%%%%%%%%%%%%
%%%%%%%%%%%%%%%%%%%%%%%%%%%%%%%%%%%%%%%%%%%%%%%%%%%%%%%%%%%%%%%%%%%%%%
\section{ Introduction}
%1%%%%%%%%%%%%%%%%%%%%%%%%%%%%%%%%%%%%%%%%%%%%%%%%%%%%%%%%%%%%%%%%%%%%%
%%%%%%%%%%%%%%%%%%%%%%%%%%%%%%%%%%%%%%%%%%%%%%%%%%%%%%%%%%%%%%%%%%%%%

The Grassmannian-like coset model \cite{CHR1306,BK1990}
plays an important role
for the matrix generalization of $AdS_3$ higher spin theory
\cite{PV1,PV2}, according to the observation of \cite{CH1812}.
For example, in this coset model
in the large $k$, the central charge
reproduces the one in the asymptotic symmetry
of the matrix generalization of $AdS_3$ higher spin theory.
The `charged' spin-$2,3$ currents and
the `neutral' spin-$2,3$ currents
are found in \cite{CH1812} explicitly
by using the coset realization (See also \cite{GKO1,GKO2,BBSS1}).
The OPEs
of the spin-$1$ currents with the charged spin-$2,3$ currents
have simple poles
while
the OPE
of the spin-$1$ currents with the neutral spin-$3$ current
does not have any singular terms.
The conjecture by \cite{GG1011} can be obtained
from this coset model by level rank
duality for the particular value of the parameter (without
matrix generalization).
At the specific 't Hooft-like coupling constant, the above coset model
provides the `rectangular' (or matrix valued) $W$ algebra \cite{AM}.
In \cite{Ahn2011},
the OPE of the
charged spin-$2$ current with itself
is obtained for generic parameters and
the large $k$ limit is determined explicitly
\footnote{The OPEs of the (charged) spin-$1$ currents (and the
charged spin-$2$ currents) with the neutral
spin-$3$ current are determined also.}.

On the other hand,
the supersymmetric Grassmannian-like coset model \cite{CHR1306}
describes the ${\cal N}=2$ supersymmetric
$AdS_3$ higher spin gravity
\cite{PV1,PV2} for the particular parameter (without matrix
generalization),
from the analysis in \cite{CHR1111}.
Moreover, the ${\cal N}=4$ supersymmetric holography \cite{GG1305}
can be studied by this supersymmetric coset model for
different value of the parameter. 
In \cite{Ahn2107},
some singlet (without an adjoint index)
and nonsinglet (having an adjoint index)
${\cal N}=2$ supersymmetric
multiplets of superspin-$1,2$
are obtained from the coset realization
by considering the fermionic operators of
spin-$\frac{1}{2}$ to the above
bosonic Grassmannian-like coset model \footnote{ The OPEs
of the lowest nonsinglet multiplet with itself are
determined also explicitly.}.
The above charged and neutral spin-$2,3$ currents
appearing in previous paragraph are generalized to this
supersymmetric coset model and
they appear at the last components of
singlet and nonsinglet
${\cal N}=2$
multiplets.
The two singlets of spin-$\frac{3}{2}$ currents
living on the singlet ${\cal N}=2$  multiplet
of superspin-$1$ are used explicitly in order to
satisfy the ${\cal N}=2$  primary conditions.

We can go one step further from the results of \cite{Ahn2011}
or \cite{Ahn2107}.
We need to describe the next higher spin-$4$ currents
which are charged or neutral. In principle, we can write down
all the possible terms from the coset fields
with undetermined coefficients
and require that they should satisfy
the regular conditions with the coset fields living on the
denominator of the coset, the primary conditions with
the spin-$2$ stress energy tensor and the primary conditions
with the spin-$1$ currents \cite{BCG}
(for the neutral higher spin current,
this leads to a trivial OPE with the spin-$1$ currents where there
are no singular terms).
However, it is 
nontrivial \cite{CG1207,CH1906} to classify all the invariant tensors
contracted with the coset fields as the spins of the currents
become larger.
See also \cite{CHU1906-1}.
For the charged or neutral spin-$3$ currents,
the number of undetermined coefficients appearing in the
possible terms is less than twenty. We expect that the number of
independent terms for the charged spin-$4$ currents is
larger than the one of independent terms for the neutral spin-$4$
current because the former should have a single free adjoint index
while the latter does not have any free adjoint index. 

Or we can extract the new charged or neutral higher spin currents
from the OPEs of the lower spin currents by using
the procedure on the OPE of \cite{Blumenhagenetal,Nahm1,Nahm2}.
From the construction of (neutral) spin-$5$ current \cite{AK1308}
in different coset model, the standard
expression for this current looks like the quintic
terms in the spin-$1$ currents where the
sum of number of derivatives and the number of spin-$1$
currents is equal to five which is the spin of the current.
If we ignore one spin-$1$ currents in the product
of five spin-$1$ currents (without a derivative)
where there exist three totally symmetric $d$ symbols \cite{BS,BBSS,NPB97}
having three adjoint indices, then
we are left with a single adjoint free index in one of the
$d$ symbols, the remaining four indices of three $d$ symbols
are contracted with the remaining four spin-$1$ currents
and the remaining four indices in the three $d$ symbols are
contracted with each other in the charged spin-$4$ current.
Moreover, the above single adjoint index can appear in the
totally antisymmetric $f$ symbols, in the generator
of the special unitary algebra, or in the spin-$1$ currents itself.

For the neutral spin-$4$ current \cite{Ahn1111}, the product of
two $d$
symbols contracted with four product of spin-$1$ currents
(without any derivatives)
can appear. Among six indices in the two $d$ symbols,
four of them is contracted with the adjoint indices
of four spin-$1$ currents and two remaining indices
are contracted with each other.
Furthermore, similarly, the two $f$ symbols can be contracted with
the four product of spin-$1$ currents
and the combination of one $d$ symbol and one $f$ symbol
is contracted with the adjoint indices of four spin-$1$ currents.
Because the Kronecker delta is an invariant tensor of
special unitary algebra, the two products of the Kronecker
deltas with four different adjoint indices
contracted with the four spin-$1$ currents can also appear
in the neutral spin-$4$ current.
Note that the totally symmetric traceless symmetric $d$
symbols with four adjoint indices \cite{Schoutens,DP} contain
the above  two products of the Kronecker
deltas.
Compared to the previous charged spin-$4$ currents
appearing in previous paragraph, the structure of the
neutral spin-$4$ current is simpler because there are no
free adjoint indices.

\begin{itemize}
\item[]
The main motivations of this paper are as follows:

1) The construction of ${\cal N}=2$ ``rectangular'' $W$-algebra

One of the reasons why we are trying to obtain the higher charged
or neutral spin currents
is to construct the generators of ${\cal N}=2$  rectangular $W$-algebra
of the $M \times M$ matrix generalization of ${\cal N}=2$ $AdS_3$
higher spin theory in three dimensions \footnote{In recent progress
on the celestial holography, the commutator between the soft
graviton is described by the wedge subalgebra of $w_{1+\infty}$ algebra.
Then it is obvious to study the different kinds of
$W$-algebra in two dimensions in order to explain four dimensional
scattering amplitudes via celestial holography.}.
Any ${\cal N}=2$ multiplet consists of
the bosonic lowest component, the fermionic second and third
components and the bosonic last component. Our charged primary
spin-$4$ current and
neutral spin-$4$ current will provide some hints for the above
bosonic last components, which include
complex fermions, for the $SU(M)$ nonsinglets and singlet
respectively.
Our goal in this direction is to determine the whole
${\cal N}=2$ $SU(M)$ nonsinglets or singlet multiplets of
spins $(3, \frac{7}{2}, \frac{7}{2},4)$
by using the two fermionic generators of ${\cal N}=2$ superconformal
algebra.
Then we will observe the structure of
${\cal N}=2$  rectangular $W$-algebra (the commutators or the
anticommutators)
for lower spin currents.
Note that ${\cal N}=2$ $SU(M)$ nonsinglets or singlet multiplets of
spins $(2, \frac{5}{2}, \frac{5}{2},3)$ are found previously.

2) The relation with a ``colored'' gravity

The nonabelian isospin extension of Einstein gravity is called
a $SU(M)$-colored gravity \cite{JKKR} in $AdS_3$ spacetimes.
The Poisson brackets between the spin-$2$
current, the spin-$1$ currents and the spin-$2$ currents are obtained.
The first is $SU(M)$ singlet which is the mode of
stress energy tensor and the last two are $SU(M)$ nonsinglets
corresponding to the spin-$1$ currents and the charged spin-$2$ currents.
There is no construction on the charged spin-$3$ currents.
Our result for the OPE of the charged spin-$2$ currents and
the charged spin-$3$ currents provides the new Poisson bracket
from colored gravity after taking the infinity limit of the level
$k$ for fixed other parameters. See also the end of Appendix for details.
It would be interesting to construct
the nonlinear Poisson algebra from colored gravity in three
dimensions by including the
spin-$3$ currents and observe our findings in two dimensions
in this paper.

3) A construction of spin-$4$ currents

The immediate answer for the reason why we
should consider the spin-$4$ currents
\footnote{It is natural to ask what happens for the
spin-$5,6,7, \cdots$ currents? For the spin-$5$ currents,
we will mention how to obtain them in the specific OPEs later.
At the moment, we do not know how to determine
all these higher spin currents systematically because we should perform
the procedures in this paper step by step successively.}
is that
so far, the charged and neutral spin-$3$ currents
are found and in order to complete
the OPE of the charged spin-$2$ currents
with the charged spin-$3$ currents, the construction of
these spin-$4$ currents is crucial.
In other words, the first order pole of this OPE
(the next simplest nontrivial OPE after adding the spin-$3$ currents)
contains the charged primary spin-$4$ currents.
By writing down the first order pole explicitly,
collecting these terms in terms of the known currents
and subtracting all the descendant terms, we are left with the
final charged primary spin-$4$ currents.
For the neutral primary spin-$4$ current, 
we should compute the OPE of the neutral primary spin-$3$ current
with itself and focus on the second order pole of this OPE.
It is rather nontrivial to obtain these charged and neutral
primary spin-$4$ currents directly without using the above procedures
because it is not obvious to exhaust all the possible terms
by using the $f$ symbol, $d$ symbol, Kronecker delta and the generators
of $SU(M)$ or $SU(N)$.

4) The role of structure constants having three generic parameters

Depending on the values of the three parameters in the
Grassmannian-like coset model (or its supersymmetric version), 
there appear different models.
In particular, the parameter $M$ is related to the number of
supersymmetry in the supersymmetric Grassmannian-like coset model.
Other parameter $N$ appears in the numerator and the denominator
of Grassmannian-like coset model. The level $k$ appears
in the second order poles of various OPEs with
the denominator currents.
One of the particular limits among
these parameters (the infinity limit of $k$ with
a fixed ratio between $k$ and $(k+N)$) is given by 
the one in the colored gravity above.
Our goal in this direction is
to find whether there exists an extension of ${\cal N}=4$ (or ${\cal N}
> 4$)
superconformal algebra
in the supersymmetric Grassmannian-like coset model or not
by changing the value of $M$.

\end{itemize}

In this paper,
in section $2$,
the construction of charged spin-$2,3$ currents
and neutral spin-$3$ current in addition to
the spin-$1$ currents and the stress energy tensor
spin-$2$ current
is reviewed.
The OPE of the charged spin-$2$ current
with the charged spin-$3$ current is reviewed
up to the second order pole for generic parameters.
In section $3$,
from the second order pole in the
OPE of the charged spin-$3$ current
with the neutral spin-$3$ current,
the primary charged spin-$4$ current
is determined for generic parameters.
Similarly,
from the second order pole in the
OPE of the neutral spin-$3$ current
with itself,
the primary neutral spin-$4$ current
is obtained for generic parameters.
In section $4$,
we summarize the main results of this paper
and list some open problems.
In Appendices $A$-$E$,
some details given in sections $2,3$ are described
and the ancillary file for the
coefficients appearing in Appendices
on the {\bf arXiv} will be submitted 
\footnote{The Thielemans package \cite{Thielemans} can be used
together with Mathematica \cite{mathematica} throughout
this paper.
The
similar coset in the recent work of \cite{EP,EP1}
where the possibility of four parameters in the specific coset
is obtained is found.
}.

%%%%%%%%%%%%%%%%%%%%%%%%%%%%%%%%%%%%%%%%%%%%%%%%%%%%%%%%%%%%%%%%%%%%%
%%%%%%%%%%%%%%%%%%%%%%%%%%%%%%%%%%%%%%%%%%%%%%%%%%%%%%%%%%%%%%%%%%%%%%
\section{ Review}
%2%%%%%%%%%%%%%%%%%%%%%%%%%%%%%%%%%%%%%%%%%%%%%%%%%%%%%%%%%%%%%%%%%%%%%
%%%%%%%%%%%%%%%%%%%%%%%%%%%%%%%%%%%%%%%%%%%%%%%%%%%%%%%%%%%%%%%%%%%%%

The Grassmannian-like coset model \cite{CHR1306} is
given by 
\bea
\frac{SU(N+M)_k}{SU(N)_k \times U(1)_{k N M (N+M)}}.
\label{coset}
\eea
Under the decomposition of $SU(N+M)$ into the
$SU(N) \times SU(M)$,
the adjoint representation of $SU(N+M)$ breaks into
\bea
({\bf N+M})^2-{\bf 1} \longrightarrow ({\bf N}^2-{\bf 1},{\bf 1})
\oplus ({\bf 1},{\bf M}^2-{\bf 1}) \oplus ({\bf 1}, {\bf 1})
\oplus ({\bf N},
\overline{{\bf M}}) \oplus (\overline{{\bf N}},{\bf M}).
\label{branching}
\eea
The generators for each factor on the right hand side
of (\ref{branching})
are denoted by
\bea
(t^{\al}, t^a, t^{u(1)},
t^{(\rho \bar{i})}, t^{(\bar{\si} j)})
\label{generators}
\eea
in the decomposition of $SU(N+M)$ of the coset model (\ref{coset}).
The metric is normalized as follows
\bea
\mbox{Tr} (t^{\al} t^{\beta}) =\de^{\al \beta},
\qquad
\mbox{Tr} (t^{a} t^{b}) =\de^{a b},
\qquad
\mbox{Tr} (t^{u(1)} t^{u(1)}) =1,
\qquad
\mbox{Tr} (t^{(\rho \bar{i})} t^{(\bar{\si} j)}) =\de^{\rho \bar{\si}} \, \de^{j \bar{i}},
\label{metric}
\eea
where
the trace (denoted by Tr) is taken over the
fundamental representation of $SU(N+M)$.
For the $SU(N), SU(M)$ and $U(1)$ indices in (\ref{generators}), where
the adjoint indices are given by $\al =1,2, \cdots, (N^2-1)$
and $a=1,2,\cdots, (M^2-1)$ respectively, they can
be raised or lowered without
modifying (\ref{metric}).
The barred index in (\ref{metric})
becomes the unbarred one  when we raise or lower
it 
and vice versa.
The fundamental indices $\rho$ of $SU(N)$ and $j$
of $SU(M)$ among (\ref{generators})
run over $\rho =1,2,\cdots, N$ and $j =1,2,\cdots, M$
respectively,
while  the antifundamental indices $\bar{\si}$
of $SU(N)$ and $\bar{i}$ of $SU(M)$
run over $\bar{\si} =1,2,\cdots, N$ and $\bar{i} =1,2,\cdots, M$
respectively.

We present the fundamental OPEs in the $SU(N+M)$ currents
together with the generators in Appendix $A$.

%%%%%%%%%%%%%%%%%%%%%%%%%%%%%%%%%%%%%%%%%%%%%
\subsection{The charged spin-$2,3$ currents}
%%%%%%%%%%%%%%%%%%%%%%%%%%%%%%%%%%%%%%%%%%%%%%

The stress energy tensor \cite{CH1812}
by the Sugawara construction
is given by
\footnote{The normal ordered product is a regularization method
for composite operators where singular short-distances
in the OPE are removed. In the normal ordered products
of three operators or more, one should specify how
the normal ordering is carried out between the successive pairs
of operators. The precise convention we are using is given by
\cite{BBSS}. This is so-called fully normal ordered denoted as
the round brackets but we do not write down them explicitly in this paper.}
\bea
T  & =  &\frac{1}{2(k+N+M)} \Bigg( J^{\al}  J^{\al} + J^a  J^a +
\de_{\rho \bar{\si}} \de_{j \bar{i}} \,
J^{(\rho \bar{i})}  J^{(\bar{\si} j)} + \de_{\rho \bar{\si}} \de_{j \bar{i}} 
J^{(\bar{\si} j)}  J^{(\rho \bar{i})} + J^{u(1)}  J^{u(1)} 
\Bigg)
\nonu \\
&-& \frac{1}{2(k+N)} \,  J^{\al} \, J^{\al} - \frac{1}{2k} \,
J^{u(1)} \, J^{u(1)}.
\label{T}
\eea
The OPE of this stress energy tensor with itself
takes the standard form where the central charge \cite{CH1812}
is given by
\bea
c  & = &  \frac{k((N+M)^2-1)}{(k+M+N)}-\frac{k(N^2-1)}{(k+N)}-1
\nonu \\
&= &
\frac{(-k^2+k^2 M^2-2k N-M N+2k^2 M N + k M^2 N-N^2 + k M N^2)}{(k+N)(k+M+N)}.
\label{charge}
\eea
Then the $SU(M)$ adjoint spin-$1$ current $J^a$
transforms as a primary operator under the above stress energy
tensor $T$.

The charged spin-$2$ current \cite{CH1812} can be summarized by
\bea
K^a & = & \de_{\rho \bar{\si}} \,
t^a_{j\bar{i}} \, (J^{(\rho \bar{i})} \,  J^{(\bar{\si} j)} +
J^{(\bar{\si} j)} \,  J^{(\rho \bar{i})})  
-\frac{N}{(M+2k)} \, d^{abc} \, J^b\, J^c
\nonu \\
& + & \frac{2N}{k} \sqrt{\frac{M+N}{M N}} \, J^a  \,  J^{u(1)}.
\label{spin2expression}
\eea
Some of the relative coefficients appearing on the right hand side
of (\ref{spin2expression})
are fixed by the regularity conditions,
with the coset fields living in the denominator of the coset,
where there are no singular terms in the OPEs
of $J^{u(1)}(z)$ with $K^a(w)$ and of $J^{\alpha}(z)$ with $K^a(w)$.
The charged spin-$2$ current should transform as a primary operator
under the above $SU(M)$ adjoint spin-$1$ current.
In other words, the first order pole of $J^a(z)$ with $K^b(w)$
has $i \, f^{ab}_{\,\,\,\,\,\,c} \, K^c(w)$ which is the
interpretation for the `charge'.
%\footnote{
%\label{JK}
%The OPE between the spin-$1$ current
%and the charged spin-$2$ current is
%given by
%\bea
%J^a(z) \, K^b(w) = \frac{1}{(z-w)}\, i\, f^{a b c}\, K^c(w) +
%\cdots.
%\nonu
%\eea}.
Moreover, the charged spin-$2$ current should transform as
a primary operator under the stress energy tensor.
All the coefficients can be fixed except the
overall normalization after imposing these two conditions.
%\footnote{
%\label{TK}
%Explicitly, we have
%\bea
%T(z) \, K^a(w) = \frac{1}{(z-w)^2}\, 2\, K^a(w) +
%\frac{1}{(z-w)}\, \pa \, K^a(w) + \cdots,
%\nonu
%\eea
%for the spin-$2$ current,
%in addition to
%\bea
%T(z) \, J^a(w) = \frac{1}{(z-w)^2}\,  J^a(w) +
%\frac{1}{(z-w)}\, \pa \, J^a(w) + \cdots,
%\nonu
%\eea
%for the spin-$1$ current.
%}.

The charged spin-$3$ current \cite{CH1812} can be described as
%\footnote{We have, for the spin-$3$ current,
%\bea
%T(z) \, P^a(w) = \frac{1}{(z-w)^2}\, 3 \, P^a(w)
%+ \frac{1}{(z-w)}\, \pa \, P^a(w) + \cdots.
%\nonu
%\eea}
\bea
P^a & = & a_1 \,
t^{\al}_{\rho \bar{\si}} \, t^a_{j\bar{i}} \, J^{\al} 
J^{(\rho \bar{i})}  \, J^{(\bar{\si} j)}
+ a_2 \, J^{\al}  \, J^{\al}  \, J^a+a_3 \, J^b  \, J^b \, J^a
+a_4 \, J^a  \, J^{u(1)}  \, J^{u(1)}
\nonu \\
&+& (a_5 \, d^{abc} + a_6 \, i\, f^{abc} ) \, \de_{\rho \bar{\rho}} \,
t^b_{j\bar{i}} \, J^c   (J^{(\rho \bar{i})}   J^{(\bar{\rho} j)} +
J^{(\bar{\rho} j)}   J^{(\rho \bar{i})})  
\nonu \\
&+& a_7 \, \de_{\rho \bar{\si}} \,
t^a_{j\bar{i}}\, J^{u(1)}   (J^{(\rho \bar{i})}  J^{(\bar{\si} j)} +
J^{(\bar{\si} j)}   J^{(\rho \bar{i})})  
+ a_8 \, \de_{\rho \bar{\si}} \, \de_{j\bar{i}}\,
J^a   \, (J^{(\rho \bar{i})} \, J^{(\bar{\si} j)} +
J^{(\bar{\si} j)}  \,  J^{(\rho \bar{i})})  
\nonu \\
&+& a_9 \,  d^{abc} \, J^b  \, J^c  \,  J^{u(1)}
+(a_{10} \, d^{abc} + a_{11} \, i \, f^{abc}) \, \pa \,   J^b  \,  J^c
\nonu \\
&+& a_{12} \, \de_{\rho \bar{\si}} \,
t^a_{j\bar{i}} \, \pa \, J^{(\rho \bar{i})} \, J^{(\bar{\si} j)}
+ a_{13} \, \de_{\rho \bar{\si}} \,
t^a_{j\bar{i}}  \, \pa  \, J^{(\bar{\si} j)} \,  J^{(\rho \bar{i})}
+a_{14}\, J^a \, \pa \, J^{u(1)} 
+a_{15}\, \pa \, J^a  \, J^{u(1)}
\nonu \\
&+ & a_{16} \, \pa^2  \, J^a
+ a_{17} \, 6 \, \mbox{Tr} \, (t^{a} \,
t^{\left(b \right.} \, t^c \, t^{\left. d \right)}) \, 
J^b  \, J^c  \, J^d.
\label{spin3exp}
\eea
The relative coefficients are given by
\bea
a_2  & = &
\frac{a_1}{k}, \qquad a_3= \frac{ N (k+2 N)}{k (k+M) (3 k+2 M)} \, a_1,
\qquad
a_4 = \frac{ (k+2 N) (M+N)}{k^2 M} \, a_1,
\nonu \\
a_5  & = & -\frac{ (k+2 N)}{4 (k+M)} \, a_1,
\qquad
a_6=0, \qquad
a_7 =   \frac{(k+2 N)}{2 k} \, \sqrt{\frac{M+N}{M N}} \, a_1,
\nonu \\
a_8  &= & -\frac{ (k+2 N)}{2 k M} \, a_1,
\nonu \\
a_9 &=& -  \frac{(k+2 N)N}{2 k (k+M)} \,
\sqrt{\frac{ (M+N)}{M N}} \, a_1,
\qquad
a_{10} = 0, \qquad
a_{11} =\frac{ (k^2-8) N (k+2 N)}{4 k (k+M) (3 k+2 M)} \, a_1,
\nonu \\
a_{12} & = & -\frac{1}{2} \, (k+2 N) \, a_1,
\qquad
a_{13}  =\frac{1}{2}  \, (k+2 N) \, a_1,
\qquad
a_{14} =0, \qquad
a_{15} =0,
\label{avalues}
\\
a_{16} & = & -\frac{ N (6 k^3+9 k^2 M+4 k M^2+12 M)
(k+2 N)}{12 k (k+M) (3 k+2 M)} \, a_1,
\qquad
a_{17}  =  \frac{ N (k+2 N)}{6 (k+M) (3 k+2 M)} a_1.
\nonu
\eea
Some of the relative coefficients appearing on the right hand side
of (\ref{spin3exp})
are determined by the regular conditions
where there are no singular terms in the OPEs
of $J^{u(1)}(z)$ with $P^a(w)$ and of $J^{\alpha}(z)$ with $P^a(w)$.
The charged spin-$3$ current should transform as a primary operator
under the above $SU(M)$ adjoint spin-$1$ current.
That is, the first order pole of $J^a(z)$ with $P^b(w)$
has the term $i \, f^{ab}_{\,\,\,\,\,\,c} \, P^c(w)$
\footnote{
\label{JP}
The OPE of the spin-$1$ current
with the charged spin-$3$ current is
described as
\bea
J^a(z) \, P^b(w) = \frac{1}{(z-w)}\, i\, f^{a b c}\, P^c(w) +
\cdots.
\nonu
\eea}.
Moreover, the charged spin-$3$ current should transform as
a primary operator under the stress energy tensor.
%\footnote{
%\label{cond1}
%At $k=-2N$, we are left with
%the coefficients $a_1$ and $a_2$.}.
After that all the coefficients can be fixed explicitly.

%%%%%%%%%%%%%%%%%%%%%%%%%%%%%%%%%%%%%%%%%%%%%
\subsection{The neutral spin-$3$ current}
%%%%%%%%%%%%%%%%%%%%%%%%%%%%%%%%%%%%%%%%%%%%%%

The neutral primary
spin-$3$ current \cite{CH1812} can be written in terms of
\bea
W^{(3)} & = &
b_1 \, d^{\alpha \beta \gamma} \, J^{\alpha} J^{\beta} J^{\gamma}
+ b_2 \, d^{a b c} \, J^a J^b J^c+
b_3 \, J^{u(1)}  J^{u(1)}  J^{u(1)} + b_4 \, J^{\al} J^{\al} J^{u(1)}
\nonu \\
&+& b_5 \, J^a \, J^a \, J^{u(1)} + b_6 \, 
t^{\al}_{\rho \bar{\si}} \, \de_{j\bar{i}}\,
J^{\al}   \, (J^{(\rho \bar{i})}  \, J^{(\bar{\si} j)} +
J^{(\bar{\si} j)} \,   J^{(\rho \bar{i})})  
\nonu \\
&+& b_7  \, \de_{\rho \bar{\si}} \,
t^a_{j\bar{i}} \, J^a  (J^{(\rho \bar{i})}  J^{(\bar{\si} j)} +
J^{(\bar{\si} j)}  J^{(\rho \bar{i})})  
+b_8 \, \de_{\rho \bar{\si}} \,
\de_{j\bar{i}} \, J^{u(1)}  (J^{(\rho \bar{i})}  J^{(\bar{\si} j)} +
J^{(\bar{\si} j)}  J^{(\rho \bar{i})})   
\nonu \\
&+& b_9 \, \pa \, J^{\alpha} \, J^{\alpha} +
b_{10}\, \pa \, J^{a} \, J^{a} +
b_{11}\, \pa \, J^{u(1)} \, J^{u(1)}
\nonu \\
&+& b_{12} \, 
\de_{\rho \bar{\si}} \,
\de_{j\bar{i}} \, \pa  \, J^{(\rho \bar{i})}  \,J^{(\bar{\si} j)}
+ b_{13} \,
\, \de_{\rho \bar{\si}} \,
\de_{j\bar{i}}  \, \pa  \, J^{(\bar{\si} j)} \,  J^{(\rho \bar{i})}
+ b_{14} \, \pa^2 \, J^{u(1)}.
\label{W}
\eea
The relative coefficients are given by
\bea
b_2  & = &  -\frac{N(k+N)(k+2N)}{M(k+M)(k+2M)} \, b_1,
\qquad
b_3 =  \frac{2(k+N)(M+N)(k+2N)}{k^2 M} \, \sqrt{\frac{M+N}{M N}} \, b_1,
\nonu \\
b_4 & = & \frac{6(k+N)}{k} \, \sqrt{\frac{M+N}{M N}} \, b_1,
\qquad
b_5 = \frac{6N(k+N)(k+2N)}{k M(k+2M)} \, \sqrt{\frac{M+N}{M N}} \, b_1,
% \frac{2 N}{k} \, \sqrt{\frac{M+N}{M N}} \, b_7,
\nonu \\
b_6 & = &  -\frac{3(k+N)}{M} \, b_1, \qquad
b_7 = \frac{3(k+N)(k+2N)}{M(k+2M)} \, b_1,
\nonu \\
b_8  & = & -\frac{3(k+N)(k+2N)}{k M} \, \sqrt{\frac{M+N}{M N}} \, b_1,
\qquad
b_9 =  0, \qquad b_{10}=0,\qquad b_{11}=0,
\nonu \\
b_{12}  & = &  \frac{3(k+N)(k+2N)}{M} \, b_1, 
\nonu \\
b_{13} &=& -\frac{3(k+N)(k+2N)}{M}\, b_1,
\qquad
b_{14} = -N (k+N)(k+2N) \, \sqrt{\frac{M+N}{M N}} \, b_1.
\label{interbvalue}
\eea
Again, some of
the relative coefficients appearing on the right hand side
of (\ref{W})
are obtained by the regular conditions
where there are no singular terms in the OPEs
of $J^{u(1)}(z)$ with $W^{(3)}(w)$ and of
$J^{\alpha}(z)$ with $W^{(3)}(w)$.
Furthermore,  the OPE of this neutral spin-$3$ current
with $SU(M)$ adjoint spin-$1$ current should not contain any
singular terms which is the interpretation of `uncharged' or neutral
\footnote{
\label{JW3}
The OPE is summarized by
\bea
J^a(z) \, W^{(3)}(w) = 0 +
\cdots,
\nonu
\eea
in addition to the primary condition
for the spin-$3$ current
\bea
T(z) \, W^{(3)}(w) = \frac{1}{(z-w)^2}\, 3 \, W^{(3)}(w)
+ \frac{1}{(z-w)}\, \pa \, W^{(3)}(w) + \cdots.
\nonu
\eea}.
Moreover, the neutral spin-$3$ current should transform as
a primary operator under the stress energy tensor.
After that, all the coefficients can be fixed completely \cite{Ahn2011}.
%\footnote{
%\label{cond2}
%At $k=-2N$, we are left with
%the coefficients $b_1, b_4$ and $b_6$.}.

The OPE of the charged spin-$2$ current
with the neutral spin-$3$ current has the nontrivial
second and first order poles. See also the equation
$(7.10)$ of \cite{Ahn2011}.
The charged spin-$3$ current (\ref{spin3exp}) appears at the
second order pole of this OPE. Other primary spin-$4$ operator,
which is written as the composite operator of
the spin-$1$ current and charged spin-$3$ current
together with a derivative of charged spin-$3$ current,
appears at the first order pole.
There is no new primary spin-$4$ current.
On the other hand, for the OPE of the charged spin-$2$ current
with the charged spin-$3$ current, we will describe
its first order pole related to the presence of
new primary charged spin-$4$ current in the
next subsection.
 
%%%%%%%%%%%%%%%%%%%%%%%%%%%%%%%%%%%%%%%%%%%%%
\subsection{The OPE of the charged spin-$2$ current
with the charged spin-$3$ current}
%%%%%%%%%%%%%%%%%%%%%%%%%%%%%%%%%%%%%%%%%%%%%%

The OPE of the charged spin-$2$ current
with the charged spin-$3$ current can be summarized by
\cite{Ahn2011}
\bea
K^a(z) \, P^b(w) & = &
\frac{1}{(z-w)^3} \,
\Bigg[\frac{(k^2-4)(2k+M)(k+2N)(3k+2M+2N)}{2k(k+M)(3k+2M)}
\Bigg]
\, a_1 \, i\, f^{a b c} \, K^c(w)
\nonu \\
&+& \frac{1}{(z-w)^2} \, \Bigg[
\frac{1}{4} \,
\frac{(k^2-4)(2k+M)(k+2N)(3k+2M+2N)}{2k(k+M)(3k+2M)}
\, a_1 \, i\, f^{a b c} \, \pa \, K^c
\nonu \\
& + & Q^{a b} - \de^{a b} \, \frac{a_1}{b_1} \, W^{(3)}
+\frac{k(3k+2M)(2k+M+2N)}{(k+M)(2k+M)} \, d^{a b c} \, P^c
\Bigg](w) 
\nonu \\
&+& \frac{1}{(z-w)} \, \Bigg[
\frac{1}{20} \,
\frac{(k^2-4)(2k+M)(k+2N)(3k+2M+2N)}{2k(k+M)(3k+2M)}
\, a_1 \, i\, f^{a b c} \, \pa^2 \, K^c
\nonu \\
& + & \frac{1}{3} \pa \, Q^{a b} -
\frac{1}{3}
\de^{a b} \, \frac{a_1}{b_1} \, \pa \, W^{(3)}
+\frac{1}{3} \,
\frac{k(3k+2M)(2k+M+2N)}{(k+M)(2k+M)} \, d^{a b c} \, \pa \, P^c
\nonu \\
&+& R^{a b} \Bigg](w) + \cdots,
\label{KPfinal1}
\eea
where $R^{a b}$ is the extra term
appearing on the first order pole in (\ref{KPfinal1})
after subtracting all the descendant terms.
Here, the spin-$3$ current appearing on the
second order pole, which is a primary
under the stress energy tensor
\footnote{That is,
\label{TQ}
the primary condition for the
spin-$3$ current
\bea
T(z) \, Q^{a b}(w) = \frac{1}{(z-w)^2} \, 3\, Q^{a b}(w)+
\frac{1}{(z-w)}\, \pa \, Q^{a b}(w) + \cdots.
\nonu
\eea}, is written in terms of
the multiple product of previous spin-$1$ and spin-$2$ currents
together with a derivative
\bea
Q^{a b} & \equiv & \frac{N (k+2N) (
3k +2M+2N)}{6 (k+M)(3k+2M)} \, a_1\, \Bigg(
-\frac{6 (4+k^2+3 k M+M^2)}{k M N} \, J^a \, K^b  \nonu \\
& + &
\frac{6 (4k +k^3 + 4 M-2 k^2 M-3 k M^2 -M^3)}{
k(k+M) M N} \,
J^b \, K^a \nonu \\
& + & \frac{3(2k+M)^2}{2(k+M)N} \,
f^{a c e} \, f^{b d e} \,
J^c \, K^d
+  \frac{3(4k +k^3+4M)}{2k(k+M)N}  \,
d^{a c e} \, d^{b d e}  \, J^d \, K^c \nonu \\
& - &  \frac{3(4+k^2+k M)}{2 k N} \,
d^{a c e} \, d^{b d e}  \, J^c \, K^d
\nonu \\
& - & \frac{6k(2k+M)}{(k+M)(k+2M)N} \,
\de^{a b} \,
J^c \, K^c
 -  \frac{3(k^2-4)M}{4k N} \,
i \, f^{a b c} \, \pa K^c \Bigg).
\label{Qab}
\eea
It is straightforward to see that the OPE
of the $J^{u(1)}(z)$ with $Q^{a b}(w)$ in (\ref{Qab})
and the OPE of the $J^{\alpha}(z)$ with
$Q^{a b}(w)$ do not contain any singular terms
\footnote{Note that there are also the five trivial regular OPEs besides the
nontrivial ten OPEs (\ref{OPEspin1spin1})
\bea
J^{\al}(z) \, J^a(w) & = & 0 + \cdots, \qquad
J^{\al}(z) \, J^{u(1)}(w) =0 + \cdots, \qquad
J^{a}(z) \, J^{u(1)}(w) =0 + \cdots,
\nonu \\
J^{(\rho \bar{i})}(z) \, J^{(\si \bar{j})}(w) & = & 0 +\cdots, \qquad
J^{(\bar{\rho} i)}(z) \, J^{(\bar{\si} j)}(w) =0 + \cdots.
\label{regular}
%\nonu
\eea
Due to the first and the third relations of
(\ref{regular}),
there are no singular terms in the OPE
of $J^{\alpha}(z)$ with $J^a(w)$ and in the OPE
of $J^{u(1)}(z)$ with $J^a(w)$.
Similar behavior for the spin-$2$ current
$K^a$
appears in the sentence just after the equation 
(\ref{spin2expression}). Note that the $Q^{ab}(z)$ depends on
$J^c(z)$ and $K^d(z)$.}.

The third order pole of (\ref{KPfinal1})
has $i\, f^{a b c} \, K^c$
and we can introduce the following expression
\cite{Ahn2011}
by multiplying the structure constant by the
first order pole
between the charged spin-$2$ current
and the charged spin-$3$ current
\bea
i \, f^{a b c} \, K^a(z) \, P^b(w)\Bigg|_{\frac{1}{(z-w)}} \equiv R^c.
\label{Rdef}
\eea
By multiplying the structure constant
by (\ref{Rdef}) further, we obtain
the right hand side multiplied by the structure constant.
There are also some terms on the first order pole
which vanish after multiplication of the structure constant.
Then we can write down the first order pole as   \cite{Ahn2011}
\bea
K^a(z) \, P^b(w)\Bigg|_{\frac{1}{(z-w)}} \equiv
-\frac{1}{2M} \, i\, f^{a b c } \, R^c + S^{a b}.
\label{KPpole1other1}
\eea
In other words, 
the $S^{a b}$ is nothing but
the first order pole subtracted by the first term of the
right hand side of (\ref{KPpole1other1}). 
After using the relation (\ref{Rdef}), we rewrite it
as
\bea
S^{a b} \equiv \Bigg( K^a(z) \, P^b(w)-
\frac{1}{2M} \,  f^{a b c } \,   f^{c d e }  \,
K^d(z) \, P^e(w)\Bigg)\Bigg|_{\frac{1}{(z-w)}}.
\label{SabSab}
\eea
For fixed $(N,M)=(5,4)$, the explicit form for the
(\ref{SabSab}) is known and can be written in terms of
the known $SU(M)$ charged spin-$1,2,3$
currents and neutral spin-$3$ current with a derivative.

By identifying the first order pole of (\ref{KPfinal1})
with the right hand side of (\ref{KPpole1other1}),
we can read off the expression for $R^{a b}$
which is given by the first four terms of the
first order pole of (\ref{KPfinal1})
plus the two terms of  the right hand side of  (\ref{KPpole1other1})
\footnote{
We can read off the first order pole in the OPE
of the charged spin-$2$ current and the charged
spin-$3$ current.
The decendant terms with the two
free indices at the first order pole are given by the first four terms
from (\ref{KPfinal1}) and the first order pole is
defined as (\ref{KPpole1other1}).
This leads to the following result
\bea
R^{a b} & = &
-\frac{1}{20} \,
\frac{(k^2-4)(2k+M)(k+2N)(3k+2M+2N)}{2k(k+M)(3k+2M)}
\, a_1 \, i\, f^{a b c} \, \pa^2 \, K^c
\nonu \\
& - & \frac{1}{3} \pa \, Q^{a b} +
\frac{1}{3}
\de^{a b} \, \frac{a_1}{b_1} \, \pa \, W^{(3)}
-\frac{1}{3} \,
\frac{k(3k+2M)(2k+M+2N)}{(k+M)(2k+M)} \, d^{a b c} \, \pa \, P^c
\nonu \\
& - & \frac{1}{2M} \, i\, f^{a b c } \, R^c + S^{a b}.
%\label{Rab}
\nonu
\eea
}.
Furthermore,
the expression of (\ref{Rdef})
is related to the second order pole of
the OPE of the charged spin-$3$ current
with the neutral spin-$3$ current together with
the above
known $SU(M)$ charged spin-$1,2,3$
currents and neutral spin-$3$ current with a derivative
for  fixed $(N,M)=(5,4)$.
In next section, we describe the structure of the first order pole
in (\ref{KPfinal1}) more detail for generic $(N,M)$.
After substituting the explicit form for the
$S^{a b}$ for generic $(N,M)$ in terms of the known currents,
we will observe that $R^{a b }$ can be written in terms of
the known currents except $i \, f^{a b c} \, R^c$ term.
This implies that
we expect that the new primary spin-$4$ current occurs
at this term, due to  the various terms which cannot be written
in terms of the known currents, from the standard two dimensional conformal
field theory analysis. See also (\ref{rtilde}).

%%%%%%%%%%%%%%%%%%%%%%%%%%%%%%%%%%%%%%%%%%%%%%%%%%%%%%%%%%%%%%%%%%%%%
%%%%%%%%%%%%%%%%%%%%%%%%%%%%%%%%%%%%%%%%%%%%%%%%%%%%%%%%%%%%%%%%%%%%%%
\section{Construction of the charged primary spin-$4$ current
and the neutral primary spin-$4$ current}
%2%%%%%%%%%%%%%%%%%%%%%%%%%%%%%%%%%%%%%%%%%%%%%%%%%%%%%%%%%%%%%%%%%%%%%
%%%%%%%%%%%%%%%%%%%%%%%%%%%%%%%%%%%%%%%%%%%%%%%%%%%%%%%%%%%%%%%%%%%%%

%%%%%%%%%%%%%%%%%%%
\subsection{The charged primary spin-$4$ current}
%%%%%%%%%%%%%%%%%%%

%%%%%%%%%%%%%%%%%%%%%%%%%%%%%%%%%%%%
\subsubsection{The OPE of the charged spin-$3$ current
with the neutral spin-$3$ current up to the second order pole}
%%%%%%%%%%%%%%%%%%%%%%%%%%%%%%%%%%%%

%\bea
%&&(common\ coeff)\equiv
%\\
%&&
%\frac{3 (-2 + k) (2 + k)(k + N) (k + M + N) (k + 2 N) (3 k + 2 M + 2 N)\, a%_{1}\, b_{1}}{2 k M (k + M) (k + 2 M)}
%\\
%&&\Big( P^{a}\,W^{(3)}\Big)_{pole-4}(w) \equiv 
%- (2 k + M) \ (common\ coeff)\,K^{a}
%\\
%&&\Big( P^{a}\,W^{(3)}\Big)_{pole-3}(w) \equiv +(common\ coeff)\Big(i f^{a %b c}J^{b}K^{c}+\frac{M}{2} K^{a}  \Big)
%\\
%\eea
It is known that
the fourth, the third and the second order poles
in 
the OPE of the charged spin-$3$ current with
the neutral spin-$3$ current for fixed
$(N,M)=(5,4)$ are obtained explicitly \cite{Ahn2011}.
One can calculate these poles for arbitrary $(N,M)$
by using (\ref{spin3exp}) and (\ref{W})
and the fourth and the third order poles
can be summarized by
\bea
&& P^a(z) \, W^{(3)}(w) = 
\frac{1}{(z-w)^4} \nonu \\
&& \times
\Bigg[-
\frac{3 (k^2-4)  (2 k + M) (k + N) (k + M + N) (k + 2 N) (3 k + 2 M + 2 N) a_{1} b_{1}}{2 k M (k + M) (k + 2 M)} \Bigg]   K^a(w)
\nonu \\
&&+
\frac{1}{(z-w)^3}
\Bigg[-
  \frac{1}{2}\,
\frac{3 (k^2-4)  (2 k + M) (k + N) (k + M + N) (k + 2 N) (3 k + 2 M + 2 N) a_{1} b_{1}}{2 k M (k + M) (k + 2 M)}  \nonu\\
&& \times   \pa \, K^a+
\frac{3 (k^2-4)  (k + N) (k + M + N) (k + 2 N) (3 k + 2 M + 2 N) a_{1} b_{1}}{2 k M (k + M) (k + 2 M)} \nonu\\
&& \times  \Bigg( i\, f^{a b c} \, J^b \, K^c + \frac{M}{2}
\, \pa K^a \Bigg) \Bigg](w)+
{\cal O}\left(\frac{1}{(z-w)^2}\right) +\cdots.
\nonu
\eea
%It turns out that
The second order pole in the OPE of the charged spin-$3$ current
(\ref{spin3exp})
with the neutral spin-$3$ current (\ref{W}),
by using the descriptions in \cite{BBSS,BBSS1,BS}
and using the rearrangement lemmas to write down
all the normal ordered products as `fully' normal ordered product,
can be summarized by
\bea
&&P^a(z) \, W^{(3)}(w)\Bigg|_{\frac{1}{(z-w)^2}} = 
p_{1} d^{d a r} d^{f h e} d^{r c e} J^{d} J^{c} J^{f} J^{h} 
+ p_{2} d^{d a r} d^{c h e} d^{r f e} J^{d} J^{c} J^{f} J^{h} \nonu\\
&&+ p_{3} d^{d a r} d^{c f e} d^{r h e} J^{d} J^{c} J^{f} J^{h} 
+ p_{4} d^{a v e} d^{c f e} d^{v h d} J^{d} J^{c} J^{f} J^{h} 
+ p_{5} d^{a f e} d^{c v e} d^{v h d} J^{d} J^{c} J^{f} J^{h} \nonu\\
&&+ p_{6} d^{a c e} d^{f v e} d^{v h d} J^{d} J^{c} J^{f} J^{h} 
+ p_{7} d^{f h e} d^{r c e} f^{a d r} J^{d} J^{c} J^{f} J^{h} 
+ p_{8} d^{c h e} d^{r f e} f^{a d r} J^{d} J^{c} J^{f} J^{h} \nonu\\
&&+ p_{9} d^{c f e} d^{r h e} f^{a d r} J^{d} J^{c} J^{f} J^{h} 
+ p_{10} d^{a c e} d^{v h d} f^{f v e} J^{d} J^{c} J^{f} J^{h} 
+ p_{11} d^{a v e} d^{c f e} f^{v h d} J^{d} J^{c} J^{f} J^{h} \nonu\\
&&+ p_{12} d^{a f e} d^{c v e} f^{v h d} J^{d} J^{c} J^{f} J^{h} 
+ p_{13} d^{a c e} d^{f v e} f^{v h d} J^{d} J^{c} J^{f} J^{h} 
+ p_{14} d^{a c e} f^{f v e} f^{v h d} J^{d} J^{c} J^{f} J^{h} \nonu\\
&&+ p_{15} d^{a c f} \delta_{d h} J^{d} J^{c} J^{f} J^{h}
+ p_{16} d^{f h d} \delta_{a c} J^{d} J^{c} J^{f} J^{h} 
+ p_{17} f^{f h d} \delta_{a c} J^{d} J^{c} J^{f} J^{h} \nonu\\
&&+ p_{18} d^{d a h} \delta_{c f} J^{d} J^{c} J^{f} J^{h} 
+ p_{19} d^{d a h} \delta_{c f} J^{d} J^{c} J^{f} J^{h} 
+ p_{20} d^{a h d} \delta_{c f} J^{d} J^{c} J^{f} J^{h} \nonu\\
&&+ p_{21} f^{a d h} \delta_{c f} J^{d} J^{c} J^{f} J^{h} 
+ p_{22} f^{a d h} \delta_{c f} J^{d} J^{c} J^{f} J^{h} 
+ p_{23} f^{a h d} \delta_{c f} J^{d} J^{c} J^{f} J^{h} \nonu\\
&&+ p_{24} d^{d a f} \delta_{c h} J^{d} J^{c} J^{f} J^{h} 
+ p_{25} f^{a d f} \delta_{c h} J^{d} J^{c} J^{f} J^{h} 
+ p_{26} d^{d a c} \delta_{f h} J^{d} J^{c} J^{f} J^{h} \nonu\\
&&+ p_{27} f^{a d c} \delta_{f h} J^{d} J^{c} J^{f} J^{h} 
+ p_{28} J^{a} J^{u(1)} J^{u(1)} J^{u(1)} 
+ p_{29} \delta_{\bar{i} k} \delta_{\rho \bar{\tau}} J^{a} J^{u(1)} J^{(\rho \bar{i})} J^{(\bar{\tau} k)} \nonu\\
&&+ p_{30} \delta_{\bar{i} k} \delta_{\rho \bar{\tau}} J^{a} J^{(\rho \bar{i})} \partial J^{(\bar{\tau} k)} 
+ p_{31} J^{a} \partial J^{u(1)} J^{u(1)} 
+ p_{32} \delta_{\bar{i} k} \delta_{\rho \bar{\tau}} J^{a} \partial J^{(\rho \bar{i})} J^{(\bar{\tau} k)} \nonu\\
&&+ p_{33} J^{a} \partial^{2} J^{u(1)}
+ p_{34} d^{a c f} J^{c} J^{f} J^{u(1)} J^{u(1)} \nonu\\
&&+ p_{35} f^{a c f} J^{c} J^{f} J^{u(1)} J^{u(1)} 
+p_{36} d^{a h e} d^{c f e} J^{c} J^{f} J^{h} J^{u(1)} 
+ p_{37} d^{a f e} d^{c h e} J^{c} J^{f} J^{h} J^{u(1)} \nonu\\
&&+ p_{38} d^{a c e} d^{f h e} J^{c} J^{f} J^{h} J^{u(1)} 
+ p_{39} d^{f h e} f^{a c e} J^{c} J^{f} J^{h} J^{u(1)} 
+ p_{40} d^{c h e} f^{a f e} J^{c} J^{f} J^{h} J^{u(1)} \nonu\\
&&+ p_{41} d^{a f e} f^{c h e} J^{c} J^{f} J^{h} J^{u(1)} 
+ p_{42} f^{a f e} f^{c h e} J^{c} J^{f} J^{h} J^{u(1)} 
+ p_{43} d^{a c e} f^{f h e} J^{c} J^{f} J^{h} J^{u(1)} \nonu\\
&&+ p_{44} \delta_{a h} \delta_{c f} J^{c} J^{f} J^{h} J^{u(1)} 
+ p_{45} \delta_{a f} \delta_{c h} J^{c} J^{f} J^{h} J^{u(1)} 
+ p_{46} \delta_{a c} \delta_{f h} J^{c} J^{f} J^{h} J^{u(1)} \nonu\\
&&+ p_{47} d^{a c f} \delta_{\bar{i} k} \delta_{\rho \bar{\tau}} J^{c} J^{f} J^{(\rho \bar{i})} J^{(\bar{\tau} k)} 
+ p_{48} d^{a c f} J^{c} J^{f} \partial J^{u(1)} 
+ p_{49} f^{a c f} J^{c} J^{f} \partial J^{u(1)} \nonu\\
&&+ p_{50} \delta_{\beta \gamma} J^{\beta} J^{\gamma} J^{a} J^{u(1)} 
+ p_{51} \delta_{a c} \delta_{\beta \gamma} J^{\beta} J^{\gamma} J^{c} J^{u(1)} 
+ p_{52} d^{a c f} \delta_{\beta \gamma} J^{\beta} J^{\gamma} J^{c} J^{f} \nonu\\
&&+ p_{53} f^{a c f} \delta_{\beta \gamma} J^{\beta} J^{\gamma} J^{c} J^{f} 
+ p_{54} d^{\beta \gamma \delta} J^{\beta} J^{\gamma} J^{\delta} J^{a} 
+ p_{55} f^{\beta \gamma \delta} J^{\beta} J^{\gamma} J^{\delta} J^{a} \nonu\\
&&+ p_{56} d^{\beta \gamma \delta} \delta_{a c} J^{\beta} J^{\gamma} J^{\delta} J^{c}
+ p_{57} f^{\beta \gamma \delta} \delta_{a c} J^{\beta} J^{\gamma} J^{\delta} J^{c} 
+ p_{58} \delta_{\beta \gamma} J^{\beta} J^{\gamma} \partial J^{a} 
+ p_{59} \partial J^{a} J^{u(1)} J^{u(1)} \nonu\\
&&+ p_{60} \delta_{\bar{i} k} \delta_{\rho \bar{\tau}} \partial J^{a} J^{(\rho \bar{i})} J^{(\bar{\tau} k)} 
+ p_{61} \partial J^{a} \partial J^{u(1)}
+p_{62} d^{a c f} \partial J^{c} J^{f} J^{u(1)} 
+ p_{63} f^{a c f} \partial J^{c} J^{f} J^{u(1)} \nonu\\
&&+ p_{64} d^{a h e} d^{c f e} \partial J^{c} J^{f} J^{h} 
+ p_{65} d^{a f e} d^{c h e} \partial J^{c} J^{f} J^{h}
+ p_{66} d^{a c e} d^{f h e} \partial J^{c} J^{f} J^{h} \nonu\\
&&+ p_{67} d^{f h e} f^{a c e} \partial J^{c} J^{f} J^{h} 
+ p_{68} d^{c h e} f^{a f e} \partial J^{c} J^{f} J^{h} 
+ p_{69} d^{c f e} f^{a h e} \partial J^{c} J^{f} J^{h} \nonu\\
&&+ p_{70} d^{a h e} f^{c f e} \partial J^{c} J^{f} J^{h} 
+ p_{71} f^{a h e} f^{c f e} \partial J^{c} J^{f} J^{h} 
+p_{72} d^{a f e} f^{c h e} \partial J^{c} J^{f} J^{h} \nonu\\
&&+ p_{73} f^{a f e} f^{c h e} \partial J^{c} J^{f} J^{h} 
+ p_{74} d^{a c e} f^{f h e} \partial J^{c} J^{f} J^{h} 
+ p_{75} f^{a c e} f^{f h e} \partial J^{c} J^{f} J^{h}\nonu\\
&&+ p_{76} \delta_{a h} \delta_{c f} \partial J^{c} J^{f} J^{h} 
+ p_{77} \delta_{a f} \delta_{c h} \partial J^{c} J^{f} J^{h} 
+ p_{78} \delta_{a c} \delta_{f h} \partial J^{c} J^{f} J^{h} \nonu\\
&&+ p_{79} \delta_{a c} \delta_{\bar{i} k} \delta_{\rho \bar{\tau}} \partial J^{c} J^{(\rho \bar{i})} J^{(\bar{\tau} k)}
+ p_{80} d^{a c f} \partial J^{c} \partial J^{f} 
+ p_{81} f^{a c f} \partial J^{c} \partial J^{f} \nonu\\
&&+ p_{82} \delta_{\beta \gamma} \partial J^{\beta} J^{\gamma} J^{a} 
+ p_{83} \partial^{2} J^{a} J^{u(1)}
+ p_{84} d^{a c f} \partial^{2} J^{c} J^{f} 
+ p_{85} f^{a c f} \partial^{2} J^{c} J^{f} \nonu\\
&&+p_{86} \delta_{\rho \bar{\tau}} t^{a}_{k \bar{i}} J^{u(1)} J^{u(1)} J^{(\rho \bar{i})} J^{(\bar{\tau} k)}
+ p_{87} \delta_{\rho \bar{\tau}} t^{a}_{k \bar{i}} J^{u(1)}
J^{(\rho \bar{i})} \pa J^{(\bar{\tau} k)}  \nonu\\
&&+ p_{88} \delta_{\rho \bar{\tau}} t^{a}_{k \bar{i}}  J^{u(1)} \pa
J^{(\rho \bar{i})} J^{(\bar{\tau} k)}  
+ p_{89} \delta_{c f} \delta_{\rho \bar{\tau}} t^{a}_{k \bar{i}} J^{c} J^{f} J^{(\rho \bar{i})} J^{(\bar{\tau} k)}  \nonu\\
&&+ p_{90} \delta_{\beta \gamma} \delta_{\rho \bar{\tau}} t^{a}_{k \bar{i}} J^{\beta} J^{\gamma} J^{(\rho \bar{i})} J^{(\bar{\tau} k)}
+
p_{91} \delta_{\bar{g} y} \delta_{\bar{\nu} \sigma} \delta_{\rho \bar{\tau}} t^{a}_{k \bar{i}} J^{(\rho \bar{i})} J^{(\sigma \bar{g})} J^{(\bar{\nu} k)} J^{(\bar{\tau} y)}  \nonu\\
&&+ p_{92} \delta_{\bar{g} y} \delta_{\bar{\nu} \sigma} \delta_{\rho \bar{\tau}} t^{a}_{k \bar{i}} J^{(\rho \bar{i})} J^{(\sigma \bar{g})} J^{(\bar{\nu} y)} J^{(\bar{\tau} k)}  + p_{93} \delta_{\rho \bar{\tau}} t^{a}_{k \bar{i}} J^{(\rho \bar{i})} \partial^{2} J^{(\bar{\tau} k)}  
+ p_{94} \delta_{\rho \bar{\tau}} t^{a}_{k \bar{i}} \partial J^{u(1)} J^{(\rho \bar{i})} J^{(\bar{\tau} k)}  \nonu\\
&&+ p_{95} \delta_{\rho \bar{\tau}} t^{a}_{k \bar{i}} \partial J^{(\rho \bar{i})} \partial J^{(\bar{\tau} k)}  
+ p_{96} \delta_{\rho \bar{\tau}} t^{a}_{k \bar{i}} \partial^{2} J^{(\rho \bar{i})} J^{(\bar{\tau} k)}  
+ p_{97} \delta_{\bar{i} k} \delta_{\bar{\nu} \sigma} \delta_{\rho \bar{\tau}} t^{a}_{y \bar{g}} J^{(\rho \bar{i})} J^{(\sigma \bar{g})} J^{(\bar{\nu} y)} J^{(\bar{\tau} k)}  \nonu\\
&&+ p_{98} \delta_{\rho \bar{\tau}} t^{f}_{k \bar{i}} d^{a c f} J^{c} J^{u(1)} J^{(\rho \bar{i})} J^{(\bar{\tau} k)}  
+ p_{99} \delta_{a c} \delta_{\rho \bar{\tau}} t^{f}_{k \bar{i}} J^{c} J^{f} J^{(\rho \bar{i})} J^{(\bar{\tau} k)}  
+ p_{100} \delta_{\rho \bar{\tau}} t^{f}_{k \bar{i}} d^{a c f} J^{c} J^{(\rho \bar{i})} \partial J^{(\bar{\tau} k)}  \nonu\\
&&+ p_{101} \delta_{\rho \bar{\tau}} t^{f}_{k \bar{i}} f^{a c f} J^{c} J^{(\rho \bar{i})} \partial J^{(\bar{\tau} k)}  + p_{102} \delta_{\rho \bar{\tau}} t^{f}_{k \bar{i}} d^{a c f}  J^{c} \pa J^{(\rho \bar{i})} J^{(\bar{\tau} k)}  
+ p_{103} \delta_{\rho \bar{\tau}} t^{f}_{k \bar{i}} f^{a c f}
J^{c} \pa J^{(\rho \bar{i})} J^{(\bar{\tau} k)}  \nonu\\
&&+ p_{104} \delta_{\rho \bar{\tau}} t^{f}_{k \bar{i}} d^{a c f} \partial J^{c} J^{(\rho \bar{i})} J^{(\bar{\tau} k)}  
+ p_{105} \delta_{\rho \bar{\tau}} t^{f}_{k \bar{i}} f^{a c f} \partial J^{c} J^{(\rho \bar{i})} J^{(\bar{\tau} k)}  \nonu\\
&&+ p_{106} \delta_{\bar{\nu} \sigma} \delta_{\rho \bar{\tau}} t^{c}_{k \bar{i}} t^{f}_{y \bar{g}} d^{a c f} J^{(\rho \bar{i})} J^{(\sigma \bar{g})} J^{(\bar{\nu} y)} J^{(\bar{\tau} k)}  
+ p_{107} \delta_{\rho \bar{\tau}} t^{h}_{k \bar{i}} d^{a h e} d^{c f e} J^{c} J^{f} J^{(\rho \bar{i})} J^{(\bar{\tau} k)}  \nonu\\
&&+ p_{108} \delta_{\rho \bar{\tau}} t^{h}_{k \bar{i}} d^{a f e} d^{c h e} J^{c} J^{f} J^{(\rho \bar{i})} J^{(\bar{\tau} k)}  
+ p_{109} \delta_{\rho \bar{\tau}} t^{h}_{k \bar{i}} d^{a c e} d^{f h e} J^{c} J^{f} J^{(\rho \bar{i})} J^{(\bar{\tau} k)}  \nonu\\
&&+ p_{110} \delta_{\rho \bar{\tau}} t^{h}_{k \bar{i}} d^{f h e} f^{a c e} J^{c} J^{f} J^{(\rho \bar{i})} J^{(\bar{\tau} k)}  
+p_{111} \delta_{\rho \bar{\tau}} t^{h}_{k \bar{i}} d^{c h e} f^{a f e} J^{c} J^{f} J^{(\rho \bar{i})} J^{(\bar{\tau} k)}  \nonu\\
&&+ p_{112} \delta_{\rho \bar{\tau}} t^{h}_{k \bar{i}} d^{c f e} f^{a h e} J^{c} J^{f} J^{(\rho \bar{i})} J^{(\bar{\tau} k)}  
+ p_{113} \delta_{\rho \bar{\tau}} t^{h}_{k \bar{i}} d^{a h e} f^{c f e} J^{c} J^{f} J^{(\rho \bar{i})} J^{(\bar{\tau} k)}  \nonu\\
&&+ p_{114} \delta_{\rho \bar{\tau}} t^{h}_{k \bar{i}} f^{a h e} f^{c f e} J^{c} J^{f} J^{(\rho \bar{i})} J^{(\bar{\tau} k)}  
+ p_{115} \delta_{\rho \bar{\tau}} t^{h}_{k \bar{i}} d^{a f e} f^{c h e} J^{c} J^{f} J^{(\rho \bar{i})} J^{(\bar{\tau} k)}  \nonu\\
&&+ p_{116} \delta_{\rho \bar{\tau}} t^{h}_{k \bar{i}} f^{a f e} f^{c h e} J^{c} J^{f} J^{(\rho \bar{i})} J^{(\bar{\tau} k)}  
+ p_{117} \delta_{\rho \bar{\tau}} t^{h}_{k \bar{i}} d^{a c e} f^{f h e} J^{c} J^{f} J^{(\rho \bar{i})} J^{(\bar{\tau} k)}  \nonu\\
&&+ p_{118} \delta_{\rho \bar{\tau}} t^{h}_{k \bar{i}} f^{a c e} f^{f h e} J^{c} J^{f} J^{(\rho \bar{i})} J^{(\bar{\tau} k)}  
+ p_{119} \delta_{a f} \delta_{c h} \delta_{\rho \bar{\tau}} t^{h}_{k \bar{i}} J^{c} J^{f} J^{(\rho \bar{i})} J^{(\bar{\tau} k)}  \nonu\\
&&+ p_{120} \delta_{a c} \delta_{f h} \delta_{\rho \bar{\tau}} t^{h}_{k \bar{i}} J^{c} J^{f} J^{(\rho \bar{i})} J^{(\bar{\tau} k)}  
+ p_{121} \delta_{\bar{i} k} t^{\beta}_{\rho \bar{\tau}} J^{\beta} J^{a} J^{(\rho \bar{i})} J^{(\bar{\tau} k)}  \nonu\\
&&+ p_{122} t^{a}_{k \bar{i}} t^{\beta}_{\rho \bar{\tau}} J^{\beta} J^{u(1)} J^{(\rho \bar{i})} J^{(\bar{\tau} k)}  
+ p_{123} t^{a}_{k \bar{i}} t^{\beta}_{\rho \bar{\tau}} J^{\beta} J^{(\rho \bar{i})} \partial J^{(\bar{\tau} k)} 
+ p_{124} t^{a}_{k \bar{i}} t^{\beta}_{\rho \bar{\tau}} J^{\beta} \partial J^{(\rho \bar{i})} J^{(\bar{\tau} k)}  \nonu\\
&&+ p_{125} t^{a}_{k \bar{i}} t^{\beta}_{\rho \bar{\tau}} \partial J^{\beta} J^{(\rho \bar{i})} J^{(\bar{\tau} k)}  
+ p_{126} t^{f}_{k \bar{i}} t^{\beta}_{\rho \bar{\tau}} d^{a c f} J^{\beta} J^{c} J^{(\rho \bar{i})} J^{(\bar{\tau} k)}  \nonu\\
&&+ p_{127} \delta_{\beta \gamma} t^{a}_{k \bar{i}} t^{\gamma}_{\rho \bar{\tau}} J^{\beta} J^{u(1)} J^{(\rho \bar{i})} J^{(\bar{\tau} k)}  
+ p_{128} t^{a}_{k \bar{i}} t^{\delta}_{\rho \bar{\tau}} d^{\beta \gamma \delta} J^{\beta} J^{\gamma} J^{(\rho \bar{i})} J^{(\bar{\tau} k)}  + p_{129} \partial^{3} J^{a}, 
\label{PWPOLE2}
\eea
where the various coefficients which depend on
the parameters $(k,N,M)$ are given by (\ref{pvalue}).
Note that
the free adjoint index $a$ appears in the $d$ symbol, the $f$ symbol,
the Kronecker delta and the generator of $SU(M)$.

%%%%%%%%%%%%%%%%%%%
\subsubsection{The charged primary spin-$4$ current}
%%%%%%%%%%%%%%%%%%%

From the result of (\ref{PWPOLE2})
after subtracting the descendant terms explicitly, we can extract
the quasi-primary spin-$4$ current for generic $(N,M)$
as follows \cite{Ahn2011}:
\bea
R_2^a & \equiv &
P^{a}(z)\,W^{(3)}(w)\Bigg|_{\frac{1}{(z-w)^2}}
\nonu \\
&+&
\frac{3}{20} \frac{3 (k^2-4)  (2 k + M) (k + N) (k + M + N) (k + 2 N) (3 k + 2 M + 2 N) a_{1} b_{1}}{2 k M (k + M) (k + 2 M)} \nonu\\
&\times& \pa^2  K^a \nonu\\
&-& \frac{1}{2}\frac{3 (k^2-4)  (k + N) (k + M + N) (k + 2 N) (3 k + 2 M + 2 N) a_{1} b_{1}}{2 k M (k + M) (k + 2 M)} \nonu\\
&\times& \pa \Bigg( i\, f^{a b c} \, J^b \, K^c + \frac{M}{2}  \, \pa K^a \Bigg),
\label{r2anew}
\eea
where the coefficient $b_1$ can be fixed later in (\ref{b1value}).
Note that the numerical values
$\frac{3}{20}$ and $\frac{1}{2}$
can be obtained from the formula \cite{Blumenhagenetal,Nahm1,Nahm2}.
These can be determined from the spins of the left hand side of
this OPE ($3$ and $3$ respectively),
the spins of the right hand side ($2$ and $3$
respectively) and the number of
derivatives ($2$ and $1$ respectively)
appearing in the descendant fields on the right hand side.
See also \cite{BS}.
Moreover,
we can associate (\ref{r2anew})
with the previous spin-$4$ current
in (\ref{Rdef}) together with other
composite operators between
the $SU(M)$ adjoint spin-$1,2,3$ currents and the neutral
spin-$3$ current for generic $(N,M)$
\bea
R_2^a & \equiv &  r_{1}\, R^a + r_{2}\, J^a W^{(3)} + r_{3}\, d^{a b c} \, J^b\, P^c + r_{4}\, d_{4SS2}^{b c d a}\, J^b \, J^c\, K^d\nonu\\
&+& r_{5}\, d_{4AA1}^{b a c d}\ J^b \, J^c \, K^d + r_{6} \,\de^{b a} \, \de^{c d} \, J^b \, J^c \, K^d + r_{7}\, \de^{b c} \, \de^{d a} \, J^b \, J^c \, K^d\nonu\\
&+&r_{8}\, \pa^2 \, K^a
+r_{9}\, i\, f^{a b c}  \,J^b \, \pa \, K^c + r_{10}\, i\, f^{a b c}  \, \pa \,J^b  \, K^c,
\label{r2a}
\eea
where the four index $d$ symbols are given in (\ref{tensor}).

The ten coefficients appearing in (\ref{r2a})
are given by
\bea
r_1 & = & \frac{6 (k+M)(k+N)(k+M+N)\ b_1}{M^2 (k+2M)}, \qquad
r_2 = -\frac{12 (k+M)(k+N)(k+M+N)\ a_1}{k M (k+2M)},
\nonu \\
r_3 & = & \frac{6 k (3k+2M)(k+N)(k+M+N)(2k+M+2N)\ b_1}{M (k+M)(2k+M)(k+2M)},
\nonu \\
r_4 & = & \frac{3 k (2k+M)(k+N)(k+M+N)(k+2N)(3k+2M+2N)\ a_1b_1}{4 M (k+M)^2(k+2M)(3k+2M)},
\nonu \\
r_5 & = & \frac{3 (8k^2+16kM+k^3M+8M^2)(k+N)(k+M+N)(k+2N)(3k+2M+2N)\ a_1b_1}{4k M^2 (k+M)^2 (k+2M)(3k+2M)},
\nonu \\
r_6 & = & \frac{3 (2k+M)(2k^2+3kM+2M^2)(k+N)(k+M+N)(k+2N)(3k+2M+2N)\ a_1b_1}{M^2 (k+M)^2 (k+2M)^2 (3k+2M)},
\nonu \\
r_7 & = & \frac{3 (2k+M)(k+N)(k+M+N)(k+2N)(3k+2M+2N)\ a_1b_1}{M^2 (k+M)(k+2M)(3k+2M)},
\nonu \\
r_8 & = & \frac{3 (k^2-4)(k+N)(k+M+N)(k+2N)(3k+2M+2N)\ a_1b_1}{20k M (k+M)(k+2M)(3k+2M)}(4k^2+3kM\nonu\\
&-&2M^2),
\nonu\\
r_9 & = & -\frac{3 (k^2-4)(k+N)(k+M+N)(k+2N)(3k+2M+2N)\ a_1b_1}{4k M (k+M)(3k+2M)},
\nonu \\
r_{10} & = & -\frac{3(k+N)(k+M+N)(k+2N)(3k+2M+2N)\ a_1b_1}{2k M^2 (k+M)(k+2M)(3k+2M)}(-4k^2-14kM+k^3M
\nonu\\
&-&8M^2+k^2M^2).
\label{rvalues}
\eea
The factor $(k+2N)$,
which decouples the charged and neutral spin-$3$ currents
\cite{CH1812},
does not appear in the coefficients,
$r_1,r_2$ and $r_3$ in (\ref{rvalues}) \footnote{By construction,
$R_2^a$ should contain the overall factor
$a_1 b_1$ where the first factor
comes from the charged spin-$3$ current while
the second factor comes from the neutral spin-$3$ current.
It is obvious that the $r_2$ term has the $b_1$ factor
and $r_3$ term has $a_1$ factor in (\ref{r2a}).
Moreover, the $r_1$ term has $a_1$ factor from $R^a$.
Therefore, the right hand side of (\ref{r2a}) has
the above $a_1 b_1$ factor.}.

The quasi primary spin-$4$ current is introduced in \cite{Ahn2011}
\bea
\hat{R}^c & \equiv &
R^c -\frac{1}{3}\, i\, f^{a b c} \, \pa \, Q^{a b} 
\nonu \\
&+ & \frac{M(k^2-4)(2k+M)(k+2N)(3k+2M+2N)}{20k(k+M)(3k+2M)}
\, a_1 \, \pa^2 \, K^c,
\label{Rhata}
\eea
where $R^a$ and $Q^{ a b}$ are given by (\ref{Rdef}) and
(\ref{Qab}) respectively \footnote{We can observe
that the right hand side of (\ref{Rhata}) contains
the overall factor $a_1$.}.
By adding the quasi primary \footnote{
In other words, the third order pole in the OPE of
$T$ with this operator vanishes. }
spin-$4$ current
\bea
&& \frac{2(k^2-4)(2k+M)(k+2N)(3k+2M+2N)}{7k(k+M)(3k+2M)}
\nonu \\
&& \times a_1
i f^{a b c}  (J^b  \pa  K^c -2 \pa  J^b  K^c-\frac{i}{10} 
f^{b c d}  \pa^2  K^d)
\label{extraterms}
\eea
into (\ref{Rhata})
which leads to the vanishing of
the pole $4$ in the OPE
of the stress energy tensor with this
spin-$4$ current
\footnote{As in the result of \cite{Ahn2011},
the OPE of $T(z)$ with $\hat{R}^a(w)$
for fixed $(N,M)=(5,4)$
contains the fourth order pole
which is proportional to $K^a(w)$, in addition to the second
and the first order poles.
In order to fix the overall coefficient
appearing in (\ref{extraterms}),
we need to calculate this
OPE explicitly. However, we can focus on
the particular fourth order pole.
Moreover, $K^a(w)$ contains the last term of
(\ref{spin2expression}).
We can observe that 
there is no contribution from
the second term in (\ref{Rhata})
according to the footnote \ref{TQ}.
The contribution from the third term
in (\ref{Rhata}) can be obtained also.
%from the footnote
%\ref{TK}.
The nontrivial part comes from the
first term in (\ref{Rhata}) which is defined in (\ref{Rdef}).
See also (\ref{KPPOLE1}).
This implies that we should calculate, by hand,
the fourth order pole having
$J^a \, J^{u(1)}(w)$ of the OPE of
$T(z)$ with the twentyeight terms (among ninetyseven terms)
of $R^a(w)$
where the nontrivial results of $(2.28)$ in \cite{Ahn2011}
are used
%the footnote \ref{TK} is used
and we have
$T(z) \, J^{\alpha}(w)=
0 + \cdots = T(z) \, J^{u(1)}(w)$.
\label{footnote13}}, we determine the primary spin-$4$ current
as follows:
\bea
&& \tilde{R}^a \equiv
\hat{R}^a +
\label{SPIN4} \\
&& \frac{2(k^2-4)(2k+M)(k+2N)(3k+2M+2N)}{7k(k+M)(3k+2M)}
a_1
i f^{a b c}  (J^b  \pa  K^c -2 \pa  J^b  K^c-\frac{i}{10} 
f^{b c d}  \pa^2  K^d).
\nonu
\eea
In other words,
the reason why we should include
(\ref{extraterms})
is to obtain the primary current rather than the quasi primary
current (\ref{Rhata}).
The determination for the overall factor
in (\ref{extraterms})
is explained in the footnote \ref{footnote13}.
We can rewrite the primary spin-$4$ current,
by using (\ref{Rdef}), (\ref{Rhata}) and (\ref{SPIN4}),
as \footnote{We expect that all the coefficients
should contain the overall factor $a_1$.
We have the following OPE
for the primary condition of the spin-$4$ current
\bea
T(z) \,  \tilde{R}^{a}(w) = \frac{1}{(z-w)^2}\, 4 \,
\tilde{R}^{a}(w) + \frac{1}{(z-w)}\, \pa \,  \tilde{R}^{a}
+ \cdots.
\nonu
\eea}
\bea
&&  \tilde{R}^{a}  = g_{1} J^{a} J^{u(1)} J^{u(1)} J^{u(1)} 
+ g_{2} \delta_{\bar{i} k} \delta_{\rho \bar{\tau}} J^{a} J^{u(1)} J^{(\rho \bar{i})}
J^{(\bar{\tau} k)} 
+ g_{3} \delta_{\bar{i} k} \delta_{\rho \bar{\tau}} J^{a} J^{(\rho \bar{i})} \partial J^{(\bar{\tau} k)}  \nonu\\
&&+ g_{4} J^{a} \partial J^{u(1)} J^{u(1)} 
+ g_{5} \delta_{\bar{i} k} \delta_{\rho \bar{\tau}} J^{a} \partial J^{(\rho \bar{i})}J^{(\bar{\tau} k)} 
+ g_{6} J^{a} \partial^{2} J^{u(1)}  \nonu\\
&&+ g_{7} d^{a v o} d^{b c o} d^{f v h} J^{b} J^{c} J^{f} J^{h} 
+ g_{8} d^{a c o} d^{b v o} d^{f v h} J^{b} J^{c} J^{f} J^{h} 
+ g_{9} d^{a b o} d^{c v o} d^{f v h} J^{b} J^{c} J^{f} J^{h}  \nonu\\
&&+ g_{10} d^{f v h} f^{a v o} f^{b c o} J^{b} J^{c} J^{f} J^{h} 
+ g_{11} d^{f v h} f^{a v o} f^{b c o} J^{b} J^{c} J^{f} J^{h} 
+ g_{12} d^{f v h} f^{a c o} f^{b v o} J^{b} J^{c} J^{f} J^{h}  \nonu\\
&&+ g_{13} d^{f v h} f^{a c o} f^{b v o} J^{b} J^{c} J^{f} J^{h} 
+ g_{14} d^{f v h} f^{a b o} f^{c v o} J^{b} J^{c} J^{f} J^{h} 
+ g_{15} d^{f v h} f^{a b o} f^{c v o} J^{b} J^{c} J^{f} J^{h}  \nonu\\
&&+ g_{16} d^{c f h} \delta^{a b} J^{b} J^{c} J^{f} J^{h} 
+ g_{17} d^{b f h} \delta^{a c} J^{b} J^{c} J^{f} J^{h} 
+ g_{18} d^{a f h} \delta^{b c} J^{b} J^{c} J^{f} J^{h} \nonu\\
&&+ g_{19} d^{a c f} J^{c} J^{f} J^{u(1)} J^{u(1)} 
+ g_{20} d^{a h e} d^{c f e} J^{c} J^{f} J^{h} J^{u(1)}  
+ g_{21} d^{a f e} d^{c h e} J^{c} J^{f} J^{h} J^{u(1)} \nonu\\
&&+ g_{22} d^{a c e} d^{f h e} J^{c} J^{f} J^{h} J^{u(1)} 
+ g_{23} f^{a h e} f^{c f e} J^{c} J^{f} J^{h} J^{u(1)}  
+ g_{24} f^{a f e} f^{c h e} J^{c} J^{f} J^{h} J^{u(1)} \nonu\\
&&+ g_{25} f^{a c e} f^{f h e} J^{c} J^{f} J^{h} J^{u(1)}  
+ g_{26} \delta^{a h} \delta^{c f} J^{c} J^{f} J^{h} J^{u(1)} 
+ g_{27} \delta^{a f} \delta^{c h} J^{c} J^{f} J^{h} J^{u(1)} \nonu\\
&&+ g_{28} \delta^{a c} \delta^{f h} J^{c} J^{f} J^{h} J^{u(1)}  
+ g_{29} d^{a c f} \delta_{\bar{i} k} \delta_{\rho \bar{\tau}} J^{c} J^{f} J^{(\rho \bar{i})} J^{(\bar{\tau} k)} 
+ g_{30} d^{a c f} J^{c} J^{f} \partial J^{u(1)} \nonu\\
&&+ g_{31} f^{a c f} J^{c} J^{f} \partial J^{u(1)} 
+ g_{32} J^{\beta} J^{\beta} J^{a} J^{u(1)} 
+ g_{33} d^{a c f} J^{\beta} J^{\beta} J^{c} J^{f} 
+ g_{34} J^{\beta} J^{\beta} \partial J^{a}  \nonu\\
&&+ g_{35} \partial J^{a} J^{u(1)} J^{u(1)} 
+ g_{36} \delta_{\bar{i} k} \delta_{\rho \bar{\tau}} \partial J^{a} J^{(\rho \bar{i})} J^{(\bar{\tau} k)} 
+ g_{37} \partial J^{a} \partial J^{u(1)}  \nonu\\
&&+ g_{38} d^{a c f} \partial J^{c} J^{f} J^{u(1)} 
+ g_{39} f^{a c f} \partial J^{c} J^{f} J^{u(1)} 
+ g_{40} d^{a h e} d^{c f e} \partial J^{c} J^{f} J^{h}  \nonu\\
&&+ g_{41} d^{a f e} d^{c h e} \partial J^{c} J^{f} J^{h} 
+ g_{42} d^{a c e} d^{f h e} \partial J^{c} J^{f} J^{h} 
+ g_{43} d^{f h e} f^{a c e} \partial J^{c} J^{f} J^{h}  \nonu\\
&&+ g_{44} d^{c h e} f^{a f e} \partial J^{c} J^{f} J^{h} 
+ g_{45} d^{c f e} f^{a h e} \partial J^{c} J^{f} J^{h} 
+ g_{46} f^{a h e} f^{c f e} \partial J^{c} J^{f} J^{h}  \nonu\\
&&+ g_{47} d^{a f e} f^{c h e} \partial J^{c} J^{f} J^{h} 
+ g_{48} f^{a f e} f^{c h e} \partial J^{c} J^{f} J^{h} 
+ g_{49} d^{a c e} f^{f h e} \partial J^{c} J^{f} J^{h}  \nonu\\
&&+ g_{50} f^{a c e} f^{f h e} \partial J^{c} J^{f} J^{h} 
+ g_{51} \delta^{a h} \delta^{c f} \partial J^{c} J^{f} J^{h} 
+ g_{52} \delta^{a f} \delta^{c h} \partial J^{c} J^{f} J^{h}  \nonu\\
&&+ g_{53} \delta^{a c} \delta^{f h} \partial J^{c} J^{f} J^{h} 
+ g_{54} d^{a c f} \partial J^{c} \partial J^{f} 
+ g_{55} f^{a c f} \partial J^{c} \partial J^{f}  \nonu\\
&&+ g_{56} \delta^{\beta \gamma} \partial J^{\beta} J^{\gamma} J^{a} 
+ g_{57} \partial^{2} J^{a} J^{u(1)} 
+ g_{58} d^{a c f} \partial^{2} J^{c} J^{f}  \nonu\\
&&+ g_{59} f^{a c f} \partial^{2} J^{c} J^{f} 
+ g_{60} \delta_{\rho \bar{\tau}} t^{a}_{k \bar{i}} J^{u(1)} J^{u(1)} J^{(\rho \bar{i})} J^{(\bar{\tau} k)}  
+ g_{61} \delta_{\rho \bar{\tau}} t^{a}_{k \bar{i}} J^{u(1)}
J^{(\rho \bar{i})} \pa J^{(\bar{\tau} k)} \nonu\\
&&+ g_{62} \delta_{\rho \bar{\tau}} t^{a}_{k \bar{i}} 
J^{u(1)} \pa J^{(\rho \bar{i})} J^{(\bar{\tau} k)} 
%+ g_{63} \delta^{c f} \delta_{\rho \bar{\tau}} t^{a}_{k \bar{i}} J^{c} J&%^{f} J^{(\rho \bar{i})} J^{(\bar{\tau} k)} 
+ g_{63} \delta^{\beta \gamma} \delta_{\rho \bar{\tau}} t^{a}_{k \bar{i}} J^{\beta} J^{\gamma} J^{(\rho \bar{i})} J^{(\bar{\tau} k)}  
+ g_{64} \delta_{\rho \bar{\tau}} t^{a}_{k \bar{i}} 
J^{(\rho \bar{i})} \pa^2 J^{(\bar{\tau} k)}
\nonu \\
&& + g_{65} \delta_{\rho \bar{\tau}} t^{a}_{k \bar{i}} 
\pa J^{u(1)}  J^{(\rho \bar{i})} J^{(\bar{\tau} k)} 
+ g_{66} \delta_{\rho \bar{\tau}} t^{a}_{k \bar{i}} \partial J^{(\rho \bar{i})} \partial J^{(\bar{\tau} k)} 
+ g_{67} \delta_{\rho \bar{\tau}} t^{a}_{k \bar{i}}
\pa^2 J^{(\rho \bar{i})} J^{(\bar{\tau} k)}  \nonu\\
&&+ g_{68} \delta_{\bar{g} y} \delta_{\bar{\nu} \sigma} \delta_{\rho \bar{\tau}} t^{a}_{k \bar{i}} J^{(\rho \bar{i})} J^{(\sigma \bar{g})} J^{(\bar{\nu} y)} J^{(\bar{\tau} k)}  
+ g_{69} d^{a c f} \delta_{\rho \bar{\tau}} t^{f}_{k \bar{i}} J^{c} J^{u(1)} J^{(\rho \bar{i})} J^{(\bar{\tau} k)} \nonu\\
&&+ g_{70} d^{a c f} \delta_{\rho \bar{\tau}} t^{f}_{k \bar{i}} J^{c} J^{(\rho \bar{i})} \partial J^{(\bar{\tau} k)} 
+ g_{71} f^{a c f} \delta_{\rho \bar{\tau}} t^{f}_{k \bar{i}} J^{c} J^{(\rho \bar{i})} \partial J^{(\bar{\tau} k)} 
+ g_{72} d^{a c f} \delta_{\rho \bar{\tau}} t^{f}_{k \bar{i}} J^{c} \partial J^{(\rho \bar{i})} J^{(\bar{\tau} k)}  \nonu\\
&&+ g_{73} f^{a c f} \delta_{\rho \bar{\tau}} t^{f}_{k \bar{i}} J^{c} \partial J^{(\rho \bar{i})} J^{(\bar{\tau} k)} 
+ g_{74} d^{a c f} \delta_{\rho \bar{\tau}} t^{f}_{k \bar{i}} \partial J^{c} J^{(\rho \bar{i})} J^{(\bar{\tau} k)} 
+ g_{75} f^{a c f} \delta_{\rho \bar{\tau}} t^{f}_{k \bar{i}} \partial J^{c} J^{(\rho \bar{i})} J^{(\bar{\tau} k)}  \nonu\\
&&+ g_{76} d^{a c f} \delta_{\bar{\nu} \sigma} \delta_{\rho \bar{\tau}} t^{c}_{y \bar{g}} t^{f}_{k \bar{i}} J^{(\sigma \bar{g})} J^{(\rho \bar{i})} J^{(\bar{\nu} y)} J^{(\bar{\tau} k)}  
+ g_{77} d^{a h e} d^{c f e} \delta_{\rho \bar{\tau}} t^{h}_{k \bar{i}} J^{c} J^{f} J^{(\rho \bar{i})} J^{(\bar{\tau} k)}  \nonu\\
&&+ g_{78} d^{a f e} d^{c h e} \delta_{\rho \bar{\tau}} t^{h}_{k \bar{i}} J^{c} J^{f} J^{(\rho \bar{i})} J^{(\bar{\tau} k)}  
+ g_{79} d^{a c e} d^{f h e} \delta_{\rho \bar{\tau}} t^{h}_{k \bar{i}} J^{c} J^{f} J^{(\rho \bar{i})} J^{(\bar{\tau} k)}  \nonu\\
&&+ g_{80} f^{a h e} f^{c f e} \delta_{\rho \bar{\tau}} t^{h}_{k \bar{i}} J^{c} J^{f}
J^{(\rho \bar{i})} J^{(\bar{\tau} k)}  
+ g_{81} f^{a f e} f^{c h e} \delta_{\rho \bar{\tau}} t^{h}_{k \bar{i}} J^{c} J^{f} J^{(\rho \bar{i})} J^{(\bar{\tau} k)}  \nonu\\
&&+ g_{82} f^{a c e} f^{f h e} \delta_{\rho \bar{\tau}} t^{h}_{k \bar{i}} J^{c} J^{f} J^{(\rho \bar{i})} J^{(\bar{\tau} k)}  
+ g_{83} \delta^{a h} \delta^{c f} \delta_{\rho \bar{\tau}} t^{h}_{k \bar{i}} J^{c} J^{f} J^{(\rho \bar{i})} J^{(\bar{\tau} k)}  \nonu\\
&&+ g_{84} \delta^{a f} \delta^{c h} \delta_{\rho \bar{\tau}} t^{h}_{k \bar{i}} J^{c} J^{f} J^{(\rho \bar{i})} J^{(\bar{\tau} k)}  
+ g_{85} \delta^{a c} \delta^{f h} \delta_{\rho \bar{\tau}} t^{h}_{k \bar{i}} J^{c} J^{f} J^{(\rho \bar{i})} J^{(\bar{\tau} k)}  \nonu\\
&&+ g_{86} \delta_{\bar{i} k} t^{\beta}_{\rho \bar{\tau}} J^{\beta} J^{a} J^{(\rho \bar{i})} J^{(\bar{\tau} k)}  
+ g_{87} t^{a}_{k \bar{i}} t^{\beta}_{\rho \bar{\tau}} J^{\beta} J^{u(1)} J^{(\rho \bar{i})} J^{(\bar{\tau} k)}  
+ g_{88} t^{a}_{k \bar{i}} t^{\beta}_{\rho \bar{\tau}} J^{\beta} J^{(\rho \bar{i})} \partial J^{(\bar{\tau} k)}  \nonu\\
&&+ g_{89} t^{a}_{k \bar{i}} t^{\beta}_{\rho \bar{\tau}} J^{\beta} \partial J^{(\rho \bar{i})} J^{(\bar{\tau} k)}  
+ g_{90} t^{a}_{k \bar{i}} t^{\beta}_{\rho \bar{\tau}} \partial J^{\beta} J^{(\rho \bar{i})} J^{(\bar{\tau} k)}  
+ g_{91} d^{a c f} t^{f}_{k \bar{i}} t^{\beta}_{\rho \bar{\tau}} J^{\beta} J^{c} J^{(\rho \bar{i})} J^{(\bar{\tau} k)}  \nonu\\
&&+ g_{92} d^{\beta \gamma \delta} t^{a}_{k \bar{i}} t^{\delta}_{\rho \bar{\tau}} J^{\beta} J^{\gamma} J^{(\rho \bar{i})} J^{(\bar{\tau} k)}  
+ g_{93} \partial^{3} J^{a},
\label{rtilde}
\eea
where the coefficients are given by
(\ref{gcoeff}).
The number of independent terms in (\ref{rtilde})
is different from the ones in (\ref{PWPOLE2}).
We can also check, for $(N,M)=(5,4)$,  that
\bea
J^a(z) \, \tilde{R}^b(w)\Bigg|_{\frac{1}{(z-w)}} =
 i \, f^{a b c}\,
\tilde{R}^c,
\label{JR}
\eea
which behaves as the one in the footnote
%\ref{JK} and
\ref{JP}.
There are also the nontrivial second and third poles
in this OPE. 
We expect that
for generic $(N,M)$,
the first order pole of this OPE (\ref{JR}) satisfies.
The nonderivative terms in (\ref{rtilde})
contain the triple product of $d$ symbols.
For example, the $g_7$ term in (\ref{rtilde})
or the $p_4$ term in (\ref{PWPOLE2})
can be described as the contribution from the OPE
of the $a_{17}$ term in (\ref{spin3exp})
with the $b_2$ term in (\ref{W}). In (\ref{PWP2}),
we do not substitute the various coefficients
appearing in (\ref{avalues}) and (\ref{interbvalue})
and this shows that we can observe how those contributions
originate from each term of (\ref{spin3exp}) and (\ref{W})
by recognizing the coefficients
in front of any operators.
These coefficients become the ones in (\ref{pvalue})
after substituting the coefficients (\ref{avalues}) and
(\ref{interbvalue}) explicitly.
Finally, by using (\ref{PWPOLE2}),
(\ref{r2anew}), (\ref{r2a}), (\ref{Rhata}),
(\ref{SPIN4}), (\ref{rtilde}) and (\ref{pvalue}),
we determine all the coefficients in (\ref{gcoeff}).
We also present the expression for the charged spin-$4$
current at $k=-2 N$ in (\ref{simple1}).

%%%%%%%%%%%%%%%%%%%
\subsection{The neutral primary spin-$4$ current}
%%%%%%%%%%%%%%%%%%%

%%%%%%%%%%%%%%%%%%%%%%%%%%%%%%%%%%%%
\subsubsection{The OPE of the neutral spin-$3$ current
with itself}
%%%%%%%%%%%%%%%%%%%%%%%%%%%%%%%%%%%%

It is well known that
the OPE of the neutral spin-$3$ current
with itself satisfies \cite{BBSS,BBSS1,Ahn2011}
\bea
W^{(3)}(z) \, W^{(3)}(w) & = &
\frac{1}{(z-w)^6} \frac{\hat{c}}{3}
+ \frac{1}{(z-w)^4} 2 \, \hat{T} (w)
+\frac{1}{(z-w)^3} \pa \, \hat{T} (w)
\nonu \\
&+& \frac{1}{(z-w)^2} \Bigg[ \frac{3}{10}\, \pa^2 \, \hat{T} +
\frac{32}{(5\hat{c} +22)} \Big(\hat{T} \, \hat{T} -
\frac{3}{10} \, \pa^2 \, \hat{T} \Big) +  W^{(4)} \Bigg](w)
\nonu \\
&+&
\frac{1}{(z-w)} \Bigg[
\frac{1}{15}\, \pa^3 \, \hat{T} +
\frac{1}{2} \,
\frac{32}{(5\hat{c} +22)} \pa \, \Big(\hat{T} \, \hat{T} -
\frac{3}{10} \, \pa^2 \, \hat{T} \Big) +
\frac{1}{2}\, \pa \,  W^{(4)} \Bigg](w)
\nonu \\
& + & \cdots.
\label{Spin3Spin3}
\eea
The overall factor of the neutral spin-$3$ current
\footnote{The contributions on this come from
$b_{1, \cdots, 8} \, b_{12}$, $b_{12}\, b_{12}$, $b_{1, \cdots, 8}\, b_{13}$,
$b_{12,13}\, b_{13}$, and $b_{12,13}\, b_{14}$
with $(k,N,M)$ dependent coefficients.}
is fixed by
\bea
b_{1}^2 &=& \frac{k M^{2} (k + 2 M)}{18(-4 + k^{2}) (k + N)^{2} (k + M + N)^{2} (k + 2 N) (3k + 2M + 2N)}.
\label{b1value}
\eea
Here the central term and the stress energy tensor
appearing on the fourth order pole are given by \cite{Ahn2011}
\bea
\hat{c} & = & c-\frac{k (M^2-1)}{(k+M)} =
\frac{(k^2-1)M N (2k +M+N)}{(k+M)(k+N)(k+M+N)},
\nonu \\
\hat{T} & = & 
T- \frac{1}{2(k+M)} \, J^a \, J^a. 
\label{modified}
\eea
The previous central charge and the
stress energy tensor $T$ are given by (\ref{charge}) and 
(\ref{T}) respectively
\footnote{As in \cite{Ahn2011},
the OPE of
the spin-$1$ current with
the stress energy tensor $\hat{T}$ (\ref{modified})
does not contain any singular terms and also there is a relation
given by the footnote \ref{JW3}. This implies that the spin-$1$
current is decoupled from the algebra constructed by
the neutral currents, $\hat{T}, W^{(3)}, W^{(4)}, \cdots$
and see also the footnote \ref{JW4}.}.

%%%%%%%%%%%%%%%%%%%
\subsubsection{The neutral primary spin-$4$ current}
%%%%%%%%%%%%%%%%%%%

%It turns out that
The second order pole in the OPE of the neutral spin-$3$ current
with itself can be described by
\bea
&& W^{(3)}(z) \, W^{(3)}(w)\Bigg|_{\frac{1}{(z-w)^2}} =  w_{1} J^{u(1)} J^{u(1)} J^{u(1)} J^{u(1)}  
+ w_{2} \delta_{\bar{i} k} \delta_{\rho \bar{\tau}} J^{u(1)} J^{u(1)} J^{(\rho \bar{i})} J^{(\bar{\tau} k)}  \nonu\\
&&+ w_{3} \delta_{\bar{i} k} \delta_{\rho \bar{\tau}} J^{u(1)} J^{(\rho \bar{i})} \partial J^{(\bar{\tau} k)}  
+ w_{4} \delta_{\bar{i} k} \delta_{\rho \bar{\tau}} J^{u(1)} \partial J^{(\rho \bar{i})} J^{(\bar{\tau} k)}  
+ w_{5} \delta_{b c} J^{b} J^{c} J^{u(1)} J^{u(1)}  \nonu\\
&&+ w_{6} d^{b c f} J^{b} J^{c} J^{f} J^{u(1)}  
+ w_{7} f^{b c f} J^{b} J^{c} J^{f} J^{u(1)}  
+ w_{8} d^{b h e} d^{c f e} J^{b} J^{c} J^{f} J^{h}  \nonu\\
&&+ w_{9} d^{c f e} f^{b h e} J^{b} J^{c} J^{f} J^{h}  
+ w_{10} d^{b h e} f^{c f e} J^{b} J^{c} J^{f} J^{h}  
+ w_{11} f^{b h e} f^{c f e} J^{b} J^{c} J^{f} J^{h}  \nonu\\
&&+ w_{12} \delta_{b h} \delta_{c f} J^{b} J^{c} J^{f} J^{h}  
+ w_{13} \delta_{b c} \delta_{\bar{i} k} \delta_{\rho \bar{\tau}} J^{b} J^{c} J^{(\rho \bar{i})} J^{(\bar{\tau} k)}  
+ w_{14} \delta_{b c} J^{b} J^{c} \partial J^{u(1)}  \nonu\\
&&+ w_{15} \delta_{\beta \gamma} J^{\beta} J^{\gamma} J^{u(1)} J^{u(1)}  
+ w_{16} \delta_{b c} \delta_{\beta \gamma} J^{\beta} J^{\gamma} J^{b} J^{c}  
+ w_{17} d^{\beta \gamma \delta} J^{\beta} J^{\gamma} J^{\delta} J^{u(1)}  \nonu\\
&&+ w_{18} f^{\beta \gamma \delta} J^{\beta} J^{\gamma} J^{\delta} J^{u(1)}  
+ w_{19} d^{\beta \omega \eta} d^{\gamma \delta \eta} J^{\beta} J^{\gamma} J^{\delta} J^{\omega}  
+ w_{20} d^{\gamma \delta \eta} f^{\beta \omega \eta} J^{\beta} J^{\gamma} J^{\delta} J^{\omega}  \nonu\\
&&+ w_{21} d^{\beta \omega \eta} f^{\gamma \delta \eta} J^{\beta} J^{\gamma} J^{\delta}J^{\omega}  
+ w_{22} f^{\beta \omega \eta} f^{\gamma \delta \eta} J^{\beta} J^{\gamma} J^{\delta} J^{\omega}  
+ w_{23} \delta_{\beta \omega} \delta_{\delta \gamma} J^{\beta} J^{\gamma} J^{\delta} J^{\omega}  \nonu\\
&&+ w_{24} \delta_{\beta \gamma} \delta_{\bar{i} k} \delta_{\rho \bar{\tau}} J^{\beta} J^{\gamma} J^{(\rho \bar{i})} J^{(\bar{\tau} k)}  
+ w_{25} \delta_{\beta \gamma} J^{\beta} J^{\gamma} \partial J^{u(1)}  \nonu\\
&&+ w_{26} \delta_{\bar{g} y} \delta_{\bar{i} k} \delta_{\bar{\nu} \sigma} \delta_{\rho \bar{\tau}} J^{(\rho \bar{g})} J^{(\sigma \bar{i})} J^{(\bar{\nu} y)} J^{(\bar{\tau} k)}  
+ w_{27} \delta_{\bar{g} y} \delta_{\bar{i} k} \delta_{\bar{\nu} \sigma} \delta_{\rho \bar{\tau}} J^{(\rho \bar{i})} J^{(\sigma \bar{g})} J^{(\bar{\nu} k)} J^{(\bar{\tau} y)}  \nonu\\
&&+ w_{28} \delta_{\bar{g} y} \delta_{\bar{i} k} \delta_{\bar{\nu} \sigma} \delta_{\rho \bar{\tau}} J^{(\rho \bar{i})} J^{(\sigma \bar{g})} J^{(\bar{\nu} y)} J^{(\bar{\tau} k)}  
+ w_{29} \delta_{\bar{i} k} \delta_{\rho \bar{\tau}}
J^{(\rho \bar{i})} \partial^2 J^{(\bar{\tau} k)}  \nonu\\
&&+ w_{30} \partial J^{u(1)} J^{u(1)} J^{u(1)}  
+ w_{31} \delta_{\bar{i} k} \delta_{\rho \bar{\tau}} \partial J^{u(1)} J^{(\rho \bar{i})} J^{(\bar{\tau} k)} 
+ w_{32} \partial J^{u(1)} \partial J^{u(1)} \nonu\\
&&+ w_{33} \delta_{b c} \partial J^{b} J^{c} J^{u(1)}  
+ w_{34} d^{b c f} \partial J^{b} J^{c} J^{f}  
+ w_{35} f^{b c f} \partial J^{b} J^{c} J^{f}  \nonu\\
&&+ w_{36} \delta_{b c} \partial J^{b} \partial J^{c}  
+ w_{37} \delta_{\beta \gamma} \partial J^{\beta} J^{\gamma} J^{u(1)}  
+ w_{38} d^{\beta \gamma \delta} \partial J^{\beta} J^{\gamma} J^{\delta} 
+ w_{39} f^{\beta \gamma \delta} \partial J^{\beta} J^{\gamma} J^{\delta}  \nonu\\
&&+ w_{40} \delta_{\beta \gamma} \partial J^{\beta} \partial J^{\gamma} 
+ w_{41} \delta_{\bar{i} k} \delta_{\rho \bar{\tau}} \partial J^{(\rho \bar{i})} \partial J^{(\bar{\tau} k)}  
+ w_{42} \partial^2 J^{u(1)} J^{u(1)}  
+ w_{43} \delta_{b c} \partial^2 J^{b} J^{c}  \nonu\\
&&+ w_{44} \delta_{\beta \gamma} \partial^2 J^{\beta} J^{\gamma}  
+ w_{45} \delta_{\bar{i} k} \delta_{\rho \bar{\tau}} \partial^2 J^{(\rho \bar{i})} J^{(\bar{\tau} k)}  
+ w_{46} \delta_{\rho \bar{\tau}} t^{b}_{k \bar{i}} J^{b} J^{u(1)} J^{(\rho \bar{i})} J^{(\bar{\tau} k)}  \nonu\\
&&+ w_{47} \delta_{\rho \bar{\tau}} t^{b}_{k \bar{i}} J^{b} 
 J^{(\rho \bar{i})} \pa J^{(\bar{\tau} k)} 
+ w_{48} \delta_{\rho \bar{\tau}} t^{b}_{k \bar{i}} J^{b} \partial J^{(\rho \bar{i})}J^{(\bar{\tau} k)}  
+ w_{49} \delta_{\rho \bar{\tau}} t^{b}_{k \bar{i}} \partial J^{b} J^{(\rho \bar{i})}J^{(\bar{\tau} k)}  \nonu\\
&&+ w_{50} \delta_{\rho \bar{\tau}} d^{b c f} t^{f}_{k \bar{i}} J^{b} J^{c} J^{(\rho \bar{i})} J^{(\bar{\tau} k)}  
+ w_{51} \delta_{\rho \bar{\tau}} f^{b c f} t^{f}_{k \bar{i}} J^{b} J^{c} J^{(\rho \bar{i})} J^{(\bar{\tau} k)}  \nonu\\
&&+ w_{52} \delta_{\bar{i} k} t^{\beta}_{\rho \bar{\tau}} J^{\beta} J^{u(1)} J^{(\rho \bar{i})} J^{(\bar{\tau} k)}  
+ w_{53} t^{\beta}_{\rho \bar{\tau}} \delta_{\bar{i} k} J^{\beta} J^{(\rho \bar{i})} \partial J^{(\bar{\tau} k)}  
+ w_{54} t^{\beta}_{\rho \bar{\tau}} \delta_{\bar{i} k} J^{\beta} \partial J^{(\rho \bar{i})} J^{(\bar{\tau} k)}  \nonu\\
&&+ w_{55} t^{\beta}_{\rho \bar{\tau}} \delta_{\bar{i} k} \partial J^{\beta} J^{(\rho \bar{i})} J^{(\bar{\tau} k)}  
+ w_{56} t^{b}_{k \bar{i}} t^{\beta}_{\rho \bar{\tau}}J^{\beta} J^{b}  J^{(\rho \bar{i})} J^{(\bar{\tau} k)}  
+ w_{57} t^{c}_{k \bar{i}} t^{\beta}_{\rho \bar{\tau}} \delta_{b c}J^{\beta} J^{b} J^{(\rho \bar{i})} J^{(\bar{\tau} k)}  \nonu\\
&&+ w_{58} t^{\delta}_{\rho \bar{\tau}} d^{\beta \gamma \delta} \delta_{\bar{i} k} J^{\beta} J^{\gamma} J^{(\rho \bar{i})} J^{(\bar{\tau} k)}  
+ w_{59} t^{\delta}_{\rho \bar{\tau}} f^{\beta \gamma \delta} \delta_{\bar{i} k} J^{\beta} J^{\gamma} J^{(\rho \bar{i})} J^{(\bar{\tau} k)}  \nonu\\
&&+ w_{60} t^{\gamma}_{\rho \bar{\tau}} \delta_{\beta \gamma} \delta_{\bar{i} k} J^{\beta} J^{(\rho \bar{i})} \partial J^{(\bar{\tau} k)}  
+ w_{61} t^{\gamma}_{\rho \bar{\tau}} \delta_{\beta \gamma} \delta_{\bar{i} k} J^{\beta} \partial J^{(\rho \bar{i})} J^{(\bar{\tau} k)}  \nonu\\
&&+ w_{62} t^{\gamma}_{\rho \bar{\tau}} \delta_{\beta \gamma} \delta_{\bar{i} k} \partial J^{\beta} J^{(\rho \bar{i})} J^{(\bar{\tau} k)}  
+ w_{63} \partial^{3} J^{u(1)},
\label{WWPOLE2}
\eea
where the coefficients are given by
(\ref{wvalues}).
%In (\ref{WWP2}), we do not substitute the coefficients
%given in (\ref{interbvalue}) because
%we keep all the contributions in order to see
%the origin of each composite operator.
%Moreover, after substituting (\ref{interbvalue}) into
%(\ref{WWP2}), we obtain the simplified coeffcients
%in (\ref{wvalues}).

Then the primary neutral spin-$4$ current,
by subtracting the descendant terms
(the first term) and the quasi primary
neutral spin-$4$ current (the second and the third terms)
from the second order pole
in the OPE of (\ref{Spin3Spin3}) (or (\ref{WWPOLE2})),
can be summarized by
\bea
&& W^{(4)} = d_{1} J^{u(1)} J^{u(1)} J^{u(1)} J^{u(1)} 
+ d_{2} \delta_{\bar{i} k} \delta_{\rho \bar{\tau}} J^{u(1)} J^{u(1)} J^{(\rho \bar{i})} J^{(\bar{\tau} k)} \nonu\\
&&+ d_{3} \delta_{\bar{i} k} \delta_{\rho \bar{\tau}} J^{u(1)} J^{(\rho \bar{i})} \partial J^{(\bar{\tau} k)} 
+ d_{4} \delta_{\bar{i} k} \delta_{\rho \bar{\tau}} J^{u(1)} \partial J^{(\rho \bar{i})} J^{(\bar{\tau} k)} 
+ d_{5} \delta_{b c} J^{b} J^{c} J^{u(1)} J^{u(1)} \nonu\\
&&+ d_{6} d^{b c f} J^{b} J^{c} J^{f} J^{u(1)} 
+ d_{7} f^{b c f} J^{b} J^{c} J^{f} J^{u(1)} 
+ d_{8} d^{b h e} d^{c f e} J^{b} J^{c} J^{f} J^{h} \nonu\\
&&+ d_{9} d^{c f e} f^{b h e} J^{b} J^{c} J^{f} J^{h} 
+ d_{10} d^{b h e} f^{c f e} J^{b} J^{c} J^{f} J^{h} 
+ d_{11} f^{b h e} f^{c f e} J^{b} J^{c} J^{f} J^{h} \nonu\\
&&+ d_{12} \delta_{b h} \delta_{c f} J^{b} J^{c} J^{f} J^{h} 
+ d_{13} \delta_{b c} \delta_{f h} J^{b} J^{c} J^{f} J^{h} 
+ d_{14} \delta_{b c} \delta_{\bar{i} k} \delta_{\rho \bar{\tau}} J^{b} J^{c} J^{(\rho \bar{i})} J^{(\bar{\tau} k)} \nonu\\
&&+ d_{15} \delta_{b c} J^{b} J^{c} \partial J^{u(1)} 
+ d_{16} \delta_{\beta \gamma} J^{\beta} J^{\gamma} J^{u(1)} J^{u(1)} 
+ d_{17} \delta_{b c} \delta_{\beta \gamma} J^{\beta} J^{\gamma} J^{b} J^{c} \nonu\\
&&+ d_{18} d^{\beta \gamma \delta} J^{\beta} J^{\gamma} J^{\delta} J^{u(1)} 
+ d_{19} f^{\beta \gamma \delta} J^{\beta} J^{\gamma} J^{\delta} J^{u(1)} 
+ d_{20} d^{\beta \omega \eta} d^{\gamma \delta \eta} J^{\beta} J^{\gamma} J^{\delta} J^{\omega} \nonu\\
&&+ d_{21} d^{\gamma \delta \eta} f^{\beta \omega \eta} J^{\beta} J^{\gamma} J^{\delta}J^{\omega} 
+ d_{22} d^{\beta \omega \eta} f^{\gamma \delta \eta} J^{\beta} J^{\gamma} J^{\delta} J^{\omega} 
+ d_{23} f^{\beta \omega \eta} f^{\gamma \delta \eta} J^{\beta} J^{\gamma} J^{\delta} J^{\omega} \nonu\\
&&+ d_{24} \delta_{\beta \omega} \delta_{\gamma \delta} J^{\beta} J^{\gamma} J^{\delta} J^{\omega} 
+ d_{25} \delta_{\beta \gamma} \delta_{\delta \omega} J^{\beta} J^{\gamma} J^{\delta}J^{\omega} 
+ d_{26} \delta_{\beta \gamma} \delta_{\bar{i} k} \delta_{\rho \bar{\tau}} J^{\beta}J^{\gamma} J^{(\rho \bar{i})} J^{(\bar{\tau} k)} \nonu\\
&&+ d_{27} \delta_{\beta \gamma} J^{\beta} J^{\gamma} \partial J^{u(1)} 
+ d_{28} \delta_{\bar{g} y} \delta_{\bar{i} k} \delta_{\bar{\nu} \sigma} \delta_{\rho \bar{\tau}} J^{(\rho \bar{g})} J^{(\sigma \bar{i})} J^{(\bar{\nu} y)} J^{(\bar{\tau} k)} \nonu\\
&&+ d_{29} \delta_{\bar{g} y} \delta_{\bar{i} k} \delta_{\bar{\nu} \sigma} \delta_{\rho \bar{\tau}} J^{(\rho \bar{i})} J^{(\sigma \bar{g})} J^{(\bar{\nu} k)} J^{(\bar{\tau} y)} 
+ d_{30} \delta_{\bar{g} y} \delta_{\bar{i} k} \delta_{\bar{\nu} \sigma} \delta_{\rho \bar{\tau}} J^{(\rho \bar{i})} J^{(\sigma \bar{g})} J^{(\bar{\nu} y)} J^{(\bar{\tau} k)} \nonu\\
&&+d_{31} \delta_{\bar{g} y} \delta_{\bar{i} k} \delta_{\bar{\nu} \sigma} \delta_{\rho \bar{\tau}} J^{(\rho \bar{i})} J^{(\sigma \bar{g})} J^{(\bar{\tau} k)} J^{(\bar{\nu} y)} 
+ d_{32} \delta_{\bar{i} k} \delta_{\rho \bar{\tau}}
J^{(\rho \bar{i})} \pa^2 J^{(\bar{\tau} k)} 
+ d_{33} \partial J^{u(1)} J^{u(1)} J^{u(1)} \nonu\\
&&+ d_{34} \delta_{\bar{i} k} \delta_{\rho \bar{\tau}} \partial J^{u(1)} J^{(\rho \bar{i})} J^{(\bar{\tau} k)} 
+ d_{35} \partial J^{u(1)} \partial J^{u(1)} 
+ d_{36} \delta_{b c} \partial J^{b} J^{c} J^{u(1)} \nonu\\
&&+ d_{37} d^{b c f} \partial J^{b} J^{c} J^{f} 
+ d_{38} f^{b c f} \partial J^{b} J^{c} J^{f} 
+ d_{39} \delta_{b c} \partial J^{b} \partial J^{c} 
+ d_{40} \delta_{\beta \gamma} \partial J^{\beta} J^{\gamma} J^{u(1)} \nonu\\
&&+ d_{41} d^{\beta \gamma \delta} \partial J^{\beta} J^{\gamma} J^{\delta} 
+ d_{42} f^{\beta \gamma \delta} \partial J^{\beta} J^{\gamma} J^{\delta} 
+ d_{43} \delta_{\beta \gamma} \partial J^{\beta} \partial J^{\gamma} 
+ d_{44} \delta_{\bar{i} k} \delta_{\rho \bar{\tau}} \partial J^{(\rho \bar{i})} \partial J^{(\bar{\tau} k)} \nonu\\
&&+ d_{45} \partial^{2} J^{u(1)} J^{u(1)} 
+ d_{46} \delta_{b c} \partial^{2} J^{b} J^{c} 
+ d_{47} \delta_{\beta \gamma} \partial^{2} J^{\beta} J^{\gamma} 
+ d_{48} \delta_{\bar{i} k} \delta_{\rho \bar{\tau}} \partial^{2} J^{(\rho \bar{i})}J^{(\bar{\tau} k)} \nonu\\
&&+ d_{49} \delta_{\rho \bar{\tau}}  t^{b}_{k \bar{i}}
J^{b} J^{u(1)} J^{(\rho \bar{i})} J^{(\bar{\tau} k)}
+ d_{50} \delta_{\rho \bar{\tau}}  t^{b}_{k \bar{i}}
J^{b} J^{(\rho \bar{i})} \partial J^{(\bar{\tau} k)}
+ d_{51} \delta_{\rho \bar{\tau}}
t^{b}_{k \bar{i}} J^{b} \partial J^{(\rho \bar{i})} J^{(\bar{\tau} k)}\nonu\\
&&+ d_{52} \delta_{\rho \bar{\tau}}
t^{b}_{k \bar{i}} \partial J^{b} J^{(\rho \bar{i})} J^{(\bar{\tau}k)} 
+ d_{53} d^{b c f} \delta_{\rho \bar{\tau}}
t^{f}_{k \bar{i}} J^{b} J^{c} J^{(\rho \bar{i})} J^{(\bar{\tau} k)}
+ d_{54} f^{b c f} \delta_{\rho \bar{\tau}}
t^{f}_{k \bar{i}} J^{b} J^{c} J^{(\rho \bar{i})} J^{(\bar{\tau} k)} \nonu\\
&&+ d_{55} \delta_{\bar{i} k}
t^{\beta}_{\rho \bar{\tau}} J^{\beta} J^{u(1)} J^{(\rho \bar{i})} J^{(\bar{\tau} k)} 
+ d_{56} \delta_{\bar{i} k}
t^{\beta}_{\rho \bar{\tau}} J^{\beta} J^{(\rho \bar{i})} \partial J^{(\bar{\tau} k)}
+ d_{57} \delta_{\bar{i} k}
t^{\beta}_{\rho \bar{\tau}} J^{\beta} \partial J^{(\rho \bar{i})} J^{(\bar{\tau} k)} \nonu\\
&&+ d_{58} \delta_{\bar{i} k}
t^{\beta}_{\rho \bar{\tau}}
\partial J^{\beta} J^{(\rho \bar{i})} J^{(\bar{\tau} k)} 
+ d_{59} t^{b}_{k \bar{i}} t^{\beta}_{\rho \bar{\tau}}
J^{\beta} J^{b} J^{(\rho \bar{i})} J^{(\bar{\tau} k)} 
+ d_{60} \delta_{b c}
t^{c}_{k \bar{i}} t^{\beta}_{\rho \bar{\tau}}
J^{\beta} J^{b} J^{(\rho \bar{i})} J^{(\bar{\tau} k)} \nonu\\
&&+ d_{61} \delta_{\beta \gamma} \delta_{\bar{i} k}
t^{\gamma}_{\rho \bar{\tau}}
J^{\beta} J^{(\rho \bar{i})} \partial J^{(\bar{\tau} k)}
+ d_{62} \delta_{\beta \gamma} \delta_{\bar{i} k}
t^{\gamma}_{\rho \bar{\tau}}
J^{\beta} \partial J^{(\rho \bar{i})} J^{(\bar{\tau} k)}
+ d_{63} \delta_{\beta \gamma} \delta_{\bar{i} k}
t^{\gamma}_{\rho \bar{\tau}}
\partial J^{\beta} J^{(\rho \bar{i})} J^{(\bar{\tau} k)} \nonu\\
&&+ d_{64} d^{\beta \gamma \delta} \delta_{\bar{i} k}
t^{\delta}_{\rho \bar{\tau}}
J^{\beta} J^{\gamma} J^{(\rho \bar{i})} J^{(\bar{\tau} k)}
+ d_{65} f^{\beta \gamma \delta} \delta_{\bar{i} k}
t^{\delta}_{\rho \bar{\tau}}
J^{\beta} J^{\gamma} J^{(\rho \bar{i})} J^{(\bar{\tau} k)}
+ d_{66} \partial^{3} J^{u(1)}, 
\label{W(4)}
\eea
where the coefficients are given by
(\ref{dvalue}) \footnote{
\label{JW4}
We can check
that
the regular condition is given by
\bea
J^a(z) \, W^{(4)}(w) = 0 + \cdots.
\nonu
\eea
}.
Note that the $d_8$ term in (\ref{W(4)})
without any derivatives is the
standard term in the neutral spin-$4$ current in different
coset model of \cite{Ahn1111}.
The $d_8$ term
can be written in terms of
the $d$ symbols with four adjoint indices \cite{Schoutens}
contracted with the multiple spin-$1$ currents,
the product of two Kronecker deltas with
spin-$1$ currents with any derivatives and other
derivative terms with spin-$1$ currents. This is also valid for
the $d_{20}$ term where the previous $M$ appeared in
the definition of $d$ symbols with four indices
is replaced by $N$.
The $d_{13}$ and $d_{25}$ terms come from the
$\hat{T}^2$ in the stress energy tensor (\ref{modified}).
The $d_{31}$ term can be written in tems of
$d_{30}$ term with derivatives by moving the last factor
to the left.
Therefore, there are three additional terms
compared to (\ref{WWPOLE2}).
We also present the expression for the neutral spin-$4$
current at $k=-2 N$ in (\ref{simple2})
\footnote{
We can check
again that the primary condition
for the spin-$4$ current is
\bea
T(z) \, W^{(4)}(w) =\frac{1}{(z-w)^2} \, 4 \, W^{(4)}(w)
+\frac{1}{(z-w)} \, \pa \, W^{(4)}(w) + \cdots.
\nonu
\eea}.

%%%%%%%%%%%%%%%%%%%%%%%%
\subsection{The OPE of the charged spin-$2$ current
with the charged spin-$3$ current}
%%%%%%%%%%%%%%%%%%%%%%%

%It turns out that
The first order pole in the  OPE of the charged spin-$2$ current
with the charged spin-$3$ current
can be described by 
(\ref{KPPOLE1}) together with (\ref{kvalue}).
We have found that the previous expression
(\ref{SabSab}) for generic $(N,M)$,
which is a generalization of the result in \cite{Ahn2011}
satisfying the fixed $(N,M)=(5,4)$,
can be summarized by
\bea
S^{a b} 
& = & s_{1}\,\de^{a b} \, \pa \, W^{(3)} + s_{2}\,
d^{a b c}\, \pa P^c 
+ s_{3}\, d_{4AA2}^{d c b a} \, J^c \, P^d
+ s_{4}\, d_{4SA}^{c a d b} \, J^c \, P^d\nonu \\
&+&  s_{5}\, d_{4SA}^{c b d a} \, J^c \, P^d
+ s_{6}\,d_{51}^{d b e c a}J^c \, J^d \, K^e + s_{7}\, d_{51}^{e  a d c b}J^c \, J^d \, K^e\nonu\\
&+& s_{8}\, d_{51}^{e d c b a}J^c \, J^d \, K^e
+  s_{9}\,d_{52}^{d b e c a}J^c \, J^d \, K^e + s_{10}\, d_{52}^{e  b d c a}J^c \,J^d \, K^e\nonu\\
&+& s_{11}\, d_{52}^{e d c b a}J^c \, J^d \, K^e
+ s_{12}\, f^{a b c} \, \de^{d e}J^c \, J^d \, K^e + s_{13}\,f^{a b e} \, \de^{d c} J^c \, J^d \, K^e\nonu\\
&+ &s_{14}\, f^{a e d} \, \de^{b c} J^c \, J^d \, K^e
+ s_{15}\,\pa \, J^a\, K^b +  s_{16}\,J^a \, \pa \, K^b  
+ s_{17}\,\pa \, J^b\, K^a \nonu\\
&+&  s_{18}\,J^b \, \pa \, K^a
+s_{19} f^{a c e} \, f^{b d e} \, \pa J^c \, K^d   +s_{20}\, f^{a c e} \, f^{b d e} \,  J^c \, \pa \, K^d \nonu\\
&+& s_{21}\, d^{a c e} \, d^{e b d} \,  \pa \, J^d  \, K^c  
+ s_{22}\,d^{a c e} \, d^{e b d} \,  \, J^d  \, \pa \, K^c  + s_{23}\,d^{a c e} \, d^{e b d} \, \pa \, J^c   \, K^d  \nonu\\
&+& s_{24}\,d^{a c e} \, d^{e b d} \, J^c  \, \pa  \, K^d  
+ s_{25}\, \de^{a b} \pa J^c\, K^c + s_{26}\,\de^{a b} J^c \pa \, K^c\nonu\\
&+& s_{27}\,f^{a b e}f^{c v e}f^{f h v} J^c \, J^f \, K^h,
\label{Sab}
\eea
where the various $d$ tensors
are given by Appendix $B$.
From the definition of (\ref{SabSab}),
the above expression can be determined
by the first order pole between
the charged spin-$2$ current and the charged spin-$3$ current
given in (\ref{KPPOLE1}) with (\ref{kvalue}).
Here the coefficients are given by
\bea
s_1 & = & -\frac{1}{3} \frac{a_1}{b_1}, \qquad
s_2 = \frac{k (2k+M+2N)}{2 (k+M)}, \qquad
s_3 = \frac{(3k+2M)(2k+M+2N)}{2 (k+M)(2k+M)},
\nonu \\
s_4 & = & -i \frac{k (2k+M+2N)}{2 (k+M)(2k+M)}, \qquad
s_5 = -i \frac{k (2k+M+2N)}{2 (k+M)(2k+M)},
\nonu \\
s_6 & = & \frac{(k+2N)(3k+2M+2N)(-6k-2M+k M N)}{16k (k+M)^2 (3k+2M)N} \, a_1,
\nonu \\
s_7 & = & \frac{(k+2N)(3k+2M+2N)(-6k-2M+k M N)}{16k (k+M)^2 (3k+2M)N} \, a_1,
\nonu \\
s_8 & = & -\frac{(k+2N)(3k+2M+2N)(6k+2M+2k^2 N+3k M N)}{32k (k+M)^2 (3k+2M)N} \, a_1,
\nonu \\
s_9 & = & \frac{(k+2N)(3k+2M+2N)(18k+6M+8k^2 N+5k M N)}{16k (k+M)^2 (3k+2M)N} \, a_1,
\nonu \\
s_{10} & = & \frac{(k+2N)(3k+2M+2N)(-18k-6M+4k^2 N+3k M N)}{16k (k+M)^2 (3k+2M)N} \, a_1,
\nonu \\
s_{11} & = & \frac{(k+2N)(3k+2M+2N)(-54k-18M+6k^2 N+5k M N)}{32k (k+M)^2 (3k+2M)N} \, a_1,
\nonu \\
s_{12} & = & -i \frac{(2k+M)(k+2N)(3k+2M+2N)}{2M (k+M)^2 (3k+2M)} \, a_1,
\nonu \\
s_{13} & = & -i \frac{(k+2N)(3k+2M+2N)}{2M (k+M)(3k+2M)} \, a_1, \qquad
s_{14}  =  -i \frac{(k+2N)(3k+2M+2N)}{k (k+M)(3k+2M)} \, a_1,
\nonu \\
s_{15} & = & -\frac{(k+2N)(3k+2M+2N)}{8k M^2 (k+M)^2 (3k+2M)N}(6kM^2+2M^3-16k^2 N-32k M N+8k^3 M N
\nonu\\
&-&16M^2 N+16k^2 M^2 N+19k M^3 N+8M^4 N) \, a_1,
\nonu \\
s_{16} & = & -\frac{(k+2N)(3k+2M+2N)}{2M (3k+2M)} \, a_1,
\nonu \\
s_{17} & = & \frac{(k+2N)(3k+2M+2N)}{16k M^2 (k+M)^2 (3k+2M)N}(-78kM^2-26M^3-32k^2 N-64k M N+16k^3 M N
\nonu\\
&-&32M^2 N-6k^2 M^2 N-7k M^3 N) \, a_1,
\nonu \\
s_{18} & = & \frac{(k^2-k M-M^2)(k+2N)(3k+2M+2N)}{2M (k+M)^2 (3k+2M)}
\, a_1,
\nonu \\
s_{19} & = & \frac{(k+2N)(3k+2M+2N)}{64k (k+M)^2 (3k+2M)N} (-66kM-22M^2+32k^3 N+66k^2 M N+31k M^2 N)\, a_1,
\nonu \\
s_{20} & = & \frac{k (2k+M)(k+2N)(3k+2M+2N)}{8 (k+M)^2 (3k+2M)} \, a_1,
\nonu \\
s_{21} & = & \frac{(k+2N)(3k+2M+2N)}{32k M (k+M)^2 (3k+2M)N}(-18kM^2-6M^3-16k^2 N-32k M N+8k^3 M N
\nonu\\
&-&16M^2 N+2k^2 M^2 N-k M^3 N) \, a_1, \qquad
s_{22}  =  \frac{k^2 (k+2N)(3k+2M+2N)}{8 (k+M)^2 (3k+2M)} \, a_1,
\nonu \\
s_{23} & = & -\frac{(k+2N)(3k+2M+2N)}{64k M (k+M)^2 (3k+2M)N}(-30kM^2-10M^3-32k^2 N-64k M N+16k^3 M N
\nonu\\
&-&32M^2 N+14k^2 M^2 N+k M^3 N) \, a_1, \qquad
s_{24}  =  -\frac{k (k+2N)(3k+2M+2N)}{8 (k+M)(3k+2M)} \, a_1,
\nonu \\
s_{25} & = & -\frac{(k+2N)(3k+2M+2N)
}{16k (k+M)^2 (k+2M)(3k+2M)N}
\nonu \\
& \times &  (6k^2+14k M+4M^2+18k^3 N+23k^2 M N+6k M^2 N)\, a_1,
\nonu \\
s_{26} & = & -\frac{k^2 (k+2N)(3k+2M+2N)}{2 (k+M)^2 (k+2M)(3k+2M)} \, a_1,
\nonu \\
s_{27} & = & i \frac{(k+2N)(3k+2M+2N)(-30kM-10M^2+8k N+8M N+k M^2 N)}{16k M (k+M)^2 (3k+2M)N} \, a_1.
\label{svalues}
\eea
Note that the last coefficient $s_{27}$
vanishes at $(N,M)=(5,4)$ and this is the reason
why this coefficient does not appear in \cite{Ahn2011}
\footnote{There were some
typos of overall signs in the coefficients,
$s_{6, \cdots, 14}$. The $S^{a b}$ has an overall factor
$a_1$ and the coefficients $s_{2, \cdots, 5}$ terms
have the charged spin-$3$ currents and contain the
coefficient $a_i$. For the $s_1$ coefficient,
the neutral spin-$3$ current has an overall factor
$b_1$ and this $s_1$ term has the coefficient $a_1$. }.
Among twenty seven coefficients, most of them
have the factor $(k+2N)$ and there are five terms
where the coefficients do not have
this factor \footnote{
We are left with the following operators 
\bea
s_{1}\,\de^{a b} \, \pa \, W^{(3)} + s_{2}\,
d^{a b c}\, \pa P^c 
+ s_{3}\, d_{4AA2}^{d c b a} \, J^c \, P^d
+ s_{4}\, d_{4SA}^{c a d b} \, J^c \, P^d
+  s_{5}\, d_{4SA}^{c b d a} \, J^c \, P^d,
\nonu
\eea
where the coefficients are given by (\ref{svalues}).
Moreover,
%according to the footnotes \ref{cond1} and
%\ref{cond2},
the neutral and charged spin-$3$ currents
can be simplified further.}.

Then we can further write down the
first order pole in the OPE
of the charged spin-$2$ current with the
charged spin-$3$ current
\bea
K^a(z) \, P^b(w) \Bigg|_{\frac{1}{(z-w)}} & = &
-\frac{1}{2M} \, i\, f^{a b c } \, \Bigg[
\tilde{R}^c+\frac{1}{3}\, i\, f^{c d e} \, \pa \, Q^{d e} 
\nonu \\
&- & \frac{M(k^2-4)(2k+M)(k+2N)(3k+2M+2N)}{20k(k+M)(3k+2M)}
\, a_1 \, \pa^2 \, K^c
\nonu \\
& - &
\frac{2(k^2-4)(2k+M)(k+2N)(3k+2M+2N)}{7k(k+M)(3k+2M)}
\nonu \\
& \times & a_1
i f^{c d e}  \Bigg(J^d  \pa  K^e -2 \pa  J^d  K^e-\frac{i}{10} 
f^{d e f}  \pa^2  K^f \Bigg)
\Bigg] + S^{a b},
\label{KPfirstorder}
\eea
where $Q^{a b}$, which depends on
the spin-$1$ current and the charged spin-$2$ current,
and $S^{a b}$, which depends on
the spin-$1$ current, the charged spin-$2, 3$ currents
and the neutral spin-$3$ current,
are given by (\ref{Qab}) and (\ref{Sab})
respectively.
In (\ref{KPfinal}) with (\ref{COEFF}),
we present the OPE
of the charged spin-$2$ current with the
charged spin-$3$ current in terms of
the composite operators among the known currents,
the spin-$1$ currents, the charged spin-$2, 3$ currents
and the neutral spin-$3$ current
in addition to the charged spin-$4$ currents.
Also its infinity limit of $k$ is given by (\ref{kpinfinityvalues}).

Therefore, we have found
the charged spin-$4$ current in (\ref{rtilde})
with the coefficients (\ref{gcoeff})
and
the neutral spin-$4$ current in (\ref{W(4)})
with the coefficients appearing in (\ref{dvalue})
together with (\ref{Ddenom}).
In particular, the former appears in the
first order pole between the
charged spin-$2$ current and the charged spin-$3$ current
from (\ref{KPfirstorder}).
We expect that the latter will appear in the second order pole
in the OPE of the charged spin-$3$ current with itself
at fixed $(N,M)=(5,4)$
\footnote{
We have the following expression,
corresponding to the last sentence of the abstract
(that is, the first two terms), 
\bea
&&
P^a(z) \, P^b(w)\Bigg|_{\frac{1}{(z-w)^2}}=
\nonu \\
&& t_{1}\,\delta_{a b}\,W^{(4)} + t_{2}\,d^{a b c}\,\tilde{R}^c +
t_{3}\,f^{a b c}\,\pa P^c + t_{4}\,\delta_{a b}\,\pa^2 T 
+ t_{5}\,d^{a b c}\,\pa^2 K^c +t_{6}\,\delta_{a b}\,T\,T + t_{7}\,f^{a b c}\,T\,\pa J^c
\nonu\\
&&+ t_{8}\,d^{a b c}\,J^c\,W^{(3)} +
t_{9}\,d^{a b e}\,d^{c f e}\,J^c\,J^f\,T + t_{10}\,f^{a f e}\,f^{b c e}\,J^c\,J^f\,T
+t_{11}\,\delta_{a c}\,\delta_{b f}\,J^c\,J^f\,T + t_{12}\,\delta_{a b}\,\delta_{c f}\,J^c\,J^f\,T
\nonu\\
&&+ t_{13}\,d^{a f e}\,d^{b c e}\,J^c\,P^f + t_{14}\,f^{a c e}\,f^{b f e}\,J^c\,P^f + t_{15}\,f^{a b e}\,f^{c f e}\,J^c\,P^f 
+ t_{16}\,\delta_{a f}\,\delta_{b c}\,J^c\,P^f +
t_{17}\,\delta_{a c}\,\delta_{b f}\,J^c\,P^f \nonu \\
&& + t_{18}\,\delta_{a b}\,\delta_{c f}\,J^c\,P^f
+ t_{19}\,f^{a b c}\,\pa T\,J^c
+ t_{20}\,d^{a c e}\,d^{b f v}\,d^{h v e}\,J^c\,J^f\,K^h +
t_{21}\,d^{h v e}\,f^{a c e}\,f^{b f v}\,J^c\,J^f\,K^h
\nonu \\
&& +t_{22}\,d^{b v e}\,f^{a h e}\,f^{f c v}\,J^c\,J^f\,K^h 
+ t_{23}\,d^{a h e}\,f^{b v e}\,f^{f c v}\,J^c\,J^f\,K^h + t_{24}\,d^{b v e}\,f^{a c e}\,f^{f h v}\,J^c\,J^f\,K^h \nonu \\
&& +t_{25}\,d^{a v e}\,f^{c b e}\,f^{f h v}\,J^c\,J^f\,K^h
+t_{26}\,d^{c f h}\,\delta_{a b}\,J^c\,J^f\,K^h + t_{27}\,d^{h b c}\,\delta_{a f}\,J^c\,J^f\,K^h
+t_{28}\,d^{f b c}\,\delta_{a h}\,J^c\,J^f\,K^h
\nonu \\
&& + t_{29}\,d^{a c h}\,\delta_{b f}\,J^c\,J^f\,K^h
+t_{30}\,d^{a c f}\,\delta_{b h}\,J^c\,J^f\,K^h + t_{31}\,d^{a b h}\,\delta_{c f}\,J^c\,J^f\,K^h + t_{32}\,d^{a b c}\,\delta_{f h}\,J^c\,J^f\,K^h
\nonu\\
&&+t_{33}\,d^{a f e}\,d^{b c e}\,K^c\,K^f + t_{34}\,f^{a f e}\,f^{b c e}\,K^c\,K^f + t_{35}\,f^{a b e}\,f^{c f e}\,K^c\,K^f  
+ t_{36}\,\delta_{a f}\,\delta_{b c}\,K^c\,K^f + t_{37}\,\delta_{a c}\,\delta_{b f}\,K^c\,K^f \nonu \\
&& + t_{38}\,\delta_{a b}\,\delta_{c f}\,K^c\,K^f
+t_{39}\,d^{c f e}\,f^{a b e}\,J^c\,\pa K^f + t_{40}\,d^{b c e}\,f^{a f e}\,J^c\,\pa K^f + t_{41}\,d^{a f e}\,f^{b c e}\,J^c\,\pa K^f
\nonu\\
&&+t_{42}\,d^{c e v}\,d^{f i e}\,f^{d a i}\,f^{b v h}\,J^d\,J^c\,J^f\,J^h  +  t_{43}\,d^{d h v}\,d^{i e v}\,f^{a b i}\,f^{c f e}\,J^d\,J^c\,J^f\,J^h
+t_{44}\,f^{a b d}\,f^{f c h}\,J^d\,J^c\,J^f\,J^h
\nonu \\
&& +  t_{45}\,d^{a i e}\,d^{b i v}\,f^{d c e}\,f^{f h v}\,J^d\,J^c\,J^f\,J^h
+t_{46}\,d^{d c e}\,d^{a i e}\,f^{b i v}\,f^{f h v}\,J^d\,J^c\,J^f\,J^h  \nonu \\
&& +  t_{47}\,f^{d c e}\,f^{a i e}\,f^{b i v}\,f^{f h v}\,J^d\,J^c\,J^f\,J^h
+t_{48}\,f^{d a i}\,f^{b v h}\,f^{c e v}\,f^{f i e}\,J^d\,J^c\,J^f\,J^h
\nonu \\
&& +  t_{49}\,f^{d i v}\,f^{a b i}\,f^{c f e}\,f^{h e v}\,J^d\,J^c\,J^f\,J^h
+t_{50}\,d^{b e v}\,d^{c f e}\,f^{a d i}\,f^{h i v}\,J^d\,J^c\,J^f\,J^h
\nonu \\
&& +  t_{51}\,d^{d h v}\,d^{a b i}\,f^{c f e}\,f^{i e v}\,J^d\,J^c\,J^f\,J^h
+t_{52}\,d^{a d i}\,d^{b f e}\,f^{c h v}\,f^{i e v}\,J^d\,J^c\,J^f\,J^h  +  t_{53}\,f^{a b i}\,f^{i h f}\,\delta_{d c}\,J^d\,J^c\,J^f\,J^h
\nonu\\
&&+t_{54}\,d^{d c v}\,d^{f h v}\,\delta_{a b}\,J^d\,J^c\,J^f\,J^h  +  t_{55}\,d^{d c v}\,d^{b f v}\,\delta_{a h}\,J^d\,J^c\,J^f\,J^h
+t_{56}\,d^{a d i}\,d^{c f i}\,\delta_{b h}\,J^d\,J^c\,J^f\,J^h
\nonu \\
&& +  t_{57}\,f^{a d i}\,f^{c f i}\,\delta_{b h}\,J^d\,J^c\,J^f\,J^h
+t_{58}\,\delta_{d c}\,\delta_{a b}\,\delta_{f h}\,J^d\,J^c\,J^f\,J^h +\,t_{59}\,d^{a d c}\,d^{b f h}\,J^d\,J^c\,J^f\,J^h
\nonu\\
&& + t_{60}\,d^{a b d}\,d^{c h f}\,J^d\,J^c\,J^f\,J^h + t_{61}\,d^{d a i}\,d^{b v h}\,d^{c e v}\,d^{f i e}\,J^d\,J^c\,J^f\,J^h
+t_{62}\,d^{a v h}\,d^{f e v}\,f^{d c i}\,f^{b i e}\,J^d\,J^c\,J^f\,J^h
\nonu\\
&&+t_{63}\,f^{a b e}\,f^{c v e}\,f^{f h v}\,\pa J^c\,J^f\,J^h + t_{64}\,f^{a h e}\,f^{b c v}\,f^{f v e}\,\pa J^c\,J^f\,J^h
+t_{65}\,f^{f b h}\,\delta_{a c}\,\pa J^c\,J^f\,J^h
\nonu \\
&& + t_{66}\,f^{c b h}\,\delta_{a f}\,\pa J^c\,J^f\,J^h
+t_{67}\,f^{a h c}\,\delta_{b f}\,\pa J^c\,J^f\,J^h + t_{68}\,d^{a f e}\,d^{b c e}\,\pa^2 J^c\,J^f
+t_{69}\,d^{a c e}\,d^{b f e}\,\pa^2 J^c\,J^f \nonu \\
&& + t_{70}\,\delta_{a f}\,\delta_{b c}\,\pa^2 J^c\,J^f
+t_{71}\,\delta_{a b}\,\delta_{c f}\,\pa^2 J^c\,J^f,
\nonu 
%\label{papb}
\eea
where the coefficients are complicated function of $k$
and we don't present them in this paper. Note that
the above expression depends on the known charged spin-$2,3,4$ currents
and the neutral spin-$3,4$ currents (as well as
the stress energy tensor and spin-$1$ currents) with derivatives.}.

%%%%%%%%%%%%%%%%%%%%%%%%%%%%%%%%%%%%%%%%%%%%%%%%%%%%%%%%%%%%%%%%%%%%%
%%%%%%%%%%%%%%%%%%%%%%%%%%%%%%%%%%%%%%%%%%%%%%%%%%%%%%%%%%%%%%%%%%%%%%
\section{ Conclusions and outlook}
%9%%%%%%%%%%%%%%%%%%%%%%%%%%%%%%%%%%%%%%%%%%%%%%%%%%%%%%%%%%%%%%%%%%%%%
%%%%%%%%%%%%%%%%%%%%%%%%%%%%%%%%%%%%%%%%%%%%%%%%%%%%%%%%%%%%%%%%%%%%%

We have constructed the charged primary spin-$4$ currents
(\ref{rtilde}) with the coefficients
(\ref{gcoeff}) and the neutral primary spin-$4$ current
(\ref{W(4)}) with the coefficients (\ref{dvalue}) and (\ref{Ddenom})
by calculating the particular poles in the OPEs
of the previously known charged or neutral currents
and using the fundamental OPEs (\ref{OPEspin1spin1})
with the help of rearrangement lemmas for the operators in \cite{BBSS}.
The coefficients appearing in the coset fields of these new currents
depend on the three generic parameters appearing in the
Grassmannian-coset model (\ref{coset}).

It is an immediate question to ask
how we can construct 
the  ${\cal N}=2$ supersymmetric extension
for the charged and neutral spin-$4$ currents
by regarding them as the highest components
of ${\cal N}=2$ multiplets.
So far, we did not calculate 
the first order pole for the OPE
of the charged spin-$3$ current
with the neutral spin-$3$ current.
It is an open problem to check whether there exists
a new quasi primary operator or not.
It is also nontrivial to obtain
the complete OPE of the charged spin-$3$ current
with itself because this contains the adjoint two free indices.
As long as the total spin appearing on the left hand sides,
which is less than or equal to six,
is concerned, then it is obvious to determine
the complete OPEs of the charged spin-$2$ current
with the charged (or neutral) spin-$4$ currents
by using the results of this paper.
The next open problem is to obtain
the charged spin-$5$ currents
which can be determined by the previous OPE
of the charged spin-$3$ current
with itself by focusing on the
first order pole
and the neutral spin-$5$ current
which can be fixed by the OPE
of the neutral spin-$3$ current with the
neutral spin-$4$ current by extracting the second order pole.
It is also open problem to check whether the unknown coefficients
appearing in the charged and neutral spin-$4$ currents
can be fixed by the regular conditions with
the spin-$1$ currents living on the denominator of the coset model and
the primary conditions with the stress energy tensor.

\vspace{.7cm}

%%%%%%%%%%%%%%%%%%%%%%%%%%%%%%%%%%%%%%%%%%%%%%%%%%%%%%%%%%%%%%
%%%%%%%%%%%%%%%%%%%%%%%%%%%%%%%%%%%%%%%%%%%%%%%%%%%%%%%%%%%%%%%
\centerline{\bf Acknowledgments}
%%%%%%%%%%%%%%%%%%%%%%%%%%%%%%%%%%%%%%%%%%%%%%%%%%%%%%%%%%%%%%%
%%%%%%%%%%%%%%%%%%%%%%%%%%%%%%%%%%%%%%%%%%%%%%%%%%%%%%%%%%%%%%%

CA thanks Y. Hikida for  the discussions.
%the general aspects of the celestial holography 
%and 
%A. Tropper for the discussion on
%the module of the Lie algebra,
%the global superconformal subalgebra and 
%the three point coefficients in the context of 
%\cite{Tropper2412}.
This work was supported by the National
Research Foundation of Korea(NRF) grant funded by the
Korea government(MSIT) 
(No. 2023R1A2C1003750).
%and RS-2026-25468734).
%This research was supported by Basic Science Research
%Program through the National Research Foundation of Korea(NRF)
%funded by the Ministry of Education(RS-2024-00446084).
CA
acknowledges warm hospitality from
the College of Liberal Arts,
Seoul National University of Science and Technology.

\newpage

\appendix

\renewcommand{\theequation}{\Alph{section}\mbox{.}\arabic{equation}}

\vspace{.7cm}

%%%%%%%%%%%%%%%%%%%%%%%%%%%%%%%%%%%%%%%%%%%%%%%%%%%%%%%%%%%%%%
%%%%%%%%%%%%%%%%%%%%%%%%%%%%%%%%%%%%%%%%%%%%%%%%%%%%%%%%%%%%%%%
%\centerline{\bf Acknowledgments}
%%%%%%%%%%%%%%%%%%%%%%%%%%%%%%%%%%%%%%%%%%%%%%%%%%%%%%%%%%%%%%%
%%%%%%%%%%%%%%%%%%%%%%%%%%%%%%%%%%%%%%%%%%%%%%%%%%%%%%%%%%%%%%%

\newpage

\appendix

\renewcommand{\theequation}{\Alph{section}\mbox{.}\arabic{equation}}

%%%%%%%%%%%%%%%%%%%%%%%%%%%%%%%%%%%%%%%%%%%%%%%%%%%%%%%%%%%%%%%%
%%%%%%%%%%%%%%%%%%%%%%%%%%%%%%%%%%%%%%%%%%%%%%%%%%%%%%%%%%%%%%%%%%%%%
\section{ The OPEs in the $SU(N+M)$ currents }
%%%%%AAA%%%%%%%%%%%%%%%%%%%%%%%%%%%%%%%%%%%%%%%%%%%%%%%%%%%%%%%%%%%%%%%%
%%%%%%%%%%%%%%%%%%%%%%%%%%%%%%%%%%%%%%%%%%%%%%%%%%%%%%%%%%%%

The OPEs for the $SU(N+M)$ currents \cite{CH1812} (in the ordering of
(\ref{branching})) are 
described by
\bea
J^{\al}(z) \, J^{\beta}(w) & = & \frac{1}{(z-w)^2} \, k \, \de^{\al \beta}+
\frac{1}{(z-w)} \, i \, f^{\al\beta}_{\,\,\,\,\,\,\,\,\ga} \, J^{\ga}(w) + \cdots,
\nonu \\
J^{\al}(z) \, J^{(\rho \bar{i})}(w) & = & 
\frac{1}{(z-w)} \, i \,
f^{\al (\rho \bar{i})}_{\,\,\,\,\,\,\,\,\,\,\,\,(\si \bar{j})} \,
J^{(\si \bar{j})}(w) + \cdots,
\nonu \\
J^{\al}(z) \, J^{(\bar{\rho} i)}(w) & = & 
\frac{1}{(z-w)} \, i \,
f^{\al (\bar{\rho} i)}_{\,\,\,\,\,\,\,\,\,\,\,\,(\bar{\si} j)} \,
J^{(\bar{\si} j)}(w) + \cdots,
\nonu \\
J^{a}(z) \, J^{b}(w) & = & \frac{1}{(z-w)^2} \, k \, \de^{a b}+
\frac{1}{(z-w)} \, i \, f^{a b}_{\,\,\,\,\,\,\,\,c} \, J^{c}(w) + \cdots,
\nonu \\
J^{a}(z) \, J^{(\rho \bar{i})}(w) & = & 
\frac{1}{(z-w)} \, i \,
f^{a (\rho \bar{i})}_{\,\,\,\,\,\,\,\,\,\,\,\,(\si \bar{j})} \,
J^{(\si \bar{j})}(w) + \cdots,
\nonu \\
J^{a}(z) \, J^{(\bar{\rho} i)}(w) & = & 
\frac{1}{(z-w)} \, i \,
f^{a (\bar{\rho} i)}_{\,\,\,\,\,\,\,\,\,\,\,\,(\bar{\si} j)} \,
J^{(\bar{\si} j)}(w) + \cdots,
\nonu \\
J^{u(1)}(z) \, J^{u(1)}(w) & = & \frac{1}{(z-w)^2} \, k + \cdots,
\nonu \\
J^{u(1)}(z) \, J^{(\rho \bar{i})}(w) & = & 
\frac{1}{(z-w)} \, i \,
f^{u(1) (\rho \bar{i})}_{\,\,\,\,\,\,\,\,\,\,\,\,\,\,\,\,\,\,\,\,(\si \bar{j})} \,
J^{(\si \bar{j})}(w) + \cdots,
\nonu \\
J^{u(1)}(z) \, J^{(\bar{\rho} i)}(w) & = & 
\frac{1}{(z-w)} \, i \,
f^{u(1) (\bar{\rho} i)}_{\,\,\,\,\,\,\,\,\,\,\,\,\,\,\,\,\,\,\,\,(\bar{\si} j)} \,
J^{(\bar{\si} j)}(w) + \cdots,
\nonu \\
J^{(\rho \bar{i})}(z) \, J^{(\bar{\si} j)}(w) & = & \frac{1}{(z-w)^2} \, k \,
\de^{\rho \bar{\si}} \, \de^{j \bar{i}}
\label{OPEspin1spin1}
\\
& + & 
\frac{1}{(z-w)} \, \Bigg[
i \,
f^{(\rho  \bar{i}) (\bar{\si} j)}_{\,\,\,\,\,\,\,\,\,\,\,\,\,\,\,\,\,\,\,\,\,\,u(1)}
\, J^{u(1)}+
i \,
f^{(\rho  \bar{i}) (\bar{\si} j)}_{\,\,\,\,\,\,\,\,\,\,\,\,\,\,\,\,\,\,\,\,\,\,\al}
\, J^{\al}+
i \,
f^{(\rho  \bar{i}) (\bar{\si} j)}_{\,\,\,\,\,\,\,\,\,\,\,\,\,\,\,\,\,\,\,\,\,\,a} \, J^{a}
\Bigg](w) + \cdots.
\nonu
\eea
Here the structure constants
appearing on the right hand sides are determined
by the above generators in (\ref{generators})
and are given by the trace for the triple product
in the generators.
From the last OPE in (\ref{OPEspin1spin1}),
the contraction (or singular OPE) between the spin-$1$ current
transforming as $({\bf N}, \overline{\bf M})$ and its conjugated one
in the various OPEs 
provides the remaining three different
kinds of spin-$1$ currents appearing on the first order pole
of the right hand side.
  
%%%%%%%%%%%%%%%%%%%%%%%%%%%%%%%%%%%%%%%%%%%%%%%%%%%%%%%%%%%%%%%%
%%%%%%%%%%%%%%%%%%%%%%%%%%%%%%%%%%%%%%%%%%%%%%%%%%%%%%%%%%%%%%%%%%%%%
\section{ The $SU(M)$ invariant tensors in terms of Kronecker
  delta, $f$ and $d$ symbols }
%%%%%BBB%%%%%%%%%%%%%%%%%%%%%%%%%%%%%%%%%%%%%%%%%%%%%%%%%%%%%%%%%%%%%%%%
%%%%%%%%%%%%%%%%%%%%%%%%%%%%%%%%%%%%%%%%%%%%%%%%%%%%%%%%%%%%

We list
the various invariant tensors
used in \cite{CH1812,Ahn2011} as follows:
\bea
d_{4SS1}^{a b c d} & = & \frac{4}{M} \, \de^{a b} \, \de^{c d} + d^{a b e} \,
d^{e c d},
\nonu \\
d_{4SS2}^{a b c d} & = & \frac{2}{M} \, \de^{a d} \, \de^{b c}
+\frac{2}{M} \, \de^{a c} \, \de^{b d} - \frac{1}{2} \,f^{a c e} \,
f^{e b d} + \frac{i}{2}  \, f^{a c e} \, d^{e b d} + \frac{i}{2} \,
 d^{a c e} \, f^{e b d}+ \frac{1}{2} \, d^{a c e} \,
d^{e b d}
\nonu \\
&- & \frac{1}{2} \,f^{b c e} \,
f^{e a d} + \frac{i}{2}  \, f^{b c e} \, d^{e a d} + \frac{i}{2} \,
d^{b c e} \, f^{e a d}+ \frac{1}{2} \, d^{b c e} \,
d^{e a d},
\nonu \\
d_{4SA}^{a b c d} & = & d^{a b e}\, f^{e c d},
\qquad
d_{4AA1}^{a b c d}  =  f^{a b e} \, f^{e cd},
\nonu \\
d_{4AA2}^{a b c d} & = & \frac{2}{M} \, \de^{a c} \, \de^{b d}-
\frac{2}{M} \, \de^{a d} \, \de^{b c} -\frac{1}{2} \, f^{a c e} \, f^{e b d} +
\frac{1}{2} \, f^{a d e} \, f^{e b c} + \frac{i}{2} \, f^{a c e} \, d^{e b d}
\nonu \\
&-& \frac{i}{2} \, f^{a d e} \, d^{e b c} +
\frac{i}{2} \, d^{a c e} \, f^{e b d} - \frac{i}{2} \, d^{a d e} \, f^{e b c}
+
\frac{1}{2} \, d^{a c e} \, d^{e b d} - \frac{1}{2} \, d^{a d e} \, d^{e b c},
\nonu \\
d_{51}^{a b c d e} & = &
\Big(\frac{2}{M} \, \de^{f c} \, \de^{d e}
+ 
\frac{2}{M} \, \de^{f e} \, \de^{c d}
+
\frac{2}{M} \, \de^{f d} \, \de^{c e} + \frac{i}{2} \, f^{f c g} \, d^{g d e}
+ 
\frac{1}{2} \, d^{f c g} \, d^{g d e}
\nonu \\
& + & \frac{i}{2} \, f^{f e g} \, d^{g c d}
+ 
\frac{1}{2} \, d^{f e g} \, d^{g c d}  + \frac{i}{2} \, f^{f d g} \, d^{g c e}
+ 
\frac{1}{2} \, d^{f d g} \, d^{g c e} \Big) i \, f^{a b f},
\nonu \\
d_{52}^{a b c d e} & = & d_{51}^{a b c d e} + i \, f^{c b f} \, \Big(
\frac{1}{M} \de^{a f} \, \de^{d e} + \frac{1}{4} (i f +d)^{a f g}(i f +d)^{g d e}\Big) \nonu \\
& - & i \, f^{c a f} \, \Big(
\frac{1}{M} \de^{b f} \, \de^{d e} + \frac{1}{4}
(i f +d)^{b f g}(i f +d)^{g d e}\Big)\nonu \\
&+& i \, f^{c b f} \, \Big(
\frac{1}{M} \de^{a f} \, \de^{e d} +
\frac{1}{4} (i f +d)^{a f g}(i f +d)^{g e d}\Big)
\nonu \\
& - & i \, f^{c a f} \, \Big(
\frac{1}{M} \de^{b f} \, \de^{e d} + \frac{1}{4}
(i f +d)^{b f g}(i f +d)^{g e d}\Big)\nonu \\
&+& i \, f^{e b f} \, \Big(
\frac{1}{M} \de^{a f} \, \de^{c d} +
\frac{1}{4} (i f +d)^{a f g}(i f +d)^{g c d}\Big)
\nonu \\
&-& i \, f^{e a f} \, \Big(
\frac{1}{M} \de^{b f} \, \de^{c d} +
\frac{1}{4} (i f +d)^{b f g}(i f +d)^{g c d}\Big)\nonu \\
&+& i \, f^{d b f} \, \Big(
\frac{1}{M} \de^{a f} \, \de^{c e} +
\frac{1}{4} (i f +d)^{a f g}(i f +d)^{g c e}\Big)
\nonu \\
&-& i \, f^{d a f} \, \Big(
\frac{1}{M} \de^{b f} \, \de^{c e} +
\frac{1}{4} (i f +d)^{b f g}(i f +d)^{g c e}\Big)
\nonu \\
&+& i \, f^{d b f} \, \Big(
\frac{1}{M} \de^{a f} \, \de^{e c} +
\frac{1}{4} (i f +d)^{a f g}(i f +d)^{g e c}\Big)
\nonu \\
&-& i \, f^{d a f} \, \Big(
\frac{1}{M} \de^{b f} \, \de^{e c} +
\frac{1}{4} (i f +d)^{b f g}(i f +d)^{g e c}\Big)
\nonu \\
&+& i \, f^{e b f} \, \Big(
\frac{1}{M} \de^{a f} \, \de^{d c} +
\frac{1}{4} (i f +d)^{a f g}(i f +d)^{g d c}\Big)
\nonu \\
&-& i \, f^{e a f} \, \Big(
\frac{1}{M} \de^{b f} \, \de^{d c} +
\frac{1}{4} (i f +d)^{b f g}(i f +d)^{g d c}\Big).
\label{tensor}
\eea
The typos in \cite{Ahn2011} are corrected here.

%%%%%%%%%%%%%%%%%%%%%%%%%%%%%%%%%%%%%%%%%%%%%%%%%%%%
%%%%%%%%%%%%%%%%%%%%%%%%%%%%%%%%%%%%%%%%%%%%%%%%%%%%%%%%%%%%%%%%%%%%%
\section{ The second order pole in the OPE
of the charged spin-$3$ current $P^a$ with
the neutral spin-$3$ current $W^{(3)}$}
%%%%CCC%%%%%%%%%%%%%%%%%%%%%%%%%%%%%%%%%%%%%%%%%%%%%%%%%%%%%%%%%%%%%%%%%A%
%%%%%%%%%%%%%%%%%%%%%%%%%%%%%%%%%%%%%%%%%%%%%%%%%%%%%%

By using the explicit expressions of (\ref{spin3exp}) and (\ref{W}),
we can extract the corresponding second order pole
where we do not substitute the final results for the
coefficients in (\ref{avalues}) and (\ref{interbvalue}) as follows:
\bea
&& P^{a}(z)\,W^{(3)}(w)\Bigg|_{\frac{1}{(z-w)^2}} =
\frac{3}{2} M a_{17} b_{2} d^{d a r} d^{f h e} d^{r c e} 
J^{d} J^{c} J^{f} J^{h} 
+ \frac{3}{2} M a_{17} b_{2} d^{d a r} d^{c h e} d^{r f e} 
J^{d} J^{c} J^{f} J^{h} \nonu\\
&&+ \frac{3}{2} M a_{17} b_{2} d^{d a r} d^{c f e} d^{r h e} 
J^{d} J^{c} J^{f} J^{h} 
+ \frac{3}{2} N a_{17} b_{7} d^{a v e} d^{c f e} d^{v h d} 
J^{d} J^{c} J^{f} J^{h} \nonu\\
&&+ \frac{3}{2} N a_{17} b_{7} d^{a f e} d^{c v e} d^{v h d} 
J^{d} J^{c} J^{f} J^{h} 
- \frac{1}{2} N \Bigg(2 a_{5} - 3 a_{17}\Bigg) b_{7} d^{a c e} d^{f v e} d^{v h d} 
J^{d} J^{c} J^{f} J^{h} \nonu\\
&&+ \frac{1}{2} i N a_{17} b_{7} d^{f h e} d^{r c e} f^{a d r} 
J^{d} J^{c} J^{f} J^{h} 
+ \frac{1}{2} i N a_{17} b_{7} d^{c h e} d^{r f e} f^{a d r} 
J^{d} J^{c} J^{f} J^{h} \nonu\\
&&+ \frac{1}{2} i N a_{17} b_{7} d^{c f e} d^{r h e} f^{a d r} 
J^{d} J^{c} J^{f} J^{h} 
+ i N a_{5} b_{7} d^{a c e} d^{v h d} f^{f v e} 
J^{d} J^{c} J^{f} J^{h} \nonu\\
&&+ \frac{3}{2} i N a_{17} b_{7} d^{a v e} d^{c f e} f^{v h d} 
J^{d} J^{c} J^{f} J^{h} 
+ \frac{3}{2} i N a_{17} b_{7} d^{a f e} d^{c v e} f^{v h d} 
J^{d} J^{c} J^{f} J^{h} \nonu\\
&&- \frac{1}{2} i N \Bigg(2 a_{5} - 3 a_{17}\Bigg) b_{7} d^{a c e} 
d^{f v e} f^{v h d} J^{d} J^{c} J^{f} J^{h} 
- N a_{5} b_{7} d^{a c e} f^{f v e} f^{v h d} 
J^{d} J^{c} J^{f} J^{h} \nonu\\
&&- \frac{2}{M} N \Bigg(2 a_{5} + M \sqrt{\frac{M + N}{M N}} a_{9}\Bigg) 
b_{7} d^{a c f} \delta^{d h} J^{d} J^{c} J^{f} J^{h} \nonu\\
&&+ \frac{2}{M} N \Bigg(M a_{3} - M a_{8} + 6 a_{17}\Bigg) b_{7} 
d^{f h d} \delta^{a c} J^{d} J^{c} J^{f} J^{h} \nonu\\
&&+ \frac{2 i N}{M} \Bigg(M a_{3} - M a_{8} + 6 a_{17}\Bigg) b_{7} 
f^{f h d} \delta^{a c} J^{d} J^{c} J^{f} J^{h}
+ 3 M a_{3} b_{2} d^{d a h} \delta^{c f} 
J^{d} J^{c} J^{f} J^{h} \nonu\\
&&+ 6 a_{17} b_{2} d^{d a h} \delta^{c f} 
J^{d} J^{c} J^{f} J^{h} 
+ \frac{N}{M} \Bigg(M a_{3} + 6 a_{17}\Bigg) b_{7} 
d^{a h d} \delta^{c f} J^{d} J^{c} J^{f} J^{h} \nonu\\
&&+ i N a_{3} b_{7} f^{a d h} \delta^{c f} 
J^{d} J^{c} J^{f} J^{h} 
+ \frac{2 i N a_{17} b_{7}}{M} f^{a d h} \delta^{c f} 
J^{d} J^{c} J^{f} J^{h} \nonu\\
&&+ \frac{i N}{M} \Bigg(M a_{3} + 6 a_{17}\Bigg) b_{7} f^{a h d} 
\delta^{c f} J^{d} J^{c} J^{f} J^{h} 
+ 6 a_{17} b_{2} d^{d a f} \delta^{c h} 
J^{d} J^{c} J^{f} J^{h} \nonu\\
&&+ \frac{2 i N a_{17} b_{7}}{M} f^{a d f} \delta^{c h} 
J^{d} J^{c} J^{f} J^{h}
+ 6 a_{17} b_{2} d^{d a c} \delta^{f h} 
J^{d} J^{c} J^{f} J^{h}
+ \frac{2 i N a_{17} b_{7}}{M} f^{a d c} \delta^{f h} 
J^{d} J^{c} J^{f} J^{h} \nonu\\
&&+ \frac{2}{M} \Bigg(M^2 a_{4} b_{5} - M N \sqrt{\frac{M + N}{M N}} a_{4} b_{7} 
- 2 M a_{7} b_{7} - 2 N a_{7} b_{7} 
+ 2 M^2 a_{4} b_{8} + 3 M N a_{4} b_{8}\nonu\\
&&+ 4 M N \sqrt{\frac{M + N}{M N}} 
a_5{7} b_{8} - 2 M^2 a_{8} b_{8} - 2 M N a_{8} b_{8}\Bigg) 
J^{a} J^{u(1)} J^{u(1)} J^{u(1)} \nonu\\
&&+ \frac{2}{M} \Bigg(2 M^2 a_{8} b_{5} + 4 k a_{7} b_{7} + 
4 M a_{7} b_{7} + 4 N a_{7} b_{7} - 2 M N 
\sqrt{\frac{M + N}{M N}} a_{8} b_{7} 
- \sqrt{\frac{M + N}{M N}} a_{12} b_{7}\nonu\\
&&+ \sqrt{\frac{M + N}{M N}} a_{13} b_{7} + 
4 k M a_{8} b_{8} + 4 M^2 a_{8} b_{8} 
+ 6 M N a_{8} b_{8} + a_{12} b_{8} - a_{13} b_{8} + 
M \sqrt{\frac{M + N}{M N}} a_{4} b_{12}\nonu\\
&&+ a_{7} b_{12} 
- M \sqrt{\frac{M + N}{M N}} a_{8} b_{12} - 
M \sqrt{\frac{M + N}{M N}} a_{4} b_{13} - a_{7} b_{13} + 
M \sqrt{\frac{M + N}{M N}} a_{8} b_{13}\Bigg) \nonu\\
&& \times \delta_{\bar{\imath} k} \delta_{\rho \bar{\tau}} 
J^{a} J^{u(1)} J^{(\rho \bar{\imath})} 
J^{(\bar{\tau} k)} \nonu\\
&&- \frac{1}{M^2 N} \Bigg(-16 M N a_{5} b_{7} + 4 M^3 N a_{5} b_{7} + 
8 M^2 N a_{8} b_{7} + 4 k M^2 N^2 a_{8} b_{7} \nonu\\
&&+ 4 k M N a_{12} b_{7} + 2 M^2 N a_{12} b_{7} + 
2 M N^2 a_{12} b_{7} - 2 k M N a_{13} b_{7} - 
2 M^2 N a_{13} b_{7} \nonu\\
&&- M a_{1} b_{12} + M N^2 a_{1} b_{12} + 
8 N a_{5} b_{12} - 2 M^2 N a_{5} b_{12} + 
2 M N \sqrt{\frac{M + N}{M N}} a_{7} b_{12} \nonu\\
&&- 2 M N a_{8} b_{12} - 2 M^2 N^2 a_{8} b_{12} + 
M a_{1} b_{13} - M N^2 a_{1} b_{13} - 
8 N a_{5} b_{13} + 2 M^2 N a_{5} b_{13} \nonu\\
&&- 2 M N \sqrt{\frac{M + N}{M N}} a_{7} b_{13} + 
2 M N a_{8} b_{13} - 6 k M^2 N a_{8} b_{13} - 
6 M^3 N a_{8} b_{13} \nonu\\
&&- 8 M^2 N^2 a_{8} b_{13} - 2 M N a_{12} b_{13} + 
2 M N a_{13} b_{13}\Bigg) \delta_{\bar{\imath} k} \delta_{\rho \bar{\tau}} 
J^{a} J^{(\rho \bar{\imath})} 
\partial J^{(\bar{\tau} k)} \nonu\\
&&+ \frac{1}{M} \Bigg(-M^3 N \sqrt{\frac{M + N}{M N}} a_{8} b_{5} 
- 2 k M N a_{4} b_{7} - 
2 k M N \sqrt{\frac{M + N}{M N}} a_{7} b_{7} \nonu\\
&&- 2 M^2 N \sqrt{\frac{M + N}{M N}} a_{7} b_{7} + 
2 M N^2 \sqrt{\frac{M + N}{M N}} a_{7} b_{7} + 
M^2 N a_{8} b_{7} + M N^2 a_{8} b_{7} \nonu\\
&&+ M a_{12} b_{7} + N a_{12} b_{7} + 
2 k M^2 N \sqrt{\frac{M + N}{M N}} a_{4} b_{8} - 
2 M^2 N a_{7} b_{8} - 2 M N^2 a_{7} b_{8} \nonu\\
&&- 2 k M^2 N \sqrt{\frac{M + N}{M N}} a_{8} b_{8} - 
M^2 N^2 \sqrt{\frac{M + N}{M N}} a_{8} b_{8} - 
M N \sqrt{\frac{M + N}{M N}} a_{12} b_{8} \nonu\\
&&+ M^2 a_{4} b_{12} + 2 M N a_{4} b_{12} + 
3 M N \sqrt{\frac{M + N}{M N}} a_{7} b_{12} - 
M^2 a_{8} b_{12} - M N a_{8} b_{12} \nonu\\
&&+ 2 M^2 a_{4} b_{13} + 3 M N a_{4} b_{13} + 
4 M N \sqrt{\frac{M + N}{M N}} a_{7} b_{13} - 
2 M^2 a_{8} b_{13} - 2 M N a_{8} b_{13}\Bigg) 
J^{a} 
\partial J^{u(1)} J^{u(1)} \nonu\\
&&+ \frac{1}{M^2 N} \Bigg(-16 M N a_{5} b_{7} + 4 M^3 N a_{5} b_{7} + 
8 M^2 N a_{8} b_{7} - 4 k M^2 N^2 a_{8} b_{7} \nonu\\
&&+ 2 k M N a_{12} b_{7} + 2 M^2 N a_{12} b_{7} + 
4 M N^2 a_{12} b_{7} - 4 k M N a_{13} b_{7} \nonu\\
&&- 2 M^2 N a_{13} b_{7} - 6 M N^2 a_{13} b_{7} - 
M a_{1} b_{12} + M N^2 a_{1} b_{12} + 
8 N a_{5} b_{12} \nonu\\
&&- 2 M^2 N a_{5} b_{12} + 2 M N \sqrt{\frac{M + N}{M N}} 
a_{7} b_{12} - 2 M N a_{8} b_{12} + 
6 k M^2 N a_{8} b_{12} \nonu\\
&&+ 6 M^3 N a_{8} b_{12} + 8 M^2 N^2 a_{8} b_{12} + 
2 M N a_{12} b_{12} - 2 M N a_{13} b_{12} + 
M a_{1} b_{13} \nonu\\
&&- M N^2 a_{1} b_{13} - 8 N a_{5} b_{13} + 
2 M^2 N a_{5} b_{13} - 2 M N \sqrt{\frac{M + N}{M N}} 
a_{7} b_{13} \nonu\\
&&+ 2 M N a_{8} b_{13} + 2 M^2 N^2 a_{8} b_{13}\Bigg) 
\delta_{\bar{\imath} k} \delta_{\rho \bar{\tau}} J^{a} 
\partial J^{(\rho \bar{\imath})} J^{(\bar{\tau} k)} \nonu\\
&&+ \frac{1}{2 M} \Bigg(-4 k M N \sqrt{\frac{M + N}{M N}} a_{4} b_{7} - 
8 M^2 N \sqrt{\frac{M + N}{M N}} a_{4} b_{7} \nonu\\
&&- 16 M N \sqrt{\frac{M + N}{M N}} a_{5} b_{7} + 
4 M^3 N \sqrt{\frac{M + N}{M N}} a_{5} b_{7} + 
16 N^2 \sqrt{\frac{M + N}{M N}} a_{5} b_{7} \nonu\\
&&- 4 M^2 N^2 \sqrt{\frac{M + N}{M N}} a_{5} b_{7} + 
4 M N a_{7} b_{7} + 4 k^2 M N a_{7} b_{7} + 
8 k M^2 N a_{7} b_{7} \nonu\\
&&+ 4 N^2 a_{7} b_{7} + 2 k M N^2 a_{7} b_{7} + 
4 M^2 N^2 a_{7} b_{7} + 
4 M^2 N \sqrt{\frac{M + N}{M N}} a_{8} b_{7} \nonu\\
&&- 4 M N^2 \sqrt{\frac{M + N}{M N}} a_{8} b_{7} + 
4 k M^2 N^2 \sqrt{\frac{M + N}{M N}} a_{8} b_{7} - 
8 k N a_{9} b_{7} \nonu\\
&&- 16 M N a_{9} b_{7} + 2 k M^2 N a_{9} b_{7} + 
4 M^3 N a_{9} b_{7} + 
4 k M N \sqrt{\frac{M + N}{M N}} a_{12} b_{7} \nonu\\
&&+ 2 M^2 N \sqrt{\frac{M + N}{M N}} a_{12} b_{7} - 
2 k M N \sqrt{\frac{M + N}{M N}} a_{13} b_{7} - 
2 M^2 N \sqrt{\frac{M + N}{M N}} a_{13} b_{7} \nonu\\
&&- M \sqrt{\frac{M + N}{M N}} a_{1} b_{12} + 
M N^2 \sqrt{\frac{M + N}{M N}} a_{1} b_{12} + 
2 M N \sqrt{\frac{M + N}{M N}} a_{4} b_{12} \nonu\\
&&+ 2 k M^2 N \sqrt{\frac{M + N}{M N}} a_{4} b_{12} + 
8 N \sqrt{\frac{M + N}{M N}} a_{5} b_{12} - 
2 M^2 N \sqrt{\frac{M + N}{M N}} a_{5} b_{12} \nonu\\
&&+ 2 M a_{7} b_{12} + 2 N a_{7} b_{12} - 
2 k M N a_{7} b_{12} - 
2 M^2 N a_{7} b_{12} - 3 M N^2 a_{7} b_{12} \nonu\\
&&- 2 M N \sqrt{\frac{M + N}{M N}} a_{8} b_{12} - 
2 k M^2 N \sqrt{\frac{M + N}{M N}} a_{8} b_{12} - 
2 M^2 N^2 \sqrt{\frac{M + N}{M N}} a_{8} b_{12} \nonu\\
&&+ 4 N a_{9} b_{12} - M^2 N a_{9} b_{12} - 
M N \sqrt{\frac{M + N}{M N}} a_{12} b_{12} + 
M \sqrt{\frac{M + N}{M N}} a_{1} b_{13} \nonu\\
&&- M N^2 \sqrt{\frac{M + N}{M N}} a_{1} b_{13} - 
2 M N \sqrt{\frac{M + N}{M N}} a_{4} b_{13} + 
4 k M^2 N \sqrt{\frac{M + N}{M N}} a_{4} b_{13} \nonu\\
&&- 8 N \sqrt{\frac{M + N}{M N}} a_{5} b_{13} + 
2 M^2 N \sqrt{\frac{M + N}{M N}} a_{5} b_{13} - 
2 M a_{7} b_{13} - 2 N a_{7} b_{13} \nonu\\
&&+ 2 k M N a_{7} b_{13} - 
4 M^2 N a_{7} b_{13} - 
3 M N^2 a_{7} b_{13} + 
2 M N \sqrt{\frac{M + N}{M N}} a_{8} b_{13} \nonu\\
&&- 4 k M^2 N \sqrt{\frac{M + N}{M N}} a_{8} b_{13} - 
2 M^2 N^2 \sqrt{\frac{M + N}{M N}} a_{8} b_{13} - 
4 N a_{9} b_{13} + M^2 N a_{9} b_{13} \nonu\\
&&- 2 M N \sqrt{\frac{M + N}{M N}} a_{12} b_{13} + 
M N \sqrt{\frac{M + N}{M N}} a_{13} b_{13}\Bigg) 
J^{a} \partial^2 J^{u(1)} \nonu\\
&&+ \frac{1}{M} \Bigg(3 M^2 a_{4} b_{2} + 2 M^2 a_{9} b_{5} + M N a_{4} b_{7} - 4 M a_{5} b_{7} - 4 N a_{5} b_{7} \nonu \\
&&+ 4 M N \sqrt{\frac{M + N}{M N}} a_{7} b_{7} - 4 M N \sqrt{\frac{M + N}{M N}} a_{9} b_{7} + 8 M N \sqrt{\frac{M + N}{M N}} a_{5} b_{8} \nonu \\
&&- 2 M N a_{7} b_{8} + 2 M^2 a_{9} b_{8} + 6 M N a_{9} b_{8}\Bigg) d^{a c f} J^{c} J^{f} J^{u(1)} J^{u(1)} \nonu \\
&&+ 2 i N a_{7} b_{8} f^{a c f} J^{c} J^{f} J^{u(1)} J^{u(1)} 
+a_{17} \Bigg(M b_{5} - 3 N \sqrt{\frac{M + N}{M N}} b_{7} + 3 N b_{8}\Bigg) d^{a h e} d^{c f e} J^{c} J^{f} J^{h} J^{u(1)} \nonu \\
&&+ \Bigg(M a_{17} b_{5} + 4 N \sqrt{\frac{M + N}{M N}} a_{5} b_{7} - N a_{7} b_{7} + 2 N a_{9} b_{7} 
- 3 N \sqrt{\frac{M + N}{M N}} a_{17} b_{7}\nonu\\
&& + 3 N a_{17} b_{8}\Bigg) d^{a f e} d^{c h e} J^{c} J^{f} J^{h} J^{u(1)} \nonu \\
&&+ \Bigg(3 M a_{9} b_{2} + M a_{17} b_{5} - 3 N \sqrt{\frac{M + N}{M N}} a_{17} b_{7} - 2 N a_{5} b_{8} 
+ 3 N a_{17} b_{8}\Bigg) d^{a c e} d^{f h e} J^{c} J^{f} J^{h} J^{u(1)} \nonu \\
&&+ i N a_{9} b_{7} d^{f h e} f^{a c e} J^{c} J^{f} J^{h} J^{u(1)} 
+ i N a_{7} b_{7} d^{c h e} f^{a f e} J^{c} J^{f} J^{h} J^{u(1)} \nonu \\
&&+ i N \Bigg(a_{7} - 2 a_{9}\Bigg)
b_{7} d^{a f e} f^{c h e} J^{c} J^{f} J^{h} J^{u(1)} 
+ N a_{7} b_{7} f^{a f e} f^{c h e} J^{c} J^{f} J^{h} J^{u(1)} \nonu \\
&&+ 2 i N a_{5} b_{8} d^{a c e} f^{f h e} J^{c} J^{f} J^{h} J^{u(1)} \nonu \\ 
&&+ \frac{2}{M} \Bigg(M^{2} a_{3} b_{5} + 2 M a_{17} b_{5} + M N a_{3} b_{8} + 6 N a_{17} b_{8}\Bigg) 
\delta^{a h} \delta^{c f} J^{c} J^{f} J^{h} J^{u(1)} \nonu \\ 
&&+ \frac{4}{M} \Bigg(M a_{17} b_{5} - M N \sqrt{\frac{M + N}{M N}} a_{3} b_{7} - 
M N \sqrt{\frac{M + N}{M N}} a_{4} b_{7} - N a_{7} b_{7} + 
2 M N \sqrt{\frac{M + N}{M N}} a_{8} b_{7}\nonu\\
&&- 6 N \sqrt{\frac{M + N}{M N}} a_{17} b_{7}\Bigg) 
\delta^{a f} \delta^{c h} J^{c} J^{f} J^{h} J^{u(1)} \nonu \\ 
&&+ \frac{2}{M} \Bigg(2 M a_{17} b_{5} - 
M N \sqrt{\frac{M + N}{M N}} a_{3} b_{7} - 
6 N \sqrt{\frac{M + N}{M N}} a_{17} b_{7} + 
2 M N a_{3} b_{8} - 2 M N a_{8} b_{8} \nonu\\
&&+ 12 N a_{17} b_{8}\Bigg) 
\delta^{a c} \delta^{f h} J^{c} J^{f} J^{h} J^{u(1)} \nonu \\
&&+ \frac{1}{M} \Bigg(6 M^{2} a_{8} b_{2} + 8 k a_{5} b_{7} + 8 M a_{5} b_{7} + 8 N a_{5} b_{7} + 2 M N a_{8} b_{7} 
+ a_{12} b_{7} - a_{13} b_{7} + 2 a_{5} b_{12}\nonu\\
&&+ M \sqrt{\frac{M + N}{M N}} a_{9} b_{12} 
- 2 a_{5} b_{13} - M \sqrt{\frac{M + N}{M N}} a_{9} b_{13}\Bigg) 
d^{a c f} \delta_{\bar{\imath} k} \delta_{\rho \bar{\tau}} J^{c} J^{f} J^{(\rho \bar{\imath})} J^{(\bar{\tau} k)} \nonu \\ 
&&- \frac{i}{M} \Bigg(a_{12} + a_{13}\Bigg) b_{7} f^{a c f} \delta_{\bar{\imath} k} \delta_{\rho \bar{\tau}} J^{c} J^{f} J^{(\rho \bar{\imath})} J^{(\bar{\tau} k)} \nonu \\
&&- \Bigg(3 M^{2} N \sqrt{\frac{M + N}{M N}} a_{8} b_{2} + 4 k N \sqrt{\frac{M + N}{M N}} a_{5} b_{7} 
+ 4 M N \sqrt{\frac{M + N}{M N}} a_{5} b_{7} + 2 N^{2} a_{7} b_{7}\nonu\\
&&+ M N^{2} \sqrt{\frac{M + N}{M N}} a_{8} b_{7} 
+ 4 k N a_{9} b_{7} + 2 M N a_{9} b_{7} + N \sqrt{\frac{M + N}{M N}} a_{12} b_{7} 
- 2 N \sqrt{\frac{M + N}{M N}} a_{5} b_{12}\nonu\\
&&+ N a_{7} b_{12} 
- 2 N a_{9} b_{12} - 4 N \sqrt{\frac{M + N}{M N}} a_{5} b_{13} + N a_{7} b_{13} 
- M a_{9} b_{13} - 3 N a_{9} b_{13}\Bigg) 
d^{a c f} J^{c} J^{f} \partial J^{u(1)} \nonu \\
&&+ i N \Bigg(4 k a_{7} b_{7} + 2 N a_{7} b_{7} + \sqrt{\frac{M + N}{M N}} a_{12} b_{7} 
+ a_{7} b_{12} + a_{7} b_{13}\Bigg) 
f^{a c f} J^{c} J^{f} \partial J^{u(1)} \nonu \\
&&+ 2 \Bigg(M a_{2} b_{5} + 2 \sqrt{\frac{M + N}{M N}} a_{1} b_{6} 
+ 2 M \sqrt{\frac{M + N}{M N}} a_{2} b_{6} + 2 M \sqrt{\frac{M + N}{M N}} a_{4} b_{6} 
+ 4 a_{7} b_{6} \nonu\\
&&- 4 M \sqrt{\frac{M + N}{M N}} a_{8} b_{6} 
+ 2 a_{1} b_{8} + 2 M a_{2} b_{8} + N a_{2} b_{8} - 2 M a_{8} b_{8}\Bigg) 
\delta^{\beta \gamma} J^{\beta} J^{\gamma} J^{a} J^{u(1)} \nonu \\
&&- 2 \Bigg(2 \sqrt{\frac{M + N}{M N}} a_{1} + N \sqrt{\frac{M + N}{M N}} a_{2} + 2 a_{7}\Bigg) b_{7} 
\delta^{a c} \delta^{\beta \gamma} J^{\beta} J^{\gamma} J^{c} J^{u(1)} \nonu \\
&&+ \Bigg(3 M a_{2} b_{2} - a_{1} b_{6} + 8 a_{5} b_{6} 
+ 2 M \sqrt{\frac{M + N}{M N}} a_{9} b_{6} 
+ 2 a_{1} b_{7} + N a_{2} b_{7} - 4 a_{5} b_{7}\Bigg)
\nonu \\
&& \times
d^{a c f} \delta^{\beta \gamma} J^{\beta} J^{\gamma} J^{c} J^{f} 
+ i a_{1} b_{6} f^{a c f} \delta^{\beta \gamma} J^{\beta} J^{\gamma} J^{c} J^{f}
\nonu \\
&& +2 \Bigg(a_{1} + M a_{2} - M a_{8}\Bigg) b_{6} 
d^{\beta \gamma \delta} J^{\beta} J^{\gamma} J^{\delta} J^{a} \nonu\\
&&- 2 i M \Bigg(a_{2} - a_{8}\Bigg) b_{6} 
f^{\beta \gamma \delta} J^{\beta} J^{\gamma} J^{\delta} J^{a} 
- a_{1} b_{7} d^{\beta \gamma \delta} \delta^{a c} J^{\beta} J^{\gamma} J^{\delta} J^{c} 
- i a_{1} b_{7} f^{\beta \gamma \delta} \delta^{a c} J^{\beta} J^{\gamma} J^{\delta} J^{c} \nonu \\
&&+ \Bigg(2 k a_{1} b_{6} + 2 N a_{1} b_{6} + 4 k M a_{2} b_{6} 
+ 4 M N a_{2} b_{6} + 2 M N \sqrt{\frac{M + N}{M N}} a_{7} b_{6} \nonu \\
&&+ 4 M^{2} a_{8} b_{6} - 2 a_{12} b_{6} 
+ 2 M a_{1} b_{7} + 2 M N a_{2} b_{7} 
+ a_{1} b_{12} + 2 M a_{2} b_{12} - 2 M a_{8} b_{12} \nonu \\
&&+ 2 a_{1} b_{13} + 2 M a_{2} b_{13} + N a_{2} b_{13} 
- 2 M a_{8} b_{13}\Bigg) \delta^{\beta \gamma} J^{\beta} J^{\gamma} \partial J^{a} \nonu \\
&&+ \Bigg(2 M N a_{7} b_{5} + 2 M N a_{4} b_{7} 
+ 4 M N \sqrt{\frac{M + N}{M N}} a_{7} b_{7} 
+ 4 k M N \sqrt{\frac{M + N}{M N}} a_{4} b_{8} \nonu \\
&&+ 4 k N a_{7} b_{8} + 2 M N a_{7} b_{8} + 2 N^{2} a_{7} b_{8} 
+ 4 M^{2} N \sqrt{\frac{M + N}{M N}} a_{8} b_{8} 
+ 4 M N^{2} \sqrt{\frac{M + N}{M N}} a_{8} b_{8} \nonu \\
&&- 2 N \sqrt{\frac{M + N}{M N}} a_{12} b_{8} + 2 M a_{4} b_{12} 
+ 2 N a_{4} b_{12} + 2 N \sqrt{\frac{M + N}{M N}} a_{7} b_{12} 
- 2 M a_{8} b_{12} - 2 N a_{8} b_{12} \nonu \\
&&+ 2 M a_{4} b_{13} + 3 N a_{4} b_{13} 
+ 4 N \sqrt{\frac{M + N}{M N}} a_{7} b_{13} 
- 2 M a_{8} b_{13} - 2 N a_{8} b_{13}\Bigg) 
\partial J^{a} J^{u(1)} J^{u(1)} \nonu \\
&&+ \frac{1}{M^{2} N} \Bigg(-16 M N a_{5} b_{7} + 4 M^{3} N a_{5} b_{7} 
+ 8 M^{2} N a_{8} b_{7} + 4 M^{3} N^{2} a_{8} b_{7} - M a_{1} b_{12} 
\nonu \\
&&+ M N^{2} a_{1} b_{12} + 4 M N a_{3} b_{12} + 16 N a_{5} b_{12} 
- 4 M^{2} N a_{5} b_{12} + 2 M N \sqrt{\frac{M + N}{M N}} a_{7} b_{12} 
\nonu \\
&&+ M^{2} N^{2} \sqrt{\frac{M + N}{M N}} a_{7} b_{12} - 4 M N a_{8} b_{12} 
+ 4 k M^{2} N a_{8} b_{12} + 4 M^{3} N a_{8} b_{12} + 4 M^{2} N^{2} a_{8} b_{12} 
\nonu \\
&&- M N a_{13} b_{12} - 24 N a_{17} b_{12} + 12 M^{2} N a_{17} b_{12} + M a_{1} b_{13} 
- M N^{2} a_{1} b_{13} + 4 M N a_{3} b_{13} 
\nonu \\
&&+ 8 N a_{5} b_{13} - 2 M^{2} N a_{5} b_{13} - 2 M N \sqrt{\frac{M + N}{M N}} a_{7} b_{13} 
- M^{2} N^{2} \sqrt{\frac{M + N}{M N}} a_{7} b_{13} 
\nonu \\
&&- 2 M N a_{8} b_{13} + 4 k M^{2} N a_{8} b_{13} + 4 M^{3} N a_{8} b_{13} + 6 M^{2} N^{2} a_{8} b_{13} 
+ M N a_{12} b_{13} - 24 N a_{17} b_{13} 
\nonu \\
&&+ 12 M^{2} N a_{17} b_{13}\Bigg) \delta_{\bar{\imath} k} \delta_{\rho \bar{\tau}} 
\partial J^{a} J^{(\rho \bar{\imath})} J^{(\bar{\tau} k)} \nonu \\
&&+ N \Bigg(4 k M a_{7} b_{7} + 2 M N a_{7} b_{7} 
- 2 M^{2} N \sqrt{\frac{M + N}{M N}} a_{8} b_{7} 
+ 2 k M \sqrt{\frac{M + N}{M N}} a_{4} b_{12} 
\nonu \\
&&+ 2 k a_{7} b_{12} - 2 k M \sqrt{\frac{M + N}{M N}} a_{8} b_{12} 
+ 4 k M \sqrt{\frac{M + N}{M N}} a_{4} b_{13} 
+ 2 k a_{7} b_{13} - M a_{7} b_{13} 
\nonu \\
&&- N a_{7} b_{13} - 2 k M \sqrt{\frac{M + N}{M N}} a_{8} b_{13} 
+ 2 M^{2} \sqrt{\frac{M + N}{M N}} a_{8} b_{13} 
+ M N \sqrt{\frac{M + N}{M N}} a_{8} b_{13} 
\nonu \\
&&- 2 \sqrt{\frac{M + N}{M N}} a_{12} b_{13}\Bigg) 
\partial J^{a} \partial J^{u(1)} \nonu \\
&&+ \Bigg(3 M N a_{7} b_{2} + 2 M N a_{5} b_{5} 
+ 4 M N \sqrt{\frac{M + N}{M N}} a_{5} b_{7} 
+ 2 N^{2} \sqrt{\frac{M + N}{M N}} a_{5} b_{7} 
\nonu \\
&&+ 2 k N a_{7} b_{7} - 4 k N a_{9} b_{7} 
+ 2 M N a_{9} b_{7} - N \sqrt{\frac{M + N}{M N}} a_{12} b_{7} 
+ 4 k N a_{5} b_{8} - 4 M N a_{5} b_{8} 
\nonu \\
&&- 2 N^{2} a_{5} b_{8} + 4 k M N \sqrt{\frac{M + N}{M N}} a_{9} b_{8} 
+ N a_{12} b_{8} + 6 N \sqrt{\frac{M + N}{M N}} a_{5} b_{12} 
\nonu \\
&&- N a_{7} b_{12} + 2 M a_{9} b_{12} + 4 N a_{9} b_{12} 
+ 8 N \sqrt{\frac{M + N}{M N}} a_{5} b_{13} - 2 N a_{7} b_{13} 
+ 2 M a_{9} b_{13} 
\nonu \\
&&+ 6 N a_{9} b_{13}\Bigg) d^{a c f} 
\partial J^{c} J^{f} J^{u(1)} \nonu \\
&&+ \frac{i}{M} \Bigg(2 M^{2} a_{11} b_{5} 
- 8 M N \sqrt{\frac{M + N}{M N}} a_{3} b_{7} 
+ 4 M N \sqrt{\frac{M + N}{M N}} a_{4} b_{7} 
+ 16 N \sqrt{\frac{M + N}{M N}} a_{5} b_{7} 
\nonu \\
&&- 4 M^{2} N \sqrt{\frac{M + N}{M N}} a_{5} b_{7} 
- 2 k M N a_{7} b_{7} 
- M N^{2} a_{7} b_{7} 
- 8 M N \sqrt{\frac{M + N}{M N}} a_{8} b_{7} 
\nonu \\
&&+ 8 N a_{9} b_{7} 
- 2 M^{2} N a_{9} b_{7} 
- 6 M N \sqrt{\frac{M + N}{M N}} a_{11} b_{7} 
+ M N \sqrt{\frac{M + N}{M N}} a_{12} b_{7} 
\nonu \\
&&- 12 N \sqrt{\frac{M + N}{M N}} a_{17} b_{7} 
+ 8 M N a_{3} b_{8} 
+ 6 M N a_{11} b_{8} 
- M N a_{12} b_{8} 
+ 12 N a_{17} b_{8} 
\nonu \\
&&+ M N a_{7} b_{12} \Bigg) 
f^{a c f} \partial J^{c} J^{f} J^{u(1)} \nonu \\
&&- \frac{N}{2 M} \Bigg(-2 M a_{3} b_{7} 
+ 8 a_{5} b_{7} 
- 4 k M a_{5} b_{7} 
+ 2 M a_{8} b_{7} 
- 36 a_{17} b_{7} 
+ 12 k M a_{17} b_{7} 
\nonu \\
&&+ 6 M^{2} a_{17} b_{7} 
- 6 M a_{17} b_{12} 
- 9 M a_{17} b_{13} \Bigg) 
d^{a h e} d^{c f e} \partial J^{c} J^{f} J^{h}  \nonu \\
&&+ \frac{N}{2 M} \Bigg(6 M^{2} a_{5} b_{2} 
+ 2 M a_{3} b_{7} 
+ 8 a_{5} b_{7} 
- 36 a_{17} b_{7} 
- 12 k M a_{17} b_{7} 
+ 6 M^{2} a_{17} b_{7} 
\nonu \\
&&- 2 M a_{5} b_{12} 
+ 6 M a_{17} b_{12} 
- 4 M a_{5} b_{13} 
+ 9 M a_{17} b_{13} \Bigg) 
d^{a f e} d^{c h e} \partial J^{c} J^{f} J^{h}  \nonu \\
&&- \frac{N}{2} \Bigg(4 a_{3} b_{7} 
+ 2 N a_{5} b_{7} 
- 2 a_{8} b_{7} 
- a_{12} b_{7} 
+ 12 k a_{17} b_{7} 
+ 6 M a_{17} b_{7} 
\nonu \\
&&+ 2 a_{5} b_{12} 
- 6 a_{17} b_{12} 
+ 2 a_{5} b_{13} 
- 9 a_{17} b_{13} \Bigg) 
d^{a c e} d^{f h e} \partial J^{c} J^{f} J^{h} \nonu \\
&&+ \frac{i N}{4 M} \Bigg(8 M a_{3} 
- 44 a_{5} 
+ M^{2} a_{5} 
+ 12 M a_{8} 
+ 12 M a_{11} 
- 2 M a_{12} 
+ 72 a_{17} \nonu\\
&&- 6 M^{2} a_{17} \Bigg) 
b_{7} d^{f h e} f^{a c e} \partial J^{c} J^{f} J^{h} \nonu \\
&&- \frac{i N}{2 M} \Bigg(6 M a_{3} 
+ 14 a_{5} 
- 2 M N a_{5} 
- 4 M a_{8} 
- 12 a_{17} 
+ 3 M^{2} a_{17} \Bigg) 
b_{7} d^{c h e} f^{a f e} \partial J^{c} J^{f} J^{h} \nonu \\
&&+ \frac{i N}{4 M} \Bigg(-4 M a_{3} 
- 20 a_{5} 
+ M^{2} a_{5} 
+ 4 M a_{8} 
- 16 M \sqrt{\frac{M + N}{M N}} a_{9} 
- 120 a_{17} \nonu\\
&&+ 12 M^{2} a_{17} \Bigg) 
b_{7} d^{c f e} f^{a h e} \partial J^{c} J^{f} J^{h} \nonu \\
&&- \frac{i N}{4 M} \Bigg(4 M a_{3} 
+ 8 k M a_{5} 
+ M^{2} a_{5} 
- 4 M a_{8} 
+ 24 a_{17} \Bigg) 
b_{7} d^{a h e} f^{c f e} \partial J^{c} J^{f} J^{h} \nonu \\
&&- \frac{N}{M} \Bigg(M a_{3} 
- M a_{8} 
+ 6 a_{17} \Bigg) 
b_{7} f^{a h e} f^{c f e} \partial J^{c} J^{f} J^{h} \nonu \\
&&+ \frac{i}{2 M} \Bigg(6 M^{2} a_{11} b_{2} 
- 2 M N a_{3} b_{7} 
+ 4 N a_{5} b_{7} 
- 12 N a_{17} b_{7} 
- 3 M^{2} N a_{17} b_{7} \nonu\\
&&+ 2 M N a_{5} b_{12} \Bigg) 
d^{a f e} f^{c h e} \partial J^{c} J^{f} J^{h} \nonu \\
&&- \frac{N}{M} \Bigg(-M a_{3} 
+ 2 M a_{8} 
+ M a_{11} 
- 6 a_{17} \Bigg) 
b_{7} f^{a f e} f^{c h e} \partial J^{c} J^{f} J^{h} \nonu \\
&&+ \frac{i N}{4 M} \Bigg(8 M a_{3} b_{7} 
+ 8 a_{5} b_{7} 
+ 16 k M a_{5} b_{7} 
+ M^{2} a_{5} b_{7} 
+ 4 M N a_{5} b_{7} 
- 4 M a_{8} b_{7} 
\nonu \\
&& - 16 M \sqrt{\frac{M + N}{M N}} a_{9} b_{7} 
- 2 M a_{12} b_{7} 
- 144 a_{17} b_{7} 
+ 18 M^{2} a_{17} b_{7} 
+ 4 M a_{5} b_{12} 
+ 4 M a_{5} b_{13} \Bigg)
\nonu \\
&& \times 
d^{a c e} f^{f h e} \partial J^{c} J^{f} J^{h} \nonu \\
&&+ \frac{N}{2 M} \Bigg(12 M a_{3} 
- 2 M a_{8} 
+ 6 M a_{11} 
- M a_{12} 
+ 36 a_{17} \Bigg) 
b_{7} f^{a c e} f^{f h e} \partial J^{c} J^{f} J^{h} \nonu \\
&&- \frac{2 N}{M^{2}} \Bigg(-2 M a_{3} b_{7} 
+ 2 k M^{2} a_{3} b_{7} 
+ 8 a_{5} b_{7} 
- 2 M^{2} a_{5} b_{7} 
+ 2 M a_{8} b_{7} 
- 2 k M^{2} a_{8} b_{7} 
- 36 a_{17} b_{7} 
\nonu \\
&&+ 12 k M a_{17} b_{7} 
+ 12 M^{2} a_{17} b_{7} 
- M^{2} a_{3} b_{12} 
- 6 M a_{17} b_{12} 
- M^{2} a_{3} b_{13} 
- 6 M a_{17} b_{13} \Bigg) 
\delta^{a h} \delta^{c f} \partial J^{c} J^{f} J^{h} \nonu \\
&&- \frac{2 N}{M^{2}} \Bigg(-2 M a_{3} b_{7} 
+ 2 k M^{2} a_{3} b_{7} 
- 8 a_{5} b_{7} 
+ 2 M^{2} a_{5} b_{7} 
+ 36 a_{17} b_{7} 
+ 12 k M a_{17} b_{7} 
- 18 M^{2} a_{17} b_{7} 
\nonu \\
&&- M^{2} a_{3} b_{12} 
+ M^{2} a_{8} b_{12} 
- 6 M a_{17} b_{12} 
- 2 M^{2} a_{3} b_{13} 
+ 2 M^{2} a_{8} b_{13} 
- 12 M a_{17} b_{13} \Bigg) 
\delta^{a f} \delta^{c h} \partial J^{c} J^{f} J^{h} \nonu \\
&&- \frac{N}{M} \Bigg(8 a_{3} b_{7} 
+ 4 k M a_{3} b_{7} 
+ 2 M^{2} a_{3} b_{7} 
+ 2 M N \sqrt{\frac{M + N}{M N}} a_{7} b_{7} 
- 4 a_{8} b_{7} 
+ 4 M N a_{8} b_{7} 
- 2 a_{12} b_{7} 
\nonu \\
&&+ 24 k a_{17} b_{7} 
+ 24 M a_{17} b_{7} 
- 2 M a_{3} b_{12} 
+ 2 M a_{8} b_{12} 
- 12 a_{17} b_{12} 
- 3 M a_{3} b_{13} 
+ 2 M a_{8} b_{13} \nonu\\
&&- 18 a_{17} b_{13} \Bigg) 
\delta^{a c} \delta^{f h} \partial J^{c} J^{f} J^{h} \nonu \\
&&+ \frac{2}{M^{2}} \Bigg(-2 M a_{3} b_{12} 
- 4 a_{5} b_{12} 
+ M^{2} a_{5} b_{12} 
+ M a_{8} b_{12} 
+ 12 a_{17} b_{12} 
- 6 M^{2} a_{17} b_{12} 
- 2 M a_{3} b_{13} 
\nonu \\
&& - 8 a_{5} b_{13} 
 + 2 M^{2} a_{5} b_{13} 
+ 2 M a_{8} b_{13} 
+ 12 a_{17} b_{13} 
- 6 M^{2} a_{17} b_{13} \Bigg) 
\delta^{a c} \delta_{\bar{\imath} k} \delta_{\rho \bar{\tau}} \partial J^{c} J^{\bar{\imath}} J^{\rho} J^{\bar{\tau}} \nonu \\
&&+ \frac{N}{2} \Bigg(4 k M a_{5} b_{7} 
+ 4 M N a_{5} b_{7} 
+ 4 k a_{5} b_{12} 
+ 2 k M \sqrt{\frac{M + N}{M N}} a_{9} b_{12} 
+ 4 k a_{5} b_{13} 
- 4 M a_{5} b_{13} 
\nonu \\
&& - 2 N a_{5} b_{13} 
+ 4 k M \sqrt{\frac{M + N}{M N}} a_{9} b_{13} 
+ a_{12} b_{13} \Bigg) 
d^{a c f} \partial J^{c} \partial J^{f}\nonu\\
&&+ \frac{i}{2 M^{2}} \Bigg(16 M^{2} a_{2} b_{7} 
- 16 M^{2} N^{2} a_{2} b_{7} 
+ 16 M N a_{3} b_{7} 
+ 8 k M^{2} N a_{3} b_{7} 
+ 16 M^{3} N a_{3} b_{7} 
\nonu \\
&& - 16 M^{2} a_{4} b_{7} 
- 16 M N a_{4} b_{7} 
+ 48 k M N a_{5} b_{7} 
- 12 k M^{3} N a_{5} b_{7} 
- 24 k M^{2} N a_{8} b_{7} 
\nonu \\
&& - 16 M^{3} N a_{8} b_{7} 
- 64 M N \sqrt{\frac{M + N}{M N}} a_{9} b_{7} 
+ 16 M^{3} N \sqrt{\frac{M + N}{M N}} a_{9} b_{7} 
+ 4 M^{3} N a_{11} b_{7} 
\nonu \\
&& - 288 N a_{17} b_{7} 
+ 48 k M N a_{17} b_{7} 
+ 192 M^{2} N a_{17} b_{7} 
 + 4 M^{2} a_{1} b_{12} 
- 4 M^{2} N^{2} a_{1} b_{12} 
\nonu \\
&&
- 8 M^{2} N \sqrt{\frac{M + N}{M N}} a_{7} b_{12} 
- 8 M^{2} N a_{8} b_{12} 
- 2 M^{2} N a_{12} b_{12} 
+ 2 M^{2} N a_{13} b_{12} 
\nonu \\
&& + 8 M^{2} N a_{3} b_{13} 
+ 6 M^{2} N a_{11} b_{13} 
- M^{2} N a_{12} b_{13} 
+ 12 M N a_{17} b_{13} \Bigg) 
f^{a c f} \partial J^{c} \partial J^{f} 
\nonu \\
&&- \Bigg(2 M a_{1} b_{6} 
+ 2 N a_{1} b_{6} 
- 4 k M a_{2} b_{6} 
+ 4 k M a_{8} b_{6} 
+ 2 a_{12} b_{6} 
+ 2 k a_{1} b_{7} 
+ 2 M a_{1} b_{7} 
- 2 N a_{1} b_{7} 
\nonu \\
&&+ 4 k N a_{2} b_{7} 
- 2 a_{12} b_{7} 
- 3 a_{1} b_{12} 
- 2 M a_{2} b_{12} 
- 2 N a_{2} b_{12} 
+ 2 M a_{8} b_{12} 
- 4 a_{1} b_{13} 
- 4 M a_{2} b_{13} \nonu \\
&&- 2 N a_{2} b_{13} 
+ 4 M a_{8} b_{13} \Bigg) 
\delta^{\beta \gamma} \partial J^{\beta} J^{\gamma} J^{a}\nonu\\
&&+ \frac{1}{2 M} \Bigg(2 M^{2} N a_{13} b_{5} 
+ 4 M^{2} a_{16} b_{5} 
- 16 M N \sqrt{\frac{M + N}{M N}} a_{5} b_{7} 
+ 4 M^{3} N \sqrt{\frac{M + N}{M N}} a_{5} b_{7} 
\nonu \\
&& + 2 M^{2} N^{2} a_{7} b_{7} 
+ 8 M^{2} N \sqrt{\frac{M + N}{M N}} a_{8} b_{7} 
+ 4 M^{2} N \sqrt{\frac{M + N}{M N}} a_{11} b_{7} 
+ 2 M^{2} N \sqrt{\frac{M + N}{M N}} a_{13} b_{7} 
\nonu \\
&& - 24 M N \sqrt{\frac{M + N}{M N}} a_{17} b_{7} 
+ 2 M^{2} a_{1} b_{8} 
- 2 M^{2} N^{2} a_{1} b_{8} 
- 2 k M^{2} a_{2} b_{8} 
\nonu \\
&&
+ 2 k M^{2} N^{2} a_{2} b_{8} 
+ 2 k M N a_{3} b_{8} 
+ 2 k M^{3} N a_{3} b_{8} 
+ 2 k M^{2} a_{4} b_{8} 
+ 2 k M N a_{4} b_{8} 
\nonu \\
&& - 16 N^{2} a_{5} b_{8} 
+ 4 M^{2} N^{2} a_{5} b_{8} 
- 4 M^{2} N \sqrt{\frac{M + N}{M N}} a_{7} b_{8} 
- 4 M N^{2} \sqrt{\frac{M + N}{M N}} a_{7} b_{8} \nonu\\
&&+ 4 k M^{2} N^{2} \sqrt{\frac{M + N}{M N}} a_{7} b_{8} 
+ 4 k^{2} M^{2} N a_{8} b_{8} 
+ 2 k M^{3} N a_{8} b_{8} \
+ 4 M N^{2} a_{8} b_{8} 
+ 2 k M^{2} N^{2} a_{8} b_{8} \nonu\\
&&- 4 k M N a_{12} b_{8} 
+ 2 k M N a_{13} b_{8} 
- 36 k N a_{17} b_{8} 
+ 24 k M^{2} N a_{17} b_{8} 
- M \sqrt{\frac{M + N}{M N}} a_{1} b_{12} \nonu\\
&&+ M N^{2} \sqrt{\frac{M + N}{M N}} a_{1} b_{12} 
- M^{2} \sqrt{\frac{M + N}{M N}} a_{2} b_{12} 
+ M^{2} N^{2} \sqrt{\frac{M + N}{M N}} a_{2} b_{12} 
+ M N \sqrt{\frac{M + N}{M N}} a_{3} b_{12} \nonu\\
&&+ M^{3} N \sqrt{\frac{M + N}{M N}} a_{3} b_{12} 
+ M^{2} \sqrt{\frac{M + N}{M N}} a_{4} b_{12} 
+ M N \sqrt{\frac{M + N}{M N}} a_{4} b_{12} 
\nonu \\
&& + 2 k M^{2} N \sqrt{\frac{M + N}{M N}} a_{4} b_{12} 
+ 8 N \sqrt{\frac{M + N}{M N}} a_{5} b_{12} 
- 2 M^{2} N \sqrt{\frac{M + N}{M N}} a_{5} b_{12} 
\nonu \\
&& + 2 M a_{7} b_{12} 
+ 2 N a_{7} b_{12} 
+ 2 k M N a_{7} b_{12}\nonu\\
&&+ 2 M^{2} N a_{7} b_{12} 
+ 2 M N^{2} a_{7} b_{12} 
- 2 M N \sqrt{\frac{M + N}{M N}} a_{8} b_{12} 
+ 2 k M^{2} N \sqrt{\frac{M + N}{M N}} a_{8} b_{12} \nonu\\
&&+ 3 M^{3} N \sqrt{\frac{M + N}{M N}} a_{8} b_{12} 
+ 3 M^{2} N^{2} \sqrt{\frac{M + N}{M N}} a_{8} b_{12} 
- M N \sqrt{\frac{M + N}{M N}} a_{12} b_{12} 
\nonu \\
&& - 18 N \sqrt{\frac{M + N}{M N}} a_{17} b_{12} 
+  12 M^{2} N \sqrt{\frac{M + N}{M N}} a_{17} b_{12} \Bigg) 
\partial^{2} J^{a} J^{u(1)}  \nonu \\
&&- \frac{1}{4 M} \Bigg(6 M^{3} a_{11} b_{2} 
- 6 M^{2} N a_{13} b_{2} 
- 12 M^{2} a_{16} b_{2} 
- 4 k M N a_{3} b_{7} 
- 8 M^{2} N a_{3} b_{7} 
- 24 M N a_{5} b_{7} 
\nonu \\
&&- 8 k^{2} M N a_{5} b_{7} 
- 20 k M^{2} N a_{5} b_{7} 
+ 16 N^{2} a_{5} b_{7} 
+ 4 k M N^{2} a_{5} b_{7} 
- 8 M^{2} N^{2} a_{5} b_{7} 
\nonu \\
&& + 4 M N^{2} \sqrt{\frac{M + N}{M N}} a_{7} b_{7} 
+ 8 M^{2} N a_{8} b_{7} 
- 4 M N^{2} a_{8} b_{7} 
+ 8 k M N \sqrt{\frac{M + N}{M N}} a_{9} b_{7} 
\nonu \\
&&
+ 16 M^{2} N \sqrt{\frac{M + N}{M N}} a_{9} b_{7} 
+ 2 M^{2} N a_{11} b_{7} 
+ 4 k M N a_{12} b_{7} 
+ 2 M^{2} N a_{12} b_{7} 
\nonu \\
&& - 2 k M N a_{13} b_{7} 
+ 72 k N a_{17} b_{7} 
+ 132 M N a_{17} b_{7} 
- 12 k M^{2} N a_{17} b_{7} 
\nonu \\
&&- 24 M^{3} N a_{17} b_{7} 
- M a_{1} b_{12} 
+ M N^{2} a_{1} b_{12} 
+ 2 M N a_{3} b_{12} 
+ 8 N a_{5} b_{12} 
+ 6 M^{2} N a_{5} b_{12} 
\nonu \\
&& + 2 M N^{2} a_{5} b_{12} 
+ 2 M N \sqrt{\frac{M + N}{M N}} a_{7} b_{12} 
- 2 M N a_{8} b_{12} 
- 4 M N \sqrt{\frac{M + N}{M N}} a_{9} b_{12} 
\nonu \\
&&
- 4 k M^{2} N \sqrt{\frac{M + N}{M N}} a_{9} b_{12} 
- M N a_{12} b_{12} 
- 36 N a_{17} b_{12} 
+ 6 M^{2} N a_{17} b_{12} 
+ M a_{1} b_{13} 
\nonu \\
&& - M N^{2} a_{1} b_{13} 
- 2 M N a_{3} b_{13} 
- 8 N a_{5} b_{13} 
- 12 k M N a_{5} b_{13} 
+ 12 M^{2} N a_{5} b_{13} 
\nonu \\
&& + 2 M N^{2} a_{5} b_{13} 
- 2 M N \sqrt{\frac{M + N}{M N}} a_{7} b_{13} 
+ 2 M N a_{8} b_{13} 
+ 4 M N \sqrt{\frac{M + N}{M N}} a_{9} b_{13} 
\nonu \\
&&- 8 k M^{2} N \sqrt{\frac{M + N}{M N}} a_{9} b_{13} 
- 2 M N a_{12} b_{13} 
+ M N a_{13} b_{13} 
+ 36 N a_{17} b_{13} 
- 6 M^{2} N a_{17} b_{13} \Bigg)
\nonu \\
&& \times 
d^{a c f} \partial^{2} J^{c} J^{f} \nonu\\
&&- \frac{i}{4 M} \Bigg(-24 M N a_{5} b_{2} 
+ 6 M^{3} N a_{5} b_{2} 
+ 20 k M N a_{3} b_{7} 
+ 8 M^{2} N a_{3} b_{7} 
- 16 k N a_{5} b_{7} 
\nonu \\
&& + 4 k M^{2} N a_{5} b_{7} 
- 8 N^{2} a_{5} b_{7} 
+ 2 M^{2} N^{2} a_{5} b_{7} 
- 8 M N^{2} \sqrt{\frac{M + N}{M N}} a_{7} b_{7} 
\nonu \\
&&
+ 8 k M N a_{8} b_{7} 
+ 8 M^{2} N a_{8} b_{7} 
- 4 M N^{2} a_{8} b_{7} 
\nonu \\
&&+ 20 k M N a_{11} b_{7} 
+ 8 M^{2} N a_{11} b_{7} 
- 4 k M N a_{12} b_{7} 
- 2 M^{2} N a_{12} b_{7} 
+ 2 k M N a_{13} b_{7} 
+ 2 M N^{2} a_{13} b_{7} 
\nonu \\
&&+ 4 M N a_{16} b_{7} 
+ 12 M N a_{17} b_{7} 
+ M a_{1} b_{12} 
- M N^{2} a_{1} b_{12} 
- 10 M N a_{3} b_{12} 
+ 8 N a_{5} b_{12} 
\nonu \\
&& - 2 M^{2} N a_{5} b_{12} 
- 2 M N \sqrt{\frac{M + N}{M N}} a_{7} b_{12} 
- 6 M N a_{8} b_{12} 
- 10 M N a_{11} b_{12} 
\nonu \\
&&
+ M N a_{12} b_{12} 
- M a_{1} b_{13} 
+ M N^{2} a_{1} b_{13} \nonu\\
&&- 10 M N a_{3} b_{13} 
+ 16 N a_{5} b_{13} 
- 4 M^{2} N a_{5} b_{13} 
+ 2 M N \sqrt{\frac{M + N}{M N}} a_{7} b_{13} 
- 6 M N a_{8} b_{13} 
\nonu \\
&& - 10 M N a_{11} b_{13}
+ 2 M N a_{12} b_{13} 
- M N a_{13} b_{13} \Bigg) 
f^{a c f} \partial^{2} J^{c} J^{f} \nonu\\
&&+ \Bigg(4 M a_{7} b_{5} 
+ 4 M a_{4} b_{7} 
+ 8 M \sqrt{\frac{M + N}{M N}} a_{7} b_{7} 
+ 8 k a_{7} b_{8} 
+ 8 M a_{7} b_{8} 
+ 12 N a_{7} b_{8} 
\nonu \\
&&- 2 \sqrt{\frac{M + N}{M N}} a_{12} b_{8} 
+ 2 \sqrt{\frac{M + N}{M N}} a_{13} b_{8} 
- a_{4} b_{12} 
- 2 \sqrt{\frac{M + N}{M N}} a_{7} b_{12} 
+ a_{4} b_{13} \nonu\\
&&+ 2 \sqrt{\frac{M + N}{M N}} a_{7} b_{13} \Bigg) 
\delta_{\rho \bar{\tau}} t^{a}_{k \bar{\imath}} J^{u(1)} J^{u(1)} J^{(\rho \bar{\imath})} J^{(\bar{\tau} k)} \nonu\\
&&+ \frac{1}{M N} \Bigg(2 M^{2} N a_{13} b_{5} 
- 16 M N \sqrt{\frac{M + N}{M N}} a_{5} b_{7} 
+ 4 M^{3} N \sqrt{\frac{M + N}{M N}} a_{5} b_{7} 
+ 4 k M^{2} N a_{7} b_{7} 
\nonu \\
&& + 8 M^{2} N^{2} a_{7} b_{7} 
+ 8 M^{2} N \sqrt{\frac{M + N}{M N}} a_{8} b_{7} 
- 2 M^{2} N \sqrt{\frac{M + N}{M N}} a_{12} b_{7} 
+ 2 M^{2} N \sqrt{\frac{M + N}{M N}} a_{13} b_{7} 
\nonu \\
&& + 4 k M^{2} N^{2} \sqrt{\frac{M + N}{M N}} a_{7} b_{8} 
- 4 k M N a_{12} b_{8} 
- 2 M N^{2} a_{12} b_{8} 
\nonu \\
&&
+ 2 k M N a_{13} b_{8} 
- M \sqrt{\frac{M + N}{M N}} a_{1} b_{12} 
+ M N^{2} \sqrt{\frac{M + N}{M N}} a_{1} b_{12} 
\nonu \\
&& + 8 N \sqrt{\frac{M + N}{M N}} a_{5} b_{12} 
- 2 M^{2} N \sqrt{\frac{M + N}{M N}} a_{5} b_{12} 
+ 2 M a_{7} b_{12} 
+ 2 N a_{7} b_{12} 
+ 2 M^{2} N a_{7} b_{12} 
\nonu \\
&&+ 2 M N^{2} a_{7} b_{12} 
- 2 M N \sqrt{\frac{M + N}{M N}} a_{8} b_{12} 
+ M \sqrt{\frac{M + N}{M N}} a_{1} b_{13} 
- M N^{2} \sqrt{\frac{M + N}{M N}} a_{1} b_{13} 
\nonu \\
&&- 8 N \sqrt{\frac{M + N}{M N}} a_{5} b_{13} 
+ 2 M^{2} N \sqrt{\frac{M + N}{M N}} a_{5} b_{13} 
- 2 M a_{7} b_{13} 
- 2 N a_{7} b_{13} 
+ 6 k M N a_{7} b_{13} 
\nonu \\
&&+ 4 M^{2} N a_{7} b_{13} 
+ 8 M N^{2} a_{7} b_{13} 
+ 2 M N \sqrt{\frac{M + N}{M N}} a_{8} b_{13} 
- 2 M N \sqrt{\frac{M + N}{M N}} a_{12} b_{13} \nonu\\
&&+ 2 M N \sqrt{\frac{M + N}{M N}} a_{13} b_{13} \Bigg) 
 \delta_{\rho \bar{\tau}} t^{a}_{k \bar{\imath}} J^{u(1)} J^{(\rho \bar{\imath})} \partial J^{(\bar{\tau} k)} \nonu\\
 &&- \frac{1}{M N} \Bigg(-2 M^{2} N a_{12} b_{5} 
- 16 M N \sqrt{\frac{M + N}{M N}} a_{5} b_{7} 
+ 4 M^{3} N \sqrt{\frac{M + N}{M N}} a_{5} b_{7} 
+ 4 k M^{2} N a_{7} b_{7} \nonu \\
&&+ 8 M^{2} N^{2} a_{7} b_{7} 
+ 8 M^{2} N \sqrt{\frac{M + N}{M N}} a_{8} b_{7} 
- 2 M^{2} N \sqrt{\frac{M + N}{M N}} a_{12} b_{7} 
+ 2 M^{2} N \sqrt{\frac{M + N}{M N}} a_{13} b_{7} \nonu\\
&&- 4 k M^{2} N^{2} \sqrt{\frac{M + N}{M N}} a_{7} b_{8} 
- 2 k M N a_{12} b_{8} 
- 4 M^{2} N a_{12} b_{8} 
- 4 M N^{2} a_{12} b_{8} 
+ 4 k M N a_{13} b_{8}  \nonu\\
&&+ 4 M^{2} N a_{13} b_{8} 
+ 6 M N^{2} a_{13} b_{8} 
- M \sqrt{\frac{M + N}{M N}} a_{1} b_{12} 
+ M N^{2} \sqrt{\frac{M + N}{M N}} a_{1} b_{12} 
\nonu \\
&& + 8 N \sqrt{\frac{M + N}{M N}} a_{5} b_{12} 
- 2 M^{2} N \sqrt{\frac{M + N}{M N}} a_{5} b_{12} 
+ 2 M a_{7} b_{12} 
+ 2 N a_{7} b_{12} 
\nonu \\
&& - 6 k M N a_{7} b_{12} 
- 4 M^{2} N a_{7} b_{12} 
- 8 M N^{2} a_{7} b_{12} \nonu\\
&&- 2 M N \sqrt{\frac{M + N}{M N}} a_{8} b_{12} 
+ 2 M N \sqrt{\frac{M + N}{M N}} a_{12} b_{12} 
- 2 M N \sqrt{\frac{M + N}{M N}} a_{13} b_{12} 
\nonu \\
&& + M \sqrt{\frac{M + N}{M N}} a_{1} b_{13} 
- M N^{2} \sqrt{\frac{M + N}{M N}} a_{1} b_{13} 
\nonu \\
&&
- 8 N \sqrt{\frac{M + N}{M N}} a_{5} b_{13} 
+ 2 M^{2} N \sqrt{\frac{M + N}{M N}} a_{5} b_{13} 
- 2 M a_{7} b_{13} \nonu\\
&&- 2 N a_{7} b_{13} 
- 2 M^{2} N a_{7} b_{13} 
- 2 M N^{2} a_{7} b_{13} 
+ 2 M N \sqrt{\frac{M + N}{M N}} a_{8} b_{13} \Bigg) 
 \delta_{\rho \bar{\tau}} t^{a}_{k \bar{\imath}} J^{u(1)} \partial J^{(\rho \bar{\imath})} J^{(\bar{\tau} k)}\nonu\\
 &&+ \frac{1}{M^{2}} \Bigg(-4 M a_{3} b_{7} 
+ 4 M^{3} a_{3} b_{7} 
- 4 M^{2} N \sqrt{\frac{M + N}{M N}} a_{7} b_{7} 
+ 4 M a_{8} b_{7} 
- 72 a_{17} b_{7} 
+ 36 M^{2} a_{17} b_{7} \nonu\\
&&- M^{2} a_{3} b_{12} 
- 6 M a_{17} b_{12} 
+ M^{2} a_{3} b_{13} 
+ 6 M a_{17} b_{13} \Bigg) 
\delta^{c f} \delta_{\rho \bar{\tau}} t^{a}_{k \bar{\imath}} 
J^{c} J^{f} J^{(\rho \bar{\imath})} J^{(\bar{\tau} k)} \nonu \\
&&+ \frac{1}{N} \Bigg(4 k a_{1} b_{6} 
+ 2 M a_{1} b_{6} 
+ 4 N a_{1} b_{6} 
+ 4 M N \sqrt{\frac{M + N}{M N}} a_{7} b_{6} 
- 2 a_{12} b_{6} 
+ 2 a_{13} b_{6} 
+ 4 M a_{1} b_{7} \nonu\\
&&+ 4 M N a_{2} b_{7} 
- a_{1} b_{12} 
- N a_{2} b_{12} 
+ a_{1} b_{13} 
+ N a_{2} b_{13} \Bigg) 
\delta^{\beta \gamma} \delta_{\rho \bar{\tau}} t^{a}_{k \bar{\imath}} 
J^{\beta} J^{\gamma} J^{(\rho \bar{\imath})} J^{(\bar{\tau} k)} \nonu \\
&&+ a_{1} b_{12} 
\delta_{\bar{g} y} \delta_{\bar{\nu} \sigma} \delta_{\rho \bar{\tau}} t^{a}_{k \bar{\imath}}
J^{(\rho \bar{\imath})} J^{(\sigma \bar{g})} J^{(\bar{\nu} k)} J^{(\bar{\tau} y)} 
- a_{1} b_{13} 
\delta_{\bar{g} y} \delta_{\bar{\nu} \sigma} \delta_{\rho \bar{\tau}} t^{a}_{k \bar{\imath}}
J^{(\rho \bar{\imath})} J^{(\sigma \bar{g})} J^{(\bar{\nu} k)} J^{(\bar{\tau} y)} 
\nonu \\
&&+ 2 a_{8} \Bigg(4 M b_{7} 
- b_{12} 
+ b_{13} \Bigg) 
\delta_{\bar{g} y} \delta_{\bar{\nu} \sigma} \delta_{\rho \bar{\tau}} t^{a}_{k \bar{\imath}}
J^{(\rho \bar{\imath})} J^{(\sigma \bar{g})} J^{(\bar{\nu} y)} J^{(\bar{\tau} k)}  \nonu \\
&&+ \frac{1}{2 M N} \Bigg(
-16 k M N a_{5} b_{7} 
+ 4 k M^{3} N a_{5} b_{7} 
- 32 M N^{2} a_{5} b_{7} 
+ 8 M^{3} N^{2} a_{5} b_{7}
\nonu \\
&&
- 8 M^{2} N^{2} \sqrt{\frac{M + N}{M N}} a_{7} b_{7} 
+ 8 k M^{2} N a_{8} b_{7} 
+ 8 M^{3} N a_{8} b_{7} 
+ 8 M^{2} N^{2} a_{8} b_{7} 
\nonu \\
&&
- 4 k M^{2} N a_{12} b_{7} 
- 2 M^{2} N^{2} a_{12} b_{7} 
+ 4 k M^{2} N a_{13} b_{7} \nonu\\
&&+ 6 M^{2} N^{2} a_{13} b_{7} 
- k M a_{1} b_{12} 
+ k M N^{2} a_{1} b_{12} 
+ 8 k N a_{5} b_{12} 
- 2 k M^{2} N a_{5} b_{12} 
+ 8 N^{2} a_{5} b_{12} \nonu\\
&&- 2 M^{2} N^{2} a_{5} b_{12} 
+ 2 k M N \sqrt{\frac{M + N}{M N}} a_{7} b_{12} 
+ 2 M N^{2} \sqrt{\frac{M + N}{M N}} a_{7} b_{12} 
\nonu \\
&& + 2 k M^{2} N^{2} \sqrt{\frac{M + N}{M N}} a_{7} b_{12} 
- 2 k M N a_{8} b_{12} 
- 2 M^{2} N a_{8} b_{12} 
- 2 M N^{2} a_{8} b_{12} 
\nonu \\
&&
+ k M a_{1} b_{13} 
- k M N^{2} a_{1} b_{13} 
- 8 k N a_{5} b_{13} + 2 k M^{2} N a_{5} b_{13} 
- 8 N^{2} a_{5} b_{13}
\nonu \\
&&
+ 2 M^{2} N^{2} a_{5} b_{13} 
- 2 k M N \sqrt{\frac{M + N}{M N}} a_{7} b_{13} 
- 2 M N^{2} \sqrt{\frac{M + N}{M N}} a_{7} b_{13} 
\nonu \\
&&
+ 4 k M^{2} N^{2} \sqrt{\frac{M + N}{M N}} a_{7} b_{13} 
+ 2 k M N a_{8} b_{13} 
+2 M^{2} N a_{8} b_{13} 
+ 2 M N^{2} a_{8} b_{13} 
\nonu \\
&&
- 6 k M N a_{12} b_{13} 
- 3 M N^{2} a_{12} b_{13} 
+ 4 k M N a_{13} b_{13} 
+ M N^{2} a_{13} b_{13}\Bigg) 
\delta_{\rho \bar{\tau}} t^{a}_{k \bar{\imath}} 
J^{(\rho \bar{\imath})} \partial^2 J^{(\bar{\tau} k)}  \nonu \\
&&- \frac{1}{M N} \Bigg(
-16 M N \sqrt{\frac{M + N}{M N}} a_{5} b_{7} 
+ 4 M^{3} N \sqrt{\frac{M + N}{M N}} a_{5} b_{7} 
+ 8 M^{2} N \sqrt{\frac{M + N}{M N}} a_{8} b_{7} \nonu \\
&&+ 4 M^{3} N^{2} \sqrt{\frac{M + N}{M N}} a_{8} b_{7} 
- M \sqrt{\frac{M + N}{M N}} a_{1} b_{12} 
+ M N^{2} \sqrt{\frac{M + N}{M N}} a_{1} b_{12} 
+ 8 N \sqrt{\frac{M + N}{M N}} a_{5} b_{12} \nonu \\
&&- 2 M^{2} N \sqrt{\frac{M + N}{M N}} a_{5} b_{12} 
+ 2 M a_{7} b_{12} 
+ 2 N a_{7} b_{12} 
- 4 k M N a_{7} b_{12} 
- 2 M^{2} N a_{7} b_{12} 
\nonu \\
&& - 4 M N^{2} a_{7} b_{12} 
- 2 M N \sqrt{\frac{M + N}{M N}} a_{8} b_{12} 
\nonu \\
&&
- M^{2} N^{2} \sqrt{\frac{M + N}{M N}} a_{8} b_{12} 
- M N \sqrt{\frac{M + N}{M N}} a_{13} b_{12} 
+ M \sqrt{\frac{M + N}{M N}} a_{1} b_{13} \nonu \\
&&- M N^{2} \sqrt{\frac{M + N}{M N}} a_{1} b_{13} 
- 8 N \sqrt{\frac{M + N}{M N}} a_{5} b_{13} 
+ 2 M^{2} N \sqrt{\frac{M + N}{M N}} a_{5} b_{13} 
- 2 M a_{7} b_{13} 
\nonu \\
&& - 2 N a_{7} b_{13} - 4 k M N a_{7} b_{13} 
- 4 M^{2} N a_{7} b_{13} 
- 6 M N^{2} a_{7} b_{13} 
+ 2 M N \sqrt{\frac{M + N}{M N}} a_{8} b_{13} 
\nonu \\
&& + M^{2} N^{2} \sqrt{\frac{M + N}{M N}} a_{8} b_{13} 
+ M N \sqrt{\frac{M + N}{M N}} a_{12} b_{13} \Bigg) 
 \delta_{\rho \bar{\tau}} t^{a}_{k \bar{\imath}} 
\partial J^{u(1)} J^{(\rho \bar{\imath})} J^{(\bar{\tau} k)}  \nonu \\
&&- \frac{1}{M N} \Bigg(
8 M^{2} N^{2} \sqrt{\frac{M + N}{M N}} a_{7} b_{7} 
+ 8 M^{2} N^{2} a_{8} b_{7} 
- 4 M^{2} N^{2} a_{12} b_{7} 
- M^{2} a_{1} b_{12} 
+ M^{2} N^{2} a_{1} b_{12} \nonu \\
&&+ 8 N^{2} a_{5} b_{12} 
- 2 M^{2} N^{2} a_{5} b_{12} 
+ 2 M^{2} N \sqrt{\frac{M + N}{M N}} a_{7} b_{12} 
+ 2 M N^{2} \sqrt{\frac{M + N}{M N}} a_{7} b_{12} \nonu \\
&&- 2 k M^{2} N^{2} \sqrt{\frac{M + N}{M N}} a_{7} b_{12} 
- 2 M N^{2} a_{8} b_{12} 
+ 2 k M N a_{12} b_{12} 
+ M N^{2} a_{12} b_{12} 
- k M N a_{13} b_{12} \nonu \\
&&+ M^{2} a_{1} b_{13} 
- M^{2} N^{2} a_{1} b_{13} 
- 8 N^{2} a_{5} b_{13} 
+ 2 M^{2} N^{2} a_{5} b_{13} 
- 2 M^{2} N \sqrt{\frac{M + N}{M N}} a_{7} b_{13} \nonu \\
&&- 2 M N^{2} \sqrt{\frac{M + N}{M N}} a_{7} b_{13}
- 4 k M^{2} N^{2} \sqrt{\frac{M + N}{M N}} a_{7} b_{13} 
+ 2 M N^{2} a_{8} b_{13} 
- k M N a_{12} b_{13} 
\nonu \\
&& - 3 M^{2} N a_{12} b_{13} - 3 M N^{2} a_{12} b_{13} 
+ 2 k M N a_{13} b_{13} 
\nonu \\
&&
+ 3 M^{2} N a_{13} b_{13} 
+ 4 M N^{2} a_{13} b_{13} \Bigg) 
\delta_{\rho \bar{\tau}} t^{a}_{k \bar{\imath}} 
\partial J^{(\rho \bar{\imath})} \partial J^{(\bar{\tau} k)}  \nonu \\
&&+ \frac{1}{2 M N} \Bigg(
-16 k M N a_{5} b_{7} 
+ 4 k M^{3} N a_{5} b_{7} 
- 64 M N^{2} a_{5} b_{7} 
+ 16 M^{3} N^{2} a_{5} b_{7} 
\nonu \\
&& - 8 M^{2} N^{2} \sqrt{\frac{M + N}{M N}} a_{7} b_{7} 
+ 8 k M^{2} N a_{8} b_{7} 
+ 16 M^{3} N a_{8} b_{7} 
+ 24 M^{2} N^{2} a_{8} b_{7} 
- 4 k M^{2} N a_{12} b_{7}
\nonu \\
&& - 2 M^{2} N^{2} a_{12} b_{7} 
+ 4 k M^{2} N a_{13} b_{7} 
+ 6 M^{2} N^{2} a_{13} b_{7} 
- k M a_{1} b_{12} 
- M^{2} a_{1} b_{12} 
\nonu \\
&&
+ k M N^{2} a_{1} b_{12} 
+ M^{2} N^{2} a_{1} b_{12} 
+ 8 k N a_{5} b_{12} 
- 2 k M^{2} N a_{5} b_{12} \nonu \\
&&+ 24 N^{2} a_{5} b_{12} 
- 6 M^{2} N^{2} a_{5} b_{12} 
+ 2 k M N \sqrt{\frac{M + N}{M N}} a_{7} b_{12} 
+ 2 M^{2} N \sqrt{\frac{M + N}{M N}} a_{7} b_{12} \nonu \\
&&+ 6 M N^{2} \sqrt{\frac{M + N}{M N}} a_{7} b_{12} 
+ 2 k M^{2} N^{2} \sqrt{\frac{M + N}{M N}} a_{7} b_{12} 
- 2 k M N a_{8} b_{12} 
- 4 M^{2} N a_{8} b_{12} 
\nonu \\
&& - 6 M N^{2} a_{8} b_{12}
+ 4 k M N a_{12} b_{12} 
+ 3 M^{2} N a_{12} b_{12}
+ 4 M N^{2} a_{12} b_{12} 
- 6 k M N a_{13} b_{12}
\nonu \\
&&
- 3 M^{2} N a_{13} b_{12} 
- 6 M N^{2} a_{13} b_{12} 
+ k M a_{1} b_{13} 
+ M^{2} a_{1} b_{13}
- k M N^{2} a_{1} b_{13}
\nonu \\
&& - M^{2} N^{2} a_{1} b_{13} 
- 8 k N a_{5} b_{13} 
+ 2 k M^{2} N a_{5} b_{13} - 24 N^{2} a_{5} b_{13} 
+ 6 M^{2} N^{2} a_{5} b_{13} 
\nonu \\
&&
- 2 k M N \sqrt{\frac{M + N}{M N}} a_{7} b_{13} 
- 2 M^{2} N \sqrt{\frac{M + N}{M N}} a_{7} b_{13} \nonu \\
&&- 6 M N^{2} \sqrt{\frac{M + N}{M N}} a_{7} b_{13} 
+ 4 k M^{2} N^{2} \sqrt{\frac{M + N}{M N}} a_{7} b_{13} 
+ 2 k M N a_{8} b_{13} 
+ 4 M^{2} N a_{8} b_{13} \nonu \\
&&+ 6 M N^{2} a_{8} b_{13} 
+ 3 M^{2} N a_{12} b_{13} 
+ 3 M N^{2} a_{12} b_{13} 
- 3 M^{2} N a_{13} b_{13} 
- 3 M N^{2} a_{13} b_{13} \Bigg) 
\nonu \\
&& \times \delta_{\rho \bar{\tau}} t^{a}_{k \bar{\imath}} 
\partial^{2} J^{(\rho \bar{\imath})} J^{(\bar{\tau} k)}  \nonu \\
&&+ \frac{1}{N} \Bigg(
-a_{1} 
+ 2 N \sqrt{\frac{M + N}{M N}} a_{7} \Bigg) b_{12} 
\delta_{\bar{\imath} k} \delta_{\bar{\nu} \sigma} \delta_{\rho \bar{\tau}}  t^{a}_{y \bar{g}}
J^{(\rho \bar{\imath})} J^{(\sigma \bar{g})} J^{(\bar{\nu} y)} J^{(\bar{\tau} k)}  \nonu \\
&&- \frac{1}{N} \Bigg(-a_{1} + 2 N \sqrt{\frac{M + N}{M N}} a_{7} \Bigg) b_{13} 
\delta_{\bar{\imath} k} \delta_{\bar{\nu} \sigma} \delta_{\rho \bar{\tau}}  t^{a}_{y \bar{g}}
J^{(\rho \bar{\imath})} J^{(\sigma \bar{g})} J^{(\bar{\nu} y)} J^{(\bar{\tau} k)}  \nonu \\
&&+ \Bigg(
6 M a_{7} b_{2} 
+ 4 M a_{5} b_{5} 
+ 8 M \sqrt{\frac{M + N}{M N}} a_{5} b_{7} 
- 4 N \sqrt{\frac{M + N}{M N}} a_{5} b_{7} 
+ 4 k a_{7} b_{7} \nonu \\
&&+ 4 N a_{7} b_{7} 
+ 6 M a_{9} b_{7} 
- \sqrt{\frac{M + N}{M N}} a_{12} b_{7} 
+ \sqrt{\frac{M + N}{M N}} a_{13} b_{7} 
+ 8 k a_{5} b_{8} 
+ 4 M a_{5} b_{8} \nonu \\
&&+ 12 N a_{5} b_{8} 
+ a_{12} b_{8} 
- a_{13} b_{8} 
- 2 \sqrt{\frac{M + N}{M N}} a_{5} b_{12} 
+ a_{7} b_{12} 
- 2 a_{9} b_{12} 
+ 2 \sqrt{\frac{M + N}{M N}} a_{5} b_{13} \nonu \\
&&- a_{7} b_{13} 
+ 2 a_{9} b_{13}
\Bigg) 
d^{a c f} \delta_{\rho \bar{\tau}} t^{f}_{k \bar{\imath}} 
J^{c} J^{u(1)} J^{(\rho \bar{\imath})} J^{(\bar{\tau} k)}  \nonu \\
&&- i \Bigg( 
\sqrt{\frac{M + N}{M N}} a_{12} b_{7} 
+ \sqrt{\frac{M + N}{M N}} a_{13} b_{7} 
- a_{12} b_{8} 
- a_{13} b_{8} 
+ a_{7} b_{12} 
+ a_{7} b_{13} 
\Bigg) 
f^{a c f} \delta_{\rho \bar{\tau}} t^{f}_{k \bar{\imath}} \nonu \\
&& \times J^{c} J^{u(1)} J^{(\rho \bar{\imath})} J^{(\bar{\tau} k)}  \nonu \\
&& - \frac{4}{M^{2}} \Bigg( 
M a_{3} - M a_{8} - 18 a_{17} + 6 M^{2} a_{17} 
\Bigg) b_{7} \delta_{a c} \delta_{f h} \delta_{\rho \bar{\tau}} 
t^{h}_{k \bar{\imath}} J^{c} J^{f} J^{(\rho \bar{\imath})} J^{(\bar{\tau} k)} \nonu \\
&&+ \frac{1}{2 M N} \Bigg(
6 M^{2} N a_{13} b_{2} 
+ 16 M N a_{5} b_{7} 
+ 8 k M^{2} N a_{5} b_{7} 
- 8 k M N^{2} a_{5} b_{7} 
+ 8 M^{2} N^{2} a_{5} b_{7} 
\nonu \\
&& - 8 M^{2} N a_{8} b_{7} - 4 k M N a_{12} b_{7} 
- 2 M N^{2} a_{12} b_{7} 
+ 2 k M N a_{13} b_{7} 
+ M a_{1} b_{12} 
- M N^{2} a_{1} b_{12} 
\nonu \\
&&
- 8 N a_{5} b_{12} 
+ 4 M N^{2} a_{5} b_{12} 
- 2 M N \sqrt{\frac{M + N}{M N}} a_{7} b_{12} 
+ 2 M N a_{8} b_{12} 
\nonu \\
&&
- M a_{1} b_{13} 
+ M N^{2} a_{1} b_{13} 
+ 8 N a_{5} b_{13} 
+ 12 k M N a_{5} b_{13}  \nonu \\
&&+ 4 M^{2} N a_{5} b_{13}
+ 16 M N^{2} a_{5} b_{13} 
+ 2 M N \sqrt{\frac{M + N}{M N}} a_{7} b_{13} 
- 2 M N a_{8} b_{13} 
+ 2 M N a_{12} b_{13}  \nonu \\
&&- 2 M N a_{13} b_{13} 
\Bigg) d^{a c f} \delta_{\rho \bar{\tau}} 
t^{f}_{k \bar{\imath}} J^{c} J^{(\rho \bar{\imath})} \partial J^{(\bar{\tau} k)} \nonu \\
&&- \frac{i}{2 M N} \Bigg( 
32 N^{2} a_{5} b_{7} 
- 8 M^{2} N^{2} a_{5} b_{7} 
+ 16 M N^{2} \sqrt{\frac{M + N}{M N}} a_{7} b_{7} 
- 8 M^{2} N a_{8} b_{7} 
+ 4 k M N a_{12} b_{7} \nonu \\
&&+ 2 M N^{2} a_{12} b_{7} 
- 2 k M N a_{13} b_{7} 
- 2 M N^{2} a_{13} b_{7} 
- M a_{1} b_{12} 
+ M N^{2} a_{1} b_{12} 
\nonu \\
&& + 2 M N \sqrt{\frac{M + N}{M N}} a_{7} b_{12} 
+ 2 M N a_{8} b_{12} 
+ M a_{1} b_{13} 
- M N^{2} a_{1} b_{13}
\nonu \\
&&
- 2 M N \sqrt{\frac{M + N}{M N}} a_{7} b_{13} 
- 2 M N a_{8} b_{13} 
- 2 M N a_{12} b_{13} 
\Bigg) f^{a c f} \delta_{\rho \bar{\tau}} 
t^{f}_{k \bar{\imath}} J^{c} J^{(\rho \bar{\imath})} \partial J^{(\bar{\tau} k)} \nonu \\
&&- \frac{1}{2 M N} \Bigg(
-6 M^{2} N a_{12} b_{2} 
+ 16 M N a_{5} b_{7} 
+ 8 k M^{2} N a_{5} b_{7} 
+ 8 k M N^{2} a_{5} b_{7} 
+ 24 M^{2} N^{2} a_{5} b_{7} \nonu \\
&&- 8 M^{2} N a_{8} b_{7} 
- 2 k M N a_{12} b_{7} 
- 4 M N^{2} a_{12} b_{7} 
+ 4 k M N a_{13} b_{7} 
+ 6 M N^{2} a_{13} b_{7} 
+ M a_{1} b_{12} 
\nonu \\
&& - M N^{2} a_{1} b_{12} 
- 8 N a_{5} b_{12} 
- 12 k M N a_{5} b_{12} 
- 4 M^{2} N a_{5} b_{12} 
- 16 M N^{2} a_{5} b_{12} 
\nonu \\
&&
- 2 M N \sqrt{\frac{M + N}{M N}} a_{7} b_{12} 
+ 2 M N a_{8} b_{12} 
- 2 M N a_{12} b_{12} 
+ 2 M N a_{13} b_{12}
\nonu \\
&&
- M a_{1} b_{13} 
+ M N^{2} a_{1} b_{13} 
+ 8 N a_{5} b_{13} 
- 4 M N^{2} a_{5} b_{13} \nonu\\
&&+ 2 M N \sqrt{\frac{M + N}{M N}} a_{7} b_{13} 
- 2 M N a_{8} b_{13} 
\Bigg) d^{a c f} \delta_{\rho \bar{\tau}} 
t^{f}_{k \bar{\imath}} J^{c} \partial J^{(\rho \bar{\imath})} J^{(\bar{\tau} k)} \nonu \\
&&- \frac{i}{2 M N} \Bigg(
32 N^{2} a_{5} b_{7} 
- 8 M^{2} N^{2} a_{5} b_{7} 
+ 16 M N^{2} \sqrt{\frac{M + N}{M N}} a_{7} b_{7} 
- 8 M^{2} N a_{8} b_{7} 
+ 4 k M N a_{12} b_{7}  \nonu \\
&&+ 2 M N^{2} a_{12} b_{7}
+ 4 k M N a_{13} b_{7} 
+ 6 M N^{2} a_{13} b_{7} 
+ M a_{1} b_{12} 
- M N^{2} a_{1} b_{12} 
\nonu \\
&& - 2 M N \sqrt{\frac{M + N}{M N}} a_{7} b_{12} 
- 2 M N a_{8} b_{12} 
+ 2 M N a_{13} b_{12}
\nonu \\
&&
+ M a_{1} b_{13} 
- M N^{2} a_{1} b_{13} 
+ 2 M N \sqrt{\frac{M + N}{M N}} a_{7} b_{13} 
+ 2 M N a_{8} b_{13}
\Bigg) f^{a c f} \delta_{\rho \bar{\tau}} t^{f}_{k \bar{\imath}} 
J^{c} \partial J^{(\rho \bar{\imath})} J^{(\bar{\tau} k)}  \nonu \\
&&+ \frac{1}{2 M N} \Bigg(
-16 M N a_{5} b_{7} 
+ 8 M^{2} N^{2} a_{5} b_{7} 
+ 8 M^{2} N a_{8} b_{7} 
+ M a_{1} b_{12} 
- M N^{2} a_{1} b_{12} 
+ 8 N a_{5} b_{12} \nonu \\
&&+ 8 k M N a_{5} b_{12} 
+ 4 M^{2} N a_{5} b_{12} 
+ 6 M N^{2} a_{5} b_{12} 
+ 2 M N \sqrt{\frac{M + N}{M N}} a_{7} b_{12} 
- 2 M N a_{8} b_{12} 
\nonu \\
&& - M N a_{13} b_{12} + M a_{1} b_{13} 
- M N^{2} a_{1} b_{13} 
\nonu \\
&&
- 8 N a_{5} b_{13} 
+ 8 k M N a_{5} b_{13} 
+ 4 M^{2} N a_{5} b_{13} 
+ 14 M N^{2} a_{5} b_{13} \nonu \\
&&- 2 M N \sqrt{\frac{M + N}{M N}} a_{7} b_{13} 
+ 2 M N a_{8} b_{13} 
+ M N a_{12} b_{13} 
\Bigg) d^{a c f} \delta_{\rho \bar{\tau}} t^{f}_{k \bar{\imath}} 
\partial J^{c} J^{(\rho \bar{\imath})} J^{(\bar{\tau} k)}  \nonu \\
&&- \frac{i}{2 M N} \Bigg(
8 M^{2} N a_{3} b_{7} 
- 32 k N a_{5} b_{7} 
+ 8 k M^{2} N a_{5} b_{7} 
- 32 N^{2} a_{5} b_{7} 
+ 8 M^{2} N^{2} a_{5} b_{7} 
\nonu \\
&& + 16 k M N a_{8} b_{7} 
+ 16 M N^{2} a_{8} b_{7} 
- 8 M^{2} N a_{11} b_{7} 
+ 48 M N a_{17} b_{7} 
+ 12 M^{3} N a_{17} b_{7} 
\nonu \\
&&
+ M a_{1} b_{12} 
- M N^{2} a_{1} b_{12} 
+ 8 M N a_{3} b_{12} 
+ 2 M N \sqrt{\frac{M + N}{M N}} a_{7} b_{12}
\nonu \\
&&
+ 2 M N a_{8} b_{12} 
+ 6 M N a_{11} b_{12} 
- M N a_{13} b_{12} 
+ 12 N a_{17} b_{12} 
+ M a_{1} b_{13} 
\nonu \\
&&
- M N^{2} a_{1} b_{13} 
- 8 M N a_{3} b_{13} 
- 2 M N \sqrt{\frac{M + N}{M N}} a_{7} b_{13} 
- 2 M N a_{8} b_{13} \nonu \\
&&- 6 M N a_{11} b_{13} 
- M N a_{12} b_{13} 
- 12 N a_{17} b_{13}
\Bigg) f^{a c f} \delta_{\rho \bar{\tau}} t^{f}_{k \bar{\imath}} 
\partial J^{c} J^{(\rho \bar{\imath})} J^{(\bar{\tau} k)}  \nonu \\
&&+ 2 a_{5} \big(2 M b_{7} - b_{12} 
+ b_{13}\big) d^{a c f} 
\delta_{\bar{\nu} \sigma} \delta_{\rho \bar{\tau}} 
t^{c}_{k \bar{\imath}} t^{f}_{y \bar{g}} 
J^{(\rho \bar{\imath})} J^{(\sigma \bar{g})} 
J^{(\bar{\nu} y)} J^{(\bar{\tau} k)} \nonu \\
&&+ \frac{1}{2 M} \Bigg(
-2 M a_{3} b_{7} 
+ 4 M N a_{5} b_{7} 
+ 2 M a_{8} b_{7} 
- 36 a_{17} b_{7} 
+ 12 M^{2} a_{17} b_{7} 
- 3 M a_{17} b_{12} \nonu\\
&&+ 3 M a_{17} b_{13}\Bigg) 
d^{a h e} d^{c f e} 
\delta_{\rho \bar{\tau}} 
t^{h}_{k \bar{\imath}} 
J^{c} J^{f} J^{(\rho \bar{\imath})} 
J^{(\bar{\tau} k)} \nonu \\
&&+ \frac{1}{2} \Bigg(
4 a_{3} b_{7} 
+ 8 k a_{5} b_{7} 
+ 8 N a_{5} b_{7} 
- 4 a_{8} b_{7} 
+ a_{12} b_{7} 
- a_{13} b_{7} 
+ 12 M a_{17} b_{7} 
- 3 a_{17} b_{12}\nonu\\ 
&&+ 3 a_{17} b_{13}\Bigg) 
d^{a f e} d^{c h e} 
\delta_{\rho \bar{\tau}} 
t^{h}_{k \bar{\imath}} 
J^{c} J^{f} J^{(\rho \bar{\imath})} 
J^{(\bar{\tau} k)} \nonu \\
&&+ \frac{1}{2 M} \Bigg(
12 M^{2} a_{5} b_{2} 
- 2 M a_{3} b_{7} 
+ 2 M a_{8} b_{7} 
+ 36 a_{17} b_{7} 
+ 2 M a_{5} b_{12} 
- 3 M a_{17} b_{12} 
- 2 M a_{5} b_{13}\nonu\\ 
&&+ 3 M a_{17} b_{13}\Bigg) 
d^{a c e} d^{f h e} 
\delta_{\rho \bar{\tau}} 
t^{h}_{k \bar{\imath}} 
J^{c}J^{f} J^{(\rho \bar{\imath})} 
J^{(\bar{\tau} k)} 
\nonu \\
&&- \frac{i}{4 M} \Bigg(
-4 M a_{3} 
+ M^{2} a_{5} 
- 8 M N a_{5} 
+ 4 M a_{8} 
- 24 a_{17}\Bigg) 
b_{7} d^{f h e} f^{a c e} 
\delta_{\rho \bar{\tau}} 
t^{h}_{k \bar{\imath}} 
J^{c}J^{f} J^{(\rho \bar{\imath})} 
J^{(\bar{\tau} k)} 
\nonu \\
&&+ \frac{i}{4} \Bigg(
M a_{5} 
+ 2 a_{12} 
+ 2 a_{13}\Bigg) 
b_{7} d^{c h e} f^{a f e} 
\delta_{\rho \bar{\tau}} 
t^{h}_{k \bar{\imath}}  
J^{c}J^{f} J^{(\rho \bar{\imath})} 
J^{(\bar{\tau} k)} 
\nonu \\
&&- \frac{i}{2 M} \Bigg(
-2 M a_{3} 
+ M^{2} a_{5} 
+ 2 M a_{8} 
- 12 a_{17}\Bigg) 
b_{7} d^{c f e} f^{a h e} 
\delta_{\rho \bar{\tau}} 
t^{h}_{k \bar{\imath}} 
J^{c} J^{f} J^{(\rho \bar{\imath})} 
J^{(\bar{\tau} k)} 
\nonu \\
&&+ \frac{i}{2 M} \Bigg(
-2 M a_{3} 
+ M^{2} a_{5} 
- 4 M N a_{5} 
+ 2 M a_{8} 
- 12 a_{17}\Bigg) 
b_{7} d^{a h e} f^{c f e} 
\delta_{\rho \bar{\tau}} 
t^{h}_{k \bar{\imath}} 
J^{c} J^{f} J^{(\rho \bar{\imath})} 
J^{(\bar{\tau} k)} 
\nonu \\
&&- \frac{1}{M} \Bigg(
M a_{3} 
- M a_{8} 
+ 6 a_{17}\Bigg) 
b_{7} f^{a h e} f^{c f e} 
\delta_{\rho \bar{\tau}} 
t^{h}_{k \bar{\imath}} 
J^{c} J^{f} J^{(\rho \bar{\imath})} 
J^{(\bar{\tau} k)} 
\nonu \\
&&- \frac{i}{4} \Bigg(
M a_{5} 
+ 8 N a_{5} 
- 2 a_{12} 
- 2 a_{13}\Bigg) 
b_{7} d^{a f e} f^{c h e} 
\delta_{\rho \bar{\tau}} 
t^{h}_{k \bar{\imath}} 
J^{c} J^{f} J^{(\rho \bar{\imath})} 
J^{(\bar{\tau} k)} 
\nonu \\
&&- \frac{1}{2} \Bigg(
4 a_{3} 
+ 4 a_{8} 
+ 4 a_{11} 
+ a_{12} 
- a_{13}\Bigg) 
b_{7} f^{a f e} f^{c h e} 
\delta_{\rho \bar{\tau}} 
t^{h}_{k \bar{\imath}} 
J^{c} J^{f} J^{(\rho \bar{\imath})} 
J^{(\bar{\tau} k)} 
\nonu \\
&&- \frac{i}{4 M} \Bigg(
-4 M a_{3} b_{7} 
+ 3 M^{2} a_{5} b_{7} 
+ 4 M a_{8} b_{7} 
- 24 a_{17} b_{7} 
+ 4 M a_{5} b_{12} 
+ 4 M a_{5} b_{13}\Bigg) \nonu\\
&& \times  d^{a c e} f^{f h e} 
\delta_{\rho \bar{\tau}} 
t^{h}_{k \bar{\imath}} 
J^{c} J^{f} J^{(\rho \bar{\imath})} 
J^{(\bar{\tau} k)} 
\nonu \\
&&+ \frac{3}{M} \Bigg(
-M a_{3} b_{7} 
+ M a_{8} b_{7} 
- 6 a_{17} b_{7}\Bigg) 
f^{a c e} f^{f h e} 
\delta_{\rho \bar{\tau}} 
t^{h}_{k \bar{\imath}} 
J^{c} J^{f} J^{(\rho \bar{\imath})} 
J^{(\bar{\tau} k)} 
\nonu \\
&&+ \frac{2}{M} \Bigg(
4 a_{3} b_{7} 
+ 2 M^{2} a_{3} b_{7} 
- 4 a_{8} b_{7} 
+ 4 k M a_{8} b_{7} 
+ 2 M^{2} a_{8} b_{7} 
+ 4 M N a_{8} b_{7} 
+ a_{12} b_{7} 
- a_{13} b_{7} \nonu\\
&&+ 18 M a_{17} b_{7}\Bigg) 
\delta_{a f} \delta_{c h} 
\delta_{\rho \bar{\tau}} 
t^{h}_{k \bar{\imath}} 
J^{c} J^{f} J^{(\rho \bar{\imath})} 
J^{(\bar{\tau} k)} 
\nonu \\
&&- \frac{2}{M} \Bigg(
M a_{3} b_{12} 
- M a_{8} b_{12} 
+ 6 a_{17} b_{12} 
- M a_{3} b_{13} 
+ M a_{8} b_{13} 
- 6 a_{17} b_{13}\Bigg) 
\nonu \\
&& \times \delta_{a c} \delta_{f h} 
\delta_{\rho \bar{\tau}} 
t^{h}_{k \bar{\imath}} 
J^{c} J^{f} J^{(\rho \bar{\imath})} 
J^{(\bar{\tau} k)} 
\nonu \\
&&+ 
\frac{1}{M N} \Bigg(
2 k M^2 N a_{1} b_{6} 
- 2 k M N^2 a_{1} b_{6} 
+ 2 M^2 N^2 a_{1} b_{6} 
- 4 k M N a_{12} b_{6} \nonu\\
&&- 
2 M N^2 a_{12} b_{6} 
+ 2 k M N a_{13} b_{6} 
+ 2 k M^2 N a_{1} b_{7} 
+ 2 M^2 N^2 a_{1} b_{7} 
- 16 M N a_{5} b_{7} \nonu\\
&&+ 
4 M^3 N a_{5} b_{7} 
+ 8 M^2 N a_{8} b_{7} 
- 2 M^2 N a_{12} b_{7} 
+ 2 M^2 N a_{13} b_{7} 
- M a_{1} b_{12} \nonu\\
&&+ 
M^2 N a_{1} b_{12} 
+ 8 N a_{5} b_{12} 
- 2 M^2 N a_{5} b_{12} 
+ 2 M N \sqrt{\frac{M + N}{M N}} a_{7} b_{12} 
- 2 M N a_{8} b_{12} \nonu\\
&&+ 
M a_{1} b_{13} 
+ 3 k M N a_{1} b_{13} 
+ 2 M^2 N a_{1} b_{13} 
+ M N^2 a_{1} b_{13} 
- 8 N a_{5} b_{13} \nonu\\
&&+ 
2 M^2 N a_{5} b_{13} 
- 2 M N \sqrt{\frac{M + N}{M N}} a_{7} b_{13} 
+ 2 M N a_{8} b_{13} 
- 2 M N a_{12} b_{13} 
+ 2 M N a_{13} b_{13}
\Bigg) \nonu\\
&& \times 
t^{a}_{k \bar{i}} t^{\beta}_{\rho \bar{\tau}} 
J^{\beta} J^{(\rho \bar{i})} \partial J^{(\bar{\tau} k)} \nonu\\
&&- \frac{1}{M N} \Bigg(
-2 k M^2 N a_{1} b_{6} 
- 2 k M N^2 a_{1} b_{6} 
- 6 M^2 N^2 a_{1} b_{6} 
- 2 k M N a_{12} b_{6} \nonu\\
&&- 4 M^2 N a_{12} b_{6} 
+ 4 k M N a_{13} b_{6} 
+ 4 M^2 N a_{13} b_{6} 
+ 2 M N^2 a_{13} b_{6} \nonu\\
&&+ 2 k M^2 N a_{1} b_{7} 
+ 2 M^2 N^2 a_{1} b_{7} 
- 16 M N a_{5} b_{7} 
+ 4 M^3 N a_{5} b_{7} 
+ 8 M^2 N a_{8} b_{7} \nonu\\
&&- 2 M^2 N a_{12} b_{7} 
+ 2 M^2 N a_{13} b_{7} 
- M a_{1} b_{12} 
- 3 k M N a_{1} b_{12} 
- 2 M^2 N a_{1} b_{12} \nonu\\
&&- M N^2 a_{1} b_{12} 
+ 8 N a_{5} b_{12} 
- 2 M^2 N a_{5} b_{12} 
+ 2 M N \sqrt{\frac{M + N}{M N}} a_{7} b_{12} 
- 2 M N a_{8} b_{12} \nonu\\
&&+ 2 M N a_{12} b_{12} 
- 2 M N a_{13} b_{12} 
+ M a_{1} b_{13} 
- M^2 N a_{1} b_{13} 
- 8 N a_{5} b_{13} 
+ 2 M^2 N a_{5} b_{13} \nonu\\
&&- 2 M N \sqrt{\frac{M + N}{M N}} a_{7} b_{13} 
+ 2 M N a_{8} b_{13}
\Bigg) 
t^{a}_{k \bar{i}} t^{\beta}_{\rho \bar{\tau}} 
J^{\beta} \partial J^{(\rho \bar{i})} J^{(\bar{\tau} k)} 
\nonu\\
&&- \frac{1}{M N} \Bigg(
-16 M N a_{5} b_{7} + 4 M^3 N a_{5} b_{7} + 8 M^2 N a_{8} b_{7} 
- M a_{1} b_{12} - 2 k M N a_{1} b_{12} - M^2 N a_{1} b_{12} \nonu\\
&&- M N^2 a_{1} b_{12} + 8 N a_{5} b_{12} - 2 M^2 N a_{5} b_{12} 
+ 2 M N \sqrt{\frac{M + N}{M N}} a_{7} b_{12} - 2 M N a_{8} b_{12} 
- M N a_{13} b_{12}\nonu\\
&&+ M a_{1} b_{13} 
- 2 k M N a_{1} b_{13} - 2 M^2 N a_{1} b_{13} - M N^2 a_{1} b_{13} 
- 8 N a_{5} b_{13} + 2 M^2 N a_{5} b_{13} \nonu\\
&&- 2 M N \sqrt{\frac{M + N}{M N}} a_{7} b_{13} 
+ 2 M N a_{8} b_{13} + M N a_{12} b_{13}
\Bigg) 
t^{a}_{k \bar{i}} t^{\beta}_{\rho \bar{\tau}} 
\partial J^{\beta} J^{(\rho \bar{i})} J^{(\bar{\tau} k)} \nonu\\
&&+ 
\frac{1}{2} \Bigg(
6 M a_{1} b_{2} 
+ 4 N a_{1} b_{6} 
+ 16 k a_{5} b_{6} 
+ 8 M a_{5} b_{6}+ 
8 N a_{5} b_{6} 
+ 2 a_{12} b_{6} 
- 2 a_{13} b_{6} 
+ 4 k a_{1} b_{7} 
\nonu \\
&& + 2 N a_{1} b_{7} + 
16 M a_{5} b_{7} 
- 2 a_{12} b_{7} 
+ 2 a_{13} b_{7} 
\nonu \\
&&
+ a_{1} b_{12} 
- 4 a_{5} b_{12} 
- a_{1} b_{13} 
+ 4 a_{5} b_{13}
\Bigg) 
d^{a c f} t^{f}_{k \bar{i}} t^{\beta}_{\rho \bar{\tau}} 
J^{\beta} J^{c} J^{(\rho \bar{i})} J^{(\bar{\tau} k)} \nonu\\
&&+ 
\frac{i}{2} \Bigg(
2 a_{12} b_{6} 
+ 2 a_{13} b_{6} 
- 2 a_{12} b_{7} 
- 2 a_{13} b_{7} 
- a_{1} b_{12} 
- a_{1} b_{13}
\Bigg) f^{a c f} t^{f}_{k \bar{i}} t^{\beta}_{\rho \bar{\tau}} 
J^{\beta} J^{c} J^{(\rho \bar{i})} J^{(\bar{\tau} k)} \nonu\\
&&+ 
\frac{1}{M} \Bigg(
4 N a_{1} b_{6} 
+ 4 M N a_{2} b_{6} 
+ 8 k M a_{8} b_{6} 
+ 8 M^2 a_{8} b_{6} 
+4 M N a_{8} b_{6} 
+ 2 a_{12} b_{6} 
- 2 a_{13} b_{6} \nonu\\
&&+4 k a_{1} b_{7} 
+ 4 M a_{1} b_{7}
+2 N a_{1} b_{7} 
- 2 a_{12} b_{7} 
+ 2 a_{13} b_{7} 
+ a_{1} b_{12} 
+ 2 M a_{2} b_{12} \nonu\\
&&- 2 M a_{8} b_{12} 
- a_{1} b_{13} 
- 2 M a_{2} b_{13} 
+ 2 M a_{8} b_{13}
\Bigg) \delta_{\bar{i} k} t^{\beta}_{\rho \bar{\tau}} 
J^{\beta} J^{a} J^{(\rho \bar{i})} J^{(\bar{\tau} k)} \nonu\\
&&+ \Bigg(
2 M \sqrt{\frac{M + N}{M N}} a_{1} b_{6} 
- 4 N \sqrt{\frac{M + N}{M N}} a_{1} b_{6} 
+ 8 k a_{7} b_{6}
+ 4 M a_{7} b_{6} 
+ 4 N a_{7} b_{6} 
\nonu \\
&& - 2 \sqrt{\frac{M + N}{M N}} a_{12} b_{6} 
+ 2 \sqrt{\frac{M + N}{M N}} a_{13} b_{6} 
+ 4 M \sqrt{\frac{M + N}{M N}} a_{1} b_{7} 
\nonu \\
&&
+ 8 M a_{7} b_{7} 
+ 4 k a_{1} b_{8} + 4 M a_{1} b_{8}
+ 2 N a_{1} b_{8} \nonu\\
&&- 2 a_{12} b_{8} 
+ 2 a_{13} b_{8} 
- \sqrt{\frac{M + N}{M N}} a_{1} b_{12} 
- 2 a_{7} b_{12}
+ \sqrt{\frac{M + N}{M N}} a_{1} b_{13} 
+ 2 a_{7} b_{13} 
\Bigg) \nonu \\
&& \times
t^{a}_{k \bar{i}} t^{\beta}_{\rho \bar{\tau}} 
J^{\beta} J^{u(1)} J^{(\rho \bar{i})} J^{(\bar{\tau} k)}
\nonu\\
&&+ 
2 M a_{1} b_{5} \delta_{\beta \gamma} t^{a}_{k \bar{i}} t^{\gamma}_{\rho \bar{\tau}} 
J^{\beta} J^{u(1)} J^{(\rho \bar{i})} J^{(\bar{\tau} k)} \nonu\\
&&+ 
\frac{1}{2} \Bigg(
4 k a_{1} b_{6} + 4 M a_{1} b_{6} - 2 a_{12} b_{6} + 2 a_{13} b_{6} 
+ 4 M a_{1} b_{7} - a_{1} b_{12} + a_{1} b_{13}
\Bigg) 
\nonu \\
&& \times d^{\beta \gamma \delta} t^{a}_{k \bar{i}} t^{\delta}_{\rho \bar{\tau}} 
J^{\beta} J^{\gamma} J^{(\rho \bar{i})} J^{(\bar{\tau} k)} \nonu\\
&&- 
\frac{i}{2} \Bigg(
2 a_{12} b_{6} + 2 a_{13} b_{6} + a_{1} b_{12} + a_{1} b_{13}
\Bigg) 
f^{\beta \gamma \delta} t^{a}_{k \bar{i}} t^{\delta}_{\rho \bar{\tau}} 
J^{\beta} J^{\gamma} J^{(\rho \bar{i})} J^{(\bar{\tau} k)} 
\nonu\\
&&+ 
\frac{1}{6 M} \Bigg(
8 M^2 a_{2} b_{7} - 8 M^2 N^2 a_{2} b_{7} + 8 M N a_{3} b_{7} 
+ 8 k M^2 N a_{3} b_{7} + 12 M^3 N a_{3} b_{7} \nonu\\
&&- 
8 M^2 a_{4} b_{7} - 8 M N a_{4} b_{7} + 16 k M N a_{5} b_{7} 
- 4 k M^3 N a_{5} b_{7} - 8 k M^2 N a_{8} b_{7} - 8 M^3 N a_{8} b_{7} \nonu\\
&&- 
32 M N \sqrt{\frac{M + N}{M N}} a_{9} b_{7} 
+ 8 M^3 N \sqrt{\frac{M + N}{M N}} a_{9} b_{7} 
+ 4 k M^2 N a_{11} b_{7} + 2 k M^2 N a_{13} b_{7} \nonu\\
&&+ 
4 M^2 N^2 a_{13} b_{7} + 8 M^2 N a_{16} b_{7} 
- 144 N a_{17} b_{7} + 24 k M N a_{17} b_{7} 
+ 96 M^2 N a_{17} b_{7} \nonu\\
&&- 
k M a_{1} b_{12} + 3 M^2 a_{1} b_{12} + k M N^2 a_{1} b_{12} 
- 3 M^2 N^2 a_{1} b_{12} - 2 k M^2 a_{2} b_{12} 
+ 2 k M^2 N^2 a_{2} b_{12} \nonu\\
&&+ 
2 k M N a_{3} b_{12} + 2 k M^3 N a_{3} b_{12} + 2 k M^2 a_{4} b_{12} 
+ 2 k M N a_{4} b_{12} + 8 k N a_{5} b_{12} - 2 k M^2 N a_{5} b_{12} \nonu\\
&&- 
8 N^2 a_{5} b_{12} + 2 M^2 N^2 a_{5} b_{12} 
+ 2 k M N \sqrt{\frac{M + N}{M N}} a_{7} b_{12} 
- 6 M^2 N \sqrt{\frac{M + N}{M N}} a_{7} b_{12} \nonu\\
&&- 
2 M N^2 \sqrt{\frac{M + N}{M N}} a_{7} b_{12} 
+ 3 k M^2 N^2 \sqrt{\frac{M + N}{M N}} a_{7} b_{12} \nonu\\
&&- 
2 k M N a_{8} b_{12} - 4 M^2 N a_{8} b_{12} 
+ 4 k^2 M^2 N a_{8} b_{12} + 2 k M^3 N a_{8} b_{12} 
+ 2 M N^2 a_{8} b_{12}\nonu\\ 
&&+ 2 k M^2 N^2 a_{8} b_{12} 
- 2 k M N a_{12} b_{12} - M^2 N a_{12} b_{12} 
+ k M N a_{13} b_{12} + M^2 N a_{13} b_{12} \nonu\\
&&- 
36 k N a_{17} b_{12} + 24 k M^2 N a_{17} b_{12} 
+ k M a_{1} b_{13} + 3 M^2 a_{1} b_{13} - k M N^2 a_{1} b_{13} \nonu\\
&&- 
3 M^2 N^2 a_{1} b_{13} - 2 k M^2 a_{2} b_{13} 
+ 2 k M^2 N^2 a_{2} b_{13} + 2 k M N a_{3} b_{13} 
+ 2 M^2 N a_{3} b_{13} + 2 k M^3 N a_{3} b_{13} \nonu\\
&&+ 
2 k M^2 a_{4} b_{13} + 2 k M N a_{4} b_{13} - 8 k N a_{5} b_{13} 
+ 2 k M^2 N a_{5} b_{13} - 24 N^2 a_{5} b_{13}
+ 6 M^2 N^2 a_{5} b_{13}\nonu\\
&&- 2 k M N \sqrt{\frac{M + N}{M N}} a_{7} b_{13} 
- 6 M^2 N \sqrt{\frac{M + N}{M N}} a_{7} b_{13}
- 6 M N^2 \sqrt{\frac{M + N}{M N}} a_{7} b_{13} \nonu\\
&&+ 6 k M^2 N^2 \sqrt{\frac{M + N}{M N}} a_{7} b_{13} 
+
2 k M N a_{8} b_{13} + 4 k^2 M^2 N a_{8} b_{13} 
+ 2 k M^3 N a_{8} b_{13} \nonu\\
&&+ 6 M N^2 a_{8} b_{13} 
+ 2 k M^2 N^2 a_{8} b_{13} 
- 
6 k M N a_{12} b_{13} + 4 k M N a_{13} b_{13} 
- 36 k N a_{17} b_{13} \nonu\\
&&+ 12 M N a_{17} b_{13} 
+ 24 k M^2 N a_{17} b_{13}
\Bigg)\partial^3 J^{a}.
\label{PWP2}
\eea
By substituting the relations (\ref{avalues})
and (\ref{interbvalue}) into (\ref{PWP2}),
the explicit coefficients appearing given in (\ref{PWPOLE2})
are given by
\bea
p_1 & = & -\frac{N^2 (k+N) (k+2N)^2 a_1 b_1}{4 (k+M)^2 (k+2M) (3k+2M)}, \qquad
p_2  =  -\frac{N^2 (k+N) (k+2N)^2 a_1 b_1}{4 (k+M)^2 (k+2M) (3k+2M)}, \nonu\\
p_3 & = & -\frac{N^2 (k+N) (k+2N)^2 a_1 b_1}{4 (k+M)^2 (k+2M) (3k+2M)}, \qquad
p_4  =  \frac{3N^2 (k+N) (k+2N)^2 a_1 b_1}{4 M (k+M) (k+2M) (3k+2M)}, \nonu\\
p_5 & = & \frac{3N^2 (k+N) (k+2N)^2 a_1 b_1}{4 M (k+M) (k+2M) (3k+2M)}, \qquad \nonu \\
p_6  & = &
\frac{3N (k+N) (3k+2M+N) (k+2N)^2 a_1 b_1}{4 M (k+M) (k+2M) (3k+2M)}, \nonu\\
p_7 & = & \frac{i N^2 (k+N) (k+2N)^2 a_1 b_1}{4 M (k+M) (k+2M) (3k+2M)}, \qquad
p_8  =  \frac{i N^2 (k+N) (k+2N)^2 a_1 b_1}{4 M (k+M) (k+2M) (3k+2M)}, \nonu\\
p_9 & = & \frac{i N^2 (k+N) (k+2N)^2 a_1 b_1}{4 M (k+M) (k+2M) (3k+2M)}, \qquad
p_{10}  =  -\frac{3 i N (k+N) (k+2N)^2 a_1 b_1}{4 M (k+M) (k+2M)}, \nonu\\
p_{11} & = & \frac{3 i N^2 (k+N) (k+2N)^2 a_1 b_1}{4 M (k+M) (k+2M) (3k+2M)}, \qquad
p_{12}  =  \frac{3 i N^2 (k+N) (k+2N)^2 a_1 b_1}{4 M (k+M) (k+2M) (3k+2M)}, \nonu\\
p_{13} & = & \frac{3 i N (k+N) (3k+2M+N) (k+2N)^2 a_1 b_1}{4 M (k+M) (k+2M) (3k+2M)}, \qquad
p_{14}  =  \frac{3 N (k+N) (k+2N)^2 a_1 b_1}{4 M (k+M) (k+2M)}, \nonu\\
p_{15} & = & \frac{3 N (k+N) (k+M+N) (k+2N)^2 a_1 b_1}{k M^2 (k+M) (k+2M)}, \nonu\\
p_{16}  &=&  \frac{3 N (k+N) (k+2N)^2 (3k+2M+2N) a_1 b_1}{k M^2 (k+2M) (3k+2M)}, \nonu\\
p_{17} & = & \frac{3 i N (k+N) (k+2N)^2 (3k+2M+2N) a_1 b_1}{k M^2 (k+2M) (3k+2M)}, \qquad
\nonu \\
p_{18} & = &  -\frac{3 N^2 (k+N) (k+2N)^2 a_1 b_1}{k (k+M)^2 (k+2M) (3k+2M)}, \nonu\\
p_{19} & = & -\frac{N^2 (k+N) (k+2N)^2 a_1 b_1}{M (k+M)^2 (k+2M) (3k+2M)}, \qquad
p_{20}  =  \frac{3 N^2 (k+N) (k+2N)^2 a_1 b_1}{k M^2 (k+2M) (3k+2M)},
\nonu\\
p_{21} & = & \frac{i N^2 (k+N) (k+2N)^2 a_1 b_1}{M^2 (k+M) (k+2M) (3k+2M)}, \qquad
p_{22}  =  \frac{3 i N^2 (k+N) (k+2N)^2 a_1 b_1}{k M (k+M) (k+2M) (3k+2M)}, \nonu\\
p_{23} &=&  \frac{3 i N^2 (k+N) (k+2N)^2 a_1 b_1}{k M^2 (k+2M) (3k+2M)}, \qquad
p_{24}  =  -\frac{N^2 (k+N) (k+2N)^2 a_1 b_1}{M (k+M)^2 (k+2M) (3k+2M)}, \nonu\\
p_{25}  &=&  \frac{i N^2 (k+N) (k+2N)^2 a_1 b_1}{M^2 (k+M) (k+2M) (3k+2M)}, \qquad
p_{26}  =  -\frac{N^2 (k+N) (k+2N)^2 a_1 b_1}{M (k+M)^2 (k+2M) (3k+2M)}, \nonu\\
p_{27}  &=&  \frac{i N^2 (k+N) (k+2N)^2 a_1 b_1}{M^2 (k+M) (k+2M) (3k+2M)}, \nonu\\
p_{28} & = & -\frac{24 (k+M) (k+N) (M+N) \sqrt{\frac{M+N}{M N}} (k+M+N) (k+2N)^2 a_1 b_1}{k^3 M^2 (k+2M)}, \nonu\\
p_{29} & = & \frac{48 (k+M) (k+N) \sqrt{\frac{M+N}{M N}} (k+M+N) (k+2N)^2 a_1 b_1}{k^2 M^2 (k+2M)}, \nonu\\
p_{30} & = & \frac{6 (k+N) (k+M+N) (k+2N) (-4 + 4k^2 + 3kM + 6kN + 6MN) a_1 b_1}{k M^2 (k+2M)}, \nonu\\
p_{31} & = & -\frac{24 (k+M) (k+N) (M+N) (k+M+N) (k+2N)^2 a_1 b_1}{k^2 M^2 (k+2M)}, \nonu\\
p_{32} & = & -\frac{6 (k+N) (k+M+N) (k+2N) (-4 + 4k^2 + 3kM + 6kN + 6MN) a_1 b_1}{k M^2 (k+2M)}, \nonu\\
p_{33} & = & -\frac{3 N (k+N) \sqrt{\frac{M+N}{M N}} 
(k+M+N) (k+2N) a_1 b_1}{2 k M (k+M) (k+2M)}
\nonu \\
& \times & (-12k + 5k^3 - 8M + 6k^2M + 2kM^2 
- 8N + 6k^2N + 8kMN + 4M^2N), \nonu\\
p_{34} & = & \frac{6 (3k+2M) (k+N) (M+N) (k+M+N) (k+2N)^2 a_1 b_1}{k^2 M^2 (k+M) (k+2M)}, \nonu\\
p_{35} & = & -\frac{3 i (k+N) (M+N) (k+2N)^2 a_1 b_1}{k^2 M^2}, \qquad
p_{36}  =  -\frac{N^2 (k+N) \sqrt{\frac{M+N}{M N}} (k+2N)^2 a_1 b_1}{k M (k+M) (k+2M)}, \nonu\\
p_{37} & = & -\frac{N (k+N) \sqrt{\frac{M+N}{M N}} (k+2N)^2 (9k + 3M + 8N) a_1 b_1}{2 k M (k+M) (k+2M)}, \nonu\\
p_{38} & = & -\frac{N (k+N) \sqrt{\frac{M+N}{M N}} (k+2N)^2 (3k^2 + 9kM + 6M^2 + 2kN - MN) a_1 b_1}{2 k M (k+M)^2 (k+2M)}, \nonu\\
p_{39} & = & -\frac{3 i N^2 (k+N) \sqrt{\frac{M+N}{M N}} (k+2N)^2 a_1 b_1}{2 k M (k+M) (k+2M)}, \qquad
p_{40}  =  \frac{3 i N (k+N) \sqrt{\frac{M+N}{M N}} (k+2N)^2 a_1 b_1}{2 k M (k+2M)},\nonu\\
p_{41} & = & \frac{3 i N (k+N) \sqrt{\frac{M+N}{M N}} (k+2N)^2 (k+M+2N) a_1 b_1}{2 k M (k+M) (k+2M)}, \nonu\\
p_{42} & = & \frac{3 N (k+N) \sqrt{\frac{M+N}{M N}} (k+2N)^2 a_1 b_1}{2 k M (k+2M)}, \qquad
p_{43}  =  \frac{3 i N (k+N) \sqrt{\frac{M+N}{M N}} (k+2N)^2 a_1 b_1}{2 k M (k+M)}, \nonu\\
p_{44} & = & -\frac{2 (3k+7M) N^2 (k+N) \sqrt{\frac{M+N}{M N}} (k+2N)^2 a_1 b_1}{k M^2 (k+M) (k+2M) (3k+2M)}, \nonu\\
p_{45} & = & -\frac{2 N (k+N) \sqrt{\frac{M+N}{M N}} (k+2N)^2 (9k^2 + 15kM + 6M^2 + 8kN + 6MN) a_1 b_1}{k^2 M^2 (k+M) (k+2M)}, \nonu\\
p_{46} & = & -\frac{2 N (k+N)
\sqrt{\frac{M+N}{M N}} (k+2N)^2  a_1 b_1}{k^2 M^2 (k+M) (k+2M) (3k+2M)}
(9k^3 + 33k^2M + 36kM^2 + 12M^3 + 9k^2N \nonu\\
&+& 19kMN + 12M^2N), \nonu\\
p_{47} & = & -\frac{6 (2k+M) (k+N) (k+M+N) (k+2N)^2 a_1 b_1}{k M^2 (k+M) (k+2M)}, \nonu\\
p_{48} & = & \frac{3 (2k+M) N (k+N) \sqrt{\frac{M+N}{M N}} (k+M+N) (k+2N)^2 a_1 b_1}{k M (k+M) (k+2M)}, \nonu\\
p_{49} & = & \frac{3 i N (k+N) \sqrt{\frac{M+N}{M N}} (k+2N)^2 (3k+2N) a_1 b_1}{2 k M (k+2M)}, \nonu\\
p_{50} & = & -\frac{6 (k+N) \sqrt{\frac{M+N}{M N}} a_1 b_1}
{k^2 M (k+2M)}(9k^3 + 24k^2M + 12kM^2 + 19k^2N + 44kMN + 16M^2N \nonu\\
&+& 10kN^2 + 16MN^2), \nonu\\
p_{51} & = & -\frac{18 (k+N)^2 \sqrt{\frac{M+N}{M N}} (k+2N) a_1 b_1}{k M (k+2M)}, \nonu\\
p_{52} & = & \frac{6 (k+N) (k+M+N) (3k^2 + 2kM + 4kN + 2MN) a_1 b_1}{k M (k+M) (k+2M)}, \nonu\\
p_{53} & = & -\frac{3 i (k+N) a_1 b_1}{M}, \qquad
p_{54}  =  -\frac{3 (k+N) (3k+2M+2N) a_1 b_1}{k M}, \nonu\\
p_{55} & = & \frac{3 i (k+N) (k+2M+2N) a_1 b_1}{k M}, \nonu\\
p_{56} & = & -\frac{3 (k+N) (k+2N) a_1 b_1}{M (k+2M)}, \qquad
p_{57}  =  -\frac{3 i (k+N) (k+2N) a_1 b_1}{M (k+2M)}, \nonu\\
p_{58} & = & -\frac{3 (k+N)}{k M (k+2M)}\nonu \\
& \times &  (4k^3 + 9k^2M + 6kM^2 + 8k^2N + 12kMN + 4M^2N + 4kN^2 + 4MN^2) a_1 b_1, \nonu\\
p_{59} & = & -\frac{3 (k+N) (M+N) (k+2N)^2 (4k^2 + 9kM + 6M^2 + 4kN + 4MN) a_1 b_1}{k^2 M^2 (k+2M)},\nonu\\
p_{60} & = & \frac{3 (k+N) (k+2N) a_1 b_1}{2 k M^3 (k+M) (k+2M)}(2k^3 + 18k^2M + 28kM^2 + 3k^3M^2 + 16M^3 + 6k^2M^3 \nonu\\
&+& 4kM^4 + 4k^2N + 20kMN + 8k^3MN + 8M^2N + 26k^2M^2N + 24kM^3N +
8M^4N \nonu\\
&+& 8k^2MN^2 + 16kM^2N^2 + 8M^3N^2), \nonu\\
p_{61} & = & -\frac{3 N (k+N) \sqrt{\frac{M+N}{M N}} (k+2N)^2 (2k^2 + kM + 2M^2 + 2kN) a_1 b_1}{2 k M (k+2M)}, \nonu\\
p_{62} & = & \frac{3 N (k+N) \sqrt{\frac{M+N}{M N}} (k+2N)^2 (8k^2 + 15kM + 8M^2 + 8kN + 5MN) a_1 b_1}{2 k M (k+M) (k+2M)}, \nonu\\
p_{63} & = & \frac{3 i N (k+N) \sqrt{\frac{M+N}{M N}} 
(k+2N)^2 a_1 b_1}{2 k^2 M^2 (k+M) (k+2M) (3k+2M)}\nonu \\
& \times & (48k^2M - 9k^4M + 56kM^2 - 9k^3M^2 + 16M^3 - 2k^2M^3 \nonu\\
&-& 8k^2N + 32kMN - 9k^3MN + 16M^2N - 3k^2M^2N + 2kM^3N), \nonu\\
p_{64} & = & -\frac{3 N (k+N) (k+2N)^2 a_1 b_1}{4 k M^2 (k+M) (k+2M) (3k+2M)}\nonu \\
& \times & (-18k^2 - 18kM + 6k^3M - 4M^2 + 4k^2M^2 - 12kN \nonu\\
&-& 4MN + 5k^2MN + 4kM^2N), \nonu\\
p_{65} & = & -\frac{3 N (k+N) (k+2N)^2  a_1 b_1}{4 k M^2 (k+M)^2 (k+2M) (3k+2M)}(12k^3 + 20k^2M + 3k^4M + 8kM^2 + 11k^3M^2 \nonu\\
&+& 12k^2M^3 + 4kM^4 + 12k^2N + 8kMN + 5k^3MN - 4M^2N + 2k^2M^2N - 2kM^3N
), \nonu\\
p_{66} & = & -\frac{3 N (k+N) (k+2N)^2  a_1 b_1}{4 k M^2 (k+M) (k+2M) (3k+2M)}\nonu \\
& \times & (6k^2 + 10kM + 3k^3M + 4M^2 + 5k^2M^2 + 2kM^3 
+ 8MN + 2k^2MN + 2kM^2N), \nonu\\
p_{67} & = & \frac{3 i N (k+N) (k+2N)^2  a_1 b_1}{16 k M^2 (k+M) (k+2M) (3k+2M)}(60k^2 - 32kM + 12k^3M - 48M^2 + 17k^2M^2 \nonu\\
&+& 6kM^3 + 48kN - 64MN + 12k^2MN - 4kM^2N), \nonu\\
p_{68} & = & -\frac{3 i N (k+N) (k+2N)^2 }{4 k M^2 (k+M) (k+2M) (3k+2M)}
\nonu \\
& \times & (-9k^2 + 6kM + 8M^2 - 4kN + 12MN + 3k^2MN + 3kM^2N) a_1 b_1,
\nonu\\
p_{69} & = & -\frac{3 i N (k+N) (k+2N)^2  a_1 b_1}{16 k M^2 (k+M) (k+2M) (3k+2M)}(-36k^2 - 96kM - 48M^2 + 3k^2M^2 + 2kM^3 \nonu\\
&-& 16kN - 48MN - 8kM^2N), \nonu\\
p_{70} & = & \frac{3 i N (k+N) (k+2N)^2 a_1 b_1}{16 k M^2 (k+M) (k+2M) (3k+2M)}(-24k^2 - 40kM + 24k^3M - 16M^2 + 19k^2M^2 \nonu\\
&+& 2kM^3 - 16kN - 16MN), \nonu\\
p_{71} & = & -\frac{3 N (k+N) (k+2N)^2 (3k+2M+2N) a_1 b_1}{2 k M^2 (k+2M) (3k+2M)}, \nonu\\
p_{72} & = & -\frac{3 i N (k+N) (k+2N)^2 a_1 b_1}{4 k M^2 (k+M)^2 (k+2M) (3k+2M)}(6k^3 + 10k^2M + 3k^4M + 4kM^2 + 11k^3M^2 \nonu\\
&+& 12k^2M^3 + 4kM^4 + 4k^2N + 8kMN - 4M^2N + 2k^2M^2N + kM^3N), \nonu\\
p_{73} & = & \frac{3 N (k+N) (k+2N)^2 (12k^2 + 20kM + 8M^2 + 4kN + 12MN - k^2MN) a_1 b_1}{4 k M^2 (k+M) (k+2M) (3k+2M)}, \nonu\\
p_{74} & = & \frac{3 i N (k+N) (k+2N)^2  a_1 b_1}{16 k M (k+M) (k+2M) (3k+2M)}(120k - 36k^3 + 80M - 15k^2M + 6kM^2 \nonu\\
&+& 96N - 12k^2N + 4kMN), \nonu\\
p_{75} & = & \frac{3 N (k+N) (k+2N)^2 a_1 b_1}{4k M^2 (k+M) (k+2M) (3k+2M)} 
(6k^2 + 10kM + 3k^3M + 4M^2 + 5k^2M^2 + 2kM^3 \nonu\\
&+& 12kN + 3k^2MN), \nonu\\
p_{76} & = & -\frac{3 N (k+N) (k+2N)^2 a_1 b_1}{k M^3 (k+M) (k+2M)
(3k+2M)} (-18k^2 - 18kM + 6k^3M - 4M^2 + 13k^2M^2\nonu\\
&+& 6kM^3 - 12kN - 4MN + 4k^2MN + 8kM^2N), \nonu\\
p_{77} & = & -\frac{3 N (k+N) (k+2N)^2 a_1 b_1}{k M^3 (k+M) (k+2M) (3k+2M)} 
\nonu \\
& \times & (12k^2 + 8kM + 3k^3M + 8k^2M^2 + 10kM^3 + 4M^4 \nonu\\
&+& 12kN - 4MN + 6k^2MN + 4kM^2N + 4M^3N), \nonu\\
p_{78} & = & -\frac{3 N (k+N) (k+2N)^2 a_1 b_1}{k M^3 (k+M) (k+2M) (3k+2M)} 
(6k^2 + 10kM + 3k^3M + 4M^2 + 8k^2M^2 + 7kM^3 \nonu\\
&+& 2M^4 + 8MN + 2k^2MN + 6kM^2N + 2M^3N), \nonu\\
p_{79} & = & \frac{3 (-2k + 2M + kM^2) (k+N) (k+2N)^2 a_1 b_1}{2k M^3 (k+M)}, \nonu\\
p_{80} & = & \frac{3 N (k+N) (k+2N)^2 (k^2 + kM + 2M^2 + kN) a_1 b_1}{4M (k+M) (k+2M)}, \nonu\\
p_{81} & = & \frac{3 i N (k+N) (k+2N) a_1 b_1}{4k^2 M^2 (k+M) (k+2M) (3k+2M)} 
(-384k^2M + 72k^4M + 9k^6M - 448kM^2 \nonu\\
&-& 96k^3M^2 + 51k^5M^2 - 128M^3 - 192k^2M^3 + 48k^4M^3 
- 64kM^4 \nonu\\
&+& 12k^3M^4 + 12k^4N - 448kMN + 176k^3MN - 9k^5MN - 256M^2N \nonu\\
&+& 10k^4M^2N - 64kM^3N - 8k^2M^4N 
+ 24k^3N^2 - 128MN^2 \nonu\\
&+& 64k^2MN^2 - 6k^4MN^2 - 8k^3M^2N^2), \nonu\\
p_{82} & = & -\frac{12 (k+N) (k+M+N) (2k^2 + 2kM + 3kN + 2MN) a_1 b_1}{k M (k+2M)}, \nonu\\
p_{83} & = & -\frac{N (k+N) \sqrt{\frac{M+N}{M N}} (k+2N) a_1 b_1}{2k^2 M (k+M) (k+2M) (3k+2M)} 
(-108k^4 + 45k^6 - 360k^3M + 165k^5M \nonu\\
& -& 444k^2M^2 + 216k^4M^2 - 240kM^3 + 120k^3M^3 
-  48M^4 + 24k^2M^4 - 168k^3N + 99k^5N \nonu\\
& -& 456k^2MN + 318k^4MN - 360kM^2N + 351k^3M^2N - 96M^3N 
+ 158k^2M^3N \nonu\\
& +& 24kM^4N - 48k^2N^2 + 54k^4N^2 - 144kMN^2 + 144k^3MN^2 - 48M^2N^2 \nonu\\
&+& 114k^2M^2N^2 + 28kM^3N^2), \nonu\\
p_{84} & = & \frac{N (k+N) (k+2N) a_1 b_1}{8k M (k+M)^2 (k+2M) (3k+2M)} 
\nonu \\
& \times & (-216k^4 + 90k^6 - 684k^3M + 303k^5M - 828k^2M^2 \nonu\\
& +& 387k^4M^2 - 456kM^3 + 222k^3M^3
-96M^4 + 48k^2M^4 - 348k^3N + 198k^5N \nonu \\
&- & 780k^2MN 
+ 555k^4MN - 576kM^2N + 582k^3M^2N - 144M^3N \nonu \\
& + & 272k^2M^3N 
+48kM^4N 
- 120k^2N^2 + 108k^4N^2 - 168kMN^2 \nonu \\
& + & 234k^3MN^2 - 48M^2N^2 + 168k^2M^2N^2 + 40kM^3N^2), 
\nonu \\
p_{85} & = & -\frac{i N (k+N) (k+2N) a_1 b_1}{8k M (k+M)^2 (k+2M) (3k+2M)} 
\nonu \\
& \times & (-324k^4 + 72k^6 - 972k^3M + 219k^5M - 1080k^2M^2 \nonu\\
& +& 249k^4M^2 - 528kM^3 
+126k^3M^3 - 96M^4 + 24k^2M^4 - 684k^3N + 144k^5N \nonu\\
& -& 1536k^2MN + 321k^4MN -  1152kM^2N + 244k^3M^2N - 288M^3N + 64k^2M^3N \nonu\\
& -& 360k^2N^2 + 72k^4N^2 - 528kMN^2 
+102k^3MN^2 - 192M^2N^2 \nonu \\
& + &  44k^2M^2N^2 + 8kM^3N^2), \nonu\\
p_{86} & = & -\frac{24 (k+M) (k+N) (M+N) (k+M+N) (k+2N)^2 a_1 b_1}{k^2 M^2 (k+2M) N}, \nonu\\
p_{87} & = & -\frac{6 (k+N) \sqrt{\frac{M+N}{M N}} (k+M+N) (k+2N) a_1 b_1}{k M (k+2M)} 
(-4 + 4k^2 + 3kM + 6kN + 6MN), \nonu\\
p_{88} & = & \frac{6 (k+N) \sqrt{\frac{M+N}{M N}} (k+M+N) (k+2N) a_1 b_1}{k M (k+2M)} 
(-4 + 4k^2 + 3kM + 6kN + 6MN), \nonu\\
p_{89} & = & -\frac{6 (k+N) (k+2N)^2 a_1 b_1}{k M^3 (k+M) (k+2M) (3k+2M)} 
(3k^2 + 5kM + 2M^2 + 3k^2M^2 + 5kM^3 + 2M^4 \nonu\\
& +& 6kN + 2MN 
+ 4k^2MN + 5kM^2N + 2M^3N), \nonu\\
p_{90} & = & -\frac{24 (k+M) (k+N)^2 (k+M+N) a_1 b_1}{k M (k+2M) N}, \qquad
p_{91}  =  \frac{6 (k+N) (k+2N) a_1 b_1}{M}, \nonu\\
p_{92} & = & \frac{6 (k+N) (k+2N)^2 a_1 b_1}{M^2 (k+2M)}, \nonu\\
p_{93} & = & -\frac{3 (k+N) (k+M+N) (k+2N) a_1 b_1}{2 M (k+M) (k+2M)} 
(-12k + 5k^3 - 8M + 6k^2M + 2kM^2 - 8N \nonu\\
& +& 6k^2N + 8kMN + 4M^2N), \nonu\\
p_{94} & = & -\frac{6 (k+N) \sqrt{\frac{M+N}{M N}} (k+M+N) (k+2N) a_1 b_1}{k M (k+2M)} 
(4 + kM + 2kN + 2MN), \nonu\\
p_{95} & = & \frac{3 (k+N) (k+M+N) (k+2N) a_1 b_1}{k M (k+M) (k+2M)} 
(-4k^2 + 3k^4 - 16kM + 8k^3M - 8M^2 + 4k^2M^2 \nonu\\
& -& 8kN + 6k^3N - 16MN +12k^2MN + 4kM^2N), \nonu\\
p_{96} & = & -\frac{3 (-2 + k) (2 + k) (k+N) (k+M+N) (k+2N)}{2k M (k+M) (k+2M)} \nonu \\
& \times &  (5k^2 + 6kM + 2M^2 + 6kN + 4MN) a_1 b_1, \nonu\\
p_{97} & = & \frac{6 (k+N) (k+2N) (k+2M+2N) a_1 b_1}{k M^2}, \nonu\\
p_{98} & = & \frac{6 (4k + 3M) (k+N) \sqrt{\frac{M+N}{M N}} (k+M+N) (k+2N)^2 a_1 b_1}{k M (k+M) (k+2M)}, \nonu\\
p_{99} & = & -\frac{6 (k+N) (k+2N)^2 a_1 b_1}{k M^3 (k+M) (k+2M) (3k+2M)} 
(3k^2 + 5kM + 2M^2 - 6kN + 2MN + 2kM^2N), \nonu\\
p_{100} & = & \frac{3 (k+N) (k+M+N) (k+2N) a_1 b_1}{k M (k+M) (k+2M)} 
(-4k + 4k^3 - 4M + 3k^2M + 6k^2N + 4kMN), \nonu\\
p_{101} & = & \frac{3 i (-2 + k) (2 + k) (k+N) (k+M+N) (k+2N) (2k + M + 2N) a_1 b_1}{k M (k+M) (k+2M)}, \nonu\\
p_{102} & = & -\frac{3 (k+N) (k+M+N) (k+2N) a_1 b_1}{k M (k+M) (k+2M)} 
\nonu \\
& \times & (-4k + 4k^3 - 4M + 3k^2M + 6k^2N + 4kMN), \nonu\\
p_{103} & = & \frac{3 i (-2 + k) (2 + k) (k+N) (k+M+N) (k+2N) (2k + M + 2N) a_1 b_1}{k M (k+M) (k+2M)}, \nonu\\
p_{104} & = & \frac{6 (k+N) (k+M+N) (k+2N) (2 + kN) a_1 b_1}{k M (k+2M)}, \nonu\\
p_{105} & = & \frac{3 i (k+N) (k+2N) a_1 b_1}{2 k M^2 (k+M) (k+2M) (3k+2M)} (12k^3M + 3k^5M - 28k^2M^2 + 17k^4M^2 - 48kM^3 \nonu\\
& +& 16k^3M^3 - 16M^4 + 4k^2M^4 - 4k^3N + 64k^2MN - 3k^4MN + 40kM^2N + 14k^3M^2N \nonu\\
& +& 10k^2M^3N - 8k^2N^2 + 32kMN^2 - 6k^3MN^2 + 16M^2N^2 + 4k^2M^2N^2 + 4kM^3N^2 ), \nonu\\
p_{106} & = & \frac{3 (k+N) (k+2N)^2 a_1 b_1}{M (k+2M)}, \nonu\\
p_{107} & = & -\frac{3 (k+N) (k+2N)^2 a_1 b_1}{2 k M^2 (k+M) (k+2M) (3k+2M)}(3k^2 + 5kM + 2M^2 + 6kN + 2MN \nonu \\
&+ & 4k^2MN + 2kM^2N ), \nonu\\
p_{108} & = & -\frac{3 (k+N) (k+2N)^2 a_1 b_1}{2 k M^2 (k+M) (k+2M) (3k+2M)}(-6k^2 - 10kM + 9k^3M - 4M^2 + 9k^2M^2 \nonu\\
& +& 2kM^3 - 4MN + 7k^2MN + 4kM^2N ), \nonu\\
p_{109} & = & -\frac{3 (k+N) (k+2N)^2 a_1 b_1}{2 k M^2 (k+M)^2 (k+2M) (3k+2M)} (3k^3 + 8k^2M + 3k^4M + 7kM^2 + 11k^3M^2 \nonu \\
&+& 2M^3 
+ 12k^2M^3 + 4kM^4 - 6k^2N - 4kMN + k^3MN + 2M^2N ), \nonu\\
p_{110} & = & \frac{3 i (k+N) (k+2N)^2 a_1 b_1}{16 k M^2 (k+M) (k+2M) (3k+2M)} \nonu \\
& \times &
(24k^2 + 40kM + 16M^2 + 3k^2M^2 + 2kM^3 + 16kN 
+ 16MN \nonu \\
&- & 24k^2MN - 16kM^2N ), \nonu\\
p_{111} & = & -\frac{3 i (k+N) (k+2N)^2 a_1 b_1}{16 (k+M) (k+2M)},
\nonu \\
p_{112}  & = &
\frac{3 i (k+N) (k+2N)^2 a_1 b_1}{8 k M^2 (k+M) (k+2M) (3k+2M)} (12k^2 + 20kM \nonu\\
& +& 8M^2 + 3k^2M^2 + 2kM^3 + 8kN + 8MN ), \nonu\\
p_{113} & = & -\frac{3 i (k+N) (k+2N)^2 a_1 b_1}{8 k M^2 (k+M) (k+2M) (3k+2M)} \nonu \\
& \times & (12k^2 + 20kM + 8M^2 + 3k^2M^2 + 2kM^3 + 8kN 
+ 8MN - 12k^2MN - 8kM^2N ), \nonu\\
p_{114} & = & -\frac{3 (k+N) (k+2N)^2 (3k+2M+2N) a_1 b_1}{2 k M^2 (k+2M) (3k+2M)}, \nonu \\
p_{115}  & = &
\frac{3 i (k+N) (k+2N)^2 (M+8N) a_1 b_1}{16 M (k+M) (k+2M)}, \nonu\\
p_{116} & = & \frac{3 (k+N) (k+2N)^2}{2 k M^2 (k+M) (k+2M) (3k+2M)}
\nonu \\
& \times &  (6k^2 + 10kM + 3k^3M + 4M^2 + 5k^2M^2 + 2kM^3 + 4MN - k^2MN) a_1 b_1, \nonu\\
p_{117} & = & \frac{3 i (k+N) (k+2N)^2 }{16 k M^2 (k+M) (k+2M) (3k+2M)}
\nonu \\
& \times & (24k^2 + 40kM + 16M^2 + 9k^2M^2 + 6kM^3 + 16kN + 16MN) a_1 b_1, \nonu\\
p_{118} & = & -\frac{9 (k+N) (k+2N)^2 (3k+2M+2N) a_1 b_1}{2 k M^2 (k+2M) (3k+2M)}, \nonu\\
p_{119} & = & -\frac{6 (k+N) (k+2N)^2 a_1 b_1}{k M^3 (k+M) (k+2M) (3k+2M)} 
(-6k^2 - 10kM + 9k^3M - 4M^2 + 18k^2M^2 \nonu\\
& +& 11kM^3 + 2M^4 - 4MN + 6k^2MN + 7kM^2N + 2M^3N ), \nonu\\
p_{120} & = & -\frac{6 (k+N) (k+2N)^2 a_1 b_1}{k M^2 (3k+2M)} 
(3k+2M+2N ), \nonu\\
p_{121} & = & \frac{24 (k+M) (k+N) (k+M+N) (2k+3N) a_1 b_1}{k M^2 (k+2M)}, \nonu\\
p_{122} & = & -\frac{12 (k+N) \sqrt{\frac{M+N}{M N}} a_1 b_1}{k M (k+2M)} 
(4k^3 + 8k^2M + 4kM^2 + 10k^2N + 17kMN + 6M^2N\nonu\\
& +& 6kN^2 + 8MN^2 ), \nonu\\
p_{123} & = & -\frac{6 (k+N) (k+M+N) a_1 b_1}{k M (k+2M)} 
(-4k + 4k^3 + 3k^2M - 8N + 6k^2N + 4kMN ), \nonu\\
p_{124} & = & \frac{6 (k+N) (k+M+N) a_1 b_1}{k M (k+2M)} 
(-4k + 4k^3 + 3k^2M - 8N + 6k^2N + 4kMN ), \nonu\\
p_{125} & = & -\frac{6 (4+kM) (k+N) (k+M+N) (k+2N) a_1 b_1}{k M (k+2M)}, \nonu\\
p_{126} & = & \frac{6 (k+N) (k+M+N) a_1 b_1}{M (k+M) (k+2M)} 
(4k^2 + 3kM + 6kN + 4MN ), \nonu\\
p_{127} & = & \frac{12 N (k+N) \sqrt{\frac{M+N}{M N}} (k+2N) a_1 b_1}{k (k+2M)}, \nonu\\
p_{128} & = & -\frac{12 (k+M) (k+N) (k+M+N) a_1 b_1}{M (k+2M)}, \nonu\\
p_{129} & = & -\frac{N (k+N) (k+2N) a_1 b_1}{12 k^2 M (k+M) (k+2M) (3k+2M)} 
\nonu \\
& \times &
(-216k^5 + 63k^7 + 576k^2M - 864k^4M + 204k^6M \nonu\\
& +& 672kM^2 - 768k^3M^2 + 189k^5M^2 + 192M^3 - 192k^2M^3 + 72k^4M^3 + 12k^3M^4 \nonu \\
& - &  372k^4N 
+ 117k^6N + 672kMN - 1236k^3MN + 339k^5MN + 384M^2N \nonu \\
&- &  768k^2M^2N + 270k^4M^2N 
-96kM^3N + 100k^3M^3N + 24k^2M^4N - 168k^3N^2 \nonu \\
& + & 54k^5N^2 + 192MN^2 - 360k^2MN^2 \nonu\\
&+& 126k^4MN^2 - 96kM^2N^2 + 72k^3M^2N^2 + 8k^2M^3N^2 ). 
\label{pvalue}
\eea
The coefficient $b_1$ is given by (\ref{b1value}).

We list the coefficients appearing in (\ref{rtilde})
as follows:
\bea
g_{1} &=& -\frac{4 N (M + N) \sqrt{\frac{M + N}{M N}} (k + 2 N) a_{1}}{k^{3}}, \quad
g_{2} = \frac{4 (-k + N) \sqrt{\frac{M + N}{M N}} (k + 2 N) a_{1}}{k^{2}}, \nonu \\
g_{3} &=& -\frac{(4 + 2 k^{2} + 3 k M) (k + 2 N) a_{1}}{k (k + M)}, \quad
g_{4} = -\frac{2 (-k + N) (M + N) (k + 2 N) a_{1}}{k^{2}}, \nonu \\
g_{5} &=& \frac{(4 + 2 k^{2} + 3 k M) (k + 2 N) a_{1}}{k (k + M)},
\nonu \\
g_{6} & = &
\frac{M N \sqrt{\frac{M + N}{M N}} (k + 2 N) a_{1}}{42 k^{2} (k + M) (3 k + 2 M)} (36 k^{2} + 54 k^{4} + 105 k^{3} M \nonu \\
& -& 16 M^{2} + 46 k^{2} M^{2} + 24 k N - 6 k^{3} N - 16 M N +
4 k^{2} M N ), \nonu\\
g_{7} &=& \frac{(-2 + M) (2 + M) N^{2} (k + 2 N) a_{1}}{2 M (k + M) (2 k + M) (3 k + 2 M)}, \quad
g_{8} = -\frac{(-12 + M^{2}) N^{2} (k + 2 N) a_{1}}{6 M (k + M) (2 k + M) (3 k + 2 M)}, \nonu \\
g_{9} &=& \frac{M N^{2} (k + 2 N) a_{1}}{6 (k + M) (2 k + M) (3 k + 2 M)}, \quad
g_{10} = -\frac{k M N^{2} a_{1}}{6 (k + M) (2 k + M) (3 k + 2 M)}, \nonu \\
g_{11} &=& -\frac{M N^{3} a_{1}}{3 (k + M) (2 k + M) (3 k + 2 M)}, \quad
g_{12} = -\frac{2 N^{2} a_{1}}{(k + M) (2 k + M) (3 k + 2 M)}, \nonu \\
g_{13} &=& -\frac{4 N^{3} a_{1}}{k (k + M) (2 k + M) (3 k + 2 M)}, \quad
g_{14} = -\frac{2 k N^{2} a_{1}}{M (k + M) (2 k + M) (3 k + 2 M)}, \nonu \\
g_{15} &=& -\frac{4 N^{3} a_{1}}{M (k + M) (2 k + M) (3 k + 2 M)}, \quad
g_{16} = -\frac{N^{2} (k + 2 N) a_{1}}{3 (k + M) (2 k + M) (3 k + 2 M)}, \nonu \\
g_{17} &=& -\frac{(-24 + 5 M^{2}) N^{2} (k + 2 N) a_{1}}{3 M^{2} (k + M) (2 k + M) (3 k + 2 M)}, \quad
g_{18} = \frac{2 (-4 k + 2 k M^{2} + M^{3}) N^{2} (k + 2 N) a_{1}}{k M^{2} (k + M) (2 k + M) (3 k + 2 M)}, \nonu \\
g_{19} &=& \frac{2 (3 k + 2 M) N (M + N) (k + 2 N) a_{1}}{k^{2} (k + M) (2 k + M)}, \nonu \\
g_{20} & = &
-\frac{(-2 + M) (2 + M) N^{2} \sqrt{\frac{M + N}{M N}} (k + 2 N) a_{1}}{k M (k + M) (3 k + 2 M)}, \nonu\\
g_{21} &=& \frac{(-12 + M^{2}) N^{2} \sqrt{\frac{M + N}{M N}} (k + 2 N) a_{1}}{3 k M (k + M) (3 k + 2 M)}, \quad
g_{22} = -\frac{M (11 k + 7 M) N^{2} \sqrt{\frac{M + N}{M N}} (k + 2 N) a_{1}}{3 k (k + M) (2 k + M) (3 k + 2 M)}, \nonu \\
g_{23} &=& \frac{M N^{2} \sqrt{\frac{M + N}{M N}} (k + 2 N) a_{1}}{3 k (k + M) (3 k + 2 M)}, \quad
g_{24} = \frac{4 N^{2} \sqrt{\frac{M + N}{M N}} (k + 2 N) a_{1}}{k^{2} (k + M) (3 k + 2 M)}, \nonu \\
g_{25} &=& \frac{4 N^{2} \sqrt{\frac{M + N}{M N}} (k + 2 N) a_{1}}{k M (k + M) (3 k + 2 M)}, \quad
g_{26} = -\frac{4 (-4 k + 2 k M^{2} + M^{3}) N^{2} \sqrt{\frac{M + N}{M N}} (k + 2 N) a_{1}}{k^{2} M^{2} (k + M) (3 k + 2 M)}, \nonu \\
g_{27} &=& \frac{2 (-24 + 5 M^{2}) N^{2} \sqrt{\frac{M + N}{M N}} (k + 2 N) a_{1}}{3 k M^{2} (k + M) (3 k + 2 M)}, \quad
g_{28} = \frac{2 N^{2} \sqrt{\frac{M + N}{M N}} (k + 2 N) a_{1}}{3 k (k + M) (3 k + 2 M)}, \nonu \\
g_{29} &=& \frac{(k + 2 N) (2 k^{2} + k M - 2 k N - 2 M N) a_{1}}{k (k + M) (2 k + M)}, \nonu\\
g_{30} &=& \frac{M N \sqrt{\frac{M + N}{M N}} (k + 2 N) (-2 k^{2} - k M + 2 k N + 2 M N) a_{1}}{2 k (k + M) (2 k + M)}, \nonu \\
g_{31} &=& \frac{i N \sqrt{\frac{M + N}{M N}} (k + 2 N) a_{1}}{42 k^{2} (k + M) (3 k + 2 M)} ( -324 k^{2} + 144 k^{4} - 336 k M + 189 k^{3} M \nonu \\
& -& 80 M^{2} + 62 k^{2} M^{2} - 384 k N + 96 k^{3} N - 416 M N + 104 k^{2} M N ), \nonu \\
g_{32} &=& -\frac{4 M N \sqrt{\frac{M + N}{M N}} a_{1}}{k^{2}}, \quad
g_{33} = \frac{2 M N a_{1}}{k (2 k + M)},\quad
g_{34} = -\frac{2 M (k + N) a_{1}}{k}, \nonu\\
g_{35} &=& -\frac{2 (M + N) (k + 2 N)^{2} a_{1}}{k^{2}},\quad
g_{36} = \frac{(k + 2 N) (4 + k M + 2 k N + 2 M N) a_{1}}{k (k + M)},
\nonu\\
g_{37} &=& -\frac{M N \sqrt{\frac{M + N}{M N}} (k + 2 N) a_{1}}{21 k^{2} (k + M) (3 k + 2 M)} ( 216 k^{2} - 54 k^{4} + 168 k M 
 - 42 k^{3} M + 16 M^{2} - 4 k^{2} M^{2}\nonu\\
 &+& 228 k N + 6 k^{3} N + 184 M N + 59 k^{2} M N + 42 k M^{2} N ), \nonu\\
 g_{38} &=& \frac{M N \sqrt{\frac{M + N}{M N}} (k + 2 N) (2 k^{2} + k M + 10 k N + 6 M N) a_{1}}{2 k (k + M) (2 k + M)}, \nonu\\
g_{39} &=& -\frac{i N \sqrt{\frac{M + N}{M N}} (k + 2 N) a_{1}}{14 k^{2} (k + M) (3 k + 2 M)} ( 324 k^{2} - 144 k^{4} 
+ 336 k M - 189 k^{3} M + 80 M^{2} - 62 k^{2} M^{2}\nonu \\
& + & 300 k N - 96 k^{3} N + 304 M N - 104 k^{2} M N 
- 14 k M^{2} N ), \nonu \\
g_{40} &=& -\frac{N (k + 2 N) (72 k^{2} + 84 k M + 24 M^{2} + 48 k N + 24 M N + 5 k M^{2} N + 4 M^{3} N) a_{1}}{12 M (k + M) (2 k + M) (3 k + 2 M)}, \nonu\\
g_{41} &=& -\frac{M (13 k + 8 M) N^{2} (k + 2 N) a_{1}}{12 (k + M) (2 k + M)(3 k + 2 M)}, \quad
g_{42} = \frac{N (k + 2 N) (6 k + 4 M + 4 N - M^{2} N) a_{1}}{2 M (k + M) (3 k + 2 M)}, \nonu\\
g_{43} &=& \frac{i N (k + 2 N) a_{1}}{168 k M (k + M) (2 k + M) (3 k + 2 M)}( 504 k^{3} - 708 k^{2} M \nonu \\
& +& 576 k^{4} M - 1680 k M^{2} + 630 k^{3} M^{2} - 656 M^{3} + 101 k^{2} M^{3} - 42 k M^{4} + 504 k^{2} N \nonu \\
& - & 696 k M N + 384 k^{3} M N - 992 M^{2} N + 290 k^{2} M^{2} N ), \nonu \\
g_{44} &=& -\frac{i N (k + 2 N) a_{1}}{168 k M (k + M) (2 k + M) (3 k + 2 M)} ( 504 k^{3} + 1884 k^{2} M - 576 k^{4} M + 1008 k M^{2} \nonu \\
& -& 630 k^{3} M^{2} - 16 M^{3} - 227 k^{2} M^{3} - 42 k M^{4} + 504 k^{2} N +1872 k M N - 384 k^{3} M N \nonu \\
& +& 1664 M^{2} N - 290 k^{2} M^{2} N + 84 k M^{3} N ), \nonu \\
g_{45} &=& \frac{i N (k + 2 N) (2 k + M + 2 N) a_{1}}{2 M (k + M)
(2 k + M)}, \quad
g_{46} = -\frac{2 N^{2} (k + 2 N) a_{1}}{k (k + M) (3 k + 2 M)}, \nonu \\
g_{47} &=& \frac{i M N (k + 2 N) (2 k + M + 2 N) a_{1}}{8 (k + M)
(2 k + M)}, \nonu \\
g_{48} & = &
-\frac{N (k + 2 N) (3 k^{2} + 5 k M + 2 M^{2} + 2 k N) a_{1}}{k M (k + M) (3 k + 2 M)}, \nonu \\
g_{49} &=& \frac{i M N (k + 2 N) a_{1}}{8 (k + M)}, \quad
g_{50} = \frac{M N^{2} (k + 2 N) a_{1}}{6 (k + M) (3 k + 2 M)}, \nonu\\
g_{51} &=& \frac{N (k + 2 N) (-36 k - 24 M + 9 k M^{2} + 6 M^{3} - 24 N + 5 M^{2} N) a_{1}}{3 M^{2} (k + M) (3 k + 2 M)}, \nonu \\
g_{52}  & = & \frac{N^{2} (k + 2 N) a_{1}}{3 (k + M) (3 k + 2 M)}, \nonu \\
g_{53} &=& -\frac{N (k + 2 N) a_{1}}{k M^{2} (k + M) (3 k + 2 M)}
\nonu \\
& \times &
(-12 k^{2} - 8 k M + 3 k^{2} M^{2} + 2 k M^{3} - 8 k N + 4 k M^{2} N + 2 M^{3} N ),\nonu\\
g_{54} &=& -\frac{M N (k + 2 N) a_{1}}{84 k (k + M) (2 k + M) (3 k + 2 M)}
(-432 k^{2} + 108 k^{4} - 336 k M + 84 k^{3} M - 32 M^{2} \nonu\\
&+& 8 k^{2} M^{2} - 456 k N - 12 k^{3} N - 368 M N - 55 k^{2} M N - 42 k M^{2} N ), \nonu \\
g_{55} &=& -\frac{i N (k + 2 N) a_{1}}{84 k (k + M) (2 k + M) (3 k + 2 M)}
(-1800 k^{3} + 576 k^{5} - 2412 k^{2} M + 876 k^{4} M \nonu\\
&-& 964 k M^{2} + 430 k^{3} M^{2} - 104 M^{3} + 68 k^{2} M^{3} - 1284 k^{2} N + 384 k^{4} N - 1228 k M N \nonu\\
&+& 496 k^{3} M N - 272 M^{2} N + 299 k^{2} M^{2} N + 84 k M^{3} N ), \quad
g_{56} = 2 M a_{1}, \nonu \\
g_{57} &=& -\frac{M N \sqrt{\frac{M + N}{M N}} (k + 2 N) a_{1}}{21 k^{2} (k + M) (3 k + 2 M)} (-180 k^{2} + 108 k^{4} - 168 k M + 147 k^{3} M - 32 M^{2} + 50 k^{2} M^{2} \nonu\\
&-& 204 k N + 114 k^{3} N - 200 M N + 155 k^{2} M N + 42 k M^{2} N ),
\nonu\\
g_{58} &=& \frac{M N (k + 2 N) a_{1}}{28 k (k + M) (2 k + M) (3 k + 2 M)}
(-132 k^{2} + 54 k^{4} - 154 k M + 63 k^{3} M - 44 M^{2} \nonu\\
&+& 18 k^{2} M^{2} - 144 k N + 78 k^{3} N - 128 M N + 81 k^{2} M N + 14 k M^{2} N ), \nonu \\
g_{59} &=& -\frac{i N (k + 2 N) a_{1}}{84 k (k + M) (2 k + M) (3 k + 2 M)}
(-396 k^{3} + 288 k^{5} - 450 k^{2} M + 522 k^{4} M \nonu\\
&-& 118 k M^{2} + 313 k^{3} M^{2} + 4 M^{3} + 62 k^{2} M^{3} - 12 k^{2} N + 192 k^{4} N + 44 k M N + 304 k^{3} M N \nonu\\
&+& 4 M^{2} N + 125 k^{2} M^{2} N ), \quad
g_{60} = -\frac{4 (M + N) (k + 2 N)^{2} a_{1}}{k^{2} N}, \nonu\\
g_{61} &=& -\frac{M \sqrt{\frac{M + N}{M N}} (k + 2 N) a_{1}}{k (k + M)}
(-4 + 4 k^{2} + 3 k M + 6 k N + 6 M N ), \nonu\\
g_{62} &=& \frac{M \sqrt{\frac{M + N}{M N}} (k + 2 N) a_{1}}{k (k + M)}
(-4 + 4 k^{2} + 3 k M + 6 k N + 6 M N ), \quad
g_{63} = -\frac{4 M (k + N) a_{1}}{k N}, \nonu\\
g_{64} &=& -\frac{M (k + 2 N) a_{1}}{42 k (k + M) (3 k + 2 M)}
(-36 k^{2} + 72 k^{4} + 105 k^{3} M + 16 M^{2} + 38 k^{2} M^{2} - 24 k N \nonu\\
&+& 132 k^{3} N + 16 M N + 206 k^{2} M N + 84 k M^{2} N ), \nonu \\
g_{65} &=& -\frac{M \sqrt{\frac{M + N}{M N}} (k + 2 N) (4 + k M + 2 k N + 2 M N) a_{1}}{k (k + M)}, \nonu\\
g_{66} &=& \frac{2 M (k + 2 N) a_{1}}{21 k (k + M) (3 k + 2 M)} (-234 k^{2} + 90 k^{4} - 294 k M + 126 k^{3} M - 92 M^{2} + 44 k^{2} M^{2} \nonu\\
&-& 240 k N + 123 k^{3} N - 176 M N + 149 k^{2} M N + 42 k M^{2} N ), \nonu \\
g_{67} &=& -\frac{(-2 + k) (2 + k) M (k + 2 N) a_{1}}{42 k (k + M) (3 k + 2 M)} (72 k^{2} + 105 k M + 38 M^{2} + 132 k N + 80 M N ), \nonu\\
g_{68} &=& \frac{4 (k + 2 N) a_{1}}{k}, \quad
g_{69} = \frac{M \sqrt{\frac{M + N}{M N}} (k + 2 N) a_{1}}{k (k + M) (2 k + M)} (2 k^{2} + k M + 10 k N + 6 M N ), \nonu\\
g_{70} &=& \frac{M (k + 2 N) a_{1}}{2 k (k + M) (2 k + M)} (-8 k + 2 k^{3} - 4 M + k^{2} M + 6 k^{2} N + 4 k M N ), \nonu \\
g_{71} &=& \frac{i (-2 + k) (2 + k) (k + 2 N) a_{1}}{42 k (k + M) (3 k + 2 M)} (144 k^{2} + 189 k M + 62 M^{2} + 96 k N + 104 M N ), \nonu\\
g_{72} &=& -\frac{M (k + 2 N) a_{1}}{2 k (k + M) (2 k + M)} (-8 k + 2 k^{3} - 4 M + k^{2} M + 6 k^{2} N + 4 k M N ), \nonu \\
g_{73} &=& \frac{i (-2 + k) (2 + k) (k + 2 N) a_{1}}{42 k (k + M) (3 k + 2 M)} (144 k^{2} + 189 k M + 62 M^{2} + 96 k N + 104 M N ), \nonu\\
g_{74} &=& \frac{M (k + 2 N) (2 + k N) a_{1}}{k (k + M)}, \nonu \\
g_{75} &=& -\frac{i (k + 2 N) a_{1}}{42 k (k + M) (2 k + M) (3 k + 2 M)}
(-1296 k^{3} + 576 k^{5} - 1992 k^{2} M + 1044 k^{4} M \nonu\\
&-& 992 k M^{2} + 626 k^{3} M^{2} - 160 M^{3} + 124 k^{2} M^{3} - 780 k^{2} N + 384 k^{4} N - 1340 k M N \nonu\\
&+& 608 k^{3} M N - 496 M^{2} N + 229 k^{2} M^{2} N ), \nonu\\
g_{76} &=& \frac{M (k + 2 N) a_{1}}{k + M}, \quad
g_{77} = \frac{(k + 2 N) (6 k + 4 M + 4 N - M^{2} N) a_{1}}{M (k + M) (3 k +2 M)}, \nonu \\
g_{78} &=& -\frac{(k + 2 N) a_{1}}{6 M (k + M) (2 k + M) (3 k + 2 M)}
\nonu \\
& \times &
(72 k^{2} + 84 k M + 24 M^{2} + 48 k N + 24 M N + 5 k M^{2} N \nonu\\
&+& 4 M^{3} N ), \quad
g_{79} = -\frac{M (13 k + 8 M) N (k + 2 N) a_{1}}{6 (k + M) (2 k + M) (3 k +2 M)},\quad
g_{80} = \frac{M N (k + 2 N) a_{1}}{3 (k + M) (3 k + 2 M)}, \nonu\\
g_{81} &=& \frac{4 N (k + 2 N) a_{1}}{k (k + M) (3 k + 2 M)}, \quad
g_{82} = \frac{2 (k + 2 N) a_{1}}{k M (k + M) (3 k + 2 M)} (3 k^{2} + 5 k M + 2 M^{2} + 2 k N ), \nonu\\
g_{83} &=& -\frac{2 (k + 2 N) a_{1}}{k M^{2} (k + M) (3 k + 2 M)}
\nonu \\
& \times &
(-12 k^{2} - 8 k M + 3 k^{2} M^{2} + 2 k M^{3} - 8 k N + 4 k M^{2} N + 2 M^{3}N ), \nonu \\
g_{84} &=& \frac{2 (k + 2 N) a_{1}}{3 M^{2} (k + M) (3 k + 2 M)} (-36 k - 24M + 9 k M^{2} + 6 M^{3} - 24 N + 5 M^{2} N ), \nonu\\
g_{85} &=& \frac{2 N (k + 2 N) a_{1}}{3 (k + M) (3 k + 2 M)}, \quad
g_{86} = -4 a_{1}, \quad
g_{87} = -\frac{4 M \sqrt{\frac{M + N}{M N}} (2 k + 3 N) a_{1}}{k}, \nonu \\
g_{88} &=& -\frac{M a_{1}}{k (k + M)} (-4 k + 4 k^{3} + 3 k^{2} M - 8 N + 6 k^{2} N + 4 k M N ), \nonu\\
g_{89} &=& \frac{M a_{1}}{k (k + M)} (-4 k + 4 k^{3} + 3 k^{2} M - 8 N + 6 k^{2} N + 4 k M N ), \nonu \\
g_{90} &=& -\frac{M (4 + k M) (k + 2 N) a_{1}}{k (k + M)}, \quad
g_{91} = \frac{M (2 k^{2} + k M + 6 k N + 4 M N) a_{1}}{(k + M) (2 k + M)},
\nonu \\
g_{92} & = & -2 M a_{1}, \nonu\\
g_{93} &=& -\frac{M N (k + 2 N) a_{1}}{12 k (k + M) (3 k + 2 M)} (-24 k^{2} + 12 k^{4} - 16 k M + 14 k^{3} M + 4 k^{2} M^{2} - 16 k N \nonu\\
&+& 10 k^{3} N - 8 M N + 11 k^{2} M N + 4 k M^{2} N ).
\label{gcoeff}
\eea

At the particular value of the level
\bea
k = - 2 N,
\label{kcondition}
\eea
where the previous charged and neutral spin-$3$ currents
are decoupled,
the charged spin-$4$ current becomes 
\bea
&& g_{10} d^{f v h} f^{a v o} f^{b c o} J^{b} J^{c} J^{f} J^{h} 
+ g_{11} d^{f v h} f^{a v o} f^{b c o} J^{b} J^{c} J^{f} J^{h} 
+ g_{12} d^{f v h} f^{a c o} f^{b v o} J^{b} J^{c} J^{f} J^{h}  \nonu\\
&&+ g_{13} d^{f v h} f^{a c o} f^{b v o} J^{b} J^{c} J^{f} J^{h} 
+ g_{14} d^{f v h} f^{a b o} f^{c v o} J^{b} J^{c} J^{f} J^{h} 
+ g_{15} d^{f v h} f^{a b o} f^{c v o} J^{b} J^{c} J^{f} J^{h}  \nonu\\
&& + g_{32} J^{\beta} J^{\beta} J^{a} J^{u(1)} 
+ g_{33} d^{a c f} J^{\beta} J^{\beta} J^{c} J^{f} 
+ g_{34} J^{\beta} J^{\beta} \partial J^{a}
+ g_{63} \delta^{c f} \delta_{\rho \bar{\tau}} t^{a}_{k \bar{i}} J^{c} J^{f} J^{(\rho \bar{i})} J^{(\bar{\tau} k)} \nonu \\
&&+ g_{86} \delta_{\bar{i} k} t^{\beta}_{\rho \bar{\tau}} J^{\beta} J^{a} J^{(\rho \bar{i})} J^{(\bar{\tau} k)}  
+ g_{87} t^{a}_{k \bar{i}} t^{\beta}_{\rho \bar{\tau}} J^{\beta} J^{u(1)} J^{(\rho \bar{i})} J^{(\bar{\tau} k)}  
+ g_{88} t^{a}_{k \bar{i}} t^{\beta}_{\rho \bar{\tau}} J^{\beta} J^{(\rho \bar{i})} \partial J^{(\bar{\tau} k)}  \nonu\\
&&+ g_{89} t^{a}_{k \bar{i}} t^{\beta}_{\rho \bar{\tau}} J^{\beta} \partial J^{(\rho \bar{i})} J^{(\bar{\tau} k)}  
+ g_{91} d^{a c f} t^{f}_{k \bar{i}} t^{\beta}_{\rho \bar{\tau}} J^{\beta} J^{c} J^{(\rho \bar{i})} J^{(\bar{\tau} k)}.
\label{simple1}
\eea
We expect that the eighth order pole of the OPE
of (\ref{simple1}) with itself vanishes.

%%%%%%%%%%%%%%%%%%%%%%%%%%%%%%%%%%%%%%%%%%%%%%%%%%%%%%%%%%%%%%
%%%%%%%%%%%%%%%%%%%%%%%%%%%%%%%%%%%%%%%%%%%%%%%%%%%%%%%%%%%%%%%%%%%%%
\section{ The  OPE
of the neutral spin-$3$ current $W^{(3)}$ with itself}
%%%%%%%%%%%%%%%%%%%%%%%%%%%%%%%%%%%%%%%%%%%%%%%%%%%%%%%%%%%%%%%%%%%%%A%
%%%%DDD%%%%%%%%%%%%%%%%%%%%%%%%%%%%%%%%%%%%%%%%%%%%%%%%%%%%%%%%%%%%%%%%

%%%%%%%%%%%%%%%%%%%%%%%%%%%%%%%%%%%%%%%%%%%%%%%%%%%%%%%%%%%%%%%%%%%%%%%%%%%
%After substituting (\ref{interbvalue})
%into (\ref{WWP2}),
The explicit coefficients appearing in (\ref{WWPOLE2})
are given by
\bea
w_{1} & = & -\frac{36 (k+N)^2 (M+N)^2 (k+M+N) (k+2N)^2 b_1^2}{k^3 M^3 N}, \nonu\\
w_{2} & = & \frac{144 (k+N)^2 (M+N) (k+M+N) (k+2N)^2 b_1^2}{k^2 M^3 N}, \nonu\\
w_{3} & = & \frac{72 (k+N)^2 \sqrt{\frac{M+N}{M N}} (k+M+N) (k+2N) b_1^2}{k M^2 (k+2M)} (-2 + 2k^2 + 3kM + 3kN + 6MN), \nonu\\
w_{4} & = & -\frac{72 (k+N)^2 \sqrt{\frac{M+N}{M N}} (k+M+N) (k+2N) b_1^2}{k M^2 (k+2M)} (-2 + 2k^2 + 3kM + 3kN + 6MN), \nonu\\
w_{5} & = & -\frac{72 (3k + 4M) (k+N)^2 (M+N) (k+M+N) (k+2N)^2 b_1^2}{k^2 M^3 (k+2M)^2}, \nonu\\
w_{6} & = & \frac{72 N (k+N)^2 \sqrt{\frac{M+N}{M N}} (k+M+N) (k+2N)^2 b_1^2}{k M^2 (k+2M)^2}, \nonu\\
w_{7} & = & \frac{36 i N (k+N)^2 \sqrt{\frac{M+N}{M N}} (2k + 2M + N) (k+2N)^2 b_1^2}{k M^2 (k+2M)^2}, \nonu\\
w_{8} & = & -\frac{9 N (k+N)^2 (k+M+N) (k+2N)^2 b_1^2}{M^2 (k+M) (k+2M)^2}, \nonu\\
w_{9} & = & \frac{9 i N (k+N)^2 (k+M+N) (k+2N)^2 b_1^2}{M^2 (k+M) (k+2M)^2}, \nonu\\
w_{10} & = & -\frac{9 i N (k+N)^2 (k+2N)^2 b_1^2}{M^2 (k+2M)^2}, \qquad
w_{11}  =  -\frac{9 N (k+N)^2 (k+2N)^2 b_1^2}{M^2 (k+2M)^2}, \nonu\\
w_{12} & = & -\frac{36 N (k+N)^2 (k+M+N) (k+2N)^2 b_1^2}{k M^3 (k+2M)^2}, \nonu\\
w_{13} & = & \frac{144 (k+M) (k+N)^2 (k+M+N) (k+2N)^2 b_1^2}{k M^3 (k+2M)^2}, \nonu\\
w_{14} & = & -\frac{72 (k+M) N (k+N)^2 \sqrt{\frac{M+N}{M N}} (k+M+N) (k+2N)^2 b_1^2}{k M^2 (k+2M)^2}, \nonu\\
w_{15} & = & -\frac{72 (k+N)^2 (M+N) (k+M+N) (3k + 4N) b_1^2}{k^2 M^2 N}, \nonu\\
w_{16} & = & -\frac{72 (k+N)^2 (k+M+N) b_1^2}{k M^2 (k+2M)^2} (3k^2 + 4kM + 4kN + 4MN), \nonu\\
w_{17} & = & -\frac{72 (k+N)^2 \sqrt{\frac{M+N}{M N}} (k+M+N) b_1^2}{k M}, \qquad
w_{18}  =  \frac{36 i (k+N)^2 \sqrt{\frac{M+N}{M N}} b_1^2}{k}, \nonu\\
w_{19} & = & -\frac{9 (k+N) (k+M+N) b_1^2}{M}, \qquad
w_{20}  =  -\frac{9 i (k+N) (k+M+N) b_1^2}{M}, \nonu\\
w_{21} & = & -\frac{9 i (k+N)^2 b_1^2}{M}, \qquad
w_{22} =  \frac{9 (k+N)^2 b_1^2}{M}, \qquad
w_{23}  =  -\frac{36 (k+N)^2 (k+M+N) b_1^2}{k M N}, \nonu\\
w_{24} & = & \frac{144 (k+N)^3 (k+M+N) b_1^2}{k M^2 N},  \qquad
w_{25}  =  -\frac{72 (k+N)^3 \sqrt{\frac{M+N}{M N}} (k+M+N) b_1^2}{k M}, \nonu\\
w_{26} & = & -\frac{36 (k+N)^2 (k+M+N) (k+2N) b_1^2}{M^2 (k+2M)},  \qquad
\nonu \\
w_{27}  & = &  -\frac{36 (k+N)^2 (k+M+N) (k+2N) b_1^2}{M^2 (k+2M)}, \nonu\\
w_{28} & = & -\frac{144 (k+N)^2 (k+M+N) (k+2N) b_1^2}{k M^2 (k+2M)}, \nonu\\
w_{29} & = & \frac{9 (k+N)^2 (k+M+N) (k+2N) b_1^2}{k M^2 (k+2M)} \nonu \\
& \times & (-12k + 5k^3 - 8M + 6k^2M - 8N + 6k^2N + 8kMN), \nonu\\
w_{30} & = & -\frac{72 (k+N)^2 (M+N) \sqrt{\frac{M+N}{M N}} (k+M+N) (k+2N)^2 b_1^2}{k^2 M^2}, \nonu\\
w_{31} & = & \frac{72 (k+N)^2 \sqrt{\frac{M+N}{M N}} (k+M+N) (k+2N) b_1^2}{k M^2 (k+2M)} (2 + kM + kN + 2MN), \nonu\\
w_{32} & = & -\frac{9 (k+N)^2 (M+N) (k+M+N) (k+2N)^2 b_1^2}{k M^2}, \nonu\\
w_{33} & = & -\frac{36 N (k+N)^2 \sqrt{\frac{M+N}{M N}} (k+2N)^2 b_1^2}{k M^2 (k+2M)^2} (4k^2 + 8kM + 4M^2 + 4kN + 5MN\Big), \nonu\\
w_{34} & = & \frac{9 N (k+N)^2 (k+2N)^2 b_1^2}{M^2 (k+M) (k+2M)^2} \Big(4k^2 + 6kM + 2M^2 + 4kN + 3MN), \nonu\\
w_{35} & = & \frac{9 i N (k+N)^2 (k+2N)^2 b_1^2}{k M^2 (k+M) (k+2M)^2}
\nonu \\
& \times &
(-8k + 2k^3 - 8M + 4k^2M + 2kM^2 - 8N + k^2N - kMN), \nonu\\
w_{36} & = & -\frac{9 N (k+N)^3 (k+2N)^2 b_1^2}{M^2 (k+2M)}, \nonu\\
w_{37} & = & -\frac{36 (k+N)^2 \sqrt{\frac{M+N}{M N}} b_1^2}{k M} (4k^2 + 4kM + 10kN + 5MN + 6N^2), \nonu\\
w_{38} & = & -\frac{9 (k+N) b_1^2}{M} (4k^2 + 4kM + 8kN + 3MN + 4N^2), \nonu\\
w_{39} & = & -\frac{9 i (k+N) b_1^2}{k M} (8k + 8M - k^2M + 8N + 4k^2N + kMN + 4kN^2), \nonu\\
w_{40} & = & -\frac{9 (k+N)^2 (k+2N) (k+M+2N) b_1^2}{M}, \nonu\\
w_{41} & = & -\frac{18 (k+N)^2 (k+M+N) (k+2N) b_1^2}{k M^2 (k+2M)}
\nonu \\
& \times & (-4k + 3k^3 - 8M + 6k^2M - 8N + 6k^2N + 8kMN), \nonu\\
w_{42} & = & -\frac{9 (k+N)^2 (M+N) (k+M+N) (k+2N) b_1^2}{k^2 M^2 (k+2M)}
(-12k + 5k^3 - 8M \nonu\\
&+& 6k^2M - 8N + 6k^2N + 8kMN),\nonu\\
w_{43} & = & -\frac{9 N (k+N)^2 (k+2N) b_1^2}{k M^2 (k+2M)^2}
\nonu \\
& \times & (-12k^2 + 5k^4 - 28kM + 12k^3M - 16M^2 + 6k^2M^2 - 20kN \nonu\\
&+& 11k^3N - 24MN + 19k^2MN + 4kM^2N - 8N^2 + 6k^2N^2 + 6kMN^2), \nonu\\
w_{44} & = & -\frac{9 (k+N)^2 b_1^2}{k M (k+2M)} (-12k^2 + 5k^4 - 20kM + 11k^3M - 8M^2 + 6k^2M^2 - 28kN \nonu\\
&+& 12k^3N - 24MN + 19k^2MN + 6kM^2N - 16N^2 + 8k^2N^2 + 8kMN^2), \nonu\\
w_{45} & = & \frac{9 (-2+k)(2+k)(k+N)^2 (k+M+N) (k+2N) (5k + 6M + 6N) b_1^2}{k M^2 (k+2M)}, \nonu\\
w_{46} & = & -\frac{144 (2k+3M) (k+N)^2 \sqrt{\frac{M+N}{M N}} (k+M+N) (k+2N)^2 b_1^2}{k M^2 (k+2M)^2}, \nonu\\
w_{47} & = & -\frac{72 (k+N)^2 (k+M+N) (k+2N) b_1^2}{k M^2 (k+2M)^2}
\nonu \\
& \times & (-2k + 2k^3 - 4M + 3k^2M + 3k^2N + 4kMN), \nonu\\
w_{48} & = & \frac{72 (k+N)^2 (k+M+N) (k+2N) b_1^2}{k M^2 (k+2M)^2} (-2k + 2k^3 - 4M + 3k^2M + 3k^2N + 4kMN), \nonu\\
w_{49} & = & -\frac{18 (k+N)^2 (k+2N) b_1^2}{k M^2 (k+2M)} (8k + 8M - k^2M + 8N + 4k^2N + 2kMN + 4kN^2), \nonu\\
w_{50} & = & \frac{72 (k+N)^2 (k+M+N) (k+2N)^2 b_1^2}{M^2 (k+2M)^2}, \qquad
w_{51}  =  \frac{18 i (k+N)^2 (k+2N)^2 b_1^2}{M^2 (k+2M)}, \nonu\\
w_{52} & = & \frac{144 (k+N)^2 \sqrt{\frac{M+N}{M N}} (k+M+N) (2k+3N) b_1^2}{k M^2}, \nonu\\
w_{53} & = & \frac{36 (k+N)^2 (k+M+N) b_1^2}{k M^2 (k+2M)} (-2k + 4k^3 + 7k^2M - 4N + 6k^2N + 10kMN), \nonu\\
w_{54} & = & -\frac{18 (k+N)^2 (k+M+N) b_1^2}{k M^2 (k+2M)} (-4k + 3k^3 + 4k^2M - 8N + 2k^2N), \nonu\\
w_{55} & = & \frac{9 (k+N)^2 (k+2N) b_1^2}{k M^2 (k+2M)} \nonu \\
& \times & (8k + 5k^3 + 8M + 19k^2M + 14kM^2 + 8N + 6k^2N + 16kMN), \nonu\\
w_{56} & = & -\frac{36 (k+N)^2 (k+M+N) b_1^2}{M^2 (k+2M)^2} (7k^2 + 10kM + 10kN + 12MN), \nonu\\
w_{57} & = & -\frac{36 (k+N)^2 (k+M+N) (k+2N) b_1^2}{M^2 (k+2M)}, \qquad
w_{58}  =  \frac{72 (k+N)^2 (k+M+N) b_1^2}{M^2}, \nonu\\
w_{59} & = & \frac{18 i (k+N)^2 (k+2N) b_1^2}{M^2}, \qquad
\nonu \\
w_{60}  & =&  -\frac{36 (2+kM) (k+N)^2 (k+M+N) (k+2N) b_1^2}{k M^2 (k+2M)}, \nonu\\
w_{61} & = & -\frac{18 (-4 + 5k^2 + 8kM) (k+N)^2 (k+M+N) (k+2N) b_1^2}{k M^2 (k+2M)}, \nonu\\
w_{62} & = & -\frac{9 (k+N)^2 (k+2N) b_1^2}{k M^2 (k+2M)} \nonu \\
& \times & (-8k + 5k^3 - 8M + 11k^2M + 6kM^2 - 8N + 4k^2N + 4kMN), \nonu\\
w_{63} & = & -\frac{3 N (k+N)^2 \sqrt{\frac{M+N}{M N}} (k+M+N) (k+2N) b_1^2}{2k M (k+2M)} (-24k + 7k^3 - 16M + 6k^2M - 16N\nonu\\
&+& 6k^2N + 4kMN).
\label{wvalues}
\eea
Again, the coefficient $b_1$ is given by
(\ref{b1value}).

The coefficients appearing in (\ref{W(4)})
are determined by
\bea
d_{1} &=& -\frac{36 (k + N)^2 (M + N)^2 (k + M + N) (k + 2 N)}{k^3 M^3 (k + 2 M) N \, D(k,N,M)}
\nonu \\
& \times & \Bigg(22 k^5+ 88 k^4 M + 110 k^3 M^2 + 44 k^2 M^3
\nonu \\
&+& 88 k^4 N + 272 k^3 M N + 22 k^5 M N + 269 k^2 M^2 N + 45 k^4 M^2 N 
+ 90 k M^3 N + 18 k^3 M^3 N
\nonu \\
&+& 110 k^3 N^2 + 269 k^2 M N^2 
+ 45 k^4 M N^2 + 180 k M^2 N^2 + 88 k^3 M^2 N^2 + 36 M^3 N^2
\nonu \\
&+& 28 k^2 M^3 N^2 + 44 k^2 N^3 + 90 k M N^3 + 18 k^3 M N^3 
+ 36 M^2 N^3 + 28 k^2 M^2 N^3 \Bigg) b_{1}^2,
\nonu \\
%%%%%%%%%%%%%%%%%%%%%%%%%%%%%%%%%%%%%%%%%
d_{2} &=& \frac{144 (k + N)^2 (M + N) (k + M + N) (k + 2 N)}{k^2 M^3 (k + 2 M) N \, D(k,N,M)}
\nonu \\
&\times & \Bigg(22 k^5 
+ 88 k^4 M + 110 k^3 M^2 + 44 k^2 M^3
\nonu \\
&+& 88 k^4 N 
+ 272 k^3 M N + 22 k^5 M N + 269 k^2 M^2 N + 45 k^4 M^2 N 
+ 90 k M^3 N + 18 k^3 M^3 N
\nonu\\
&+& 110 k^3 N^2 + 269 k^2 M N^2 + 45 k^4 M N^2 + 180 k M^2 N^2 + 88 k^3 M^2 N^2 + 36 M^3 N^2
\nonu \\
&+& 28 k^2 M^3 N^2 + 44 k^2 N^3 + 90 k M N^3 + 18 k^3 M N^3 
+ 36 M^2 N^3 + 28 k^2 M^2 N^3 \Bigg) b_{1}^2,
\nonu \\
%%%%%%%%%%%%%%%%%%%%%%%%%%%%%%%%%%%%%%%%%%%%%%%%%%%
d_{3} &=& \frac{72 (k + N)^2 \sqrt{\frac{M + N}{M N}} (k + M + N) (k + 2 N)}{k M^2 (k + 2 M) \, D(k,N,M)}
\nonu \\
& \times &
\Bigg(52 k^3 + 20 k^5 + 72 k^2 M + 114 k^4 M
\nonu \\
&+& 20 k M^2 + 160 k^3 M^2 + 66 k^2 M^3 + 72 k^2 N + 114 k^4 N + 112 k M N + 432 k^3 M N
\nonu \\
&+ & 20 k^5 M N
+ 30 M^2 N + 506 k^2 M^2 N
\nonu \\
&+ & 40 k^4 M^2 N + 183 k M^3 N + 15 k^3 M^3 N + 20 k N^2 + 160 k^3 N^2
\nonu \\
&+& 30 M N^2 + 506 k^2 M N^2 + 40 k^4 M N^2 + 438 k M^2 N^2 + 90 k^3 M^2 N^2 + 102 M^3 N^2
\nonu \\
&+& 30 k^2 M^3 N^2
+ 66 k^2 N^3 + 183 k M N^3 + 15 k^3 M N^3 + 102 M^2 N^3
+  30 k^2 M^2 N^3 \Bigg) b_{1}^2,
\nonu \\
%%%%%%%%%%%%%%%%%%%%%%%%%%%%%%%%
d_{4} &=& -\frac{72 (k + N)^2 \sqrt{\frac{M + N}{M N}} (k + M + N) (k + 2 N)}{k M^2 (k + 2 M) \, D(k,N,M)}
\Bigg(52 k^3 + 20 k^5 + 72 k^2 M + 114 k^4 M
\nonu \\
&+& 20 k M^2 + 160 k^3 M^2 + 66 k^2 M^3 + 72 k^2 N + 114 k^4 N + 112 k M N + 432 k^3 M N
\nonu \\
&+& 20 k^5 M N
+ 30 M^2 N + 506 k^2 M^2 N + 40 k^4 M^2 N
\nonu \\
&+& 183 k M^3 N + 15 k^3 M^3 N + 20 k N^2 + 160 k^3 N^2
\nonu \\
&+& 30 M N^2 + 506 k^2 M N^2 + 40 k^4 M N^2 + 438 k M^2 N^2 + 90 k^3 M^2 N^2 + 102 M^3 N^2
\nonu \\
&+& 30 k^2 M^3 N^2
+ 66 k^2 N^3 + 183 k M N^3 + 15 k^3 M N^3 + 102 M^2 N^3 + 30 k^2 M^2 N^3 \Bigg) b_{1}^2,
\nonu \\
%%%%%%%%%%%%%%%%%%%%%%%%%%%%%
d_{5} &=& -\frac{72 (k + N)^2 (M + N) (k + M + N) (k + 2 N)}{k^2 M^3 (k + 2 M)^2 \, D(k,N,M)}
\nonu \\
& \times &
\Bigg(66 k^5 + 220 k^4 M + 242 k^3 M^2 + 88 k^2 M^3
\nonu \\
&+& 264 k^4 N + 736 k^3 M N + 42 k^5 M N + 631 k^2 M^2 N + 87 k^4 M^2 N + 180 k M^3 N
\nonu \\
&+& 36 k^3 M^3 N
+ 330 k^3 N^2 + 747 k^2 M N^2
\nonu \\
&+& 95 k^4 M N^2 + 426 k M^2 N^2 + 178 k^3 M^2 N^2 + 72 M^3 N^2
\nonu \\
&+& 56 k^2 M^3 N^2
+ 132 k^2 N^3 + 246 k M N^3 + 38 k^3 M N^3 + 72 M^2 N^3 + 56 k^2 M^2 N^3 \Bigg) b_{1}^2,
\nonu \\
%%%%%%%%%%%%%%%%%%%%%%%%%%%%%%%%%%%%%
d_{6} &=& \frac{72 N (k + N)^2 \sqrt{\frac{M + N}{M N}} (k + M + N) (k + 2 N)^2}{k M^2 (k + 2 M)^2} b_{1}^2,
\nonu \\
%\nonu \\
d_{7} &=& \frac{36 i N (k + N)^2 \sqrt{\frac{M + N}{M N}} (2 k + 2 M + N) (k + 2 N)^2}{k M^2 (k + 2 M)^2} b_{1}^2,
\nonu \\
d_{8} &=& -\frac{9 N (k + N)^2 (k + M + N) (k + 2 N)^2}{M^2 (k + M) (k + 2 M)^2} b_{1}^2,
\nonu \\
d_{9} & = &
\frac{9 i N (k + N)^2 (k + M + N) (k + 2 N)^2}{M^2 (k + M) (k + 2 M)^2}
b_{1}^2, \nonu \\
d_{10} &=& -\frac{9 i N (k + N)^2 (k + 2 N)^2}{M^2 (k + 2 M)^2} b_{1}^2,
\quad
d_{11} = -\frac{9 N (k + N)^2 (k + 2 N)^2}{M^2 (k + 2 M)^2} b_{1}^2,
\nonu \\
d_{12} &=& -\frac{36 N (k + N)^2 (k + M + N) (k + 2 N)^2}{k M^3 (k + 2 M)^2} b_{1}^2,
\nonu \\
d_{13} &=& -\frac{144 (k^2-4) N^2 (k + N)^3 (k + M + N) (k + 2 N) (3 k + 2 M + 2 N)}{k M^2 (k + M) (k + 2 M) \, D(k,N,M)}b_{1}^2,
\nonu \\
%%%%%%%%%%%%%%%%%%%%%%%%%%%%%%%%%%%%%%%%%%%%%%
d_{14} &=& \frac{144 (k + N)^2 (k + M + N) (k + 2 N)}{k M^3 (k + 2 M)^2
\, D(k,N,M)} \Bigg(22 k^5 + 66 k^4 M + 66 k^3 M^2 + 22 k^2 M^3
\nonu \\
&+& 88 k^4 N + 184 k^3 M N + 22 k^5 M N + 77 k^2 M^2 N + 47 k^4 M^2 N - 3 k M^3 N + 21 k^3 M^3 N
\nonu \\
&+& 110 k^3 N^2 + 159 k^2 M N^2 + 45 k^4 M N^2 - 29 k M^2 N^2 + 83 k^3 M^2 N^2 - 30 M^3 N^2
\nonu \\
&+& 26 k^2 M^3 N^2
+ 44 k^2 N^3 + 46 k M N^3 + 18 k^3 M N^3 - 30 M^2 N^3 + 26 k^2 M^2 N^3 \Bigg) b_{1}^2,
\nonu \\
%%%%%%%%%%%%%%%%%%%%%%%%%%%%%%%%%%%%%%%%%%%%%
d_{15} &=& -\frac{72 N (k + N)^2 \sqrt{\frac{M + N}{M N}} (k + M + N) (k + 2 N)}{k M^2 (k + 2 M)^2 \, D(k,N,M)} \Bigg(22 k^5 + 66 k^4 M + 66 k^3 M^2 + 22 k^2 M^3
\nonu \\
&+& 88 k^4 N + 184 k^3 M N + 22 k^5 M N + 77 k^2 M^2 N + 47 k^4 M^2 N - 3 k M^3 N + 21 k^3 M^3 N
\nonu \\
&+& 110 k^3 N^2 + 159 k^2 M N^2 + 45 k^4 M N^2 - 29 k M^2 N^2 + 83 k^3 M^2 N^2 - 30 M^3 N^2
\nonu \\
& + & 26 k^2 M^3 N^2
+ 44 k^2 N^3 + 46 k M N^3 + 18 k^3 M N^3 - 30 M^2 N^3 + 26 k^2 M^2 N^3 \Bigg) b_{1}^2,
\nonu \\
%%%%%%%%%%%%%%%%%%%%%%%%%%%%%%%%%%%%%%%%%%%%%%%%%%
d_{16} &=& -\frac{72 (k + N)^2 (M + N) (k + M + N)}{k^2 M^2 (k + 2 M) N
\, D(k,N,M)} \Bigg(66 k^5 + 264 k^4 M + 330 k^3 M^2 + 132 k^2 M^3
\nonu \\
&+& 220 k^4 N + 736 k^3 M N + 42 k^5 M N + 747 k^2 M^2 N + 95 k^4 M^2 N + 246 k M^3 N
\nonu \\
& + & 38 k^3 M^3 N
+ 242 k^3 N^2 + 631 k^2 M N^2
\nonu \\
&+& 87 k^4 M N^2 + 426 k M^2 N^2 + 178 k^3 M^2N^2 + 72 M^3 N^2
\nonu \\
&+& 56 k^2 M^3 N^2
+ 88 k^2 N^3 + 180 k M N^3 + 36 k^3 M N^3 + 72 M^2 N^3
+ 56 k^2 M^2 N^3 \Bigg) b_{1}^2,
\nonu \\
%%%%%%%%%%%%%%%%%%%%%%%%%%%%%%%%%%%%%%%%%%%%%
d_{17} &=& -\frac{72 (k + N)^2 (k + M + N)}{k M^2 (k + 2 M)^2
\, D(k,N,M)} \Bigg(66 k^5 + 220 k^4 M + 242 k^3 M^2 + 88 k^2 M^3
\nonu \\
&+& 220 k^4 N + 560 k^3 M N + 42 k^5 M N + 411 k^2 M^2 N + 87 k^4 M^2 N + 92k M^3 N
\nonu \\
& + & 36 k^3 M^3 N
+ 242 k^3 N^2 + 411 k^2 M N^2
\nonu \\
&+& 87 k^4 M N^2 + 40 k M^2 N^2 + 160 k^3 M^2 N^2 - 60 M^3 N^2
\nonu \\
& + & 52 k^2 M^3 N^2
+ 88 k^2 N^3 + 92 k M N^3 + 36 k^3 M N^3 - 60 M^2 N^3 + 52 k^2 M^2 N^3 \Bigg) b_{1}^2,
\nonu \\
%%%%%%%%%%%%%%%%%%%%%%%%%%%%%%%%%%
d_{18} &=& -\frac{72 (k + N)^2 \sqrt{\frac{M + N}{M N}} (k + M + N)}{k M} b_{1}^2 ,\quad
d_{19} = \frac{36 i (k + N)^2 \sqrt{\frac{M + N}{M N}}}{k} b_{1}^2,
\nonu \\
%\nonu \\
d_{20} &=& -\frac{9 (k + N) (k + M + N)}{M} b_{1}^2 ,\quad
d_{21} = -\frac{9 i (k + N) (k + M + N)}{M} b_{1}^2,
\nonu \\
%\nonu \\
d_{22} &=& -\frac{9 i (k + N)^2}{M} b_{1}^2,\quad
d_{23} = \frac{9 (k + N)^2}{M} b_{1}^2 ,\quad
d_{24} = -\frac{36 (k + N)^2 (k + M + N)}{k M N} b_{1}^2,
\nonu \\
%\nonu \\
d_{25} &=& -\frac{144 (k^2-4) (k + M) (k + N) (k + M + N) (k + 2 N) (3 k + 2 M + 2 N)}{k (k + 2 M) \, D(k,N,M)} b_{1}^2,
\nonu \\
%%%%%%%%%%%%%%%%%%%%%%%%%%%%%%%%%%%%%%%%%%%%
d_{26} &=& \frac{144 (k + N)^2 (k + M + N)}{k M^2 (k + 2 M) N
\, D(k,N,M)
} \Bigg(22 k^5 + 88 k^4 M + 110 k^3 M^2 + 44 k^2 M^3
\nonu \\
&+& 66 k^4 N + 184 k^3 M N + 22 k^5 M N + 159 k^2 M^2 N + 45 k^4 M^2 N + 46 k M^3 N + 18 k^3 M^3 N
\nonu \\
&+& 66 k^3 N^2 + 77 k^2 M N^2 + 47 k^4 M N^2 - 29 k M^2 N^2 + 83 k^3 M^2 N^2 - 30 M^3 N^2
\nonu \\
&+& 26 k^2 M^3 N^2
+ 22 k^2 N^3 - 3 k M N^3 + 21 k^3 M N^3 - 30 M^2 N^3 + 26 k^2 M^2 N^3 \Bigg) b_{1}^2, \nonu \\
%%%%%%%%%%%%%%%%%%%%%%%%%%%%%%%%%%%
d_{27} &=& -\frac{72 (k + N)^2 \sqrt{\frac{M + N}{M N}
} (k + M + N)}{k M \, D(k,N,M)} \Bigg(22 k^5 + 88 k^4 M + 110 k^3 M^2 + 44 k^2 M^3
\nonu \\
&+& 66 k^4 N + 184 k^3 M N + 22 k^5 M N + 159 k^2 M^2 N + 45 k^4 M^2 N + 46 k M^3 N + 18 k^3 M^3 N
\nonu \\
&+& 66 k^3 N^2 + 77 k^2 M N^2 + 47 k^4 M N^2 - 29 k M^2 N^2 + 83 k^3 M^2 N^2 - 30 M^3 N^2
\nonu \\
& +& 26 k^2 M^3 N^2
+ 22 k^2 N^3 - 3 k M N^3 + 21 k^3 M N^3 - 30 M^2 N^3 + 26 k^2 M^2 N^3 \Bigg) b_{1}^2,
\nonu \\
%%%%%%%%%%%%%%%%%%%%%%%%%%%%%%%%%%%%%%
d_{28} &=& -\frac{36 (k + N)^2 (k + M + N) (k + 2 N)}{M^2 (k + 2 M)}
b_{1}^2,
\nonu \\
%\nonu \\
d_{29} &=& -\frac{36 (k + N)^2 (k + M + N) (k + 2 N)}{M^2 (k + 2 M)}
b_{1}^2,
\nonu \\
d_{30} &=& -\frac{144 (k + N)^2 (k + M + N) (k + 2 N)}{k M^2 (k + 2 M)}
b_{1}^2, \nonu \\
d_{31} &=& -\frac{576 (k^2-4) (k + M) (k + N)^3 (k + M + N) (k + 2 N) (3 k + 2 M + 2 N)}{k M^2 (k + 2 M) \, D(k,N,M)} b_{1}^2,
\nonu \\
%%%%%%%%%%%%%%%%%%%%%%%%%%%%%%%%%%%%%%%
d_{32} &=& \frac{72 (k + N)^2 (k + M + N) (k + 2 N)}{k M^2 (k + 2 M)
\, D(k,N,M)} \Bigg(6 k^4 + 4 k^6 + 16 k^3 M + 18 k^5 M + 14 k^2 M^2
\nonu \\
&+& 24 k^4 M^2 + 4 k M^3 + 10 k^3 M^3 + 16 k^3 N + 18 k^5 N + 40 k^2 M N + 64 k^4 M N + 4 k^6 M N
\nonu \\
&+& 29 k M^2 N + 73 k^3 M^2 N + 8 k^5 M^2 N + 6 M^3 N + 27 k^2 M^3 N + 3 k^4M^3 N + 14 k^2 N^2
\nonu \\
& + & 24 k^4 N^2 + 29 k M N^2
\nonu \\
& +& 73 k^3 M N^2 + 8 k^5 M N^2 + 12 M^2 N^2 + 66 k^2 M^2 N^2 + 16 k^4 M^2 N^2 + 17 k M^3 N^2
\nonu \\
&+& 5 k^3 M^3 N^2 + 4 k N^3 + 10 k^3 N^3 + 6 M N^3 + 27 k^2 M N^3 + 3 k^4 M N^3 + 17 k M^2 N^3
\nonu \\
&+& 5 k^3 M^2 N^3 \Bigg) b_{1}^2,
\nonu \\
%%%%%%%%%%%%%%%%%%%%%%%%%%%%%%%%%%%%%%
d_{33} &=& -\frac{72 (k + N)^2 (M + N) \sqrt{\frac{M + N}{M N}} (k + M + N)(k + 2 N)}{k^2 M^2 (k + 2 M) \, D(k,N,M)}
\nonu \\
& \times &
\Bigg(22 k^5 + 88 k^4 M + 110 k^3 M^2 + 44 k^2 M^3
\nonu \\
&+& 88 k^4 N + 272 k^3 M N + 22 k^5 M N + 269 k^2 M^2 N + 45 k^4 M^2 N + 90 k M^3 N + 18 k^3 M^3 N
\nonu \\
&+& 110 k^3 N^2 + 269 k^2 M N^2 + 45 k^4 M N^2 + 180 k M^2 N^2 + 88 k^3 M^2 N^2 + 36 M^3 N^2
\nonu \\
& + & 28 k^2 M^3 N^2
+ 44 k^2 N^3 + 90 k M N^3 + 18 k^3 M N^3 + 36 M^2 N^3 + 28 k^2 M^2 N^3 \Bigg) b_{1}^2,
\nonu \\
%%%%%%%%%%%%%%%%%%%%%%%%%%%%%%%%%%%
d_{34} &=& \frac{72 (k + N)^2 \sqrt{\frac{M + N}{M N}} (k + M + N) (k + 2 N)}{k M^2 (k + 2 M) \, D(k,N,M)} \Bigg(-52 k^3 + 24 k^5 - 72 k^2 M + 62 k^4 M
\nonu \\
&-& 20 k M^2 + 60 k^3 M^2 + 22 k^2 M^3 - 72 k^2 N + 62 k^4 N - 112 k M N + 112 k^3 M N
\nonu \\
& + & 24 k^5 M N
- 30 M^2 N + 32 k^2 M^2 N
\nonu \\
&+& 50 k^4 M^2 N - 3 k M^3 N + 21 k^3 M^3 N - 20 k N^2 + 60 k^3 N^2
\nonu \\
&-& 30 M N^2 + 32 k^2 M N^2 + 50 k^4 M N^2 - 78 k M^2 N^2 + 86 k^3 M^2 N^2 - 30 M^3 N^2
\nonu \\
& + & 26 k^2 M^3 N^2
+ 22 k^2 N^3 - 3 k M N^3 + 21 k^3 M N^3 - 30 M^2 N^3 + 26 k^2 M^2 N^3 \Bigg) b_{1}^2,
\nonu \\
%%%%%%%%%%%%%%%%%%%%%%%%%%%%%%%%%%%%
d_{35} &=& -\frac{36 (k + N)^2 (M + N) (k + M + N) (k + 2 N)}{
k^2 M^2 (k + 2 M) \, D(k,N,M)} \Bigg(-18 k^4 + 10 k^6 - 48 k^3 M + 34 k^5 M \nonu \\
&-& 42 k^2 M^2 + 38 k^4 M^2 - 12 k M^3 + 14 k^3 M^3 - 48 k^3 N + 34 k^5 N - 120 k^2 M N + 80 k^4 M N
\nonu \\
&+& 10 k^6 M N - 87 k M^2 N + 50 k^3 M^2 N + 21 k^5 M^2 N - 18 M^3 N + 9 k^2M^3 N + 9 k^4 M^3 N
\nonu \\
&-& 42 k^2 N^2 + 38 k^4 N^2 - 87 k M N^2 + 50 k^3 M N^2 + 21 k^5 M N^2 - 36 M^2 N^2 - 18 k^2 M^2 N^2
\nonu \\
&+& 40 k^4 M^2 N^2 - 15 k M^3 N^2 + 13 k^3 M^3 N^2 - 12 k N^3 + 14 k^3 N^3 - 18 M N^3 + 9 k^2 M N^3
\nonu \\
&+& 9 k^4 M N^3 - 15 k M^2 N^3 + 13 k^3 M^2 N^3 \Bigg) b_{1}^2,
\nonu \\
%%%%%%%%%%%%%%%%%%%%%%%%%%%%%%%%%%%%%%%%%%%%
d_{36} &=& -\frac{36 N (k + N)^2 \sqrt{\frac{M + N}{M N}} (k + 2 N)^2 (4 k^2+ 8 k M + 4 M^2 + 4 k N + 5 M N)}{k M^2 (k + 2 M)^2} b_{1}^2,
\nonu\\
d_{37} &=& \frac{9 N (k + N)^2 (k + 2 N)^2 (4 k^2 + 6 k M + 2 M^2 + 4 k N + 3 M N)}{M^2 (k + M) (k + 2 M)^2} b_{1}^2,
\nonu \\
d_{38} &=& \frac{9 i N (k + N)^2 (k + 2 N)^2}{k M^2 (k + M) (k + 2 M)^2} b_{1}^2
\nonu \\
& \times &  (-8 k + 2 k^3 - 8 M + 4 k^2 M + 2 k M^2 - 8 N + k^2 N - k M N),
\nonu \\
%%%%%%%%%%%%%%%%%%%%%%%%%%%%%%%%%%%%
d_{39} &=& \frac{9 N (k + N)^2 (k + 2 N)}{k M^2 (k + M) (k + 2 M)
\, D(k,N,M)}
\nonu \\
&\times &
\Bigg(72 k^5 - 40 k^7 + 264 k^4 M - 132 k^6 M + 360 k^3 M^2 - 156 k^5 M^2
\nonu \\
&+& 216 k^2 M^3 - 76 k^4 M^3 + 48 k M^4 - 12 k^3 M^4 + 264 k^4 N - 176 k^6 N+ 864 k^3 M N
\nonu \\
& - & 586 k^5 M N + 8 k^7 M N + 996 k^2 M^2 N
\nonu \\
&- & 676 k^4 M^2 N + 24 k^6 M^2 N + 468 k M^3 N - 308 k^3 M^3 N + 22 k^5 M^3N \nonu \\
&+& 72 M^4 N - 42 k^2 M^4 N + 6 k^4 M^4 N + 360 k^3 N^2 - 288 k^5 N^2 + 996 k^2 M N^2 - 876 k^4 M N^2
\nonu \\
&+& 4 k^6 M N^2 + 840 k M^2 N^2 - 794 k^3 M^2 N^2 + 4 k^5 M^2 N^2 + 216 M^3 N^2 - 221 k^2 M^3 N^2
\nonu \\
& +& 3 k^4 M^3 N^2
+ 216 k^2 N^3 - 208 k^4 N^3
\nonu \\
&+& 468 k M N^3 - 542 k^3 M N^3 - 8 k^5 M N^3 + 216 M^2 N^3
\nonu \\
& - & 323 k^2 M^2 N^3
- 27 k^4 M^2 N^3 - 34 k M^3 N^3
\nonu \\
&-& 10 k^3 M^3 N^3 + 48 k N^4 - 56 k^3 N^4 + 72 M N^4 - 120 k^2 M N^4
\nonu \\
&-& 4 k^4 M N^4 - 34 k M^2 N^4 - 10 k^3 M^2 N^4 \Bigg) b_{1}^2,
\nonu \\
%%%%%%%%%%%%%%%%%%%%%%%%%%%%%%%%%%%%%%%
d_{40} &=& -\frac{36 (k + N)^2 \sqrt{\frac{M + N}{M N}} (4 k^2 + 4 k M + 10 k N + 5 M N + 6 N^2)}{k M} b_{1}^2,
\nonu \\
d_{41} &=& -\frac{9 (k + N) (4 k^2 + 4 k M + 8 k N + 3 M N + 4 N^2)}{M} b_{1}^2,
\nonu \\
d_{42} &=& \frac{9 i (k + N) (-8 k - 8 M + k^2 M - 8 N - 4 k^2 N - k M N - 4 k N^2)}{k M} b_{1}^2,
\nonu \\
%%%%%%%%%%%%%%%%%%%%%%%%%%%%%%%%%%%%
d_{43} &=& \frac{9 (k + N) (k + 2 N)}{k M (k + 2 M)
\, D(k,N,M)
} \nonu \\
& \times &
\Bigg(72 k^5 - 40 k^7 + 264 k^4 M - 176 k^6 M + 360 k^3 M^2 - 288 k^5 M^2
\nonu \\
&+& 216 k^2 M^3 - 208 k^4 M^3 + 48 k M^4 - 56 k^3 M^4 + 264 k^4 N - 176 k^6 N + 864 k^3 M N
\nonu \\
& - & 762 k^5 M N
+ 8 k^7 M N + 996 k^2 M^2 N
\nonu \\
&-& 1096 k^4 M^2 N + 4 k^6 M^2 N + 468 k M^3 N -630 k^3 M^3 N
\nonu \\
& - & 8 k^5 M^3 N + 72 M^4 N - 120 k^2 M^4 N
\nonu \\
& - & 4 k^4 M^4 N + 360 k^3 N^2 - 288 k^5 N^2 + 996 k^2 M N^2 - 1140 k^4 M N^2 \nonu \\
&+& 4 k^6 M N^2 + 840 k M^2 N^2 - 1272 k^3 M^2 N^2 - 46 k^5 M^2 N^2 + 216 M^3N^2 - 479 k^2 M^3 N^2
\nonu \\
&-& 47 k^4 M^3 N^2 - 34 k M^4 N^2 - 10 k^3 M^4 N^2 + 216 k^2 N^3 - 208 k^4 N^3 + 468 k M N^3
\nonu \\
& - & 718 k^3 M N^3
- 8 k^5 M N^3 + 216 M^2 N^3
\nonu \\
& - & 547 k^2 M^2 N^3 - 67 k^4 M^2 N^3 - 102 k M^3 N^3 - 30 k^3 M^3 N^3
\nonu \\
&+& 48 k N^4 - 56 k^3 N^4 + 72 M N^4 - 164 k^2 M N^4 - 4 k^4 M N^4 - 68 k M^2 N^4 - 20 k^3 M^2 N^4 \Bigg) \nonu \\
& \times & b_{1}^2,
\nonu \\
%%%%%%%%%%%%%%%%%%%%%%%%%%%%%%%%%%%%%%%%%%%%%%%%%%%
d_{44} &=& -\frac{72 (k + N)^2 (k + M + N) (k + 2 N)}{k M^2 (k + 2 M)
\, D(k,N,M)} 
\Bigg(-4 k^4 + 12 k^6 + 56 k^3 M + 30 k^5 M + 92 k^2 M^2
\nonu \\
&+& 32 k^4 M^2 
+ 32 k M^3 + 14 k^3 M^3 + 56 k^3 N + 30 k^5 N + 208 k^2 M N + 80 k^4 M N
\nonu \\
& +& 12 k^6 M N
+ 202 k M^2 N + 78 k^3 M^2 N + 24 k^5 M^2 N
\nonu \\
&+& 48 M^3 N + 33 k^2 M^3 N + 9 k^4 M^3 N
\nonu \\
&+& 92 k^2 N^2 + 32 k^4 N^2 + 202 k M N^2 + 78 k^3 M N^2 + 24 k^5 M N^2 + 96M^2 N^2
\nonu \\
&+& 90 k^2 M^2 N^2 + 38 k^4 M^2 N^2 + 34 k M^3 N^2 + 10 k^3 M^3 N^2 + 32 k N^3 + 14 k^3 N^3
\nonu \\
&+& 48 M N^3 + 33 k^2 M N^3 + 9 k^4 M N^3 + 34 k M^2 N^3 + 10 k^3 M^2 N^3 \Bigg) b_{1}^2,
\nonu \\
%%%%%%%%%%%%%%%%%%%%%%%%%%%%%%%%%%%%%
d_{45} &=& -\frac{72 (k + N)^2 (M + N) (k + M + N) (k + 2 N)}{
  k^2 M^2 (k + 2 M) \, D(k,N,M)} 
\nonu \\
& \times
& \Bigg(6 k^4 + 4 k^6 + 16 k^3 M + 18 k^5 M + 14 k^2 M^2
\nonu \\
&+& 24 k^4 M^2 
+ 4 k M^3 + 10 k^3 M^3 + 16 k^3 N + 18 k^5 N + 40 k^2 M N + 64 k^4 M N + 4 k^6 M N
\nonu \\
&+& 29 k M^2 N + 73 k^3 M^2 N + 8 k^5 M^2 N + 6 M^3 N + 27 k^2 M^3 N + 3 k^4 M^3 N
\nonu \\
&+& 14 k^2 N^2 + 24 k^4 N^2 + 29 k M N^2 + 73 k^3 M N^2 + 8 k^5 M N^2 + 12 M^2 N^2
\nonu \\
&+& 66 k^2 M^2 N^2 + 16 k^4 M^2 N^2 + 17 k M^3 N^2 + 5 k^3 M^3 N^2 + 4 k N^3 + 10 k^3 N^3
\nonu \\
&+& 6 M N^3 + 27 k^2 M N^3 + 3 k^4 M N^3 + 17 k M^2 N^3 + 5 k^3 M^2 N^3 \Bigg) b_{1}^2,
\nonu \\
%%%%%%%%%%%%%%%%%%%%%%%%%%%%%%%%%%%%
d_{46} &=& -\frac{9 N (k + N)^2 (k + 2 N)}{k M^2 (k + M) (k + 2 M)^2
\, D(k,N,M)} 
\nonu \\
& \times &
\Bigg(48 k^6 + 32 k^8 + 360 k^5 M + 152 k^7 M + 856 k^4 M^2 + 292 k^6 M^2
\nonu \\
&+& 888 k^3 M^3 + 284 k^5 M^3 + 408 k^2 M^4 + 140 k^4 M^4 + 64 k M^5 + 28 k^3 M^5 + 176 k^5 N
\nonu \\
& + & 176 k^7 N 
+ 1368 k^4 M N + 712 k^6 M N
\nonu \\
& + & 32 k^8 M N + 2920 k^3 M^2 N + 1186 k^5 M^2 N + 120 k^7 M^2 N \nonu \\
&+& 2548 k^2 M^3 N + 996 k^4 M^3 N + 160 k^6 M^3 N + 916 k M^4 N + 412 k^3 M^4 N + 90 k^5 M^4 N
\nonu \\
&+& 96 M^5 N + 66 k^2 M^5 N + 18 k^4 M^5 N + 240 k^4 N^2 + 336 k^6 N^2 + 1936 k^3 M N^2
\nonu \\
& + & 1168 k^5 M N^2 
+ 96 k^7 M N^2 + 3420 k^2 M^2 N^2
\nonu \\
&+& 1780 k^4 M^2 N^2 + 308 k^6 M^2 N^2 + 2160 k M^3 N^2 + 1382 k^3 M^3 N^2 \nonu \\
&+& 344 k^5 M^3 N^2 + 424 M^4 N^2 + 517 k^2 M^4 N^2 + 149 k^4 M^4 N^2 + 68 k M^5 N^2 + 20 k^3 M^5 N^2
\nonu \\
&+& 144 k^3 N^3 + 272 k^5 N^3 + 1216 k^2 M N^3 + 808 k^4 M N^3 + 88 k^6 M N^3 + 1596 k M^2 N^3
\nonu\\
&+& 1138 k^3 M^2 N^3 + 228 k^5 M^2 N^3 + 560 M^3 N^3 + 727 k^2 M^3 N^3 + 199 k^4 M^3 N^3
\nonu \\
& + & 170 k M^4 N^3 
+ 50 k^3 M^4 N^3 + 32 k^2 N^4
\nonu \\
& + & 80 k^4 N^4 + 288 k M N^4 + 200 k^3 M N^4 + 24 k^5 M N^4 + 232 M^2 N^4
\nonu\\
&+& 252 k^2 M^2 N^4 
+ 48 k^4 M^2 N^4 + 102 k M^3 N^4 + 30 k^3 M^3 N^4 \Bigg) b_{1}^2,
\nonu \\
%%%%%%%%%%%%%%%%%%%%%%%%%%%%%%%%%%%%%%%
d_{47} &=& -\frac{9 (k + N)}{k M (k + 2 M) \, D(k,N,M)} 
\nonu \\
& \times &
\Bigg(48 k^6 + 32 k^8 + 176 k^5 M + 176 k^7 M + 240 k^4 M^2 + 336 k^6 M^2 + 144 k^3 M^3
\nonu \\
&+& 272 k^5 M^3 + 32 k^2 M^4 + 80 k^4 M^4 + 360 k^5 N + 152 k^7 N + 1368 k^4M N + 712 k^6 M N
\nonu \\
&+& 32 k^8 M N + 1936 k^3 M^2 N + 1168 k^5 M^2 N + 96 k^7 M^2 N + 1216 k^2 M^3 N + 808 k^4 M^3 N
\nonu \\
&+& 88 k^6 M^3 N + 288 k M^4 N + 200 k^3 M^4 N + 24 k^5 M^4 N + 856 k^4 N^2 + 336 k^6 N^2
\nonu \\
&+& 2920 k^3 M N^2 
+ 1362 k^5 M N^2 + 120 k^7 M N^2
\nonu \\
&+& 3420 k^2 M^2 N^2 + 2000 k^4 M^2 N^2 + 308 k^6 M^2 N^2 + 1596 k M^3 N^2
\nonu \\
&+& 1226 k^3 M^3 N^2 + 228 k^5 M^3 N^2 + 232 M^4 N^2 + 252 k^2 M^4 N^2 + 48 k^4 M^4 N^2 + 888 k^3 N^3
\nonu \\
&+& 416 k^5 N^3 + 2548 k^2 M N^3 + 1460 k^4 M N^3 + 180 k^6 M N^3 + 2160 k M^2 N^3
\nonu \\
& + & 1860 k^3 M^2 N^3 
+ 394 k^5 M^2 N^3 + 560 M^3 N^3
\nonu \\
& +& 883 k^2 M^3 N^3 + 219 k^4 M^3 N^3 + 102 k M^4 N^3 + 30 k^3 M^4 N^3
\nonu \\
&+& 408 k^2 N^4 + 272 k^4 N^4 + 916 k M N^4 + 822 k^3 M N^4 + 120 k^5 M N^4 + 424 M^2 N^4
\nonu \\
& + & 843 k^2 M^2 N^4 
+ 219 k^4 M^2 N^4 + 238 k M^3 N^4
\nonu \\
& +& 70 k^3 M^3 N^4 + 64 k N^5 + 72 k^3 N^5 + 96 M N^5 + 188 k^2 M N^5
\nonu \\
&+& 28 k^4 M N^5 + 136 k M^2 N^5 + 40 k^3 M^2 N^5 \Bigg) b_{1}^2,
\nonu \\
%%%%%%%%%%%%%%%%%%%%%%%%%%%%%%%%%%%%%%%%%%%%
d_{48} &=& \frac{72 (k^2-4) (k + N)^2 (k + M + N) (2 k + M + N) (k + 2 N)}{k M^2 (k + 2 M) \, D(k,N,M)} 
\nonu \\
& \times & \Bigg(2 k^3 - 4 k^2 M- 6 k M^2
\nonu \\
&-& 4 k^2 N - 12 k M N + 2 k^3 M N - 9 M^2 N + 3 k^2 M^2 N - 6 k N^2 - 9 M N^2 + 3 k^2 M N^2 \Bigg) b_{1}^2,
\nonu \\
%%%%%%%%%%%%%%%%%%%%%%%%%%%%%%%%%%%%%%%
d_{49} &=& -\frac{144 (2 k + 3 M) (k + N)^2}{k M^2 (k + 2 M)^2} 
\Bigg(\sqrt{\frac{M + N}{M N}} (k + M + N) (k + 2 N)^2 \Bigg) b_{1}^2,
\nonu \\
%%%%%%%%%%%%%%%%%%%%%%%%%%%%%%%%
d_{50} &=& -\frac{72 (k + N)^2 (k + M + N) (k + 2 N)}{k M^2 (k + 2 M)^2
\, D(k,N,M)} 
\Bigg(52 k^4 + 20 k^6 + 176 k^3 M + 66 k^5 M + 164 k^2 M^2
\nonu\\
&+& 80 k^4 M^2 
+ 40 k M^3 + 34 k^3 M^3 + 72 k^3 N + 114 k^5 N + 256 k^2 M N + 308 k^4 M N
\nonu \\
&+& 20 k^6 M N
+ 254 k M^2 N + 266 k^3 M^2 N + 40 k^5 M^2 N
\nonu \\
&+& 60 M^3 N + 87 k^2 M^3 N + 15 k^4 M^3 N
\nonu \\
&+& 20 k^2 N^2 + 160 k^4 N^2 + 70 k M N^2 + 386 k^3 M N^2 + 40 k^5 M N^2 + 60 M^2 N^2
\nonu \\
&+& 274 k^2 M^2 N^2 + 70 k^4 M^2 N^2 + 68 k M^3 N^2 + 20 k^3 M^3 N^2 + 66 k^3 N^3
\nonu \\
&+& 139 k^2 M N^3 + 15 k^4 M N^3 + 68 k M^2 N^3 + 20 k^3 M^2 N^3 \Bigg)
b_{1}^2,
\nonu \\
%%%%%%%%%%%%%%%%%%%%%%%%%%%%%%%%%%%%%%
d_{51} &=& \frac{72 (k + N)^2 (k + M + N) (k + 2 N)}{k M^2 (k + 2 M)^2
\, D(k,N,M)} 
\Bigg(52 k^4 + 20 k^6 + 176 k^3 M + 66 k^5 M + 164 k^2 M^2\nonu \\
&+& 80 k^4 M^2
+ 40 k M^3 + 34 k^3 M^3 + 72 k^3 N
\nonu \\
&+& 114 k^5 N + 256 k^2 M N + 308 k^4 M N + 20 k^6 M N
\nonu \\
&+& 254 k M^2 N + 266 k^3 M^2 N + 40 k^5 M^2 N + 60 M^3 N + 87 k^2 M^3 N + 15 k^4 M^3 N
\nonu \\
&+& 20 k^2 N^2 + 160 k^4 N^2 + 70 k M N^2 + 386 k^3 M N^2 + 40 k^5 M N^2 + 60 M^2 N^2
\nonu \\
&+& 274 k^2 M^2 N^2 + 70 k^4 M^2 N^2 + 68 k M^3 N^2 + 20 k^3 M^3 N^2 + 66 k^3N^3
\nonu \\
&+& 139 k^2 M N^3 + 15 k^4 M N^3 + 68 k M^2 N^3 + 20 k^3 M^2 N^3 \Bigg)
b_{1}^2,
\nonu \\
%%%%%%%%%%%%%%%%%%%%%%%%%%%%%%%%%
d_{52} &=& -\frac{18 (k + N)^2 (k + 2 N)}{k M^2 (k + 2 M)
\, D(k,N,M)} 
\nonu \\
& \times &
\Bigg(-208 k^4 + 96 k^6 - 496 k^3 M + 234 k^5 M - 368 k^2 M^2 + 180 k^4 M^2
\nonu \\
&-& 80 k M^3 + 42 k^3 M^3 - 496 k^3 N + 344 k^5 N - 1024 k^2 M N + 800 k^4 M N - 648 k M^2 N
\nonu \\
&+& 592 k^3 M^2 N - 10 k^5 M^2 N - 120 M^3 N + 131 k^2 M^3 N - 5 k^4 M^3 N - 368 k^2 N^2
\nonu \\
&+& 488 k^4 N^2 - 648 k M N^2 + 938 k^3 M N^2 + 40 k^5 M N^2 - 240 M^2 N^2 + 459 k^2 M^2 N^2
\nonu \\
&+& 35 k^4 M^2 N^2 + 34 k M^3 N^2 + 10 k^3 M^3 N^2 - 80 k N^3 + 328 k^3 N^3 -120 M N^3
\nonu \\
&+& 440 k^2 M N^3 + 60 k^4 M N^3 + 102 k M^2 N^3 + 30 k^3 M^2 N^3 + 88 k^2 N^4 + 68 k M N^4
\nonu \\
&+& 20 k^3 M N^4 \Bigg) b_{1}^2, \nonu \\
%%%%%%%%%%%%%%%%%%%%%%%%%%%%%%%%%
d_{53} &=& \frac{72 (k + N)^2 (k + M + N) (k + 2 N)^2}{M^2 (k + 2 M)^2}
b_{1}^2,\quad
d_{54} = \frac{18 i (k + N)^2 (k + 2 N)^2}{M^2 (k + 2 M)} b_{1}^2,
\nonu \\
d_{55} &=& \frac{144 (k + N)^2}{k M^2} 
\Bigg(\sqrt{\frac{M + N}{M N}} (k + M + N) (2 k + 3 N) \Bigg) b_{1}^2,
\nonu \\
%%%%%%%%%%%%%%%%%%%%%%%%%%%%%%%
d_{56} &=& \frac{36 (k + N)^2 (k + M + N)}{k M^2 (k + 2 M)
\, D(k,N,M)} 
\nonu \\
& \times &
\Bigg(148 k^4 + 40 k^6 + 232 k^3 M + 250 k^5 M + 84 k^2 M^2 + 364 k^4 M^2
\nonu \\
&+& 528 k^3 N + 132 k^5 N + 800 k^2 M N + 724 k^4 M N + 40 k^6 M N + 262 k M^2 N
\nonu \\
& + & 926 k^3 M^2 N 
+ 90 k^5 M^2 N + 339 k^2 M^3 N
\nonu \\
&+& 35 k^4 M^3 N + 548 k^2 N^2 + 160 k^4 N^2 + 766 k M N^2 + 692 k^3 M N^2 \nonu \\
&+& 80 k^5 M N^2 + 188 M^2 N^2 + 697 k^2 M^2 N^2 + 165 k^4 M^2 N^2 + 170 k M^3 N^2 + 50 k^3 M^3 N^2
\nonu \\
& + & 154 k^3 M^3 
+ 168 k N^3 + 68 k^3 N^3
\nonu \\
& +& 188 M N^3 + 238 k^2 M N^3 + 30 k^4 M N^3 + 170 k M^2 N^3 + 50 k^3 M^2 N^3 \Bigg) b_{1}^2,
\nonu \\
%%%%%%%%%%%%%%%%%%%%%%%%%%%%%
d_{57} &=& -\frac{18 (k + N)^2 (k + M + N)}{k M^2 (k + 2 M)
\, D(k,N,M)} 
\nonu \\
& \times &
\Bigg(296 k^4 - 30 k^6 + 464 k^3 M + 60 k^5 M + 168 k^2 M^2 + 178 k^4 M^2
\nonu \\
&+& 1056 k^3 N - 176 k^5 N + 1600 k^2 M N - 152 k^4 M N + 30 k^6 M N + 524 k M^2 N \nonu \\
& + & 107 k^3 M^2 N 
+ 55 k^5 M^2 N + 68 k^2 M^3 N
\nonu \\
&+& 20 k^4 M^3 N + 1096 k^2 N^2 - 230 k^4 N^2 + 1532 k M N^2 - 361 k^3 M N^2\nonu \\
&+& 35 k^5 M N^2 + 376 M^2 N^2 - 66 k^2 M^2 N^2 + 30 k^4 M^2 N^2 + 336 k N^3 - 84 k^3 N^3 + 376 M N^3
\nonu \\
&-& 134 k^2 M N^3 + 10 k^4 M N^3 + 88 k^3 M^3\Bigg) b_{1}^2,
\nonu \\
%%%%%%%%%%%%%%%%%%%%%%%%%%%%%%%%%
d_{58} &=& \frac{9 (k + N)^2 (k + 2 N)}{k M^2 (k + 2 M)
\, D(k,N,M)} 
\nonu \\
& \times &
\Bigg(-592 k^4 + 302 k^6 - 1520 k^3 M + 1150 k^5 M - 1264 k^2 M^2 + 1702 k^4M^2
\nonu \\
&-& 336 k M^3 + 1162 k^3 M^3 + 308 k^2 M^4 - 1520 k^3 N + 864 k^5 N - 3200 k^2 M N
\nonu \\
&+& 2900 k^4 M N 
+ 50 k^6 M N - 2056 k M^2 N
\nonu \\
&+& 3425 k^3 M^2 N + 215 k^5 M^2 N - 376 M^3 N + 1627 k^2 M^3 N
\nonu \\
&+& 235 k^4 M^3 N + 238 k M^4 N + 70 k^3 M^4 N - 1264 k^2 N^2 + 822 k^4 N^2 - 2056 k M N^2
\nonu \\
&+& 2367 k^3 M N^2 + 85 k^5 M N^2 - 752 M^2 N^2 + 1965 k^2 M^2 N^2 + 285 k^4M^2 N^2 + 510 k M^3 N^2
\nonu \\
&+& 150 k^3 M^3 N^2 - 336 k N^3 + 260 k^3 N^3 - 376 M N^3 + 622 k^2 M N^3 + 30 k^4 M N^3
\nonu \\
&+& 272 k M^2 N^3 + 80 k^3 M^2 N^3 \Bigg) b_{1}^2,
\nonu \\
%%%%%%%%%%%%%%%%%%%%%%%%%%%%%%%%%%%%%%%%%%%
d_{59} &=& -\frac{36 (k + N)^2 (k + M + N)}{M^2 (k + 2 M)^2} 
\Bigg(7 k^2 + 10 k M + 10 k N + 12 M N \Bigg) b_{1}^2,
\nonu \\
d_{60} &=& -\frac{36 (k + N)^2 (k + M + N) (k + 2 N)}{M^2 (k + 2 M)} b_{1}^2,
\nonu \\
d_{61} &=& -\frac{36 (2 + k M) (k + N)^2 (k + M + N) (k + 2 N)}{k M^2 (k + 2 M)} b_{1}^2,
\nonu \\
d_{62} &=& -\frac{18 (-4 + 5 k^2 + 8 k M) (k + N)^2 (k + M + N) (k + 2 N)}{kM^2 (k + 2 M)} b_{1}^2,
\nonu \\
d_{63} &=& -\frac{9 (k + N)^2 (k + 2 N)}{k M^2 (k + 2 M)} 
\nonu \\
&\times &
\Bigg(-8 k + 5 k^3 - 8 M + 11 k^2 M + 6 k M^2 - 8 N + 4 k^2 N + 4 k M N \Bigg) b_{1}^2,
\nonu \\
d_{64} &=& \frac{72 (k + N)^2 (k + M + N)}{M^2} b_{1}^2,\quad
d_{65} = \frac{18 i (k + N)^2 (k + 2 N)}{M^2} b_{1}^2,
\nonu \\
%%%%%%%%%%%%%%%%%%%%%%%%%%%%%%%%%%%%%%%%%%
d_{66} &=& -\frac{6 N (k + N)^2}{k M \, (k+2M)\, D(k,N,M)} 
\Bigg(\sqrt{\frac{M + N}{M N}} (k + M + N) (k + 2 N) \Bigg) 
\nonu \\
& \times & \Bigg(6 k^4 + 4 k^6 + 16 k^3 M + 18 k^5 M + 14 k^2 M^2
\nonu \\
&+& 24 k^4 M^2 + 4 k M^3 + 10 k^3 M^3 + 16 k^3 N + 18 k^5 N + 40 k^2 M N + 64 k^4 M N + 4 k^6 M N
\nonu \\
&+& 29 k M^2 N + 73 k^3 M^2 N + 8 k^5 M^2 N + 6 M^3 N + 27 k^2 M^3 N + 3 k^4M^3 N + 14 k^2 N^2
\nonu \\
&+& 24 k^4 N^2 
+ 29 k M N^2 + 73 k^3 M N^2
\nonu \\
&+& 8 k^5 M N^2 + 12 M^2 N^2 + 66 k^2 M^2 N^2 + 16 k^4 M^2 N^2 + 17 k M^3 N^2
\nonu \\
&+& 5 k^3 M^3 N^2 + 4 k N^3 + 10 k^3 N^3 + 6 M N^3 + 27 k^2 M N^3 + 3 k^4 M N^3 + 17 k M^2 N^3
\nonu \\
& + & 5 k^3 M^2 N^3 \Bigg) b_{1}^2,
\label{dvalue}
\eea
where we introduce 
\bea
D(k,N,M) & \equiv & \Bigg(22 k^3 + 44 k^2 M + 22 k M^2 + 44 k^2 N + 56 k M N 
+ 10 k^3 M N + 17 M^2 N \nonu \\
& + & 5 k^2 M^2 N 
+ 22 k N^2 
+ 17 M N^2 + 5 k^2 M N^2 \Bigg).
\label{Ddenom}
\eea

For the condition of (\ref{kcondition})
on the level $k$,
we are left with the following neutral spin-$4$ current
\bea
&&
d_{16} \delta_{\beta \gamma} J^{\beta} J^{\gamma} J^{u(1)} J^{u(1)} 
+ d_{17} \delta_{b c} \delta_{\beta \gamma} J^{\beta} J^{\gamma} J^{b} J^{c}
\nonu\\
&&+ d_{18} d^{\beta \gamma \delta} J^{\beta} J^{\gamma} J^{\delta} J^{u(1)} 
+ d_{19} f^{\beta \gamma \delta} J^{\beta} J^{\gamma} J^{\delta} J^{u(1)} 
+ d_{20} d^{\beta \omega \eta} d^{\gamma \delta \eta} J^{\beta} J^{\gamma} J^{\delta} J^{\omega} \nonu\\
&&+ d_{21} d^{\gamma \delta \eta} f^{\beta \omega \eta} J^{\beta} J^{\gamma} J^{\delta} J^{\omega} 
+ d_{22} d^{\beta \omega \eta} f^{\gamma \delta \eta} J^{\beta} J^{\gamma} J^{\delta} J^{\omega} 
+ d_{23} f^{\beta \omega \eta} f^{\gamma \delta \eta} J^{\beta} J^{\gamma} J^{\delta} J^{\omega} \nonu\\
&&+ d_{24} \delta_{\beta \omega} \delta_{\gamma \delta} J^{\beta} J^{\gamma} J^{\delta} J^{\omega} 
+ d_{26} \delta_{\beta \gamma} \delta_{\bar{i} k} \delta_{\rho \bar{\tau}} J^{\beta}J^{\gamma} J^{(\rho \bar{i})} J^{(\bar{\tau} k)} 
+ d_{27} \delta_{\beta \gamma} J^{\beta} J^{\gamma} \partial J^{u(1)}
\nonu \\
&& + d_{40} \delta_{\beta \gamma} \partial J^{\beta} J^{\gamma} J^{u(1)}
+ d_{41} d^{\beta \gamma \delta} \partial J^{\beta} J^{\gamma} J^{\delta} 
+ d_{42} f^{\beta \gamma \delta} \partial J^{\beta} J^{\gamma} J^{\delta} 
+ d_{47} \delta_{\beta \gamma} \partial^{2} J^{\beta} J^{\gamma} 
\nonu\\
&&+ d_{55} \delta_{\bar{i} k} J^{\beta} J^{u(1)} J^{(\rho \bar{i})} J^{(\bar{\tau} k)}t^{\beta}_{\rho \bar{\tau}} 
+ d_{56} \delta_{\bar{i} k} J^{\beta} J^{(\rho \bar{i})} \partial J^{(\bar{\tau} k)} t^{\beta}_{\rho \bar{\tau}} 
+ d_{57} \delta_{\bar{i} k} J^{\beta} \partial J^{(\rho \bar{i})} J^{(\bar{\tau} k)} t^{\beta}_{\rho \bar{\tau}} \nonu\\
&&
+ d_{59} J^{\beta} J^{b} J^{(\rho \bar{i})} J^{(\bar{\tau} k)} t^{b}_{k \bar{i}} t^{\beta}_{\rho \bar{\tau}} 
+ d_{64} d^{\beta \gamma \delta} \delta_{\bar{i} k} J^{\beta} J^{\gamma} J^{(\rho \bar{i})} J^{(\bar{\tau} k)} t^{\delta}_{\rho \bar{\tau}}. 
\label{simple2}
\eea
The eighth order pole of the OPE
of (\ref{simple2}) with itself should vanish.

%%%%%%%%%%%%%%%%%%%%%%%%%%%%%%%%%%%%%%%%%%%%%%%%%%%%%%%%%%%%%%%%%%%%
%%%%%%%%%%%%%%%%%%%%%%%%%%%%%%%%%%%%%%%%%%%%%%%%%%%%%%%%%%%%%%%%%%%%%
\section{ The first order pole in the OPE
of the charged spin-$2$ current
$K^a$ with the charged spin-$3$ current $P^b$}
%%%%EEE%%%%%%%%%%%%%%%%%%%%%%%%%%%%%%%%%%%%%%%%%%%%%%%%%%%%%%%%%%%%%%%%%%
%%%%%%%%%%%%%%%%%%%%%%%%%%%%%%%%%%%%%%%%%%%%%%%%%%%%%%%%%%%%%%%%%%%%

Again, from (\ref{spin2expression}) and (\ref{spin3exp}),
the following first order pole can be obtained
\bea
&&K^a(z) \, P^b(w) \Bigg|_{\frac{1}{(z-w)}} = 
\frac{4 (k + N) a_{7}}{M} 
\delta_{a b} \delta_{\bar{i} k} \delta_{\rho \bar{\tau}}J^{u(1)} J^{(\rho \bar{i})} \partial J^{(\bar{\tau} k)} \nonu\\
&&+
\frac{4 (k + N) a_{7} }{M}\delta_{a b} \delta_{\bar{i} k} \delta_{\rho \bar{\tau}} 
J^{u(1)} \partial J^{(\rho \bar{i})} J^{(\bar{\tau} k)} 
-\frac{3 i N a_{17}}{2 (2 k + M)} 
d^{d c i} d^{b i e} d^{f v h} f^{a e v} 
J^{d} J^{c} J^{f} J^{h} 
\nonu\\
&&- 
\frac{3 i N a_{17}}{(2 k + M)} 
d^{d b i} d^{c i e} d^{f v h} f^{a e v} 
J^{d} J^{c} J^{f} J^{h} 
-
\frac{i N a_{17}}{2 (2 k + M)} 
d^{f v h} f^{d i e} f^{a e v} f^{b c i} 
J^{d} J^{c} J^{f} J^{h} 
\nonu\\
&&- 
\frac{i N a_{17}}{(2 k + M)} 
d^{f v h} f^{d c i} f^{a e v} f^{b i e} 
J^{d} J^{c} J^{f} J^{h} 
- 
\frac{i N a_{17}}{2 (2 k + M)} 
d^{f v h} f^{d b i} f^{a e v} f^{c i e} 
J^{d} J^{c} J^{f} J^{h}  
\nonu\\
&&- 
\frac{12 i N a_{17}}{M (2 k + M)}  d^{f h e} f^{a c e} \delta_{d b}
J^{d} J^{c} J^{f} J^{h}  
-
\frac{i N \Bigg(M a_{3} + 6 a_{17}\Bigg) }
     {M (2 k + M)} d^{f h e} f^{a b e} \delta_{d c}
J^{d} J^{c} J^{f} J^{h}  
\nonu\\
&&- 
\frac{2 i N a_{3} }{(2 k + M)} d^{f h e} f^{a d e} \delta_{b c}
J^{d} J^{c} J^{f} J^{h}  
+
\frac{4 N (k + N) \sqrt{\frac{M + N}{M N}} a_{8}}{k} 
J^{b} J^{a} \partial J^{u(1)}  
\nonu\\
&&+ 
\frac{2 i N \sqrt{\frac{M + N}{M N}} a_{4}}{k}  f^{a b c}
J^{c} J^{u(1)} J^{u(1)} J^{u(1)}  
- 
\frac{i N a_{4}}{(2 k + M)} 
d^{c f e} f^{a b e} 
J^{c} J^{f} J^{u(1)} J^{u(1)}  
\nonu\\
&&+ 
\frac{4 i}{k M} \Bigg(k a_{7} + M N \sqrt{\frac{M + N}{M N}} a_{8}\Bigg) 
f^{a b c} \delta_{\bar{i} k} \delta_{\rho \bar{\tau}} 
J^{c} J^{u(1)} J^{(\rho \bar{i})} J^{(\bar{\tau} k)}  
\nonu\\
&&- 
\frac{4 i N \sqrt{\frac{M + N}{M N}} a_{9}}{k} 
d^{b c e} f^{a f e} 
J^{c} J^{f} J^{u(1)} J^{u(1)}  
-
\frac{2 i N a_{9}}{(2 k + M)} 
d^{b c e} d^{f v h} f^{a e v} 
J^{c} J^{f} J^{h} J^{u(1)}  
\nonu\\
&&+ 
\frac{3 i N \sqrt{\frac{M + N}{M N}} a_{17}}{k} 
d^{b e v} d^{c f e} f^{a v h} 
J^{c} J^{f} J^{h} J^{u(1)}  
+ 
\frac{6 i N \sqrt{\frac{M + N}{M N}} a_{17}}{k} 
d^{b c e} d^{f e v} f^{a v h} 
J^{c} J^{f} J^{h} J^{u(1)}  
\nonu\\
&&+ 
\frac{i N \sqrt{\frac{M + N}{M N}} a_{17}}{k} 
f^{a v h} f^{b f e} f^{c e v} 
J^{c} J^{f} J^{h} J^{u(1)}  
+
\frac{2 i N \sqrt{\frac{M + N}{M N}} a_{17}}{k} 
f^{a v h} f^{b e v} f^{c f e} 
J^{c} J^{f} J^{h} J^{u(1)}  
\nonu\\
&&- 
\frac{i N \sqrt{\frac{M + N}{M N}} a_{17}}{k} 
f^{a v h} f^{b c e} f^{f e v} 
J^{c} J^{f} J^{h} J^{u(1)}  
+
\frac{24 i N \sqrt{\frac{M + N}{M N}} a_{17}}{k M} 
f^{a f h} \delta_{b c} 
J^{c} J^{f} J^{h} J^{u(1)}  
\nonu\\
&&+ 
\frac{4 i N \sqrt{\frac{M + N}{M N}} a_{3}}{k} 
f^{a c h} \delta_{b f} 
J^{c} J^{f} J^{h} J^{u(1)}  
+ 
\frac{2 i N \sqrt{\frac{M + N}{M N}} \Bigg(M a_{3} + 6 a_{17}\Bigg)}{k M} 
f^{a b h} \delta_{c f} 
J^{c} J^{f} J^{h} J^{u(1)}  
\nonu\\
&&- 
\frac{2 i N a_{8}}{(2 k + M)} 
d^{c f e} f^{a b e} \delta_{\bar{i} k} \delta_{\rho \bar{\tau}} 
J^{c} J^{f} J^{(\rho \bar{i})} J^{(\bar{\tau} k)}  
-
\frac{4 i a_{5}}{M} 
d^{b c e} f^{a f e} \delta_{\bar{i} k} \delta_{\rho \bar{\tau}} 
J^{c} J^{f} J^{(\rho \bar{i})} J^{(\bar{\tau} k)}  
\nonu\\
&&+ 
\frac{2 N \sqrt{\frac{M + N}{M N}}
  \Bigg(k M a_{5} + M N a_{5} - 12 a_{17} + 3 M^2 a_{17}\Bigg)}{k M} 
d^{a f e} d^{b c e} 
J^{c} J^{f} \partial J^{u(1)}  
\nonu\\
&&- 
\frac{2 (-12 + M^2) N \sqrt{\frac{M + N}{M N}} a_{17}}{k M} 
d^{a c e} d^{b f e} 
J^{c} J^{f} \partial J^{u(1)}  
+
\frac{2 M N \sqrt{\frac{M + N}{M N}} a_{17}}{k} 
d^{a b e} d^{c f e} 
J^{c} J^{f} \partial J^{u(1)}  
\nonu\\
&&+ 
\frac{i M N^2 \sqrt{\frac{M + N}{M N}} a_{8}}{(2 k + M)} 
d^{c f e} f^{a b e} 
J^{c} J^{f} \partial J^{u(1)}  
+
2 i N \sqrt{\frac{M + N}{M N}} a_{5} 
d^{b c e} f^{a f e} 
J^{c} J^{f} \partial J^{u(1)}  
\nonu\\
&&- 
\frac{2 N \sqrt{\frac{M + N}{M N}} \Bigg(a_{11} + M a_{17}\Bigg)}{k} 
f^{a f e} f^{b c e} 
J^{c} J^{f} \partial J^{u(1)}  
-
\frac{4 N \sqrt{\frac{M + N}{M N}} a_{3}}{k} 
f^{a c e} f^{b f e} 
J^{c} J^{f} \partial J^{u(1)}  
\nonu\\
&&- 
\frac{24 N \sqrt{\frac{M + N}{M N}} a_{17}}{k M} 
f^{a b e} f^{c f e} 
J^{c} J^{f} \partial J^{u(1)}  
+
\frac{2 (2 k + M + 2 N) a_{5}}{M} 
d^{a b c} \delta_{\bar{i} k} \delta_{\rho \bar{\tau}} 
J^{c} J^{(\rho \bar{i})} \partial J^{(\bar{\tau} k)}  
\nonu\\
&&+ 
\frac{4 N \sqrt{\frac{M + N}{M N}}
\Bigg(M^3 a_{3} - 24 a_{17} + 12 M^2 a_{17}\Bigg)}{k M^2} 
\delta_{a f} \delta_{b c} 
J^{c} J^{f} \partial J^{u(1)}  
\nonu\\
&&- 
\frac{4 (-24 + 5 M^2) N \sqrt{\frac{M + N}{M N}} a_{17}}{k M^2} 
\delta_{a c} \delta_{b f} 
J^{c} J^{f} \partial J^{u(1)}  
- 
\frac{4 N \sqrt{\frac{M + N}{M N}} a_{17}}{k} 
\delta_{a b} \delta_{c f} 
J^{c} J^{f} \partial J^{u(1)}  
\nonu\\
&&- 
\frac{2 i \Bigg(-4 a_{5} + M^2 a_{5} + 2 M a_{8} - M a_{13}\Bigg)}{M^2} 
f^{a b c} \delta_{\bar{i} k} \delta_{\rho \bar{\tau}} 
J^{c} J^{(\rho \bar{i})} \partial J^{(\bar{\tau} k)}  
\nonu\\
&&- 
\frac{2}{k M} \Bigg(
2 k M a_{5} + 2 k N a_{5} + 2 M N a_{5} + 2 N^2 a_{5} 
- k M N \sqrt{\frac{M + N}{M N}} a_{7} 
- M N^2 \sqrt{\frac{M + N}{M N}} a_{7} \nonu\\
&&- M^2 N \sqrt{\frac{M + N}{M N}} a_{9}
\Bigg) 
d^{a b c} J^{c} \partial J^{u(1)} J^{u(1)}  
\nonu\\
&&- 
\frac{2 i N}{k} \Bigg(
k \sqrt{\frac{M + N}{M N}} a_{7} + M a_{8} + N a_{8}
\Bigg) 
f^{a b c} J^{c} \partial J^{u(1)} J^{u(1)}  
\nonu\\
&&+ 
\frac{2 (2 k + M + 2 N) a_{5}}{M} 
d^{a b c} \delta_{\bar{i} k} \delta_{\rho \bar{\tau}} 
J^{c} \partial J^{(\rho \bar{i})} J^{(\bar{\tau} k)}  
\nonu\\
&&+ 
\frac{2 i}{M^2} \Bigg(
-4 a_{5} + M^2 a_{5} + 2 M a_{8} + M a_{12}
\Bigg) 
f^{a b c} \delta_{\bar{i} k} \delta_{\rho \bar{\tau}} 
J^{c} \partial J^{(\rho \bar{i})} J^{(\bar{\tau} k)}  
\nonu\\
&&- 
\frac{N \sqrt{\frac{M + N}{M N}}}{2 k} \Bigg(
4 k^2 a_{5} + 2 k M a_{5} + 4 k N a_{5} 
- k a_{12} - N a_{12} - k a_{13} - N a_{13}
\Bigg) 
d^{a b c} J^{c} \partial^2 J^{u(1)}  
\nonu\\
&&+ 
\frac{i N \sqrt{\frac{M + N}{M N}}}{2 k M} \Bigg(
4 M^2 a_{3} - 8 k a_{5} + 2 k M^2 a_{5} - 8 N a_{5} 
+ 2 M^2 N a_{5} + 4 k M a_{8} + 4 M N a_{8} \nonu\\
&&- 
2 M^2 a_{11} - k M a_{12} - M N a_{12} 
- k M a_{13} + M N a_{13} + 4 M a_{16} 
+ 12 M a_{17} + 2 M^3 a_{17}
\Bigg)\nonu \\
&& \times 
f^{a b c} J^{c} \partial^2 J^{u(1)}  
\nonu\\
&&+ 
\frac{2 i N \sqrt{\frac{M + N}{M N}} a_{2}}{k} 
f^{a b c} J^{\beta} J^{\beta} J^{c} J^{u(1)}  
-
\frac{i N a_{2}}{(2 k + M)} 
d^{c f e} f^{a b e} 
J^{\beta} J^{\beta} J^{c} J^{f}  
\nonu\\
&&- 
\frac{2 (k + N) \sqrt{\frac{M + N}{M N}} a_{1}}{k} 
\delta_{a b} J^{\beta} J^{\beta} \partial J^{u(1)}  
+
\frac{(2 k + M + 2 N) a_{1}}{(2 k + M)} 
d^{a b c} J^{\beta} J^{\beta} \partial J^{c}  
\nonu\\
&&+ 
i \Bigg(
a_{1} + N a_{2}
\Bigg) 
f^{a b c} J^{\beta} J^{\beta} \partial J^{c}  
+
\frac{2 (k + N) a_{13}}{M} 
\delta_{a b} \delta_{\bar{i} k} \delta_{\rho \bar{\tau}} 
J^{(\rho \bar{i})} \partial^2 J^{(\bar{\tau} k)}  
\nonu\\
&&- 
\frac{4 (k + N) (M + N) a_{7}}{k M} 
\partial J^{u(1)} J^{u(1)} J^{u(1)}  
- 
\frac{2 (k + N) (M + N) a_{13}}{k M} 
\partial J^{u(1)} \partial J^{u(1)}  
\nonu\\
&&- 
\frac{2 (k + N) \sqrt{\frac{M + N}{M N}} \Bigg(a_{12} - a_{13}\Bigg)}{k M} 
\delta_{a b} \delta_{\bar{i} k} \delta_{\rho \bar{\tau}} 
\partial J^{u(1)} J^{(\rho \bar{i})} J^{(\bar{\tau} k)}  
\nonu\\
&&- 
\frac{4 N (k + M + N)}{k M} 
\Bigg(
a_{7} - M \sqrt{\frac{M + N}{M N}} a_{8}
\Bigg) 
\partial J^{a} J^{b} J^{u(1)}  
-
\frac{2 N (k + M + N) a_{13}}{k M} 
\partial J^{a} \partial J^{b}  
\nonu\\
&&+ 
\frac{2 N \sqrt{\frac{M + N}{M N}}}{k (2 k + M)} 
\Bigg(
2 k^2 a_{7} + k M a_{7} + 2 k N a_{7} 
+ 2 k M a_{9} + M^2 a_{9}
\Bigg) 
d^{a b c} \partial J^{c} J^{u(1)} J^{u(1)}  
\nonu\\
&&+ 
\frac{i N}{k} 
\Bigg(
k a_{4} + 2 k \sqrt{\frac{M + N}{M N}} a_{7} 
+ 2 N \sqrt{\frac{M + N}{M N}} a_{7}
\Bigg) 
f^{a b c} \partial J^{c} J^{u(1)} J^{u(1)}  
\nonu\\
&&- 
\frac{2 (-12 + M^2) N \sqrt{\frac{M + N}{M N}} a_{17}}{k M} 
d^{a f e} d^{b c e} \partial J^{c} J^{f} J^{u(1)}  
\nonu\\
&&+ 
\frac{N}{k M (2 k + M)} 
\Bigg(
4 k^2 M \sqrt{\frac{M + N}{M N}} a_{5} 
+ 2 k M^2 \sqrt{\frac{M + N}{M N}} a_{5} 
+ 4 k M N \sqrt{\frac{M + N}{M N}} a_{5} \nonu\\
&&- 
2 k^2 M a_{7} - k M^2 a_{7} - 2 k M N a_{7} 
- 48 k \sqrt{\frac{M + N}{M N}} a_{17} 
- 24 M \sqrt{\frac{M + N}{M N}} a_{17} \nonu\\
&&+ 
12 k M^2 \sqrt{\frac{M + N}{M N}} a_{17} 
+ 6 M^3 \sqrt{\frac{M + N}{M N}} a_{17}
\Bigg) 
d^{a c e} d^{b f e} \partial J^{c} J^{f} J^{u(1)}  
\nonu\\
&&+ 
\frac{2 M N}{k (2 k + M)} 
\Bigg(
-k a_{9} + 2 k \sqrt{\frac{M + N}{M N}} a_{17} 
+ M \sqrt{\frac{M + N}{M N}} a_{17}
\Bigg) 
d^{a b e} d^{c f e} \partial J^{c} J^{f} J^{u(1)}  
\nonu\\
&&- 
\frac{2 i N^2 a_{7}}{(2 k + M)} 
d^{c f e} f^{a b e} \partial J^{c} J^{f} J^{u(1)}  
- 
\frac{2 i N^2 \sqrt{\frac{M + N}{M N}} a_{5}}{k} 
d^{b c e} f^{a f e} \partial J^{c} J^{f} J^{u(1)}  
\nonu\\
&&- 
\frac{i N}{k} 
\Bigg(
2 k \sqrt{\frac{M + N}{M N}} a_{5} + 2 N \sqrt{\frac{M + N}{M N}} a_{5} 
- k a_{7} + 2 k a_{9}
\Bigg) 
d^{b f e} f^{a c e} \partial J^{c} J^{f} J^{u(1)}  
\nonu\\
&&- 
\frac{2 N \sqrt{\frac{M + N}{M N}}}{k} 
\Bigg(
2 a_{3} - a_{11}
\Bigg) 
f^{a f e} f^{b c e} \partial J^{c} J^{f} J^{u(1)}  
\nonu \\
&& +
\frac{i N (2 k + M + 2 N) a_{7}}{(2 k + M)} 
d^{a c e} f^{b f e} \partial J^{c} J^{f} J^{u(1)}  
\nonu\\
&&+ 
\frac{N}{k} 
\Bigg(
k a_{7} - 2 \sqrt{\frac{M + N}{M N}} a_{11} 
- 2 M \sqrt{\frac{M + N}{M N}} a_{17}
\Bigg) 
f^{a c e} f^{b f e} \partial J^{c} J^{f} J^{u(1)}  
\nonu\\
&&+ 
\frac{24 N \sqrt{\frac{M + N}{M N}} a_{17}}{k M} 
f^{a b e} f^{c f e} \partial J^{c} J^{f} J^{u(1)}  
- 
\frac{4 (-24 + 5 M^2) N \sqrt{\frac{M + N}{M N}} a_{17}}{k M^2} 
\delta_{a f} \delta_{b c} \partial J^{c} J^{f} J^{u(1)}  
\nonu\\
&&+ 
\frac{4 N \sqrt{\frac{M + N}{M N}}
\Bigg(M^3 a_{3} - 24 a_{17} + 12 M^2 a_{17}\Bigg)}{k M^2} 
\delta_{a c} \delta_{b f} \partial J^{c} J^{f} J^{u(1)}  
\nonu\\
&&- 
\frac{4 N \sqrt{\frac{M + N}{M N}} a_{17}}{k} 
\delta_{a b} \delta_{c f} \partial J^{c} J^{f} J^{u(1)}  
+ 
\frac{2 (-12 + M^2) N a_{17}}{M (2 k + M)} 
d^{a f e} d^{b e v} d^{c v h} \partial J^{c} J^{f} J^{h}  
\nonu\\
&&- 
\frac{6 (-2 + M) (2 + M) N a_{17}}{M (2 k + M)} 
d^{a e v} d^{b f e} d^{c v h} \partial J^{c} J^{f} J^{h}  
-
\frac{2 M N a_{17}}{(2 k + M)} 
d^{a b e} d^{c v h} d^{f e v} \partial J^{c} J^{f} J^{h}  
\nonu\\
&&- 
\frac{N (2 k + M + 2 N) a_{5}}{(2 k + M)} 
d^{a c e} d^{b v f} d^{h e v} \partial J^{c} J^{f} J^{h}  
-
\frac{3}{2} i N a_{17} 
d^{b e v} d^{f v h} f^{a c e} \partial J^{c} J^{f} J^{h}  
\nonu\\
&&+ 
i N \Bigg(a_{5} - 3 a_{17}\Bigg) 
d^{b v f} d^{h e v} f^{a c e} 
\partial J^{c} J^{f} J^{h}  
-
\frac{2 i N^2 a_{5}}{(2 k + M)} 
d^{b f e} d^{c v h} f^{a e v} 
\partial J^{c} J^{f} J^{h}  
\nonu\\
&&- 
\frac{i N^2 a_{5}}{(2 k + M)} 
d^{b c e} d^{f v h} f^{a e v} 
\partial J^{c} J^{f} J^{h}  
+ 
\frac{N a_{11}}{(2 k + M)} 
d^{f v h} f^{a e v} f^{b c e} 
\partial J^{c} J^{f} J^{h}  
\nonu\\
&&- 
\frac{4 N a_{3}}{(2 k + M)} 
d^{c v h} f^{a f e} f^{b e v} 
\partial J^{c} J^{f} J^{h}  
- 
\frac{2 N \Bigg(a_{11} + M a_{17}\Bigg)}{(2 k + M)} 
d^{c v h} f^{a e v} f^{b f e} 
\partial J^{c} J^{f} J^{h}  
\nonu\\
&&- 
\frac{24 N a_{17}}{M (2 k + M)} 
d^{c v h} f^{a b e} f^{f e v} 
\partial J^{c} J^{f} J^{h}  
-
\frac{1}{2} i N a_{17} 
f^{a c e} f^{b v h} f^{f e v} 
\partial J^{c} J^{f} J^{h}  
\nonu\\
&&- 
i N a_{17} 
f^{a c e} f^{b e v} f^{f v h} 
\partial J^{c} J^{f} J^{h}  
+
\frac{i N (2 k + M + 2 N) a_{5}}{(2 k + M)} 
d^{a c e} d^{b v f} f^{h e v} 
\partial J^{c} J^{f} J^{h}  
\nonu\\
&&+ 
N a_{5} 
d^{b v f} f^{a c e} f^{h e v} 
\partial J^{c} J^{f} J^{h}  
+ 
\frac{1}{2} i N a_{17} 
f^{a c e} f^{b v f} f^{h e v} 
\partial J^{c} J^{f} J^{h}  
\nonu\\
&&+ 
\frac{4 N a_{17}}{(2 k + M)} 
d^{c f h} \delta_{a b} 
\partial J^{c} J^{f} J^{h}  
-
\frac{4 N (k + M + N) a_{5}}{k M} 
d^{b f h} \delta_{a c} 
\partial J^{c} J^{f} J^{h}  
\nonu\\
&&+ 
\frac{4 (-24 + 5 M^2) N a_{17}}{M^2 (2 k + M)} 
d^{b c h} \delta_{a f} 
\partial J^{c} J^{f} J^{h}  
+
2 i N a_{8} 
f^{a c h} \delta_{b f} \partial J^{c} J^{f} J^{h}  
\nonu\\
&&- 
\frac{2 N}{M^2 (2 k + M)} 
\Bigg(
2 M^3 a_{3} + 2 k M^2 a_{8} + M^3 a_{8} 
+ 2 M^2 N a_{8} - 48 a_{17} + 24 M^2 a_{17}
\Bigg) \nonu \\
&& \times 
d^{a c h} \delta_{b f} \partial J^{c} J^{f} J^{h}  
\nonu\\
&&- 
\frac{12 i N a_{17}}{M} 
f^{a c h} \delta_{b f} \partial J^{c} J^{f} J^{h}  
- 
2 i N a_{3} 
f^{a c f} \delta_{b h} \partial J^{c} J^{f} J^{h}  
\nonu \\
&& +
\frac{i N \Bigg(M a_{3} + 6 a_{17}\Bigg)}{M} 
f^{a b c} \delta_{f h} \partial J^{c} J^{f} J^{h}  
\nonu\\
&&+ 
\frac{1}{M (2 k + M)} 
\Bigg(
4 k M a_{5} + 2 M^2 a_{5} + 2 k a_{12} + M a_{12} 
+ 2 N a_{12} - 2 k a_{13} - M a_{13} \nonu\\
&&- 2 N a_{13}
\Bigg) 
d^{a b c} \delta_{\bar{i} k} \delta_{\rho \bar{\tau}} 
\partial J^{c} J^{(\rho \bar{i})} J^{(\bar{\tau} k)}  
\nonu\\
&&+ 
\frac{i}{M^2} 
\Bigg(
-8 a_{5} + 2 M^2 a_{5} + 4 M a_{8} 
+ 2 M^2 N a_{8} + M a_{12} + M a_{13}
\Bigg) 
f^{a b c} \delta_{\bar{i} k} \delta_{\rho \bar{\tau}} 
\partial J^{c} J^{(\rho \bar{i})} J^{(\bar{\tau} k)}  
\nonu\\
&&+ 
\frac{N \sqrt{\frac{M + N}{M N}}}{k (2 k + M)} 
\Bigg(
4 k M N a_{5} + 2 M^2 N a_{5} + 4 k^2 a_{13} 
+ 2 k M a_{13} + 4 k N a_{13} + M N a_{13}
\Bigg) \nonu \\
&& \times 
d^{a b c} \partial J^{c} \partial J^{u(1)}  
\nonu\\
&&+ 
\frac{i N \sqrt{\frac{M + N}{M N}}}{k} 
\Bigg(
4 M a_{3} - k M N a_{8} - 2 M a_{11} 
+ N a_{13} + 4 a_{16} + 12 a_{17} + 2 M^2 a_{17}
\Bigg) 
\nonu \\
&& \times f^{a b c} \partial J^{c} \partial J^{u(1)}  
\nonu\\
&&- 
\frac{N (2 k + M + 2 N) a_{13}}{2 (2 k + M)} 
d^{a c e} d^{b f e} \partial J^{c} \partial J^{f}  
-
\frac{M N^2 a_{5}}{(2 k + M)} 
d^{a b e} d^{c f e} \partial J^{c} \partial J^{f}  
\nonu\\
&&- 
\frac{i N}{(2 k + M)} 
\Bigg(
2 M a_{3} - M a_{11} + N a_{13} 
+ 2 a_{16} + 6 a_{17} + M^2 a_{17}
\Bigg) 
d^{c f e} f^{a b e} \partial J^{c} \partial J^{f}  
\nonu\\
&&+ 
\frac{1}{2} i N a_{13} 
d^{b f e} f^{a c e} \partial J^{c} \partial J^{f}  
-
i N^2 a_{5} 
d^{b c e} f^{a f e} \partial J^{c} \partial J^{f}  
+ 
N a_{11} 
f^{a f e} f^{b c e} \partial J^{c} \partial J^{f}  
\nonu\\
&&+ 
\frac{i N (2 k + M + 2 N) a_{13}}{2 (2 k + M)} 
d^{a c e} f^{b f e} \partial J^{c} \partial J^{f}  
+ 
\frac{1}{2} N a_{13} 
f^{a c e} f^{b f e} 
\partial J^{c} \partial J^{f}  
\nonu\\
&&- 
2 \Bigg(
\sqrt{\frac{M + N}{M N}} a_{1} + 2 a_{7}
\Bigg) 
\delta_{a b} \partial J^{\beta} J^{\beta} J^{u(1)}  
+ 
\Bigg(
a_{1} - 4 a_{5}
\Bigg) 
d^{a b c} \partial J^{\beta} J^{\beta} J^{c}  
\nonu\\
&&- 
i a_{1} 
f^{a b c} \partial J^{\beta} J^{\beta} J^{c}  
-
a_{1} 
d^{\beta \gamma \delta} \delta_{a b} 
\partial J^{\beta} J^{\gamma} J^{\delta}  
-
i a_{1} 
f^{\beta \gamma \delta} \delta_{a b} 
\partial J^{\beta} J^{\gamma} J^{\delta}  
\nonu\\
&&- 
2 a_{13} 
\delta_{a b} \partial J^{\beta} \partial J^{\beta}  
+
\frac{2 (k + N) \Bigg(a_{12} + a_{13}\Bigg)}{M} 
\delta_{a b} \delta_{\bar{i} k} \delta_{\rho \bar{\tau}} 
\partial J^{(\rho \bar{i})} \partial J^{(\bar{\tau} k)}  
\nonu\\
&&- 
\frac{(k + N)}{k M} 
\Bigg(
2 k M N \sqrt{\frac{M + N}{M N}} a_{7} 
+ M a_{12} + N a_{12} + M a_{13} + N a_{13}
\Bigg) 
\partial^2 J^{u(1)} J^{u(1)}  
\nonu\\
&&+ 
\frac{N (k + M + N)}{k M} 
\Bigg(
2 k M a_{8} - a_{12} - a_{13}
\Bigg) 
\partial^2 J^{a} J^{b}  
\nonu\\
&&+ 
\frac{N (2 k + M + 2 N)}{2 (2 k + M)} 
\Bigg(
2 k a_{7} + \sqrt{\frac{M + N}{M N}} a_{12} 
+ \sqrt{\frac{M + N}{M N}} a_{13}
\Bigg) 
d^{a b c} \partial^2 J^{c} J^{u(1)}  
\nonu\\
&&- 
\frac{i N}{2 k} 
\Bigg(
4 M \sqrt{\frac{M + N}{M N}} a_{3} - 2 k^2 a_{7} 
- 2 k N a_{7} - k \sqrt{\frac{M + N}{M N}} a_{12} 
- k \sqrt{\frac{M + N}{M N}} a_{13} \nonu\\
&&- 
2 N \sqrt{\frac{M + N}{M N}} a_{13} 
- 4 \sqrt{\frac{M + N}{M N}} a_{16} 
+ 24 \sqrt{\frac{M + N}{M N}} a_{17} 
+ 4 M^2 \sqrt{\frac{M + N}{M N}} a_{17}
\Bigg) \nonu \\
&& \times 
f^{a b c} \partial^2 J^{c} J^{u(1)}  
\nonu\\
&&+ 
\frac{1}{2 M (2 k + M)} 
N \Bigg(
-8 k a_{5} - 4 M a_{5} + 2 k M^2 a_{5} + M^3 a_{5} 
+ M^2 N a_{5} + 24 k a_{17} \nonu \\
&& + 12 M a_{17} - 
4 k M^2 a_{17} - 2 M^3 a_{17}
\Bigg) 
d^{a f e} d^{b c e} \partial^2 J^{c} J^{f}  
\nonu\\
&&+ 
\frac{1}{4 (2 k + M)} 
N \Bigg(
8 k^2 a_{5} + 8 k M a_{5} + 2 M^2 a_{5} 
+ 8 k N a_{5} + 2 M N a_{5} - 2 k a_{12} - M a_{12} \nonu\\
&&- 
2 N a_{12} - 2 k a_{13} - M a_{13} - 2 N a_{13}
\Bigg) 
d^{a c e} d^{b f e} \partial^2 J^{c} J^{f}  
\nonu\\
&&- 
\frac{1}{2 M (2 k + M)} 
N \Bigg(
-8 k a_{5} - 4 M a_{5} + 2 k M^2 a_{5} + M^3 a_{5} 
+ M^2 N a_{5} + 24 k a_{17} \nonu \\
&& + 12 M a_{17} 
- 
4 k M^2 a_{17} - 2 M^3 a_{17}
\Bigg) 
d^{a b e} d^{c f e} \partial^2 J^{c} J^{f}  
\nonu\\
&&- 
\frac{i N}{2 M (2 k + M)} 
\Bigg(
4 M^2 a_{3} - 4 k a_{5} - 2 M a_{5} - 4 N a_{5} 
+ 4 k M a_{8} + 2 M^2 a_{8} + 4 M N a_{8} \nonu\\
&&- 
M^2 a_{11} + 2 M N a_{13} + 4 M a_{16} 
- 12 M a_{17} + 2 M^3 a_{17}
\Bigg) 
d^{c f e} f^{a b e} \partial^2 J^{c} J^{f}  
\nonu\\
&&- 
\frac{i N}{4 (2 k + M)} 
\Bigg(
4 M a_{3} + 8 k^2 a_{5} + 8 k M a_{5} + 2 M^2 a_{5} 
+ 8 k N a_{5} + 4 M N a_{5} - 2 k a_{12} \nonu\\
&&- 
M a_{12} - 2 k a_{13} - M a_{13} + 16 a_{17} 
+ 2 M^2 a_{17}
\Bigg) 
d^{b f e} f^{a c e} \partial^2 J^{c} J^{f}  
\nonu\\
&&- 
\frac{i N}{M (2 k + M)} 
\Bigg(
2 k a_{5} + M a_{5} + 2 N a_{5} 
- 6 M a_{17} + M^3 a_{17}
\Bigg) 
d^{b c e} f^{a f e} \partial^2 J^{c} J^{f}  
\nonu\\
&&- 
\frac{4 (-24 + 5 M^2) a_{17}}{M^2} 
\delta_{\rho \bar{\tau}} t^{b}_{k \bar{i}} 
J^{a} J^{(\rho \bar{i})} \partial J^{(\bar{\tau} k)}  
-\frac{4 (-24 + 5 M^2) a_{17}}{M^2} 
\delta_{\rho \bar{\tau}} t^{b}_{k \bar{i}} 
J^{a} \partial J^{(\rho \bar{i})} J^{(\bar{\tau} k)}  
\nonu\\
&&- 
4 i a_{8} f^{a c f} \delta_{\rho \bar{\tau}} t^{f}_{k \bar{i}} 
J^{b} J^{c} J^{(\rho \bar{i})} J^{(\bar{\tau} k)}  
\nonu\\
&&+ 
\frac{4
\Bigg(M^3 a_{3} + k M^2 a_{8} + M^2 N a_{8} - 24 a_{17} +
12 M^2 a_{17}\Bigg)}{M^2} 
\delta_{\rho \bar{\tau}} t^{a}_{k \bar{i}} 
J^{b} J^{(\rho \bar{i})} \partial J^{(\bar{\tau} k)}  
\nonu\\
&&+ 
\frac{4 \Bigg(M^3 a_{3} + k M^2 a_{8} + M^2 N a_{8} - 24 a_{17} +
12 M^2 a_{17}\Bigg)}{M^2} 
\delta_{\rho \bar{\tau}} t^{a}_{k \bar{i}} 
J^{b} \partial J^{(\rho \bar{i})} J^{(\bar{\tau} k)}  
\nonu\\
&&+ 
4 a_{8} t^{a}_{k \bar{i}} t^{\beta}_{\rho \bar{\nu}} 
J^{\beta} J^{b} J^{(\rho \bar{i})} J^{(\bar{\nu} k)}  
- 4 a_{8} t^{a}_{k \bar{i}} t^{\beta}_{\rho \bar{\tau}} 
J^{\beta} J^{b} J^{(\rho \bar{i})} J^{(\bar{\tau} k)}  
\nonu\\
&&+ 
\frac{2 i \Bigg(a_{1} + N a_{2}\Bigg)}{N}
f^{a b c} \delta_{\rho \bar{\tau}} t^{c}_{k \bar{i}} 
J^{\beta} J^{\beta} J^{(\rho \bar{i})} J^{(\bar{\tau} k)}  
+
\frac{2 i a_{1}}{M} f^{a b c} \delta_{\bar{i} k} t^{\beta}_{\rho \bar{\tau}} 
J^{\beta} J^{c} J^{(\rho \bar{i})} J^{(\bar{\tau} k)}  
\nonu\\
&&- 
i a_{1} d^{b f e} f^{a c e} t^{\beta}_{\rho \bar{\tau}} t^{f}_{k \bar{i}} 
J^{\beta} J^{c} J^{(\rho \bar{i})} J^{(\bar{\tau} k)}  
-
4 i a_{5} d^{b c e} f^{a f e} t^{\beta}_{\rho \bar{\tau}} t^{f}_{k \bar{i}} 
J^{\beta} J^{c} J^{(\rho \bar{i})} J^{(\bar{\tau} k)}  
\nonu\\
&&- 
\frac{i (2 k + M + 2 N) a_{1}}{(2 k + M)} d^{a c e} f^{b f e} 
t^{\beta}_{\rho \bar{\tau}} t^{f}_{k \bar{i}} 
J^{\beta} J^{c} J^{(\rho \bar{i})} J^{(\bar{\tau} k)}  
+
\frac{2 i a_{1}}{M} f^{\beta \gamma \delta} \delta_{a b} \delta_{\bar{i} k} t^{\delta}_{\rho \bar{\tau}} 
J^{\beta} J^{\gamma} J^{(\rho \bar{i})} J^{(\bar{\tau} k)}  
\nonu\\
&&+ 
i a_{1} d^{\beta \gamma \delta} f^{a b c} t^{c}_{k \bar{i}}
t^{\delta}_{\rho \bar{\tau}} 
J^{\beta} J^{\gamma} J^{(\rho \bar{i})} J^{(\bar{\tau} k)}  
+ 
i a_{1} d^{a b c} f^{\beta \gamma \delta} t^{c}_{k \bar{i}} t^{\delta}_{\rho \bar{\tau}} 
J^{\beta} J^{\gamma} J^{(\rho \bar{i})} J^{(\bar{\tau} k)} \nonu\\
&&+ 
\frac{2 i}{k} \Bigg( k \sqrt{\frac{M + N}{M N}} a_{1} + N \sqrt{\frac{M + N}{M N}} a_{1} + 2 k a_{7} \Bigg) 
f^{a b c} t^{\beta}_{\rho \bar{\tau}} t^{c}_{k \bar{i}} 
J^{\beta} J^{u(1)} J^{(\rho \bar{i})} J^{(\bar{\tau} k)} 
\nonu\\
&&+ 
4 i a_{8} f^{a b c} t^{\beta}_{\rho \bar{\nu}} t^{c}_{k \bar{i}} 
J^{\beta} J^{(\rho \bar{i})} \partial J^{(\bar{\nu} k)} 
+
\frac{2 (k + N) a_{1}}{M} \delta_{a b} \delta_{\bar{i} k} t^{\beta}_{\rho \bar{\tau}} 
J^{\beta} J^{(\rho \bar{i})} \partial J^{(\bar{\tau} k)} 
\nonu\\
&&+ 
\Bigg( k a_{1} + N a_{1} + 2 M a_{5} \Bigg) 
d^{a b c} t^{\beta}_{\rho \bar{\tau}} t^{c}_{k \bar{i}} 
J^{\beta} J^{(\rho \bar{i})} \partial J^{(\bar{\tau} k)} 
\nonu\\
&&+ 
\frac{i}{M} \Bigg( k M a_{1} + M N a_{1} - 8 a_{5} + 2 M^2 a_{5} + 2 M a_{13} \Bigg) 
f^{a b c} t^{\beta}_{\rho \bar{\tau}} t^{c}_{k \bar{i}} 
J^{\beta} J^{(\rho \bar{i})} \partial J^{(\bar{\tau} k)} 
\nonu\\
&&+ 
\frac{2 (k + N) a_{1}}{M} \delta_{a b} \delta_{\bar{i} k} t^{\beta}_{\rho \bar{\tau}} 
J^{\beta} \partial J^{(\rho \bar{i})} J^{(\bar{\tau} k)} 
+ 
\Bigg( k a_{1} + N a_{1} + 2 M a_{5} \Bigg) 
d^{a b c} t^{\beta}_{\rho \bar{\tau}} t^{c}_{k \bar{i}} 
J^{\beta} \partial J^{(\rho \bar{i})} J^{(\bar{\tau} k)} 
\nonu\\
&&- 
\frac{i}{M} \Bigg( k M a_{1} + M N a_{1} - 8 a_{5} + 2 M^2 a_{5} + 4 M a_{8} -2 M a_{12} \Bigg) 
f^{a b c} t^{\beta}_{\rho \bar{\tau}} t^{c}_{k \bar{i}} 
J^{\beta} \partial J^{(\rho \bar{i})} J^{(\bar{\tau} k)} 
\nonu\\
&&+ 
\frac{2 i N a_{5}}{(2 k + M)} 
d^{b v h} d^{c f e} f^{a e v} \delta_{\rho \bar{\tau}} t^{h}_{k \bar{i}} 
J^{c} J^{f} J^{(\rho \bar{i})} J^{(\bar{\tau} k)} 
+
2 i a_{5} d^{b c e} d^{h e v} f^{a v f} \delta_{\rho \bar{\tau}} t^{h}_{k \bar{i}} 
J^{c} J^{f} J^{(\rho \bar{i})} J^{(\bar{\tau} k)} 
\nonu\\
&&+ 
3 i a_{17} d^{b e v} d^{c f e} f^{a v h} \delta_{\rho \bar{\tau}} t^{h}_{k \bar{i}} 
J^{c} J^{f} J^{(\rho \bar{i})} J^{(\bar{\tau} k)} 
+
6 i a_{17} d^{b c e} d^{f e v} f^{a v h} \delta_{\rho \bar{\tau}} t^{h}_{k \bar{i}} 
J^{c} J^{f} J^{(\rho \bar{i})} J^{(\bar{\tau} k)} 
\nonu\\
&&+ 
i a_{17} f^{a v h} f^{b f e} f^{c e v} \delta_{\rho \bar{\tau}} t^{h}_{k \bar{i}} 
J^{c} J^{f} J^{(\rho \bar{i})} J^{(\bar{\tau} k)} 
+
2 i a_{17} f^{a v h} f^{b e v} f^{c f e} \delta_{\rho \bar{\tau}} t^{h}_{k \bar{i}} 
J^{c} J^{f} J^{(\rho \bar{i})} J^{(\bar{\tau} k)} 
\nonu\\
&&- 
i a_{17} f^{a v h} f^{b c e} f^{f e v} \delta_{\rho \bar{\tau}} t^{h}_{k \bar{i}} 
J^{c} J^{f} J^{(\rho \bar{i})} J^{(\bar{\tau} k)} 
\nonu\\
&&+ 
\frac{2 i (2 k + M + 2 N) a_{5}}{(2 k + M)} 
d^{a v f} d^{b c e} f^{h e v} \delta_{\rho \bar{\tau}} t^{h}_{k \bar{i}} 
J^{c} J^{f} J^{(\rho \bar{i})} J^{(\bar{\tau} k)}\nonu\\
&&+ \frac{24 i a_{17}}{M} f^{a f h} \delta_{b c} \delta_{\rho \bar{\tau}} t^{h}_{k \bar{i}} 
J^{c} J^{f} J^{(\rho \bar{i})} J^{(\bar{\tau} k)} 
+ 4 i a_{3} f^{a c h} \delta_{b f} \delta_{\rho \bar{\tau}} t^{h}_{k \bar{i}} 
J^{c} J^{f} J^{(\rho \bar{i})} J^{(\bar{\tau} k)} 
\nonu\\
&&+ \frac{2 i \Bigg(M a_{3} + 6 a_{17}\Bigg)}{M} f^{a b h} \delta_{c f} \delta_{\rho \bar{\tau}} t^{h}_{k \bar{i}} 
J^{c} J^{f} J^{(\rho \bar{i})} J^{(\bar{\tau} k)} 
\nonu\\
&&- \frac{2 i}{k} \Bigg( 2 N \sqrt{\frac{M + N}{M N}} a_{5} + k a_{7} \Bigg)d^{b f e} f^{a c e} \delta_{\rho \bar{\tau}} t^{f}_{k \bar{i}} 
J^{c} J^{u (1)} J^{(\rho \bar{i})} J^{(\bar{\tau} k)} 
\nonu\\
&&- \frac{4 i}{k} \Bigg( k \sqrt{\frac{M + N}{M N}} a_{5} + N \sqrt{\frac{M + N}{M N}} a_{5} + k a_{9} \Bigg)d^{b c e} f^{a f e} \delta_{\rho \bar{\tau}} t^{f}_{k \bar{i}} 
J^{c} J^{u (1)} J^{(\rho \bar{i})} J^{(\bar{\tau} k)} 
\nonu\\
&&- \frac{2 i (2 k + M + 2 N) a_{7}}{(2 k + M)} d^{a c e} f^{b f e} \delta_{\rho \bar{\tau}} t^{f}_{k \bar{i}} 
J^{c} J^{u (1)} J^{(\rho \bar{i})} J^{(\bar{\tau} k)} 
\nonu\\
&&+ \frac{2}{M} \Bigg( 2 a_{5} + k M a_{5} + M N a_{5} - 12 a_{17} + 3 M^{2} a_{17} \Bigg)d^{a f e} d^{b c e} \delta_{\rho \bar{\tau}} t^{f}_{k \bar{i}} 
J^{c} J^{(\rho \bar{i})} \partial J^{(\bar{\tau} k)} 
\nonu\\
&&- \frac{2}{M} \Bigg( 2 a_{5} - 12 a_{17} + M^{2} a_{17} \Bigg)d^{a c e} d^{b f e} \delta_{\rho \bar{\tau}} t^{f}_{k \bar{i}} 
J^{c} J^{(\rho \bar{i})} \partial J^{(\bar{\tau} k)} 
\nonu\\
&&+ 2 M a_{17} d^{a b e} d^{c f e} \delta_{\rho \bar{\tau}} t^{f}_{k \bar{i}} 
J^{c} J^{(\rho \bar{i})} \partial J^{(\bar{\tau} k)} 
- \frac{2 i \Bigg(-a_{5} + M a_{8}\Bigg)}{M} d^{c f e} f^{a b e} \delta_{\rho \bar{\tau}} t^{f}_{k \bar{i}} 
J^{c} J^{(\rho \bar{i})} \partial J^{(\bar{\tau} k)} 
\nonu\\
&&- \frac{i \Bigg(2 a_{5} + M a_{13}\Bigg)}{M}
d^{b f e} f^{a c e} \delta_{\rho \bar{\tau}} t^{f}_{k \bar{i}} 
J^{c} J^{(\rho \bar{i})} \partial J^{(\bar{\tau} k)} 
\nonu \\
&& - 2 \Bigg(a_{11} + M a_{17}\Bigg)
f^{a f e} f^{b c e} \delta_{\rho \bar{\tau}} t^{f}_{k \bar{i}} 
J^{c} J^{(\rho \bar{i})} \partial J^{(\bar{\tau} k)} 
\nonu\\
&&- \frac{2 i (1 + k M + 2 M N) a_{5}}{M} d^{b c e} f^{a f e} \delta_{\rho \bar{\tau}} t^{f}_{k \bar{i}} 
J^{c} J^{(\rho \bar{i})} \partial J^{(\bar{\tau} k)} 
\nonu\\
&&- \frac{i (2 k + M + 2 N) a_{13}}{(2 k + M)} d^{a c e} f^{b f e} \delta_{\rho \bar{\tau}} t^{f}_{k \bar{i}} 
J^{c} J^{(\rho \bar{i})} \partial J^{(\bar{\tau} k)} 
- 4 a_{3} f^{a c e} f^{b f e} \delta_{\rho \bar{\tau}} t^{f}_{k \bar{i}} 
J^{c} J^{(\rho \bar{i})} \partial J^{(\bar{\tau} k)} 
\nonu\\
&&+ \frac{2 \Bigg(M a_{8} - 12 a_{17}\Bigg)}{M}
f^{a b e} f^{c f e} \delta_{\rho \bar{\tau}} t^{f}_{k \bar{i}} 
J^{c} J^{(\rho \bar{i})} \partial J^{(\bar{\tau} k)} 
- 4 a_{17} \delta_{a b} \delta_{c f} \delta_{\rho \bar{\tau}} t^{f}_{k \bar{i}} 
J^{c} J^{(\rho \bar{i})} \partial J^{(\bar{\tau} k)} \nonu\\
&&- \frac{4 (-2 + M) (2 + M) a_{5}}{M^{2}} \delta_{a f} \delta_{b c} \delta_{\rho \bar{\tau}} t^{f}_{k \bar{i}} 
J^{c} J^{(\rho \bar{i})} \partial J^{(\bar{\tau} k)} 
\nonu\\
&&+ \frac{4 (-2 + M) (2 + M) a_{5}}{M^{2}} \delta_{a c} \delta_{b f} \delta_{\rho \bar{\tau}} t^{f}_{k \bar{i}} 
J^{c} J^{(\rho \bar{i})} \partial J^{(\bar{\tau} k)} 
\nonu\\
&&- \frac{2 \Bigg(2 a_{5} - 12 a_{17} + M^{2} a_{17}\Bigg)}{M} d^{a c e} d^{b f e} \delta_{\rho \bar{\tau}} t^{f}_{k \bar{i}} 
J^{c} \partial J^{(\rho \bar{i})} J^{(\bar{\tau} k)} 
\nonu \\
&& + 2 M a_{17} d^{a b e} d^{c f e} \delta_{\rho \bar{\tau}} t^{f}_{k \bar{i}} 
J^{c} \partial J^{(\rho \bar{i})} J^{(\bar{\tau} k)} 
\nonu\\
&&+ \frac{2 i \Bigg(-a_{5} + M a_{8}\Bigg)}{M} d^{c f e} f^{a b e} \delta_{\rho\bar{\tau}} t^{f}_{k \bar{i}} 
J^{c} \partial J^{(\rho \bar{i})} J^{(\bar{\tau} k)} 
\nonu \\
&& - \frac{i \Bigg(-2 a_{5} + M a_{12}\Bigg)}{M} d^{b f e} f^{a c e} \delta_{\rho \bar{\tau}} t^{f}_{k \bar{i}} 
J^{c} \partial J^{(\rho \bar{i})} J^{(\bar{\tau} k)} 
\nonu\\
&&+ \frac{2 i (1 + k M + 2 M N) a_{5}}{M} d^{b c e} f^{a f e} \delta_{\rho \bar{\tau}} t^{f}_{k \bar{i}} 
J^{c} \partial J^{(\rho \bar{i})} J^{(\bar{\tau} k)} 
\nonu\\
&&- 2 \Bigg(a_{11} + M a_{17}\Bigg) f^{a f e} f^{b c e} \delta_{\rho \bar{\tau}} t^{f}_{k \bar{i}} 
J^{c} \partial J^{(\rho \bar{i})} J^{(\bar{\tau} k)} 
\nonu \\
&& + \frac{2 \Bigg(M a_{8} - 12 a_{17}\Bigg)}{M} f^{a b e} f^{c f e} \delta_{\rho \bar{\tau}} t^{f}_{k \bar{i}} 
J^{c} \partial J^{(\rho \bar{i})} J^{(\bar{\tau} k)} 
\nonu\\
&&- \frac{i (2 k + M + 2 N) a_{12}}{(2 k + M)} d^{a c e} f^{b f e} \delta_{\rho \bar{\tau}} t^{f}_{k \bar{i}} 
J^{c} \partial J^{(\rho \bar{i})} J^{(\bar{\tau} k)} 
- 4 a_{3} f^{a c e} f^{b f e} \delta_{\rho \bar{\tau}} t^{f}_{k \bar{i}} 
J^{c} \partial J^{(\rho \bar{i})} J^{(\bar{\tau} k)} 
\nonu\\
&&- \frac{4 (-2 + M) (2 + M) a_{5}}{M^{2}} \delta_{a f} \delta_{b c} \delta_{\rho \bar{\tau}} t^{f}_{k \bar{i}} 
J^{c} \partial J^{(\rho \bar{i})} J^{(\bar{\tau} k)} 
\nonu\\
&&+ \frac{4 (-2 + M) (2 + M) a_{5}}{M^{2}} \delta_{a c} \delta_{b f} \delta_{\rho \bar{\tau}} t^{f}_{k \bar{i}} 
J^{c} \partial J^{(\rho \bar{i})} J^{(\bar{\tau} k)} 
- 4 a_{17} \delta_{a b} \delta_{c f} \delta_{\rho \bar{\tau}} t^{f}_{k \bar{i}} 
J^{c} \partial J^{(\rho \bar{i})} J^{(\bar{\tau} k)} 
\nonu\\
&&+ \frac{2 i}{k} \Bigg( k a_{4} + 2 k \sqrt{\frac{M + N}{M N}} a_{7} + 2 N \sqrt{\frac{M + N}{M N}} a_{7} \Bigg)
f^{a b c} \delta_{\rho \bar{\tau}} t^{c}_{k \bar{i}} 
J^{u (1)} J^{u (1)} J^{(\rho \bar{i})} J^{(\bar{\tau} k)} 
\nonu\\
&&+ 2 \Bigg( M \sqrt{\frac{M + N}{M N}} a_{5} + k a_{7} + N a_{7} + M a_{9} \Bigg)
d^{a b c} \delta_{\rho \bar{\tau}} t^{c}_{k \bar{i}} 
J^{u (1)} J^{(\rho \bar{i})} \partial J^{(\bar{\tau} k)} 
\nonu\\
&&+ \frac{2 i}{k M} \Bigg( -4 k \sqrt{\frac{M + N}{M N}} a_{5} + k M^{2} \sqrt{\frac{M + N}{M N}} a_{5} + k^{2} M a_{7} 
+ 2 k M N a_{7} \nonu \\
&& + 2 k M \sqrt{\frac{M + N}{M N}} a_{8} 
+ k M \sqrt{\frac{M + N}{M N}} a_{13} 
+ M N \sqrt{\frac{M + N}{M N}} a_{13} \Bigg) 
\nonu \\
&& \times f^{a b c} \delta_{\rho \bar{\tau}} t^{c}_{k \bar{i}} 
J^{u (1)} J^{(\rho \bar{i})} \partial J^{(\bar{\tau} k)} 
\nonu\\
&&+ 2 \Bigg( M \sqrt{\frac{M + N}{M N}} a_{5} + k a_{7} + N a_{7} + M a_{9} \Bigg)
d^{a b c} \delta_{\rho \bar{\tau}} t^{c}_{k \bar{i}} 
J^{u (1)} \partial J^{(\rho \bar{i})} J^{(\bar{\tau} k)} 
\nonu\\
&&- \frac{2 i}{k M} \Bigg( -4 k \sqrt{\frac{M + N}{M N}} a_{5} + k M^{2} \sqrt{\frac{M + N}{M N}} a_{5} + k^{2} M a_{7} 
+ 2 k M N a_{7} \nonu \\
&& + 2 k M \sqrt{\frac{M + N}{M N}} a_{8}
- k M \sqrt{\frac{M + N}{M N}} a_{12} 
- M N \sqrt{\frac{M + N}{M N}} a_{12} \Bigg) 
f^{a b c} \delta_{\rho \bar{\tau}} t^{c}_{k \bar{i}} 
J^{u (1)} \partial J^{(\rho \bar{i})} J^{(\bar{\tau} k)} \nonu\\
&&- 4 i a_{5} d^{b f e} f^{a c e} \delta_{\sigma \bar{\nu}} \delta_{\rho \bar{\tau}} t^{c}_{y \bar{g}} t^{f}_{k \bar{i}} 
J^{(\sigma \bar{g})} J^{(\rho \bar{i})} J^{(\bar{\nu} y)} J^{(\bar{\tau} k)} 
+ 4 i a_{8} f^{a b c} \delta_{\bar{i} k} \delta_{\sigma \bar{\nu}} \delta_{\rho \bar{\tau}} t^{c}_{y \bar{g}} 
J^{(\rho \bar{i})} J^{(\sigma \bar{g})} J^{(\bar{\tau} k)} J^{(\bar{\nu} y)} 
\nonu\\
&&+ \Bigg(k M a_{5} + 2 M N a_{5} + k a_{13} + N a_{13}\Bigg)d^{a b c} \delta_{\rho \bar{\tau}} t^{c}_{k \bar{i}} 
J^{(\rho \bar{i})} \partial^2 J^{(\bar{\tau} k)} 
\nonu\\
&&+ \frac{i}{M} \Bigg(2 M^{2} a_{3} - 4 k a_{5} + k M^{2} a_{5} - 4 N a_{5} + M^{2} N a_{5} + 2 k M a_{8} 
+ 2 M N a_{8} - M^{2} a_{11} \nonu\\
&&+ k M a_{13} + 2 M N a_{13} + 2 M a_{16} + 6 M a_{17} + M^{3} a_{17}\Bigg) 
f^{a b c} \delta_{\rho \bar{\tau}} t^{c}_{k \bar{i}} J^{(\rho \bar{i})}
\pa^2 J^{(\bar{\tau} k)} 
\nonu\\
&&+ \frac{2 (k + M + N) \Bigg(a_{12} - a_{13}\Bigg)}{k M} \delta_{\rho \bar{\tau}} t^{b}_{k \bar{i}} 
\partial J^{a} J^{(\rho \bar{i})} J^{(\bar{\tau} k)} 
- \frac{2 \Bigg(a_{12} - a_{13}\Bigg)}{M} \delta_{a b} \delta_{\bar{i} k} t^{\beta}_{\rho \bar{\tau}} 
\partial J^{\beta} J^{(\rho \bar{i})} J^{(\bar{\tau} k)} 
\nonu\\
&&+ \Bigg(2 M a_{5} - a_{12} + a_{13}\Bigg)d^{a b c} t^{\beta}_{\rho \bar{\tau}} t^{c}_{k \bar{i}} 
\partial J^{\beta} J^{(\rho \bar{i})} J^{(\bar{\tau} k)} 
\nonu\\
&&- \frac{i}{M} \Bigg(-8 a_{5} + 2 M^{2} a_{5} + 4 M a_{8} - M a_{12} - M a_{13}\Bigg)f^{a b c} t^{\beta}_{\rho \bar{\tau}} t^{c}_{k \bar{i}} 
\partial J^{\beta} J^{(\rho \bar{i})} J^{(\bar{\tau} k)} 
\nonu\\
&&+ \frac{2 (4 k + 2 M + M^{2} N) a_{5}}{M (2 k + M)} d^{a f e} d^{b c e} \delta_{\rho \bar{\tau}} t^{f}_{k \bar{i}} 
\partial J^{c} J^{(\rho \bar{i})} J^{(\bar{\tau} k)} 
\nonu\\
&&- \frac{1}{2 M (2 k + M)} \Bigg(16 k a_{5} + 8 M a_{5} + 4 M^{2} N a_{5} - 2 k M a_{12} - M^{2} a_{12} - 2 M N a_{12} 
+ 2 k M a_{13}\nonu\\
&&+ M^{2} a_{13} + 2 M N a_{13}\Bigg) d^{a c e} d^{b f e} \delta_{\rho \bar{\tau}} t^{f}_{k \bar{i}} 
\partial J^{c} J^{(\rho \bar{i})} J^{(\bar{\tau} k)} 
- \frac{2 M N a_{5}}{(2 k + M)} d^{a b e} d^{c f e} \delta_{\rho \bar{\tau}} t^{f}_{k \bar{i}} 
\partial J^{c} J^{(\rho \bar{i})} J^{(\bar{\tau} k)} 
\nonu\\
&&+ \frac{2 i \Bigg(-a_{5} + M a_{8}\Bigg)}{M} d^{c f e} f^{a b e} \delta_{\rho \bar{\tau}} t^{f}_{k \bar{i}} 
\partial J^{c} J^{(\rho \bar{i})} J^{(\bar{\tau} k)} 
\nonu\\
&&- \frac{i \Bigg(-4 a_{5} + 4 M N a_{5} + M a_{12} + M a_{13}
  \Bigg)}{2 M} d^{b f e} f^{a c e} \delta_{\rho \bar{\tau}} t^{f}_{k \bar{i}} 
\partial J^{c} J^{(\rho \bar{i})} J^{(\bar{\tau} k)} 
\nonu\\
&&- \frac{2 i (-1 + M N) a_{5}}{M} d^{b c e} f^{a f e} \delta_{\rho \bar{\tau}} t^{f}_{k \bar{i}} 
\partial J^{c} J^{(\rho \bar{i})} J^{(\bar{\tau} k)} 
+ 2 a_{11} f^{a f e} f^{b c e} \delta_{\rho \bar{\tau}} t^{f}_{k \bar{i}} 
\partial J^{c} J^{(\rho \bar{i})} J^{(\bar{\tau} k)} 
\nonu\\
&&- \frac{i (2 k + M + 2 N) \Bigg(a_{12} + a_{13}\Bigg)}{2 (2 k + M)} d^{a c e} f^{b f e} \delta_{\rho \bar{\tau}} t^{f}_{k \bar{i}} 
\partial J^{c} J^{(\rho \bar{i})} J^{(\bar{\tau} k)} 
\nonu\\
&&- \frac{1}{2} \Bigg(a_{12} - a_{13}\Bigg) f^{a c e} f^{b f e} \delta_{\rho \bar{\tau}} t^{f}_{k \bar{i}} 
\partial J^{c} J^{(\rho \bar{i})} J^{(\bar{\tau} k)} 
+ 2 a_{8} f^{a b e} f^{c f e} \delta_{\rho \bar{\tau}} t^{f}_{k \bar{i}} 
\partial J^{c} J^{(\rho \bar{i})} J^{(\bar{\tau} k)} 
\nonu\\
&&- \frac{4 (-2 + M) (2 + M) a_{5}}{M^{2}} \delta_{a f} \delta_{b c} \delta_{\rho \bar{\tau}} t^{f}_{k \bar{i}} 
\partial J^{c} J^{(\rho \bar{i})} J^{(\bar{\tau} k)} 
\nonu\\
&&+ \frac{4 (-2 + M) (2 + M) a_{5}}{M^{2}} \delta_{a c} \delta_{b f} \delta_{\rho \bar{\tau}} t^{f}_{k \bar{i}} 
\partial J^{c} J^{(\rho \bar{i})} J^{(\bar{\tau} k)} 
\nonu\\
&&+ \frac{1}{k} \sqrt{\frac{M + N}{M N}} \Bigg(2 k M a_{5} + 4 M N a_{5} - ka_{12} - N a_{12} + k a_{13} + N a_{13}\Bigg)
\nonu \\
&& \times
d^{a b c} \delta_{\rho \bar{\tau}} t^{c}_{k \bar{i}} \partial J^{u (1)} J^{(\rho \bar{i})} J^{(\bar{\tau} k)} 
\nonu\\
&&- \frac{i}{k M} \sqrt{\frac{M + N}{M N}} \Bigg(-8 k a_{5} + 2 k M^{2} a_{5} + 4 k M a_{8} + 2 k M^{2} N a_{8} 
- k M a_{12} - M N a_{12} \nonu \\
&& - k M a_{13} 
- M N a_{13}\Bigg) f^{a b c} \delta_{\rho \bar{\tau}} t^{c}_{k \bar{i}} 
\partial J^{u (1)} J^{(\rho \bar{i})} J^{(\bar{\tau} k)} 
\nonu \\
&& +
(k + N)\Bigg(a_{12} + a_{13}\Bigg)d^{a b c} \delta_{\rho \bar{\tau}} t^{c}_{k \bar{i}} 
\partial J^{(\rho \bar{i})} \partial J^{(\bar{\tau} k)}\nonu\\
&&+ \frac{i}{M} \Bigg(4 M^2 a_{3} + 8 N a_{5} - 2 M^2 N a_{5} - 4 M^2 a_{8} 
- 4 M N a_{8} - 2 M^2 a_{11} + k M a_{12} + M N a_{12}\nonu\\ 
&&- k M a_{13} - M N a_{13} + 4 M a_{16} + 12 M a_{17} 
+ 2 M^3 a_{17}\Bigg) f^{a b c} \delta_{\rho \bar{\tau}} t^{c}_{k \bar{i}} 
\partial J^{(\rho \bar{i})} \partial J^{(\bar{\tau} k)}
\nonu\\
&&- \Bigg(k a_{1} + N a_{1} + a_{12} + a_{13}\Bigg) \delta_{a b} 
\partial^{2} J^{\beta} J^{\beta}
\nonu\\
&&+ \frac{i N}{M (2 k + M)} \Bigg(2 M^2 a_{3} - 2 k a_{5} - M a_{5} - 2 N a_{5} 
+ 2 k M a_{8} + M^2 a_{8} + 2 M N a_{8} + 2 M a_{17}\nonu\\
&&+ M^3 a_{17}\Bigg) 
d^{a f e} f^{b c e} \partial^{2} J^{c} J^{f}
- \frac{N}{M} \Bigg(M a_{3} + M a_{8} - 6 a_{17}\Bigg) f^{a f e} f^{b c e} 
\partial^{2} J^{c} J^{f}
\nonu\\
&&- \frac{i N}{4 (2 k + M)} \Bigg(-4 M a_{3} + 4 k M a_{5} + 2 M^2 a_{5} 
+ 4 M N a_{5} - 2 k a_{12} - M a_{12} - 2 N a_{12} 
\nonu \\
&& - 2 k a_{13} 
- M a_{13} - 2 N a_{13} + 24 a_{17} - 2 M^2 a_{17}\Bigg) 
d^{a c e} f^{b f e} \partial^{2} J^{c} J^{f}
\nonu\\
&&- \frac{N}{4} \Bigg(4 a_{11} - a_{12} - a_{13} + 4 M a_{17}\Bigg) 
f^{a c e} f^{b f e} \partial^{2} J^{c} J^{f}
\nonu\\
&&- \frac{i N}{2 M (2 k + M)} \Bigg(4 k a_{5} + 2 M a_{5} + 4 N a_{5} 
+ M^2 a_{11} + 16 M a_{17} - 2 M^3 a_{17}\Bigg) d^{a b e} f^{c f e} 
\partial^{2} J^{c} J^{f}
\nonu\\
&&- \frac{N}{M} \Bigg(M a_{3} + M a_{8} - 6 a_{17}\Bigg) f^{a b e} f^{c f e} 
\partial^{2} J^{c} J^{f}
+ \frac{2 (k + N) a_{12}}{M} \delta_{a b} \delta_{\bar{i} k} \delta_{\rho \bar{\tau}} 
\partial^{2} J^{(\rho \bar{i})} J^{(\bar{\tau} k)}
\nonu\\
&&+ \frac{2 N}{M^2} \Bigg(-4 a_{5} + M^2 a_{5} + 12 a_{17} - 2 M^2 a_{17}\Bigg) 
\delta_{a f} \delta_{b c} \partial^{2} J^{c} J^{f}
\nonu\\
&&- \frac{2 N}{M^2} \Bigg(-4 a_{5} + M^2 a_{5} + 12 a_{17} - 2 M^2 a_{17}\Bigg) 
\delta_{a b} \delta_{c f} \partial^{2} J^{c} J^{f}
\nonu\\
&&- \Bigg(k M a_{5} + 2 M N a_{5} - k a_{12} - N a_{12}\Bigg) d^{a b c} 
\delta_{\rho \bar{\tau}} t^{c}_{k \bar{i}} \partial^{2} J^{(\rho \bar{i})} 
J^{(\bar{\tau} k)}
\nonu\\
&&+ \frac{i}{M} \Bigg(2 M^2 a_{3} - 4 k a_{5} + k M^2 a_{5} - 12 N a_{5} 
+ 3 M^2 N a_{5} + 2 k M a_{8} + 2 M^2 a_{8} + 6 M N a_{8} \nonu\\
&&- M^2 a_{11} - k M a_{12} - 2 M N a_{12} + 2 M a_{16} 
+ 6 M a_{17} + M^3 a_{17}\Bigg) f^{a b c} \delta_{\rho \bar{\tau}} 
t^{c}_{k \bar{i}} \partial^{2} J^{(\rho \bar{i})} J^{(\bar{\tau} k)}
\nonu\\
&&+ \frac{N}{6 (2 k + M)} \Bigg(2 k M N a_{5} + M^2 N a_{5} + 2 k^2 a_{12} 
+ k M a_{12} + 2 k N a_{12} + 4 k^2 a_{13} + 2 k M a_{13} \nonu\\
&&+ 4 k N a_{13}\Bigg) d^{a b c} \partial^{3} J^{c}
- \frac{N}{3} \sqrt{\frac{M + N}{M N}} \Bigg(a_{12} + 2 a_{13}\Bigg) 
(k + N) \delta_{a b} \partial^{3} J^{u (1)}\nonu\\
&&+ \frac{i N}{6} \Bigg(2 M a_{3} - M a_{11} + k a_{12} + 2 k a_{13} + 3 N a_{13} 
+ 6 a_{16}\Bigg) f^{a b c} \partial^{3} J^{c},
\label{KPpole1other}
\eea
which corresponds to Appendix $E$ of \cite{Ahn2011}.

From the following first order pole,
\bea
&& K^a(z) \, P^b(w)\Bigg|_{\frac{1}{(z-w)}} =
k_{1} \delta_{a b} \delta_{\bar{i} k} \delta_{\rho \bar{\tau}} J^{u(1)} \partial J^{(\rho \bar{i})} J^{(\bar{\tau} k)}  
+ k_{2} \delta_{a b} \delta_{\bar{i} k} \delta_{\rho \bar{\tau}} J^{u(1)}
J^{(\bar{\tau} k)} \pa J^{(\rho \bar{i})}  \nonu\\
&&+ k_{3} d^{d c i} d^{b i e} d^{f v h} f^{a e v} J^{d} J^{c} J^{f} J^{h}  
+ k_{4} d^{d b i} d^{c i e} d^{f v h} f^{a e v} J^{d} J^{c} J^{f} J^{h}  \nonu\\
&&+ k_{5} d^{f v h} f^{d i e} f^{a e v} f^{b c i} J^{d} J^{c} J^{f} J^{h}  
+ k_{6} d^{f v h} f^{d c i} f^{a e v} f^{b i e} J^{d} J^{c} J^{f} J^{h}  \nonu\\
&&+ k_{7} d^{f v h} f^{d b i} f^{a e v} f^{c i e} J^{d} J^{c} J^{f} J^{h}  
+ k_{8} d^{f h e} f^{a c e} \delta_{d b} J^{d} J^{c} J^{f} J^{h}  \nonu\\
&&+ k_{9} d^{f h e} f^{a b e} \delta_{d c} J^{d} J^{c} J^{f} J^{h}  
+ k_{10} d^{f h e} f^{a d e} \delta_{b c} J^{d} J^{c} J^{f} J^{h}  
+ k_{11} J^{b} J^{a} \partial J^{u(1)}  \nonu\\
&&+ k_{12} f^{a b c} J^{c} J^{u(1)} J^{u(1)} J^{u(1)}  
+k_{13} f^{a b c} \delta_{\bar{i} k} \delta_{\rho \bar{\tau}} J^{c} J^{u(1)} J^{(\rho\bar{i})} J^{(\bar{\tau} k)}  \nonu\\
&&+ k_{14} d^{c f e} f^{a b e} J^{c} J^{f} J^{u(1)} J^{u(1)}  
+ k_{15} d^{b c e} f^{a f e} J^{c} J^{f} J^{u(1)} J^{u(1)}  \nonu\\
&&+ k_{16} d^{b c e} d^{f v h} f^{a e v} J^{c} J^{f} J^{h} J^{u(1)}  
+ k_{17} d^{b e v} d^{c f e} f^{a v h} J^{c} J^{f} J^{h} J^{u(1)}  \nonu\\
&&+ k_{18} d^{b c e} d^{f e v} f^{a v h} J^{c} J^{f} J^{h} J^{u(1)}  
+ k_{19} f^{a v h} f^{b f e} f^{c e v} J^{c} J^{f} J^{h} J^{u(1)}  \nonu\\
&&+ k_{20} f^{a v h} f^{b e v} f^{c f e} J^{c} J^{f} J^{h} J^{u(1)}  
+ k_{21} f^{a v h} f^{b c e} f^{f e v} J^{c} J^{f} J^{h} J^{u(1)}  \nonu\\
&&+ k_{22} f^{a f h} \delta_{b c} J^{c} J^{f} J^{h} J^{u(1)}  
+ k_{23} f^{a c h} \delta_{b f} J^{c} J^{f} J^{h} J^{u(1)}  
+ k_{24} f^{a b h} \delta_{c f} J^{c} J^{f} J^{h} J^{u(1)}  \nonu\\
&&+ k_{25} d^{c f e} f^{a b e} \delta_{\bar{i} k} \delta_{\rho \bar{\tau}} J^{c} J^{f} J^{(\rho \bar{i})} J^{(\bar{\tau} k)}  
+ k_{26} d^{b c e} f^{a f e} \delta_{\bar{i} k} \delta_{\rho \bar{\tau}} J^{c} J^{f}J^{(\rho \bar{i})} J^{(\bar{\tau} k)}  \nonu\\
&&+ k_{27} d^{a f e} d^{b c e} J^{c} J^{f} \partial J^{u(1)}  
+ k_{28} d^{a c e} d^{b f e} J^{c} J^{f} \partial J^{u(1)}  
+ k_{29} d^{a b e} d^{c f e} J^{c} J^{f} \partial J^{u(1)}  \nonu\\
&&+ k_{30} d^{c f e} f^{a b e} J^{c} J^{f} \partial J^{u(1)}  
+ k_{31} d^{b c e} f^{a f e} J^{c} J^{f} \partial J^{u(1)}  
+ k_{32} f^{a f e} f^{b c e} J^{c} J^{f} \partial J^{u(1)}  \nonu\\
&&+ k_{33} f^{a c e} f^{b f e} J^{c} J^{f} \partial J^{u(1)} 
+k_{34} f^{a b e} f^{c f e} J^{c} J^{f} \partial J^{u(1)}  
+ k_{35} \delta_{a f} \delta_{b c} J^{c} J^{f} \partial J^{u(1)}  \nonu\\
&&+ k_{36} \delta_{a c} \delta_{b f} J^{c} J^{f} \partial J^{u(1)}  
+ k_{37} \delta_{a b} \delta_{c f} J^{c} J^{f} \partial J^{u(1)}  
+ k_{38} \delta_{\bar{i} k} \delta_{\rho \bar{\tau}} d^{a b c} J^{c} J^{(\rho \bar{i})} \partial J^{(\bar{\tau} k)}  \nonu\\
&&+ k_{39} \delta_{\bar{i} k} \delta_{\rho \bar{\tau}} f^{a b c} J^{c} J^{(\rho \bar{i})} \partial J^{(\bar{\tau} k)}  
+ k_{40} d^{a b c} J^{c} \partial J^{u(1)} J^{u(1)}  
+ k_{41} f^{a b c} J^{c} \partial J^{u(1)} J^{u(1)}  \nonu\\
&&+ k_{42} \delta_{\bar{i} k} \delta_{\rho \bar{\tau}} d^{a b c} \partial J^{c} J^{(\rho \bar{i})} J^{(\bar{\tau} k)}  
+ k_{43} \delta_{\bar{i} k} \delta_{\rho \bar{\tau}} f^{a b c} \partial J^{c} J^{(\rho \bar{i})} J^{(\bar{\tau} k)}  
+ k_{44} d^{a b c} J^{c} \partial^{2} J^{u(1)}  \nonu\\
&&+ k_{45} f^{a b c} J^{c} \partial^{2} J^{u(1)} 
+ k_{46} f^{a b c} J^{\beta} J^{\beta} J^{c} J^{u(1)}  
+ k_{47} d^{c f e} f^{a b e} J^{\beta} J^{\beta} J^{c} J^{f}  \nonu\\
&&+ k_{48} \delta_{a b} J^{\beta} J^{\beta} \partial J^{u(1)}  
+ k_{49} d^{a b c} J^{\beta} J^{\beta} \partial J^{c}  
+ k_{50} f^{a b c} J^{\beta} J^{\beta} \partial J^{c}  \nonu\\
&&+ k_{51} \delta_{a b} \delta_{\bar{i} k} \delta_{\rho \bar{\tau}} J^{(\rho \bar{i})} \partial^{2} J^{(\bar{\tau} k)}  
+ k_{52} \delta_{a b} \partial J^{u(1)} J^{u(1)} J^{u(1)}  
+ k_{53} \delta_{a b} \delta_{\bar{i} k} \delta_{\rho \bar{\tau}} \partial J^{u(1)} J^{(\rho \bar{i})} J^{(\bar{\tau} k)}  \nonu\\
&&+ k_{54} \delta_{a b} \partial J^{u(1)} \partial J^{u(1)}  
+ k_{55} \partial J^{a} J^{b} J^{u(1)}  
+ k_{56} \partial J^{a} \partial J^{b} 
+ k_{57} d^{a b c} \partial J^{c} J^{u(1)} J^{u(1)}  \nonu\\
&&+ k_{58} f^{a b c} \partial J^{c} J^{u(1)} J^{u(1)}  
+ k_{59} d^{a f e} d^{b c e} \partial J^{c} J^{f} J^{u(1)}  
+ k_{60} d^{a c e} d^{b f e} \partial J^{c} J^{f} J^{u(1)}  \nonu\\
&&+ k_{61} d^{a b e} d^{c f e} \partial J^{c} J^{f} J^{u(1)}  
+ k_{62} d^{c f e} f^{a b e} \partial J^{c} J^{f} J^{u(1)}  
+ k_{63} d^{b f e} f^{a c e} \partial J^{c} J^{f} J^{u(1)}  \nonu\\
&&+ k_{64} d^{b c e} f^{a f e} \partial J^{c} J^{f} J^{u(1)}  
+ k_{65} f^{a f e} f^{b c e} \partial J^{c} J^{f} J^{u(1)}  
+ k_{66} d^{a c e} f^{b f e} \partial J^{c} J^{f} J^{u(1)}  \nonu\\
&&+ k_{67} f^{a c e} f^{b f e} \partial J^{c} J^{f} J^{u(1)}  
+ k_{68} f^{a b e} f^{c f e} \partial J^{c} J^{f} J^{u(1)}  
+ k_{69} \delta_{a f} \delta_{b c} \partial J^{c} J^{f} J^{u(1)}  \nonu\\
&&+ k_{70} \delta_{a c} \delta_{b f} \partial J^{c} J^{f} J^{u(1)}  
+ k_{71} \delta_{a b} \delta_{c f} \partial J^{c} J^{f} J^{u(1)} 
+k_{72} d^{a f e} d^{b e v} d^{c v h} \partial J^{c} J^{f} J^{h}  \nonu\\
&&+ k_{73} d^{a e v} d^{b f e} d^{c v h} \partial J^{c} J^{f} J^{h}  
+ k_{74} d^{a b e} d^{c v h} d^{f e v} \partial J^{c} J^{f} J^{h}  
+ k_{75} d^{a c e} d^{b v f} d^{h e v} \partial J^{c} J^{f} J^{h}  \nonu\\
&&+ k_{76} d^{b e v} d^{f v h} f^{a c e} \partial J^{c} J^{f} J^{h}  
+ k_{77} d^{b v f} d^{h e v} f^{a c e} \partial J^{c} J^{f} J^{h}  
+ k_{78} d^{b f e} d^{c v h} f^{a e v} \partial J^{c} J^{f} J^{h}  \nonu\\
&&+ k_{79} d^{b c e} d^{f v h} f^{a e v} \partial J^{c} J^{f} J^{h}  
+ k_{80} d^{f v h} f^{a e v} f^{b c e} \partial J^{c} J^{f} J^{h}  
+ k_{81} d^{c v h} f^{a f e} f^{b e v} \partial J^{c} J^{f} J^{h}  \nonu\\
&&+ k_{82} d^{c v h} f^{a e v} f^{b f e} \partial J^{c} J^{f} J^{h}  
+ k_{83} d^{c v h} f^{a b e} f^{f e v} \partial J^{c} J^{f} J^{h}  
+ k_{84} f^{a c e} f^{b v h} f^{f e v} \partial J^{c} J^{f} J^{h}  \nonu\\
&&+ k_{85} f^{a c e} f^{b e v} f^{f v h} \partial J^{c} J^{f} J^{h}  
+ k_{86} d^{a c e} d^{b v f} f^{h e v} \partial J^{c} J^{f} J^{h}  
+ k_{87} d^{b v f} f^{a c e} f^{h e v} \partial J^{c} J^{f} J^{h}  \nonu\\
&&+ k_{88} f^{a c e} f^{b v f} f^{h e v} \partial J^{c} J^{f} J^{h}  
+ k_{89} d^{c f h} \delta_{a b} \partial J^{c} J^{f} J^{h}  
+ k_{90} d^{b f h} \delta_{a c} \partial J^{c} J^{f} J^{h}  \nonu\\
&&+ k_{91} d^{b c h} \delta_{a f} \partial J^{c} J^{f} J^{h}  
+ k_{92} d^{a c h} \delta_{b f} \partial J^{c} J^{f} J^{h}  
+ k_{93} f^{a c h} \delta_{b f} \partial J^{c} J^{f} J^{h}  \nonu\\
&&+ k_{94} f^{a c h} \delta_{b f} \partial J^{c} J^{f} J^{h}  
+ k_{95} f^{a c f} \delta_{b h} \partial J^{c} J^{f} J^{h}  
+ k_{96} f^{a b c} \delta_{f h} \partial J^{c} J^{f} J^{h}  \nonu\\
&&+ k_{97} \delta_{\bar{i} k} \delta_{\rho \bar{\tau}} d^{a b c} \partial J^{c} J^{(\rho \bar{i})} J^{(\bar{\tau} k)}  
+ k_{98} \delta_{\bar{i} k} \delta_{\rho \bar{\tau}} f^{a b c} \partial J^{c} J^{(\rho \bar{i})} J^{(\bar{\tau} k)} 
+k_{99} d^{a b c} \partial J^{c} \partial J^{u(1)}  \nonu\\
&&+ k_{100} f^{a b c} \partial J^{c} \partial J^{u(1)}  
+ k_{101} d^{a c e} d^{b f e} \partial J^{c} \partial J^{f}  
+ k_{102} d^{a b e} d^{c f e} \partial J^{c} \partial J^{f}  \nonu\\
&&+ k_{103} d^{c f e} f^{a b e} \partial J^{c} \partial J^{f}  
+ k_{104} d^{b f e} f^{a c e} \partial J^{c} \partial J^{f}  
+ k_{105} d^{b c e} f^{a f e} \partial J^{c} \partial J^{f}  \nonu\\
&&+ k_{106} f^{a f e} f^{b c e} \partial J^{c} \partial J^{f}  
+ k_{107} d^{a c e} f^{b f e} \partial J^{c} \partial J^{f}  
+ k_{108} f^{a c e} f^{b f e} \partial J^{c} \partial J^{f}  \nonu\\
&&+ k_{109} \delta_{a b} \partial J^{\beta} J^{\beta} J^{u(1)}  
+ k_{110} d^{a b c} \partial J^{\beta} J^{\beta} J^{c}  
+ k_{111} f^{a b c} \partial J^{\beta} J^{\beta} J^{c}  \nonu\\
&&+ k_{112} \delta_{a b} d^{\beta \gamma \delta} \partial J^{\beta} J^{\gamma} J^{\delta}  
+ k_{113} \delta_{a b} f^{\beta \gamma \delta} \partial J^{\beta} J^{\gamma} J^{\delta}  
+ k_{114} \delta_{a b} \partial J^{\beta} \partial J^{\beta}  \nonu\\
&&+ k_{115} \delta_{a b} \partial^{2} J^{u(1)} J^{u(1)}  
+ k_{116} \partial^{2} J^{a} J^{b}  
+ k_{117} d^{a b c} \partial^{2} J^{c} J^{u(1)}  
+ k_{118} f^{a b c} \partial^{2} J^{c} J^{u(1)}  \nonu\\
&&+ k_{119} d^{a f e} d^{b c e} \partial^{2} J^{c} J^{f}  
+ k_{120} d^{a c e} d^{b f e} \partial^{2} J^{c} J^{f}  
+ k_{121} d^{a b e} d^{c f e} \partial^{2} J^{c} J^{f}  \nonu\\
&&+ k_{122} d^{c f e} f^{a b e} \partial^{2} J^{c} J^{f}  
+ k_{123} d^{b f e} f^{a c e} \partial^{2} J^{c} J^{f}  
+ k_{124} d^{b c e} f^{a f e} \partial^{2} J^{c} J^{f}  \nonu\\
&&+ k_{125} d^{a f e} f^{b c e} \partial^{2} J^{c} J^{f}  
+ k_{126} f^{a f e} f^{b c e} \partial^{2} J^{c} J^{f}  
+ k_{127} d^{a c e} f^{b f e} \partial^{2} J^{c} J^{f}  \nonu\\
&&+ k_{128} f^{a c e} f^{b f e} \partial^{2} J^{c} J^{f}  
+ k_{129} d^{a b e} f^{c f e} \partial^{2} J^{c} J^{f}  
+ k_{130} f^{a b e} f^{c f e} \partial^{2} J^{c} J^{f}  \nonu\\
&&+ k_{131} \delta_{a f} \delta_{b c} \partial^{2} J^{c} J^{f} 
+ k_{132} \delta_{a b} \delta_{c f} \partial^{2} J^{c} J^{f}  
+ k_{133} \delta_{a b} \partial^{2} J^{\beta} J^{\beta}  \nonu\\
&&+ k_{134} \delta_{a b} \delta_{\bar{i} k} \delta_{\rho \bar{\tau}} \partial^{2} J^{(\rho \bar{i})} J^{(\bar{\tau} k)} 
+k_{135} \delta_{\rho \bar{\tau}} t^{a}_{k \bar{i}} J^{b} J^{(\rho \bar{i})} \partial J^{(\bar{\tau} k)}  
+ k_{136} \delta_{\rho \bar{\tau}} t^{a}_{k \bar{i}} J^{b} \partial J^{(\rho \bar{i})} J^{(\bar{\tau} k)}  \nonu\\
&&+ k_{137} \delta_{\rho \bar{\tau}} t^{b}_{k \bar{i}} J^{a} J^{(\rho \bar{i})} \partial J^{(\bar{\tau} k)} 
+ k_{138} \delta_{\rho \bar{\tau}} t^{b}_{k \bar{i}} J^{a} \partial J^{(\rho \bar{i})} J^{(\bar{\tau} k)}  
+ k_{139} \delta_{\rho \bar{\tau}} t^{b}_{k \bar{i}} \partial J^{a} J^{(\rho \bar{i})} J^{(\bar{\tau} k)}  \nonu\\
&&+ k_{140} \delta_{\rho \bar{\tau}} t^{c}_{k \bar{i}} f^{a b c} J^{u(1)} J^{u(1)} J^{(\rho \bar{i})} J^{(\bar{\tau} k)}  
+ k_{141} \delta_{\rho \bar{\tau}} t^{c}_{k \bar{i}} d^{a b c} J^{u(1)} J^{(\rho \bar{i})} \partial J^{(\bar{\tau} k)}  \nonu\\
&&+ k_{142} \delta_{\rho \bar{\tau}} t^{c}_{k \bar{i}} f^{a b c} J^{u(1)} J^{(\rho \bar{i})} \partial J^{(\bar{\tau} k)}  
+ k_{143} \delta_{\rho \bar{\tau}} t^{c}_{k \bar{i}} d^{a b c} J^{u(1)} \partial J^{(\rho \bar{i})} J^{(\bar{\tau} k)}  \nonu\\
&&+ k_{144} \delta_{\rho \bar{\tau}} t^{c}_{k \bar{i}} f^{a b c} J^{u(1)} \partial J^{(\rho \bar{i})} J^{(\bar{\tau} k)}  
+ k_{145} \delta_{\rho \bar{\tau}} t^{c}_{k \bar{i}} f^{a b c} J^{\beta} J^{\beta} J^{(\rho \bar{i})} J^{(\bar{\tau} k)}  \nonu\\
&&+ k_{146} \delta_{\rho \bar{\tau}} t^{c}_{k \bar{i}} d^{a b c} J^{(\rho \bar{i})} \partial^{2} J^{(\bar{\tau} k)}  
+ k_{147} \delta_{\rho \bar{\tau}} t^{c}_{k \bar{i}} f^{a b c} J^{(\rho \bar{i})} \partial^{2} J^{(\bar{\tau} k)}  \nonu\\
&&+ k_{148} \delta_{\rho \bar{\tau}} t^{c}_{k \bar{i}} d^{a b c} \partial J^{u(1)} J^{(\rho \bar{i})} J^{(\bar{\tau} k)}  
+ k_{149} \delta_{\rho \bar{\tau}} t^{c}_{k \bar{i}} f^{a b c} \partial J^{u(1)} J^{(\rho \bar{i})} J^{(\bar{\tau} k)}  \nonu\\
&&+ k_{150} \delta_{\rho \bar{\tau}} t^{c}_{k \bar{i}} f^{a b c} \partial J^{(\rho \bar{i})} \partial J^{(\bar{\tau} k)}  
+ k_{151} \delta_{\rho \bar{\tau}} t^{c}_{k \bar{i}} d^{a b c} \partial^{2} J^{(\rho \bar{i})} J^{(\bar{\tau} k)}  \nonu\\
&&+ k_{152} \delta_{\rho \bar{\tau}} t^{c}_{k \bar{i}} f^{a b c} \partial^{2} J^{(\rho \bar{i})} J^{(\bar{\tau} k)} 
+k_{153} \delta_{\bar{i} k} \delta_{\sigma \bar{\nu}} \delta_{\rho \bar{\tau}} t^{c}_{y \bar{g}} f^{a b c} J^{(\rho \bar{i})} J^{(\sigma \bar{g})} J^{(\bar{\tau} k)} J^{(\bar{\nu} y)}  \nonu\\
&&+ k_{154} \delta_{\rho \bar{\tau}} t^{f}_{k \bar{i}} f^{a c f} J^{b} J^{c} J^{(\rho\bar{i})} J^{(\bar{\tau} k)} 
+ k_{155} \delta_{\rho \bar{\tau}} t^{f}_{k \bar{i}} d^{b f e} f^{a c e} J^{c} J^{u(1)} J^{(\rho \bar{i})} J^{(\bar{\tau} k)}  \nonu\\
&&+ k_{156} \delta_{\rho \bar{\tau}} t^{f}_{k \bar{i}} d^{b c e} f^{a f e} J^{c} J^{u(1)} J^{(\rho \bar{i})} J^{(\bar{\tau} k)}  
+ k_{157} \delta_{\rho \bar{\tau}} t^{f}_{k \bar{i}} d^{a c e} f^{b f e} J^{c} J^{u(1)} J^{(\rho \bar{i})} J^{(\bar{\tau} k)}  \nonu\\
&&+ k_{158} \delta_{\rho \bar{\tau}} t^{f}_{k \bar{i}} d^{a f e} d^{b c e} J^{c} J^{(\rho \bar{i})} \partial J^{(\bar{\tau} k)}  
+ k_{159} \delta_{\rho \bar{\tau}} t^{f}_{k \bar{i}} d^{a c e} d^{b f e} J^{c} J^{(\rho \bar{i})} \partial J^{(\bar{\tau} k)}  \nonu\\
&&+ k_{160} \delta_{\rho \bar{\tau}} t^{f}_{k \bar{i}} d^{a b e} d^{c f e} J^{c} J^{(\rho \bar{i})} \partial J^{(\bar{\tau} k)}  
+ k_{161} \delta_{\rho \bar{\tau}} t^{f}_{k \bar{i}} d^{c f e} f^{a b e} J^{c} J^{(\rho \bar{i})} \partial J^{(\bar{\tau} k)}  \nonu\\
&&+ k_{162} \delta_{\rho \bar{\tau}} t^{f}_{k \bar{i}} d^{b f e} f^{a c e} J^{c} J^{(\rho \bar{i})} \partial J^{(\bar{\tau} k)}  
+ k_{163} \delta_{\rho \bar{\tau}} t^{f}_{k \bar{i}} d^{b c e} f^{a f e} J^{c} J^{(\rho \bar{i})} \partial J^{(\bar{\tau} k)}  \nonu\\
&&+ k_{164} \delta_{\rho \bar{\tau}} t^{f}_{k \bar{i}} f^{a f e} f^{b c e} J^{c} J^{(\rho \bar{i})} \partial J^{(\bar{\tau} k)}  
+ k_{165} \delta_{\rho \bar{\tau}} t^{f}_{k \bar{i}} d^{a c e} f^{b f e} J^{c} J^{(\rho \bar{i})} \partial J^{(\bar{\tau} k)}  \nonu\\
&&+ k_{166} \delta_{\rho \bar{\tau}} t^{f}_{k \bar{i}} f^{a c e} f^{b f e} J^{c} J^{(\rho \bar{i})} \partial J^{(\bar{\tau} k)}  
+ k_{167} \delta_{\rho \bar{\tau}} t^{f}_{k \bar{i}} f^{a b e} f^{c f e} J^{c} J^{(\rho \bar{i})} \partial J^{(\bar{\tau} k)}  \nonu\\
&&+ k_{168} \delta_{a f} \delta_{b c} \delta_{\rho \bar{\tau}} t^{f}_{k \bar{i}} J^{c} J^{(\rho \bar{i})} \partial J^{(\bar{\tau} k)} 
+ k_{169} \delta_{a c} \delta_{b f} \delta_{\rho \bar{\tau}} t^{f}_{k \bar{i}} J^{c}J^{(\rho \bar{i})} \partial J^{(\bar{\tau} k)}  \nonu\\
&&+ k_{170} \delta_{a b} \delta_{c f} \delta_{\rho \bar{\tau}} t^{f}_{k \bar{i}} J^{c} J^{(\rho \bar{i})} \partial J^{(\bar{\tau} k)}  
+ k_{171} \delta_{\rho \bar{\tau}} t^{f}_{k \bar{i}} d^{a f e} d^{b c e} J^{c} \partial J^{(\rho \bar{i})} J^{(\bar{\tau} k)}  \nonu\\
&&+ k_{172} \delta_{\rho \bar{\tau}} t^{f}_{k \bar{i}} d^{a c e} d^{b f e} J^{c} \partial J^{(\rho \bar{i})} J^{(\bar{\tau} k)}  
+ k_{173} \delta_{\rho \bar{\tau}} t^{f}_{k \bar{i}} d^{a b e} d^{c f e} J^{c} \partial J^{(\rho \bar{i})} J^{(\bar{\tau} k)}  \nonu\\
&&+k_{174} \delta_{\rho \bar{\tau}} t^{f}_{k \bar{i}} d^{c f e} f^{a b e} J^{c} \partial J^{(\rho \bar{i})} J^{(\bar{\tau} k)}  
+ k_{175} \delta_{\rho \bar{\tau}} t^{f}_{k \bar{i}} d^{b f e} f^{a c e} J^{c} \partial J^{(\rho \bar{i})} J^{(\bar{\tau} k)}  \nonu\\
&&+ k_{176} \delta_{\rho \bar{\tau}} t^{f}_{k \bar{i}} d^{b c e} f^{a f e} J^{c} \partial J^{(\rho \bar{i})} J^{(\bar{\tau} k)}  
+ k_{177} \delta_{\rho \bar{\tau}} t^{f}_{k \bar{i}} f^{a f e} f^{b c e} J^{c} \partial J^{(\rho \bar{i})} J^{(\bar{\tau} k)}  \nonu\\
&&+ k_{178} \delta_{\rho \bar{\tau}} t^{f}_{k \bar{i}} d^{a c e} f^{b f e} J^{c} \partial J^{(\rho \bar{i})} J^{(\bar{\tau} k)}  
+ k_{179} \delta_{\rho \bar{\tau}} t^{f}_{k \bar{i}} f^{a c e} f^{b f e} J^{c} \partial J^{(\rho \bar{i})} J^{(\bar{\tau} k)}  \nonu\\
&&+ k_{180} \delta_{\rho \bar{\tau}} t^{f}_{k \bar{i}} f^{a b e} f^{c f e} J^{c} \partial J^{(\rho \bar{i})} J^{(\bar{\tau} k)}  
+ k_{181} \delta_{a f} \delta_{b c} \delta_{\rho \bar{\tau}} t^{f}_{k \bar{i}} J^{c}
\partial J^{(\rho \bar{i})} J^{(\bar{\tau} k)}  \nonu\\
&&+ k_{182} \delta_{a c} \delta_{b f} \delta_{\rho \bar{\tau}} t^{f}_{k \bar{i}} J^{c} \partial J^{(\rho \bar{i})} J^{(\bar{\tau} k)}  
+ k_{183} \delta_{a b} \delta_{c f} \delta_{\rho \bar{\tau}} t^{f}_{k \bar{i}} J^{c}
\partial J^{(\rho \bar{i})} J^{(\bar{\tau} k)}  \nonu\\
&&+ k_{184} \delta_{\rho \bar{\tau}} t^{f}_{k \bar{i}} d^{a f e} d^{b c e} \partial J^{c} J^{(\rho \bar{i})} J^{(\bar{\tau} k)}  
+ k_{185} \delta_{\rho \bar{\tau}} t^{f}_{k \bar{i}} d^{a c e} d^{b f e} \partial J^{c} J^{(\rho \bar{i})} J^{(\bar{\tau} k)}  \nonu\\
&&+ k_{186} \delta_{\rho \bar{\tau}} t^{f}_{k \bar{i}} d^{a b e} d^{c f e} \partial J^{c} J^{(\rho \bar{i})} J^{(\bar{\tau} k)}  
+ k_{187} \delta_{\rho \bar{\tau}} t^{f}_{k \bar{i}} d^{c f e} f^{a b e} \partial J^{c} J^{(\rho \bar{i})} J^{(\bar{\tau} k)}  \nonu\\
&&+ k_{188} \delta_{\rho \bar{\tau}} t^{f}_{k \bar{i}} d^{b f e} f^{a c e} \partial J^{c} J^{(\rho \bar{i})} J^{(\bar{\tau} k)}  
+ k_{189} \delta_{\rho \bar{\tau}} t^{f}_{k \bar{i}} d^{b c e} f^{a f e} \partial J^{c} J^{(\rho \bar{i})} J^{(\bar{\tau} k)}  \nonu\\
&&+ k_{190} \delta_{\rho \bar{\tau}} t^{f}_{k \bar{i}} f^{a f e} f^{b c e} \partial J^{c} J^{(\rho \bar{i})} J^{(\bar{\tau} k)}  
+ k_{191} \delta_{\rho \bar{\tau}} t^{f}_{k \bar{i}} f^{a c e} f^{b f e} \partial J^{c} J^{(\rho \bar{i})} J^{(\bar{\tau} k)} \nonu\\
&&+ k_{192} \delta_{\rho \bar{\tau}} t^{f}_{k \bar{i}} f^{a b e} f^{c f e} \partial J^{c} J^{(\rho \bar{i})} J^{(\bar{\tau} k)}  
+ k_{193} \delta_{a f} \delta_{b c} \delta_{\rho \bar{\tau}} t^{f}_{k \bar{i}} \partial J^{c} J^{(\rho \bar{i})} J^{(\bar{\tau} k)}  \nonu\\
&&+ k_{194} \delta_{a c} \delta_{b f} \delta_{\rho \bar{\tau}} t^{f}_{k \bar{i}} \partial J^{c} J^{(\rho \bar{i})} J^{(\bar{\tau} k)} 
+ k_{195} t^{c}_{y \bar{g}} t^{f}_{k \bar{i}} \delta_{\sigma \bar{\nu}} \delta_{\rho \bar{\tau}} d^{b f e} f^{a c e} J^{(\sigma \bar{g})} J^{(\rho \bar{i})} J^{(\bar{\nu} y)} J^{(\bar{\tau} k)}  \nonu\\
&&+ k_{196} t^{h}_{k \bar{i}} \delta_{\rho \bar{\tau}} d^{b v h} d^{c f e} f^{a e v} J^{c} J^{f} J^{(\rho \bar{i})} J^{(\bar{\tau} k)}  
+ k_{197} t^{h}_{k \bar{i}} \delta_{\rho \bar{\tau}} d^{b c e} d^{h e v} f^{a v f} J^{c} J^{f} J^{(\rho \bar{i})} J^{(\bar{\tau} k)}  \nonu\\
&&+ k_{198} t^{h}_{k \bar{i}} \delta_{\rho \bar{\tau}} d^{b e v} d^{c f e} f^{a v h} J^{c} J^{f} J^{(\rho \bar{i})} J^{(\bar{\tau} k)}  
+ k_{199} t^{h}_{k \bar{i}} \delta_{\rho \bar{\tau}} d^{b c e} d^{f e v} f^{a v h} J^{c} J^{f} J^{(\rho \bar{i})} J^{(\bar{\tau} k)}  \nonu\\
&&+ k_{200} t^{h}_{k \bar{i}} \delta_{\rho \bar{\tau}} f^{a v h} f^{b f e} f^{c e v} J^{c} J^{f} J^{(\rho \bar{i})} J^{(\bar{\tau} k)}  
+ k_{201} t^{h}_{k \bar{i}} \delta_{\rho \bar{\tau}} f^{a v h} f^{b e v} f^{c f e} J^{c} J^{f} J^{(\rho \bar{i})} J^{(\bar{\tau} k)}  \nonu\\
&&+ k_{202} t^{h}_{k \bar{i}} \delta_{\rho \bar{\tau}} f^{a v h} f^{b c e} f^{f e v} J^{c} J^{f} J^{(\rho \bar{i})} J^{(\bar{\tau} k)}  
+ k_{203} t^{h}_{k \bar{i}} \delta_{\rho \bar{\tau}} d^{a v f} d^{b c e} f^{h e v} J^{c} J^{f} J^{(\rho \bar{i})} J^{(\bar{\tau} k)}  \nonu\\
&&+ k_{204} t^{h}_{k \bar{i}} \delta_{b c} \delta_{\rho \bar{\tau}} f^{a f h} J^{c} J^{f} J^{(\rho \bar{i})} J^{(\bar{\tau} k)}  
+ k_{205} t^{h}_{k \bar{i}} \delta_{b f} \delta_{\rho \bar{\tau}} f^{a c h} J^{c} J^{f} J^{(\rho \bar{i})} J^{(\bar{\tau} k)}  \nonu\\
&&+ k_{206} t^{h}_{k \bar{i}} \delta_{c f} \delta_{\rho \bar{\tau}} f^{a b h} J^{c} J^{f} J^{(\rho \bar{i})} J^{(\bar{\tau} k)}  
+ k_{207} t^{a}_{k \bar{i}} t^{\beta}_{\rho \bar{\nu}} J^{\beta} J^{b} J^{(\rho \bar{i})} J^{(\bar{\nu} k)}  \nonu\\
&&+ k_{208} t^{c}_{k \bar{i}} t^{\beta}_{\rho \bar{\nu}} f^{a b c} J^{\beta} \partial
J^{(\rho \bar{i})} J^{(\bar{\nu} k)}  
+ k_{209} t^{\beta}_{\rho \bar{\tau}} \delta_{\bar{i} k} f^{a b c} J^{c} J^{(\rho \bar{i})} J^{(\bar{\tau} k)} \nonu\\
&&+k_{210} t^{\beta}_{\rho \bar{\tau}} \delta_{a b} \delta_{\bar{i} k} J^{\beta} J^{(\rho \bar{i})} \partial J^{(\bar{\tau} k)}  
+ k_{211} t^{\beta}_{\rho \bar{\tau}} \delta_{a b} \delta_{\bar{i} k} J^{\beta} \partial J^{(\rho \bar{i})} J^{(\bar{\tau} k)}  \nonu\\
&&+ k_{212} t^{\beta}_{\rho \bar{\tau}} \delta_{a b} \delta_{\bar{i} k} \partial J^{\beta} J^{(\rho \bar{i})} J^{(\bar{\tau} k)}  
+ k_{213} t^{a}_{k \bar{i}} t^{\beta}_{\rho \bar{\tau}} J^{\beta} J^{b} J^{(\rho \bar{i})} J^{(\bar{\tau} k)}  \nonu\\
&&+ k_{214} t^{c}_{k \bar{i}} t^{\beta}_{\rho \bar{\tau}} f^{a b c} J^{\beta} J^{u(1)}J^{(\rho \bar{i})} J^{(\bar{\tau} k)}  
+ k_{215} t^{c}_{k \bar{i}} t^{\beta}_{\rho \bar{\tau}} d^{a b c} J^{\beta} J^{(\rho \bar{i})} \partial J^{(\bar{\tau} k)}  \nonu\\
&&+ k_{216} t^{c}_{k \bar{i}} t^{\beta}_{\rho \bar{\tau}} f^{a b c} J^{\beta} J^{(\rho \bar{i})} \partial J^{(\bar{\tau} k)}  
+ k_{217} t^{c}_{k \bar{i}} t^{\beta}_{\rho \bar{\tau}} d^{a b c} J^{\beta} \partial J^{(\rho \bar{i})} J^{(\bar{\tau} k)} 
+ k_{218} t^{c}_{k \bar{i}} t^{\beta}_{\rho \bar{\tau}} f^{a b c} J^{\beta} \partial J^{(\rho \bar{i})} J^{(\bar{\tau} k)}  \nonu\\
&&+ k_{219} t^{c}_{k \bar{i}} t^{\beta}_{\rho \bar{\tau}} d^{a b c} \partial J^{\beta} J^{(\rho \bar{i})} J^{(\bar{\tau} k)}  
+ k_{220} t^{c}_{k \bar{i}} t^{\beta}_{\rho \bar{\tau}} f^{a b c} \partial J^{\beta} J^{(\rho \bar{i})} J^{(\bar{\tau} k)}  \nonu\\
&&+ k_{221} t^{f}_{k \bar{i}} t^{\beta}_{\rho \bar{\tau}} d^{b f e} f^{a c e} J^{\beta}
J^{c} J^{(\rho \bar{i})} J^{(\bar{\tau} k)}  
+ k_{222} t^{f}_{k \bar{i}} t^{\beta}_{\rho \bar{\tau}} d^{b c e} f^{a f e} J^{\beta} J^{c} J^{(\rho \bar{i})} J^{(\bar{\tau} k)}  \nonu\\
&&+ k_{223} t^{f}_{k \bar{i}} t^{\beta}_{\rho \bar{\tau}} d^{a c e} f^{b f e} J^{\beta}
J^{c} J^{(\rho \bar{i})} J^{(\bar{\tau} k)}  
+ k_{224} t^{\delta}_{\rho \bar{\tau}} \delta_{a b} \delta_{\bar{i} k} f^{\beta \gamma
\delta} J^{\beta} J^{\gamma} J^{(\rho \bar{i})} J^{(\bar{\tau} k)}  \nonu\\
&&+ k_{225} t^{c}_{k \bar{i}} t^{\delta}_{\rho \bar{\tau}} d^{\beta \gamma \delta} f^{ab c} J^{\beta} J^{\gamma} J^{(\rho \bar{i})} J^{(\bar{\tau} k)}  
+ k_{226} t^{c}_{k \bar{i}} t^{\delta}_{\rho \bar{\tau}} d^{a b c} f^{\beta \gamma \delta} J^{\beta} J^{\gamma} J^{(\rho \bar{i})} J^{(\bar{\tau} k)}  \nonu\\
&&+ k_{227}\delta_{a b} \partial^{3} J^{u(1)}  
+ k_{228}d^{a b c} \partial^{3} J^{c}  
+ k_{229}f^{a b c} \partial^{3} J^{c},
\label{KPPOLE1}
\eea
after substituting (\ref{avalues}) into
(\ref{KPpole1other}),
the explicit coefficients appearing in (\ref{KPPOLE1})
are given by
\bea
k_{1} & = & \frac{2 (k+N) \sqrt{\frac{M+N}{M N}} (k+2N) a_1}{k M}, \qquad
k_{2}  =  \frac{2 (k+N) \sqrt{\frac{M+N}{M N}} (k+2N) a_1}{k M}, \nonu\\
k_{3} & = & -\frac{i N^2 (k+2N) a_1}{4 (k+M) (2k+M) (3k+2M)}, \qquad
k_{4}  =  -\frac{i N^2 (k+2N) a_1}{2 (k+M) (2k+M) (3k+2M)}, \nonu\\
k_{5} & = & -\frac{i N^2 (k+2N) a_1}{12 (k+M) (2k+M) (3k+2M)}, \qquad
k_{6}  =  -\frac{i N^2 (k+2N) a_1}{6 (k+M) (2k+M) (3k+2M)}, \nonu\\
k_{7} & = & -\frac{i N^2 (k+2N) a_1}{12 (k+M) (2k+M) (3k+2M)}, \qquad
k_{8}  =  -\frac{2 i N^2 (k+2N) a_1}{M (k+M) (2k+M) (3k+2M)}, \nonu\\
k_{9} & = & -\frac{i N^2 (k+2N) a_1}{k M (2k+M) (3k+2M)}, \qquad
k_{10}  =  -\frac{2 i N^2 (k+2N) a_1}{k (k+M) (2k+M) (3k+2M)}, \nonu\\
k_{11} & = & -\frac{2 N (k+N) \sqrt{\frac{M+N}{M N}} (k+2N) a_1}{k^2 M},
\qquad
k_{12} =  \frac{2 i N (M+N) \sqrt{\frac{M+N}{M N}} (k+2N) a_1}{k^3 M}, \nonu\\
k_{13} & = & -\frac{2 i (-k+N) \sqrt{\frac{M+N}{M N}} (k+2N) a_1}{k^2 M},
\qquad
k_{14}  =  -\frac{i N (M+N) (k+2N) a_1}{k^2 M (2k+M)}, \nonu\\
k_{15} & = & \frac{2 i N (M+N) (k+2N) a_1}{k^2 M (k+M)}, \qquad
k_{16}  =  \frac{i N^2 \sqrt{\frac{M+N}{M N}} (k+2N) a_1}{k (k+M) (2k+M)},
\nonu\\
k_{17} & = & \frac{i N^2 \sqrt{\frac{M+N}{M N}} (k+2N) a_1}{2 k (k+M) (3k+2M)}, \qquad
k_{18} = \frac{i N^2 \sqrt{\frac{M+N}{M N}} (k+2N) a_1}{k (k+M) (3k+2M)}, \nonu\\
k_{19} & = & \frac{i N^2 \sqrt{\frac{M+N}{M N}} (k+2N) a_1}{6 k (k+M) (3k+2M)}, \qquad
k_{20}  =  \frac{i N^2 \sqrt{\frac{M+N}{M N}} (k+2N) a_1}{3 k (k+M)
(3k+2M)},
\nonu\\
k_{21} & = & -\frac{i N^2 \sqrt{\frac{M+N}{M N}} (k+2N) a_1}{6 k (k+M) (3k+2M)}, \qquad
k_{22}  =  \frac{4 i N^2 \sqrt{\frac{M+N}{M N}} (k+2N) a_1}{k M (k+M) (3k+2M)}, \nonu\\
k_{23} & = & \frac{4 i N^2 \sqrt{\frac{M+N}{M N}} (k+2N) a_1}{k^2 (k+M) (3k+2M)}, \qquad
k_{24}  =  \frac{2 i N^2 \sqrt{\frac{M+N}{M N}} (k+2N) a_1}{k^2 M (3k+2M)}, \qquad
\nonu \\
k_{25}  & = &  \frac{i N (k+2N) a_1}{k M (2k+M)}, \qquad
k_{26} =  \frac{i (k+2N) a_1}{M (k+M)}, \nonu \\
k_{27} & = &  -\frac{N \sqrt{\frac{M+N}{M N}} (k+2N) (3k^2 M+2kM^2+8N+3kM N) a_1}{2 k M (k+M) (3k+2M)}, \nonu\\
k_{28} & = & -\frac{(-12+M^2) N^2 \sqrt{\frac{M+N}{M N}} (k+2N) a_1}{3 k M (k+M) (3k+2M)}, \qquad
k_{29}  =  \frac{M N^2 \sqrt{\frac{M+N}{M N}} (k+2N) a_1}{3 k (k+M) (3k+2M)}, \nonu\\
k_{30} & = & -\frac{i N^2 \sqrt{\frac{M+N}{M N}} (k+2N) a_1}{2 k (2k+M)}, \qquad
k_{31}  =  -\frac{i N \sqrt{\frac{M+N}{M N}} (k+2N) a_1}{2 (k+M)}, \nonu\\
k_{32} & = & -\frac{(-24+3k^2+2kM) N^2 \sqrt{\frac{M+N}{M N}} (k+2N) a_1}{6 k^2 (k+M) (3k+2M)}, \qquad
k_{33}  =  -\frac{4 N^2 \sqrt{\frac{M+N}{M N}} (k+2N) a_1}{k^2 (k+M) (3k+2M)}, \nonu\\
k_{34} & = & -\frac{4 N^2 \sqrt{\frac{M+N}{M N}} (k+2N) a_1}{k M (k+M) (3k+2M)}, \qquad
k_{35}  =  \frac{4 (-4k+2kM^2+M^3) N^2 \sqrt{\frac{M+N}{M N}} (k+2N) a_1}{k^2 M^2 (k+M) (3k+2M)}, \nonu\\
k_{36} & = & -\frac{2 (-24+5M^2) N^2 \sqrt{\frac{M+N}{M N}} (k+2N) a_1}{3 k M^2 (k+M) (3k+2M)}, \qquad
k_{37}  =  -\frac{2 N^2 \sqrt{\frac{M+N}{M N}} (k+2N) a_1}{3 k (k+M) (3k+2M)}, \nonu\\
k_{38} & = & -\frac{(k+2N) (2k+M+2N) a_1}{2 M (k+M)}, \qquad
k_{39}  =  \frac{i (4+2k^2+3kM) (k+2N) a_1}{2 k M (k+M)}, \nonu\\
k_{40} & = & \frac{(M+N) (k+2N) (2k+M+2N) a_1}{k M (k+M)},\qquad
k_{41}  =  -\frac{i (k-N) (M+N) (k+2N) a_1}{k^2 M}, \nonu\\
k_{42} & = & -\frac{(k+2N) (2k+M+2N) a_1}{2 M (k+M)}, \qquad
k_{43}  =  -\frac{i (4+2k^2+3kM) (k+2N) a_1}{2 k M (k+M)}, \nonu\\
k_{44} & = & \frac{N \sqrt{\frac{M+N}{M N}} (k+2N) (2k+M+2N) a_1}{4 (k+M)}, \nonu\\
k_{45} & = & -\frac{i N \sqrt{\frac{M+N}{M N}} (k+2N) (12k+8M+3k^2 M+2kM^2+8N-2k^2 N) a_1}{4 k (k+M) (3k+2M)}, \nonu\\
k_{46} & = & \frac{2 i N \sqrt{\frac{M+N}{M N}} a_1}{k^2}, \qquad
k_{47}  =  -\frac{i N a_1}{k (2k+M)}, \qquad
k_{48}  =  -\frac{2 (k+N) \sqrt{\frac{M+N}{M N}} a_1}{k}, \nonu\\
k_{49} & = & \frac{(2k+M+2N) a_1}{2k+M}, \qquad
k_{50}  =  \frac{i (k+N) a_1}{k}, \qquad
k_{51}  =  \frac{(k+N) (k+2N) a_1}{M}, \nonu\\
k_{52} & = & -\frac{2 (k+N) (M+N) \sqrt{\frac{M+N}{M N}} (k+2N) a_1}{k^2 M},\qquad
k_{53}  =  \frac{2 (k+N) \sqrt{\frac{M+N}{M N}} (k+2N) a_1}{k M}, \nonu\\
k_{54} & = & -\frac{(k+N) (M+N) (k+2N) a_1}{k M}, \qquad
k_{55}  =  -\frac{4 N \sqrt{\frac{M+N}{M N}} (k+M+N) (k+2N) a_1}{k^2 M}, \nonu\\
k_{56} & = & -\frac{N (k+M+N) (k+2N) a_1}{k M}, \nonu\\
k_{57} & = & \frac{(M+N) (k+2N) (2k^3+3k^2 M+kM^2+2k^2 N-M^2 N) a_1}{k^2 M (k+M) (2k+M)}, \nonu\\
k_{58} & = & \frac{i (M+N) (k+2N)^2 a_1}{k^2 M}, \qquad
k_{59}  =  -\frac{(-12+M^2) N^2 \sqrt{\frac{M+N}{M N}} (k+2N) a_1}{3 k M (k+M) (3k+2M)},\nonu\\
k_{60} & = & -\frac{N \sqrt{\frac{M+N}{M N}} (k+2N) (6k^2 M+7kM^2+2M^3+8N+6kM N+2M^2 N) a_1}{2 k M (k+M) (3k+2M)}, \nonu\\
k_{61} & = & \frac{M (11k+7M) N^2 \sqrt{\frac{M+N}{M N}} (k+2N) a_1}{3 k (k+M) (2k+M) (3k+2M)}, \qquad
k_{62}  =  -\frac{i N^2 \sqrt{\frac{M+N}{M N}} (k+2N) a_1}{k (2k+M)},
\nonu\\
k_{63} & = & \frac{i N \sqrt{\frac{M+N}{M N}} (k+2N) (2k+M+3N) a_1}{2 k (k+M)}, \qquad
k_{64}  =  \frac{i N^2 \sqrt{\frac{M+N}{M N}} (k+2N) a_1}{2 k (k+M)},
\nonu\\
k_{65} & = & \frac{(-4+k) (4+k) N^2 \sqrt{\frac{M+N}{M N}} (k+2N) a_1}{2 k^2 (k+M) (3k+2M)}, \qquad
\nonu \\
k_{66} & = &  \frac{i N \sqrt{\frac{M+N}{M N}} (k+2N) (2k+M+2N) a_1}{2 k (2k+M)}, \nonu\\
k_{67} & = & -\frac{N \sqrt{\frac{M+N}{M N}} (k+2N) (-9k^3-15k^2 M-6kM^2-24N+3k^2 N+2kM N) a_1}{6 k^2 (k+M) (3k+2M)}, \nonu\\
k_{68} & = & \frac{4 N^2 \sqrt{\frac{M+N}{M N}} (k+2N) a_1}{k M (k+M) (3k+2M)}, \qquad
k_{69}  =  -\frac{2 (-24+5M^2) N^2 \sqrt{\frac{M+N}{M N}} (k+2N) a_1}{3 k M^2 (k+M) (3k+2M)}, \nonu\\
k_{70} & = & \frac{4 (-4k+2kM^2+M^3) N^2 \sqrt{\frac{M+N}{M N}} (k+2N) a_1}{k^2 M^2 (k+M) (3k+2M)}, \qquad
k_{71}  =  -\frac{2 N^2 \sqrt{\frac{M+N}{M N}} (k+2N) a_1}{3 k (k+M) (3k+2M)}, \nonu\\
k_{72} & = & \frac{(-12+M^2) N^2 (k+2N) a_1}{3 M (k+M) (2k+M) (3k+2M)},
\qquad
k_{73}  =  -\frac{(-2+M) (2+M) N^2 (k+2N) a_1}{M (k+M) (2k+M) (3k+2M)},
\nonu\\
k_{74} & = & -\frac{M N^2 (k+2N) a_1}{3 (k+M) (2k+M) (3k+2M)}, \qquad
k_{75}  =  \frac{N (k+2N) (2k+M+2N) a_1}{4 (k+M) (2k+M)}, \nonu\\
k_{76} & = & -\frac{i N^2 (k+2N) a_1}{4 (k+M) (3k+2M)}, \qquad
k_{77}  =  -\frac{i N (k+2N) (3k+2M+2N) a_1}{4 (k+M) (3k+2M)}, \nonu\\
k_{78} & = & \frac{i N^2 (k+2N) a_1}{2 (k+M) (2k+M)},\qquad
k_{79}  =  \frac{i N^2 (k+2N) a_1}{4 (k+M) (2k+M)}, \nonu\\
k_{80} & = & \frac{(-8+k^2) N^2 (k+2N) a_1}{4 k (k+M) (2k+M) (3k+2M)},
\qquad
k_{81}  =  -\frac{4 N^2 (k+2N) a_1}{k (k+M) (2k+M) (3k+2M)}, \nonu\\
k_{82} & = & -\frac{(-24+3k^2+2kM) N^2 (k+2N) a_1}{6 k (k+M) (2k+M) (3k+2M)}, \qquad
k_{83}  =  -\frac{4 N^2 (k+2N) a_1}{M (k+M) (2k+M) (3k+2M)}, \nonu\\
k_{84} & = & -\frac{i N^2 (k+2N) a_1}{12 (k+M) (3k+2M)}, \qquad
k_{85}  =  -\frac{i N^2 (k+2N) a_1}{6 (k+M) (3k+2M)}, \nonu\\
k_{86} & = & -\frac{i N (k+2N) (2k+M+2N) a_1}{4 (k+M) (2k+M)}, \qquad
k_{87}  =  -\frac{N (k+2N) a_1}{4 (k+M)}, \nonu\\
k_{88} & = & \frac{i N^2 (k+2N) a_1}{12 (k+M) (3k+2M)}, \qquad
k_{89}  =  \frac{2 N^2 (k+2N) a_1}{3 (k+M) (2k+M) (3k+2M)}, \nonu\\
k_{90} & = & \frac{N (k+M+N) (k+2N) a_1}{k M (k+M)}, \qquad
k_{91}  =  \frac{2 (-24+5M^2) N^2 (k+2N) a_1}{3 M^2 (k+M) (2k+M) (3k+2M)},
\nonu\\
k_{92} & = & \frac{N (k+2N) (6k^3 M+13k^2 M^2+9kM^3+2M^4+16kN+6k^2 M N+2kM^2 N) a_1}{k M^2 (k+M) (2k+M) (3k+2M)}, \nonu\\
k_{93} & = & -\frac{i N (k+2N) a_1}{k M}, \qquad
k_{94}  =  -\frac{2 i N^2 (k+2N) a_1}{M (k+M) (3k+2M)}, \qquad
\nonu \\
k_{95}  & = &  -\frac{2 i N^2 (k+2N) a_1}{k (k+M) (3k+2M)}, \qquad
k_{96}  =  \frac{i N^2 (k+2N) a_1}{k M (3k+2M)}, \nonu \\
k_{97} & = & -\frac{(k+2N) (4k^2+8kM+3M^2+4kN+4M N) a_1}{2 M (k+M) (2k+M)},
\nonu\\
k_{98} & = & -\frac{i (k+2N) (4+kM+2kN+2M N) a_1}{2 k M (k+M)}, \nonu\\
k_{99} & = & \frac{N \sqrt{\frac{M+N}{M N}} (k+2N) (4k^2+6kM+2M^2+4kN+3M N) a_1}{2 (k+M) (2k+M)}, \nonu\\
k_{100} & = & \frac{i (4k+2k^3+8M+3k^2 M+2kM^2) N^2 \sqrt{\frac{M+N}{M N}} (k+2N) a_1}{2 k^2 (k+M) (3k+2M)}, \nonu\\
k_{101} & = & -\frac{N (k+2N) (2k+M+2N) a_1}{4 (2k+M)}, \qquad
k_{102}  =  \frac{M N^2 (k+2N) a_1}{4 (k+M) (2k+M)}, \nonu\\
k_{103} & = & -\frac{i (4k+2k^3+8M+3k^2 M+2kM^2) N^2 (k+2N) a_1}{4 k (k+M) (2k+M) (3k+2M)}, \nonu\\
k_{104} & = & \frac{i N (k+2N) a_1}{4}, \qquad
k_{105}  =  \frac{i N^2 (k+2N) a_1}{4 (k+M)}, \qquad
k_{106}  =  \frac{(-8+k^2) N^2 (k+2N) a_1}{4 k (k+M) (3k+2M)}, \nonu\\
k_{107} & = & \frac{i N (k+2N) (2k+M+2N) a_1}{4 (2k+M)}, \qquad
k_{108}  =  \frac{N (k+2N) a_1}{4}, \nonu\\
k_{109} & = & -\frac{4 (k+N) \sqrt{\frac{M+N}{M N}} a_1}{k}, \qquad
k_{110}  =  \frac{(2k+M+2N) a_1}{k+M}, \qquad
k_{111}  =  -i a_1, \nonu\\
k_{112} & = & -a_1, \qquad
k_{113}  =  -i a_1, \qquad
k_{114}  =  -(k+2N) a_1, \qquad
\nonu \\
k_{115}  & = &  -\frac{(k+N) (M+N) (k+2N) a_1}{k M}, \nonu\\
k_{116} & = & -\frac{N (k+M+N) (k+2N) a_1}{k M}, \qquad
k_{117}  =  \frac{N \sqrt{\frac{M+N}{M N}} (k+2N) (2k+M+2N) a_1}{2 (2k+M)}, \nonu\\
k_{118} & = & \frac{i N \sqrt{\frac{M+N}{M N}} (k+2N) a_1}{2 k^2 (k+M) (3k+2M)} (3k^4 + 5k^3 M + 2k^2 M^2 - 4kN + 4k^3 N - 8M N \nonu\\
&+& 7k^2 M N + 2kM^2 N), \nonu\\
k_{119} & = & -\frac{N (k+2N) a_1}{24 M (k+M) (2k+M) (3k+2M)} (-72k^2 - 84kM - 24M^2 + 18k^2 M^2 + 21kM^3 \nonu\\
&+& 6M^4 - 48kN - 24M N + 17kM^2 N + 10M^3 N), \nonu\\
k_{120} & = & -\frac{N (k+2N) (4k^2 + 4kM + M^2 + 4kN + M N) a_1}{8 (k+M) (2k+M)}, \nonu\\
k_{121} & = & \frac{N (k+2N) a_1}{24 M (k+M) (2k+M) (3k+2M)} (-72k^2 - 84kM - 24M^2 + 18k^2 M^2 + 21kM^3 \nonu\\
&+& 6M^4 - 48kN - 24M N + 17kM^2 N + 10M^3 N), \nonu\\
k_{122} & = & -\frac{i N (k+2N) a_1}{8 k M (k+M) (2k+M) (3k+2M)}
(-12k^3 - 38k^2 M - 32kM^2 - 8M^3 - 12k^2 N \nonu\\
&-& 40kM N + 4k^3 M N - 8M^2 N + 7k^2 M^2 N + 4kM^3 N), \nonu\\
k_{123} & = & \frac{i N (k+2N) a_1}{24 k (k+M) (2k+M) (3k+2M)}
(36k^4 + 60k^3 M + 33k^2 M^2 + 6kM^3 - 16kN \nonu\\
&+& 36k^3 N - 24M N + 42k^2 M N + 10kM^2 N), \nonu\\
k_{124} & = & \frac{i N (k+2N) (18k^2 + 21kM + 6M^2 + 18kN + 24M N - 2M^3 N) a_1}{12 M (k+M) (2k+M) (3k+2M)}, \nonu\\
k_{125} & = & -\frac{i N (k+2N) a_1}{12 k M (k+M) (2k+M) (3k+2M)} (18k^3 + 57k^2 M + 48kM^2 + 12M^3 + 18k^2 N \nonu\\
&+& 44kM N - 2kM^3 N), \nonu\\
k_{126} & = & \frac{N (k+2N) (3k^2 + 5kM + 2M^2 + 2kN - 2M N) a_1}{2 k M (k+M) (3k+2M)}, \nonu\\
k_{127} & = & \frac{i N (k+2N) a_1}{24 k (k+M) (2k+M) (3k+2M)} (18k^3 M + 21k^2 M^2 + 6kM^3 - 24kN + 24M N \nonu\\
&+& 18k^2 M N + 14kM^2 N), \qquad
k_{128}  =  -\frac{(-24 + 3k^2 + 2kM) N^2 (k+2N) a_1}{12 k (k+M) (3k+2M)}, \nonu\\
k_{129} & = & \frac{i N (k+2N) a_1}{24 k M (k+M) (2k+M) (3k+2M)} (36k^3 + 42k^2 M + 12kM^2 + 36k^2 N \nonu\\
&-& 8kM N + 24M^2 N - 3k^2 M^2 N + 4kM^3 N), \nonu\\
k_{130} & = & \frac{N (k+2N) (3k^2 + 5kM + 2M^2 + 2kN - 2M N) a_1}{2 k M (k+M) (3k+2M)}, \nonu\\
k_{131} & = & -\frac{N (k+2N) (-36k - 24M + 9kM^2 + 6M^3 - 24N + 4M^2 N) a_1}{6 M^2 (k+M) (3k+2M)}, \nonu\\
k_{132} & = & \frac{N (k+2N) (-36k - 24M + 9kM^2 + 6M^3 - 24N + 4M^2 N) a_1}{6 M^2 (k+M) (3k+2M)}, \nonu\\
k_{133} & = & -(k+N) a_1, \qquad
k_{134}  =  -\frac{(k+N) (k+2N) a_1}{M}, \nonu\\
k_{135} & = & -\frac{2 (k+2N) (3k^2 M + 5kM^2 + 2M^3 + 8N + 3kM N + M^2 N) a_1}{M^2 (k+M) (3k+2M)}, \nonu\\
k_{136} & = & -\frac{2 (k+2N) (3k^2 M + 5kM^2 + 2M^3 + 8N + 3kM N + M^2 N) a_1}{M^2 (k+M) (3k+2M)}, \nonu\\
k_{137} & = & -\frac{2 (-24 + 5M^2) N (k+2N) a_1}{3 M^2 (k+M) (3k+2M)},
\qquad
k_{138}  =  -\frac{2 (-24+5M^2) N (k+2N) a_1}{3 M^2 (k+M) (3k+2M)}, \nonu\\
k_{139} & = & -\frac{2 (k+M+N) (k+2N) a_1}{k M}, \qquad
k_{140}  =  \frac{2 i (M+N) (k+2N)^2 a_1}{k^2 M N}, \nonu\\
k_{141} & = & \frac{\sqrt{\frac{M+N}{M N}} (k+2N) (2k+M+2N) a_1}{2 (k+M)},
\nonu\\
k_{142} & = & \frac{i \sqrt{\frac{M+N}{M N}} (k+2N) (-4+4k^2+3kM+6kN+6M N) a_1}{2 k (k+M)}, \nonu\\
k_{143} & = & \frac{\sqrt{\frac{M+N}{M N}} (k+2N) (2k+M+2N) a_1}{2 (k+M)},
\nonu\\
k_{144} & = & -\frac{i \sqrt{\frac{M+N}{M N}} (k+2N) (-4+4k^2+3kM+6kN+6M N) a_1}{2 k (k+M)}, \qquad
k_{145}  =  \frac{2 i (k+N) a_1}{k N}, \nonu\\
k_{146} & = & \frac{k (k+2N) (2k+M+2N) a_1}{4 (k+M)}, \nonu\\
k_{147} & = & \frac{i (k+2N) }{4 (k+M) (3k+2M)}\nonu \\
& \times & (-12k+6k^3-8M+7k^2 M+2kM^2-8N+8k^2 N+10kM N+4M^2 N) a_1, \nonu\\
k_{148} & = & \frac{\sqrt{\frac{M+N}{M N}} (k+2N) (2k+M+2N) a_1}{2 (k+M)},
\nonu\\
k_{149} & = & \frac{i \sqrt{\frac{M+N}{M N}} (k+2N) (4+kM+2kN+2M N) a_1}{2 k (k+M)}, \nonu\\
k_{150} & = & -\frac{i (k+2N) (-6k+3k^3-4M+2k^2 M-8N+5k^2 N+2kM N) a_1}{k (3k+2M)}, \nonu\\
k_{151} & = & -\frac{k (k+2N) (2k+M+2N) a_1}{4 (k+M)}, \nonu\\
k_{152} & = & \frac{i (-2+k) (2+k) (2k+M) (k+2N) (3k+2M+4N) a_1}{4 k (k+M) (3k+2M)}, \qquad
\nonu \\
k_{153} & = &  -\frac{2 i (k+2N) a_1}{k M}, \qquad
k_{154}  =  \frac{2 i (k+2N) a_1}{k M}, \nonu \\
k_{155} & = & -\frac{i (k+M-N) \sqrt{\frac{M+N}{M N}} (k+2N) a_1}{k (k+M)}, \nonu\\
k_{156} & = & \frac{i \sqrt{\frac{M+N}{M N}} (k+2N) (k+3N) a_1}{k (k+M)},
\qquad
k_{157}  =  -\frac{i \sqrt{\frac{M+N}{M N}} (k+2N) (2k+M+2N) a_1}{k (2k+M)},\nonu\\
k_{158} & = & -\frac{(k+2N) (6k+4M+3k^2 M+2kM^2+8N+3kM N) a_1}{2 M (k+M) (3k+2M)}, \nonu\\
k_{159} & = & -\frac{(k+2N) (-9k-6M-12N+M^2 N) a_1}{3 M (k+M) (3k+2M)},
\qquad
k_{160}  =  \frac{M N (k+2N) a_1}{3 (k+M) (3k+2M)}, \nonu\\
k_{161} & = & \frac{i (k+2M) (k+2N) a_1}{2 k M (k+M)}, \qquad
k_{162}  =  -\frac{i (-1+kM+M^2) (k+2N) a_1}{2 M (k+M)}, \nonu\\
k_{163} & = & \frac{i (k+2N) (1+kM+2M N) a_1}{2 M (k+M)}, \qquad
k_{164}  =  -\frac{(-24+3k^2+2kM) N (k+2N) a_1}{6 k (k+M) (3k+2M)}, \nonu\\
k_{165} & = & -\frac{i (k+2N) (2k+M+2N) a_1}{2 (2k+M)}, \qquad
k_{166}  =  -\frac{4 N (k+2N) a_1}{k (k+M) (3k+2M)}, \nonu\\
k_{167} & = & -\frac{(k+2N) (3k^2+5kM+2M^2+4kN) a_1}{k M (k+M) (3k+2M)},
\qquad
\nonu \\
k_{168}  & = &  \frac{(-2+M) (2+M) (k+2N) a_1}{M^2 (k+M)}, \nonu\\
k_{169} & = & -\frac{(-2+M) (2+M) (k+2N) a_1}{M^2 (k+M)}, \qquad
k_{170}  =  -\frac{2 N (k+2N) a_1}{3 (k+M) (3k+2M)}, \nonu\\
k_{171} & = & -\frac{(k+2N) (6k+4M+3k^2 M+2kM^2+8N+3kM N) a_1}{2 M (k+M) (3k+2M)}, \nonu\\
k_{172} & = & -\frac{(k+2N) (-9k-6M-12N+M^2 N) a_1}{3 M (k+M) (3k+2M)},
\nonu\\
k_{173} & = & \frac{M N (k+2N) a_1}{3 (k+M) (3k+2M)},\qquad
k_{174}  =  -\frac{i (k+2M) (k+2N) a_1}{2 k M (k+M)}, \nonu\\
k_{175} & = & \frac{i (-1+kM+M^2) (k+2N) a_1}{2 M (k+M)}, \qquad
k_{176}  =  -\frac{i (k+2N) (1+kM+2M N) a_1}{2 M (k+M)}, \nonu\\
k_{177} & = & -\frac{(-24+3k^2+2kM) N (k+2N) a_1}{6 k (k+M) (3k+2M)}, \qquad
k_{178}  =  \frac{i (k+2N) (2k+M+2N) a_1}{2 (2k+M)}, \nonu\\
k_{179} & = & -\frac{4 N (k+2N) a_1}{k (k+M) (3k+2M)}, \qquad
k_{180}  =  -\frac{(k+2N) (3k^2+5kM+2M^2+4kN) a_1}{k M (k+M) (3k+2M)},
\nonu\\
k_{181} & = & \frac{(-2+M) (2+M) (k+2N) a_1}{M^2 (k+M)}, \qquad
k_{182}  =  -\frac{(-2+M) (2+M) (k+2N) a_1}{M^2 (k+M)}, \nonu\\
k_{183} & = & -\frac{2 N (k+2N) a_1}{3 (k+M) (3k+2M)}, \qquad
k_{184}  =  -\frac{(k+2N) (4k+2M+M^2 N) a_1}{2 M (k+M) (2k+M)}, \nonu\\
k_{185} & = & -\frac{(k+2N) (-2+kM+M^2+M N) a_1}{2 M (k+M)}, \qquad
k_{186}  =  \frac{M N (k+2N) a_1}{2 (k+M) (2k+M)}, \nonu\\
k_{187} & = & -\frac{i (k+2M) (k+2N) a_1}{2 k M (k+M)}, \qquad
k_{188}  =  \frac{i (k+2N) (-1+M N) a_1}{2 M (k+M)}, \nonu\\
k_{189} & = & \frac{i (k+2N) (-1+M N) a_1}{2 M (k+M)}, \qquad
k_{190}  =  \frac{(-8+k^2) N (k+2N) a_1}{2 k (k+M) (3k+2M)}, \nonu\\
k_{191} & = & \frac{1}{2} (k+2N) a_1, \qquad
k_{192}  =  -\frac{(k+2N) a_1}{k M}, \qquad
k_{193}  =  \frac{(-2+M) (2+M) (k+2N) a_1}{M^2 (k+M)}, \nonu\\
k_{194} & = & -\frac{(-2+M) (2+M) (k+2N) a_1}{M^2 (k+M)}, \qquad
k_{195}  =  \frac{i (k+2N) a_1}{k+M}, \nonu\\
k_{196} & = & -\frac{i N (k+2N) a_1}{2 (k+M) (2k+M)}, \qquad
k_{197}  =  -\frac{i (k+2N) a_1}{2 (k+M)}, \qquad
k_{198}  =  \frac{i N (k+2N) a_1}{2 (k+M) (3k+2M)}, \nonu\\
k_{199} & = & \frac{i N (k+2N) a_1}{(k+M) (3k+2M)}, \qquad
k_{200}  =  \frac{i N (k+2N) a_1}{6 (k+M) (3k+2M)}, \qquad
\nonu \\
k_{201}  & = &  \frac{i N (k+2N) a_1}{3 (k+M) (3k+2M)}, \nonu\\
k_{202} & = & -\frac{i N (k+2N) a_1}{6 (k+M) (3k+2M)}, \qquad
k_{203}  =  -\frac{i (k+2N) (2k+M+2N) a_1}{2 (k+M) (2k+M)}, \nonu\\
k_{204}  & = & \frac{4 i N (k+2N) a_1}{M (k+M) (3k+2M)}, \qquad
k_{205}  =  \frac{4 i N (k+2N) a_1}{k (k+M) (3k+2M)}, \nonu\\
k_{206} & = & \frac{2 i N (k+2N) a_1}{k M (3k+2M)}, \qquad
k_{207}  =  -\frac{2 (k+2N) a_1}{k M}, \qquad
k_{208}  =  -\frac{2 i (k+2N) a_1}{k M}, \nonu\\
k_{209} & = & \frac{2 i a_1}{M}, \qquad
k_{210}  =  \frac{2 (k+N) a_1}{M}, \qquad
k_{211}  =  \frac{2 (k+N) a_1}{M}, \qquad
k_{212}  =  \frac{2 (k+2N) a_1}{M}, \nonu\\
k_{213} & = & \frac{2 (k + 2N) a_{1}}{k M}, \qquad
k_{214}  =  \frac{2 i \sqrt{\frac{M+N}{M N}} (2k + 3N) a_{1}}{k}, \qquad
k_{215}  =  \frac{k (2k + M + 2N) a_{1}}{2 (k + M)}, \nonu\\
k_{216} & = & \frac{i (4k + 4k^{2} M + 3k M^{2} + 8N + 6k M N + 4M^{2} N) a_{1}}{2M (k + M)}, \nonu\\
k_{217} & = & \frac{k (2k + M + 2N) a_{1}}{2 (k + M)}, \qquad
k_{218}  =  -\frac{i (-4k + 4k^{3} + 3k^{2} M - 8N + 6k^{2} N + 4k M N) a_{1}}{2k (k + M)}, \nonu\\
k_{219} & = & \frac{(2k + M) (k + 2N) a_{1}}{2 (k + M)}, \qquad
k_{220}  =  \frac{i (4 + kM) (k + 2N) a_{1}}{2k (k + M)}, \nonu\\
k_{221} & = & -i a_{1}, \qquad
k_{222}  =  \frac{i (k + 2N) a_{1}}{k + M}, \qquad
k_{223}  =  -\frac{i (2k + M + 2N) a_{1}}{2k + M}, \nonu\\
k_{224} & = & \frac{2 i a_{1}}{M}, \qquad
k_{225}  =  i a_{1}, \qquad
k_{226}  =  i a_{1}, \nonu\\
k_{227} & = & -\frac{1}{6} N (k + N) \sqrt{\frac{M+N}{M N}} (k + 2N) a_{1}, \nonu\\
k_{228} & = & \frac{N (k + 2N) (4k^{3} + 6k^{2} M + 2k M^{2} + 4k^{2} N + 2k M N - M^{2} N) a_{1}}{24 (k + M) (2k + M)},
\label{kvalue}
%\nonu
\\
k_{229} & = & \frac{i N (k + 2N) (6k^{4} + 10k^{3} M + 4k^{2} M^{2} + 6k^{3} N- 8M N + 11k^{2} M N + 4k M^{2} N) a_{1}}{24k (k + M) (3k + 2M)}.
\nonu
\eea

The OPE in (\ref{KPfinal1}) of the charged spin-$2$ current
with the charged spin-$3$ current
can be written as
\bea
K^a(z) \, P^b(w) & = &
\frac{1}{(z-w)^3} \,
q_{1}\, f^{a b c} \, K^c(w)
\nonu \\
&+& \frac{1}{(z-w)^2} \, \Bigg[
q_{2} \  f^{a b c} \, \pa \, K^c
+q_3 \, J^{a} K^{b} 
+q_4 \, J^{b} K^{a}  
+q_5 \, f^{a c e} f^{b f e} \, J^{c} K^{f}
\nonu\\  
&+&q_6 \, d^{a f e} d^{b c e} \, J^{c} K^{f} 
+q_7 \, d^{a c e} d^{b f e} \, J^{c} K^{f} 
+q_8 \, \delta_{ab} \, J^{c} K^{c}  
+q_9 \, f^{a b c} \, \partial K^{c}
\nonu\\
&+&q_{10} \ \de^{a b}  \, W^{(3)}
+q_{11} \, d^{a b c} \, P^c \Bigg](w) 
\nonu \\
&+& \frac{1}{(z-w)} \, \Bigg[  q_{12} J^{a} \partial K^{b}
+ q_{13} J^{b} \partial K^{a}
+ q_{14} d_{51}^{f b h c a} J^{c} J^{f} K^{h}
+ q_{15} d_{51}^{h a f c b} J^{c} J^{f} K^{h}
\nonu\\
&+& q_{16} d_{51}^{h f c b a} J^{c} J^{f} K^{h}
+ q_{17} d_{52}^{f b h c a} J^{c} J^{f} K^{h}
+ q_{18} d_{52}^{h b f c a} J^{c} J^{f} K^{h}
+ q_{19} d_{52}^{h f c b a} J^{c} J^{f} K^{h}
\nonu\\
&+& q_{20} d_{4AA1}^{c e f h} f^{a b e} J^{c} J^{f} K^{h}
+ q_{21} d_{4SS2}^{c f h e} f^{a b e} J^{c} J^{f} K^{h}
+ q_{22} d^{c f v} d^{h e v} f^{a b e} J^{c} J^{f} K^{h}
\nonu\\
&+& q_{23} f^{a b e} f^{c v e} f^{f h v} J^{c} J^{f} K^{h}
+ q_{24} f^{a b h} \delta_{c f} J^{c} J^{f} K^{h}
+ q_{25} f^{a b f} \delta_{c h} J^{c} J^{f} K^{h}
\nonu\\
&+& q_{26} f^{a b c} \delta_{f h} J^{c} J^{f} K^{h}
+ q_{27} d_{4AA2}^{f c b a} J^{c} P^{f}
+ q_{28} d_{4SA}^{c a f b} J^{c} P^{f}
+ q_{29} d_{4SA}^{c b f a} J^{c} P^{f}
\nonu\\
&+& q_{30} \delta_{a b} J^{c} \partial K^{c}
+ q_{31} d^{a c e} d^{b f e} J^{c} \partial K^{f}
+ q_{32} d^{a f e} d^{c f e} J^{c} \partial K^{f}
+ q_{33} f^{a c e} f^{b f e} J^{c} \partial K^{f}
\nonu\\
&+& q_{34} f^{a b e} f^{c f e} J^{c} \partial K^{f}
+ q_{35} \partial J^{a} K^{b}
+ q_{36} \partial J^{b} K^{a}
+ q_{37} \delta_{a b} \partial J^{c} K^{c}
\nonu \\
&+& q_{38} d^{a c e} d^{b f e} \partial J^{c} K^{f}
+ q_{39} d^{a f e} d^{c f e} \partial J^{c} K^{f} \nonu \\
& + & q_{40} f^{a c e} f^{b f e} \partial J^{c} K^{f}
+ q_{41} f^{a b e} f^{c f e} \partial J^{c} K^{f}
+ q_{42} \delta_{a b} \partial W^{(3)}
\nonu\\
&+& q_{43} d^{a b c} \partial P^{c}
+ q_{44} f^{a b c}\partial^2 K^{c}
+ q_{45} f^{a b c} \widetilde{R}^{c}
\Bigg](w) + \cdots.
\label{KPfinal}
\eea
where
the coefficients are
given by
\bea
q_{1} & = & \frac{i(k^{2} - 4)(2k + M)(k + 2N)(3k + 2M + 2N)}{2k (k + M)(3k + 2M)}a_{1},
\nonu\\
q_{2} & = & \frac{i(k^{2} - 4)(2k + M)(k + 2N)(3k + 2M + 2N)}{8k (k + M)(3k+ 2M)}a_{1},
\nonu\\
q_{3} & = & -\frac{(4 + k^{2} + 3 k M + M^{2})(k + 2 N)(3 k + 2 M + 2 N) a_{1}}{k M (k + M)(3 k + 2 M)},
\nonu\\
q_{4} & = & +\frac{(4 k + k^{3} + 4 M - 2 k^{2} M - 3 k M^{2} - M^{3})(k + 2 N)(3 k + 2 M + 2 N) a_{1}}{k M (k + M)^{2}(3 k + 2 M)},
\nonu\\
q_{5} & = & +\frac{(2 k + M)^{2}(k + 2 N)(3 k + 2 M + 2 N) a_{1}}{4 (k + M)^{2}(3 k + 2 M)},
\nonu\\
q_{6} & = & +\frac{(4 k + k^{3} + 4 M)(k + 2 N)(3 k + 2 M + 2 N) a_{1}}{4 k (k + M)^{2}(3 k + 2 M)},
\nonu\\
q_{7} & = & -\frac{(4 + k^{2} + k M)(k + 2 N)(3 k + 2 M + 2 N) a_{1}}{4 k (k + M)(3 k + 2 M)},
\nonu\\
q_{8} & = & -\frac{k (2 k + M)(k + 2 N)(3 k + 2 M + 2 N) a_{1}}{(k + M)^{2}(k + 2 M)(3 k + 2 M)},
\nonu\\
q_{9} & = & -\frac{i (-2 + k)(2 + k) M (k + 2 N)(3 k + 2 M + 2 N) a_{1}}{8 k (k + M)(3 k + 2 M)},
\nonu\\
q_{10} & =&  -\frac{a_{1}}{b_{1}}, \qquad
q_{11}  =  \frac{k (3k + 2M)(2k + M + 2N)}{(k + M)(2k + M)},
\nonu \\
q_{12} & = & -\frac{(k + 2N)(3k + 2M + 2N) a_{1}}{2M(3k + 2M)},
\nonu\\
q_{13} &=& -\frac{(-k^2 + kM + M^2)(k + 2N)(3k + 2M + 2N) a_{1}}{2M(k + M)^2(3k + 2M)},
\nonu\\
q_{14} &=& \frac{(k + 2N)(3k + 2M + 2N)(-6k - 2M + kMN) a_{1}}{16k(k + M)^2(3k + 2M)N},
\nonu\\
q_{15} &=& \frac{(k + 2N)(3k + 2M + 2N)(-6k - 2M + kMN) a_{1}}{16k(k + M)^2(3k + 2M)N},
\nonu\\
q_{16} &=& -\frac{(k + 2N)(3k + 2M + 2N)(6k + 2M + 2k^2N + 3kMN) a_{1}}{32k(k + M)^2(3k + 2M)N},
\nonu\\
q_{17} &=& \frac{(k + 2N)(3k + 2M + 2N)(18k + 6M + 8k^2N + 5kMN) a_{1}}{16k(k + M)^2(3k + 2M)N},
\nonu\\
q_{18} &=& \frac{(k + 2N)(3k + 2M + 2N)(-18k - 6M + 4k^2N + 3kMN) a_{1}}{16k(k + M)^2(3k + 2M)N},
\nonu\\
q_{19} &=& \frac{(k + 2N)(3k + 2M + 2N)(-54k - 18M + 6k^2N + 5kMN) a_{1}}{32k(k + M)^2(3k + 2M)N},
\nonu\\
q_{20} &=& -\frac{i(8k^2 + 16kM + k^3M + 8M^2)(k + 2N)(3k + 2M + 2N) a_{1}}{16kM(k + M)^3(3k + 2M)},
\nonu\\
q_{21} &=& -\frac{i k(2k + M)(k + 2N)(3k + 2M + 2N) a_{1}}{16(k + M)^3(3k + 2M)},
\nonu\\
q_{22} &=& \frac{i k(2k + M)(k + 2N)(3k + 2M + 2N) a_{1}}{16(k + M)^3(3k + 2M)},
\nonu\\
q_{23} &=& \frac{i(k + 2N)(3k + 2M + 2N) a_{1}}{16kM(k + M)^3(3k + 2M)N}(-30k^2M - 40kM^2 - 10M^3 + 16k^2N + 32kMN
\nonu\\
&+& k^3MN + 16M^2N + k^2M^2N + kM^3N), \qquad
q_{24} = -\frac{i(k + 2N)(3k + 2M + 2N) a_{1}}{2M(k + M)(3k + 2M)},
\nonu\\
q_{25} &=& -\frac{i(k + 2N)(3k + 2M + 2N) a_{1}}{k(k + M)(3k + 2M)},
\nonu \\
q_{26} & = & -\frac{i(2k + M)(k + 2N)(3k + 2M + 2N) a_{1}}{2M(k + M)^2(3k + 2M)},
\nonu\\
q_{27} &=& \frac{(3k + 2M)(2k + M + 2N)}{2(k + M)(2k + M)}, \qquad
q_{28} = -\frac{i k(2k + M + 2N)}{2(k + M)(2k + M)},
\nonu\\
q_{29} &=& -\frac{i k(2k + M + 2N)}{2(k + M)(2k + M)}, \qquad
q_{30} = -\frac{k^2(k + 2N)(3k + 2M + 2N) a_{1}}{2(k + M)^2(k + 2M)(3k + 2M)},
\nonu\\
q_{31} &=& -\frac{k(k + 2N)(3k + 2M + 2N) a_{1}}{8(k + M)(3k + 2M)}, \qquad
q_{32} = \frac{k^2(k + 2N)(3k + 2M + 2N) a_{1}}{8(k + M)^2(3k + 2M)},
\nonu\\
q_{33} &=& \frac{k(2k + M)(k + 2N)(3k + 2M + 2N) a_{1}}{8(k + M)^2(3k + 2M)},
\nonu\\
q_{34} &=& -\frac{(-2 + k)(2 + k)(24k + 5M)(k + 2N)(3k + 2M + 2N) a_{1}}{84kM(k + M)(3k + 2M)},
\nonu\\
q_{35} &=& \frac{(k + 2N)(3k + 2M + 2N) a_{1}}{8kM^2(k + M)^2(3k + 2M)N}(6kM^2 + 2M^3 - 16k^2N - 32kMN + 8k^3MN
\nonu\\
&-& 16M^2N + 16k^2M^2N + 19kM^3N + 8M^4N),
\nonu\\
q_{36} &=& \frac{(k + 2N)(3k + 2M + 2N) a_{1}}{16kM^2(k + M)^2(3k + 2M)N}(-78kM^2 - 26M^3 - 32k^2N - 64kMN
\nonu\\
&+& 16k^3MN - 32M^2N - 6k^2M^2N - 7kM^3N),
\nonu\\
q_{37} &=& -\frac{(k + 2N)(3k + 2M + 2N) a_{1}}{16k(k + M)^2(k + 2M)(3k + 2M)N}
\nonu \\
& \times & (6k^2 + 14kM + 4M^2 + 18k^3N + 23k^2MN + 6kM^2N),
\nonu\\
q_{38} &=& -\frac{(k + 2N)(3k + 2M + 2N) a_{1}}{64kM(k + M)^2(3k + 2M)N}(-30kM^2 - 10M^3 - 32k^2N - 64kMN + 16k^3MN
\nonu\\
&-& 32M^2N + 14k^2M^2N + kM^3N),
\nonu\\
q_{39} &=& \frac{(k + 2N)(3k + 2M + 2N) a_{1}}{32kM(k + M)^2(3k + 2M)N}(-18kM^2 - 6M^3 - 16k^2N - 32kMN + 8k^3MN
\nonu\\
&-& 16M^2N + 2k^2M^2N - kM^3N),
\nonu\\
q_{40} &=& \frac{(k + 2N)(3k + 2M + 2N) a_{1}}{64k(k + M)^2(3k + 2M)N}(-66kM- 22M^2 + 32k^3N + 66k^2MN + 31kM^2N),
\nonu\\
q_{41} &=& \frac{(-2 + k)(2 + k)(48k + 31M)(k + 2N)(3k + 2M + 2N) a_{1}}{84kM(k + M)(3k + 2M)}, \qquad
q_{42} = -\frac{a_{1}}{3b_{1}},
\nonu\\
q_{43} &=& \frac{k(2k + M + 2N)}{2(k + M)},
\nonu \\
q_{44} & = &
\frac{i(-2 + k)(2 + k)(9k + M)(k + 2N)(3k + 2M + 2N) a_{1}}{84k(k + M)(3k +2M)},
\nonu\\
q_{45} &=& -\frac{i}{2M}.
\label{COEFF}
\eea

By substituting the $N$ in terms of $\la, k$
with
\bea
\la \equiv \frac{k}{(k+N)},
\label{la}
\eea
and taking the infinity limit of $k$,
the above coefficients (\ref{COEFF})
in terms of $k, \la$ and $M$ with (\ref{la}) become 
\bea
q_{1} & \rightarrow & -\frac{i k^{2}(\lambda^{2} - 4)}{3\lambda^{2}}a_{1},
\qquad
q_{2} \ \rightarrow  -\frac{i k^{2}(\lambda^{2} - 4)}{12\lambda^{2}}a_{1},
\qquad
q_{3}  \ \rightarrow \ \frac{k (\lambda^2 - 4) a_{1}}{3 M \lambda^2},
\nonu \\
q_{4} &\rightarrow& -\frac{k (\lambda^2 - 4) a_{1}}{3 M \lambda^2},
\qquad
q_{5} \ \rightarrow\ -\frac{k (\lambda^2 - 4) a_{1}}{3 \lambda^2},
\qquad
q_{6} \ \rightarrow \ -\frac{k (\lambda^2 - 4) a_{1}}{12 \lambda^2},
\nonu \\
q_{7} & \rightarrow & \frac{k (\lambda^2 - 4) a_{1}}{12 \lambda^2},
\qquad
q_{8}  \rightarrow \frac{2 (\lambda^2 - 4) a_{1}}{3 \lambda^2},
\nonu \\
q_{9} & \rightarrow & \frac{i k M (\lambda^2 - 4) a_{1}}{24 \lambda^2},
\qquad
q_{10} \ \rightarrow \ \frac{3 i k^3 \sqrt{2\lambda^2 -8 } \ a_{1}}{M \lambda^3},
\nonu\\
q_{11} & \rightarrow & \frac{3k}{\lambda},
\qquad
q_{12} \rightarrow \frac{k(\lambda^2 - 4)a_{1}}{6M\lambda^2},
\qquad
q_{13} \rightarrow -\frac{k(\lambda^2 - 4)a_{1}}{6M\lambda^2},
\qquad
q_{14} \rightarrow 0,
\qquad
q_{15} \rightarrow 0,
\nonu\\
q_{16} &\rightarrow& \frac{(\lambda^2 - 4)a_{1}}{48\lambda^2},
\qquad
q_{17} \rightarrow -\frac{(\lambda^2 - 4)a_{1}}{6\lambda^2},
\qquad
q_{18} \rightarrow -\frac{(\lambda^2 - 4)a_{1}}{12\lambda^2},
\nonu \\
q_{19} & \rightarrow & -\frac{(\lambda^2 - 4)a_{1}}{16\lambda^2},
\qquad
q_{20} \rightarrow \frac{i(\lambda^2 - 4)a_{1}}{48\lambda^2},
\nonu \\
q_{21} & \rightarrow & \frac{i(\lambda^2 - 4)a_{1}}{24\lambda^2},
\qquad
q_{22} \rightarrow -\frac{i(\lambda^2 - 4)a_{1}}{24\lambda^2},
\qquad
q_{23}  \rightarrow  -\frac{i(\lambda^2 - 4)a_{1}}{48\lambda^2},
\nonu\\
q_{24} &\rightarrow& \frac{i(\lambda^2 - 4)a_{1}}{6M\lambda^2},
\qquad
q_{25} \rightarrow 0,
\qquad
q_{26} \rightarrow \frac{i(\lambda^2 - 4)a_{1}}{3M\lambda^2},
\qquad
q_{27} \rightarrow \frac{3}{2\lambda},
\nonu\\
q_{28} &\rightarrow& -\frac{i}{2\lambda},
\qquad
q_{29} \rightarrow -\frac{i}{2\lambda},
\qquad
q_{30} \rightarrow \frac{(\lambda^2 - 4)a_{1}}{6\lambda^2},
\qquad
q_{31} \rightarrow \frac{k(\lambda^2 - 4)a_{1}}{24\lambda^2},
\nonu\\
q_{32} &\rightarrow& -\frac{k(\lambda^2 - 4)a_{1}}{24\lambda^2},
\qquad
q_{33} \rightarrow -\frac{k(\lambda^2 - 4)a_{1}}{12\lambda^2},
\qquad
q_{34} \rightarrow \frac{2k^2(\lambda^2 - 4)a_{1}}{21M\lambda^2},
\nonu \\
q_{35} & \rightarrow & -\frac{k(\lambda^2 - 4)a_{1}}{3M\lambda^2},
\qquad
q_{36} \rightarrow -\frac{k(\lambda^2 - 4)a_{1}}{3M\lambda^2},
\nonu \\
q_{37} & \rightarrow & \frac{3(\lambda^2 - 4)a_{1}}{8\lambda^2},
\qquad
q_{38} \rightarrow \frac{k(\lambda^2 - 4)a_{1}}{12\lambda^2},
\qquad
q_{39}  \rightarrow  -\frac{k(\lambda^2 - 4)a_{1}}{12\lambda^2},
\nonu\\
q_{40} &\rightarrow& -\frac{k(\lambda^2 - 4)a_{1}}{6\lambda^2},
\qquad
q_{41} \rightarrow -\frac{4k^2(\lambda^2 - 4)a_{1}}{21M\lambda^2},
\qquad
q_{42} \rightarrow \frac{i k^3 \sqrt{2\lambda^2 - 8} \ a_{1}}{M\lambda^3},
\nonu \\
q_{43} & \rightarrow & \frac{k}{\lambda},
\qquad
q_{44} \rightarrow -\frac{i k^2(\lambda^2 - 4)a_{1}}{28\lambda^2},
\qquad
q_{45} \rightarrow -\frac{i}{2M},
\label{kpinfinityvalues}
\eea
which is a generalization of \cite{JKKR}
(corresponding to the OPE of the
charged spin-$2$ current with itself)
in the sense that we are calculating the OPE
of the charged spin-$2$ current with charged spin-$3$
current.

%%%%%%%%%%%%%%%%%%%%%%%%%%%%%%%%%%%%%%%%%%%%%%%%%%%%%%%%%%%%%%%%%%%%%%%%%%

\end{document}